\RequirePackage{fix-cm}
\documentclass[natbib,smallextended]{svjour3}       
\smartqed  
\usepackage{graphicx}
\usepackage{color}
\usepackage[colorlinks=true,linkcolor=blue,citecolor=blue,urlcolor=blue]{hyperref}
\usepackage{txfonts}
\usepackage{lscape}
\newcommand{\msun}{$M_\odot$}  

\newcommand{\arcsec}{$^{\prime\prime}$}

\journalname{The Astronomy and Astrophysics Review}
\begin{document}
\title{Cool outflows in galaxies and their implications}

\titlerunning{Cool outflows in galaxies}

\author{Sylvain Veilleux${}^{1,2,3,4}$, Roberto Maiolino${}^{3,5}$, Alberto D. Bolatto${}^1$, and Susanne Aalto${}^6$}

\institute{$^1$ Department of Astronomy and Joint Space-Science Institute, University of Maryland, College Park, MD 20742, USA\\ $^2$ Institute of Astronomy, University of Cambridge, Cambridge, CB3 0HA, United Kingdom;\\ $^3$ Kavli Institute for Cosmology Cambridge, University of Cambridge, Cambridge, CB3 0HA, United Kingdom; \\ $^4$ Space Telescope Science Institute, Baltimore, MD 21218, USA \\ $^5$ Cavendish Laboratory, University of Cambridge, Cambridge, CB3 0HA, United Kingdom; \\ $^6$ Department of Earth and Space Sciences, Chalmers University of Technology, Onsala Space Observatory, SE-439 92, Onsala, Sweden \\
\email{Sylvain Veilleux: veilleux@astro.umd.edu, Roberto Maiolino: r.maiolino@mrao.cam.ac.uk, Alberto Bolatto: bolatto@astro.umd.edu, Susanne Aalto: susanne.aalto@chalmers.se}
}

\date{Received: date / Accepted: date}

\maketitle

\begin{abstract}
Neutral-atomic and molecular outflows are a common occurrence in galaxies, near and far. They operate over the full extent of their galaxy hosts, from the innermost regions of galactic nuclei to the outermost reaches of galaxy halos. They carry a substantial amount of material that would otherwise have been used to form new stars. These cool outflows may have a profound impact on the evolution of their host galaxies and environments. This article provides an overview of the basic physics of cool outflows, a comprehensive assessment of the observational techniques and diagnostic tools used to characterize them, a detailed description of the best-studied cases, and a more general discussion of the statistical properties of these outflows in the local and distant universe. The remaining outstanding issues that have not yet been resolved are summarized at the end of the review to inspire new research directions.
\keywords{Galaxies: Active \and Evolution \and Halo \and Kinematics and Dynamics \and Starburst}
\end{abstract}

\setcounter{tocdepth}{3} 
\tableofcontents

\section{Introduction} \label{sec:intro}

\subsection{Setting the Scene} \label{sec:intro_scene}

\begin{figure}[htb]
\begin{center}
\includegraphics[width=1.0\textwidth]{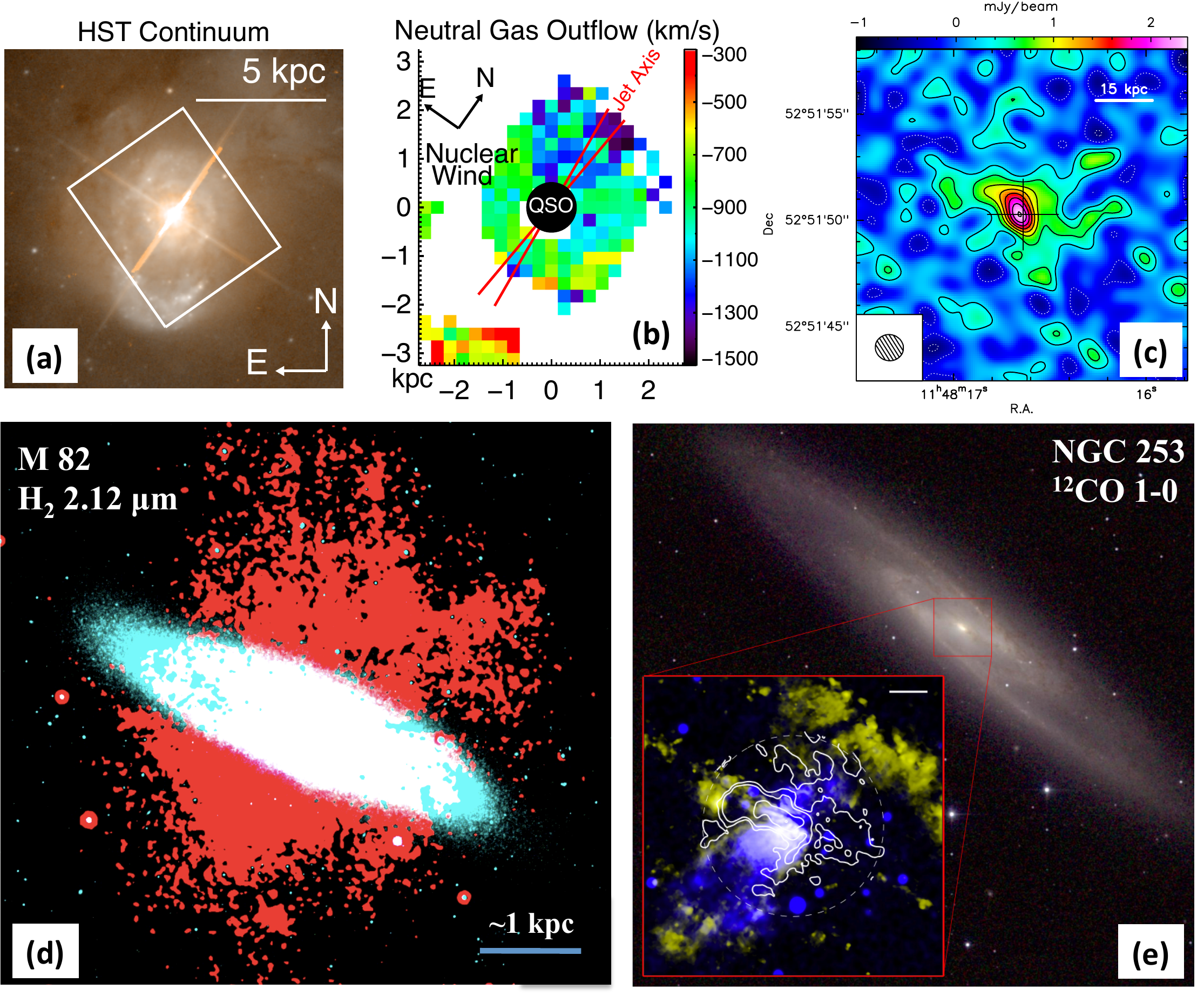}
\end{center}
\caption{Examples of cool outflows, near and far. (a) HST optical continuum image of quasar Mrk~231. The white box is the footprint of the IFS data shown in panel (b). (b) Velocity field of Na~I~D absorption line in Mrk~231. (c) [C~II] 158 $\mu$m line emission in $z$ = 6.4 quasar SDSS J1148$+$5251. (d) Starburst galaxy M~82 in H$_2$ 2.12 $\mu$m (red) and stellar light (light blue). (e) Starburst galaxy NGC 253 in CO 1$-$0 (white contours) relative to H$\alpha$ (yet), X-rays (blue), and optical continuum emission (greyscale). Images reproduced with permission from (a, b) \citet{RUPKE2011}, (d) \citet{VEILLEUX2009a}, copyright by AAS; and (c) \citet{CICONE2015}, copyright by ESO; and (e) \citet{BOLATTO2013b}, copyright by Macmillan.}
\label{fig:intro_outflows}
\end{figure}

Galaxies are not isolated ``Island Universes'' the way Thomas Wright and Immanuel Kant imagined them \citep{WRIGHT1750,KANT1755}. Instead, they are giant ecosystems where material flows in and out or is processed \textit{in-situ}. The fate of baryons in a galaxy is governed by the complex interplay between gas entering the galaxy through galaxy mergers and accretion flows, star formation within the galaxy, and gas flowing out of the galaxy. All gas phases and dust are participating in this cosmic ballet, but the focus of this review is on the cool ($T \lesssim 10^4$ K) outflowing neutral-atomic, molecular, and dusty material (Fig.\ \ref{fig:intro_outflows}). Driven by the energy released from stellar processes and gas accretion onto supermassive black holes (SMBHs) at the centers of galaxies, these cool outflows are a common feature of local and distant gas-rich systems. They often dominate the mass, and also sometimes the energetics, of gas outflows in general, and may comprise a significant fraction of the entire reservoir of cool gas in the galaxy hosts. 

The study of cool outflows is a relatively new area of research, dating back $\lesssim$ 20 years. However, in the last decade, thanks in large part to the advent of new infrared space observatories ({\em Spitzer, Herschel}), new or upgraded mm-wave and cm-wave ground-based facilities (ALMA, NOEMA, VLA, and GBT), and a suite of new-generation optical and near-infrared multi-object and integral field spectrometers on large ground-based telescopes (e.g., SAMI, KMOS, MaNGA, MOSFIRE, MUSE, KCWI),  the study of cool outflows has flourished into a full-fledged sub-discipline of its own with implications that touch on vast areas of extragalactic astronomy including galaxy formation and evolution and the synergistic connection between galaxies and supermassive black holes.

Cool outflows have deservedly received considerable attention in recent years for a number of reasons. The cool gas taking part in these outflows is the raw material from which stars are formed, so its fate may affect the evolution of the host galaxies. Fast outflows are among the leading internal negative-feedback processes to explain the rapid ($\lesssim$ 10$^9$ yrs) inside-out cessation (``quenching'') of star formation in massive galaxies \citep{SCHAWINSKI2014,ONODERA2015,TACCHELLA2015,TACCHELLA2016,SPILKER2019}. However, simply carrying the cool gas out of the galaxy and depositing it into the circumgalactic medium \citep[CGM;][]{TUMLINSON2017} may not be sufficient to quench these galaxies since the cool material, if left alone, will eventually fall back onto the galaxy and form new stars. Even if the outflows are powerful enough to eject the gas out into the intergalactic medium \citep[IGM;][]{OPPENHEIMER2019,DAVIES2019,KELLER2019}, re-accretion of the enriched material ejected by neighboring galaxies may revive the star formation activity of the host galaxy \citep{ANGLES-ALCAZAR2017a}. The duty cycle of outflows must be fine-tuned to prevent this material from infalling back onto the galaxy \citep[e.g.,][]{KIM2018}, or other processes must be at work to prevent it from forming new stars \citep{FAERMAN2017,MCQUINN2018}.

Powerful winds, driven by a central quasar or the surrounding starburst, have also been invoked to stop the growth of both the BH and spheroid component \citep{SILK1998,FABIAN1999,KING2003,MURRAY2005} and explain the tight BH-spheroid mass relation \citep[e.g.,][]{MAGORRIAN1998,GEBHARDT2000,FERRARESE2000}.
In turn, the fraction of quenched galaxies in the local universe seems strongly correlated with the properties of the spheroid component \citep[][]{TEIMOORINIA2016,BLUCK2016,BLUCK2019}, pointing back to BH-driven outflows as a means to quench massive galaxies. Fast AGN-driven cool outflows may be examples of this process in action. 

Much of our understanding of the overall impact of galactic winds and outflows on galaxy evolution comes from sophisticated cosmological galaxy formation simulations that incorporate the observable properties of the best-studied outflow systems \citep{VOGELSBERGER2014,SCHAYE2015,DUBOIS2016,CHOI2018,BIERNACKI2018,BRENNAN2018,HOPKINS2018b,DAVE2019,NELSON2019,PEEPLES2019,HAFEN2019}.
However, the cool gas phase outside of galaxies is notoriously difficult to model since it is subject to formation and destruction processes that act on scales that are unresolved in these simulations (e.g., thermal, Kelvin-Helmholtz, and Rayleigh-Taylor instabilities, turbulent boundary layers, conductive heat transport). 
We therefore expect new data on cool gas outflows to continue to inform the next generation of numerical simulations of galaxy formation and evolution for years to come.

\subsection{Scope of the Review, Key Questions, and Organization} \label{sec:intro_scope}

The focus of this review is on cool ($T \lesssim 10^4$ K) outflows comprised of neutral-atomic and molecular gas where hydrogen is mostly in neutral or molecular form, respectively. Dust is also part of our discussion since in general a significant fraction \citep[perhaps $\sim$30\%;][]{WEINGARTNER2001} of the metals in cool enriched gas is locked into dust. 
Dust grains are an essential component of these outflows, acting as anchor points for radiation forces and catalyzers for H$_2$ formation. This review is meant to provide a holistic view of cool outflows, spanning the full range of scales over which they are observed, i.e. from $\lesssim$ 10 pc in some AGN and quasars to $\gtrsim$ 100 kpc in some distant galaxies. Most of our discussion centers on the following fundamental questions:

\begin{itemize}

\item\emph{Statistical properties of cool winds in the local and distant universe:} What is their frequency of occurrence, physical extent, mass loss rate, efficiency, energetics, and evolution with look-back time? 
\item\emph{Driving mechanisms of cool winds:} What is the role of supernovae, radiation pressure, cosmic-ray pressure, and AGN?  Is there a minimum threshold of star formation or black hole activity needed to trigger cool outflows? How is the material entrained but not destroyed? 
\item\emph{Multi-scale and multi-phase nature of outflows:} What is the relative distribution of the cool, warm, and hot phases in the wind? Is the cool material produced \textit{in-situ} from the hot wind? Is the cool material transitioning to another gas phase on the way out?  Do cool winds form stars? What is the relation between large-scale cool outflows and nuclear AGN winds?
\item\emph{Impact of cool winds on the host galaxy:} What feedback effects do they exert on the host galaxy ISM and CGM? Do they help regulate the growth of the black hole? What fraction of the cool outflowing gas escapes the host galaxy? What is the role of re-accreted (recycled) gas from neighboring galaxies? 

\end{itemize}

This review is organized as follows.  Section \ref{sec:basic_physics} describes the basic physics behind cool outflows, particularly the sources of energy and momentum, driving mechanisms, physical origin of the outflowing cool gas, and minimum condition for the outflowing material to escape galaxies. Section \ref{sec:techniques} discusses the many observational techniques and diagnostic tools that have been used so far to measure the properties of cool outflows, and provides a critical assessment of their strengths and weaknesses. The best-studied cases of cool outflows are described in detail in Section \ref{sec:best_cases} to set the stage for a more general discussion of the statistical properties of cool outflows in the nearby and distant universe in Sections \ref{sec:lowz} and \ref{sec:highz}, respectively. This review concludes in Section \ref{sec:open} with a summary of the key open theoretical and observational issues, and suggestions for future research directions.  

This type of review is necessarily incomplete and biased. While our main target audience is newcomers who wish to familiarize themselves with this topic or may be searching for exciting new research ideas, we hope that this review will also be a valuable resource for more experienced scientists of all levels. Those who did not find an answer to their questions in this article may wish to consult the following recent reviews on directly relevant topics: \citet[][stellar feedback in low-mass galaxies]{ERB2015}, \citet[][AGN outflows and feedback]{KING2015}, \citet[][AGN feedback]{HARRISON2017}, \citet[][galactic winds in general]{HECKMAN2017}, \citet[][CGM]{TUMLINSON2017},  \citet[][AGN outflows and feedback]{HARRISON2018}, \citet[][galactic winds driven by star formation]{RUPKE2018}, \citet[][multiphase outflows in AGN host galaxies]{CICONE2018b}, \citet[][the driving mechanisms of outflows in AGN]{WYLEZALEK2018}, \citet[][HI 21-cm absorption as a probe of the ISM and CGM in AGN]{MORGANTI2018}, and \citet[][the theory of galactic winds driven by stellar feedback]{ZHANG2018b}. The earlier reviews by  \citet[][galactic winds in general]{VEILLEUX2005} and \citet[][observations of AGN feedback]{FABIAN2012} may also be useful. Finally, reviews on more specific topics are mentioned throughout the text. 

\section{Basic Physics} \label{sec:basic_physics}

\subsection{Energy Sources} \label{sec:energy_sources}

\subsubsection{Stellar Winds and Supernovae} \label{sec:stellar_winds_supernovae}

Stars inject mass, momentum, and energy in the surrounding environment that may prevent molecular gas from forming stars on a dynamical time scale. Given a star formation rate $\dot{M}_*$, the depletion time scale $t_{\rm dep} \equiv M_{\rm gas}/\dot{M}_*$ of the molecular gas in galaxies due to star formation is typically $\sim$1 $-$ 2 Gyr on $\gtrsim$ kpc scales \citep[e.g.,][]{BIGIEL2008}, although $t_{\rm dep}$ tends to decrease with increasing specific star formation rates, both spatially integrated  \citep[$\dot{M}_*/M_*$;][]{SAINTONGE2011} and spatially resolved \citep[$\Sigma_{\rm SFR}/\Sigma_*$;][]{ELLISON2019}, and there is considerable scatter on scales below $\sim$100 pc \citep{KENNICUTT2012,KRUIJSSEN2014}. This depletion time scale is roughly two orders of magnitude longer than the dynamical time scale of giant molecular clouds (GMCs): $t_{\rm ff}/t_{\rm dep} \equiv \epsilon_{\rm ff} \simeq 1\%$, where $t_{\rm ff} = \sqrt{3 \pi /  32 G \rho}$ is the local free-fall time and $\epsilon_{\rm ff}$ is the star formation efficiency per free-fall time. Recently, \citet{KRUIJSSEN2019} have deduced GMC lifetimes of $t_{\rm sf} \simeq$ 11 Myr resulting in a star formation efficiency per GMC lifetime $\epsilon_{\rm Sf}$ = 2.5\%. A wide range of feedback processes may be responsible for this low star formation efficiency.  Since most stars form in clusters, the emphasis of our discussion is on the dominant feedback processes in cluster environments.

\medskip
\noindent \emph{- Protostellar Outflows.}  This source of stellar feedback has been reviewed in detailed in \citet{BALLY2016}. It is clear that protostellar outflows help keep $\epsilon_{\rm ff}$ small in the dense regions of star clusters, but given the low ejection velocities, these outflows play a limited role in setting the final value of $\epsilon_{\rm ff}$ in most clusters, except those with masses less than  $\sim 100~M_\odot$ \citep{KRUMHOLZ2019}.

\medskip
\noindent \emph{- Stellar Winds from Young Massive Stars.} Winds from massive supernova (SN) progenitor stars profoundly transform the circumstellar medium around these stars \citep[e.g.,][]{VANMARLE2012,GEORGY2013}, and also likely play a significant role in clearing the denser gas from young stellar clusters, before the first core-collapse SNe occur \citep{LONGMORE2014,HOLLYHEAD2015,SOKAL2016}. Pre-existing wind bubbles at the time of the SN explosion may greatly enhance the feedback energy efficiency, although the exact boost depends on the uncertain importance of mixing, conduction, and cooling at the interface of the hot wind bubble and the compressed swept-up ambient medium \citep[e.g.,][]{FIERLINGER2016,EL-BADRY2019}.
Pre-existing wind-blown bubbles created by massive SN progenitor stars may also affect the survival or destruction of the dust generated within the ejecta or initially present in the ambient ISM \citep{MARTINEZ-GONZALEZ2019}.

\medskip
\noindent \emph{- Core Collapse Supernovae.} Stars with masses between about 8 and 25 M$_\odot$ end their lives after $\sim$3 Myr as core collapse supernovae ejecting $\sim 1-10~M_\odot$ of metal-enriched material with kinetic energy $E_{\rm SN} \simeq 10^{51}$ erg $\equiv$ $E_{51}$ and momentum $p_{\rm SN} = \sqrt{2 E_{\rm SN} M_{\rm SN}}$, where $M_{\rm SN}$ is the mass of the ejecta per SN. If the supernova rate $\Gamma_{\rm SN}$ scales with the star formation rate as $\Gamma_{\rm SN} = \alpha_{\rm SN} \dot{M_*}$, where $\alpha_{\rm SN} \simeq 0.01$--$0.02$ \citep[depending on the Initial Mass Function, IMF; $\alpha_{\rm SN} \simeq 0.02$ for a Salpeter IMF and continuous star formation;][]{LEITHERER1999}, then $\dot{E}_{\rm SN} = 7 \times 10^{41}~(\alpha_{\rm SN}/0.02)~(\dot{M}_*/M_\odot~{\rm yr}^{-1})$ erg s$^{-1}$ and $\dot{p}_{\rm SN} = 5 \times 10^{33}~(\alpha_{\rm SN}/0.02)~(\dot{M}_*/M_\odot~{\rm yr}^{-1})~{\rm dyne}$ \citep{VEILLEUX2005}. What happens next to this kinetic energy depends on the local pre-SN environment. In very low density environments, the ejecta simply remain in a state of free expansion keeping the same kinetic energy and momentum. In more typical environments, some fraction of the kinetic energy ($\xi$, the supernova thermalization or feedback efficiency) is quickly transformed into thermal energy through shocks heating the SN ejecta and ambient medium to a temperature $T = 0.4 \mu m_H (\dot{E} / k \dot{M}) \simeq 3 \times 10^8~\xi \Lambda^{-1}$ K, where $\Lambda$ is the ratio of the total mass of heated gas to the mass that is directly ejected by the SN.  Once the mass of swept up material is comparable to the mass of the ejecta, the SN remnant enters an adiabatic Sedov-Taylor phase of expansion where the outer shock radius expands as $t^{2/5}$. Once the temperature drops below $\sim10^6$ K, radiative losses become important, and the SN remnant enters a pressure-driven snowplow phase where the pressure of the interior hot gas is still larger than that of the outside pressure. Gradually, the interior pressure decreases to eventually match the outside pressure, and the remnant continues to expand and sweep up ISM gas as a momentum-conserving snowplow. Several studies have shown that the input momentum from SNe can be boosted by an order of magnitude during the Sedov-Taylor phase due to work done by the hot shocked gas \citep[e.g.,][]{KIM2015,KIMM2015,MARTIZZI2015,WALCH2015}: $p_{\rm SN} = 3 \times 10^5$ km s$^{-1}$ M$_\odot~E_{51}^{16/17} n_{\rm H}^{-2/17} Z^{-0.14}$, where $n_{\rm H}$ is the hydrogen number density in cm$^{-3}$ and $Z$ is the metallicity in solar units (formally this expression is valid only if $Z \ge 0.01$). Note that $p_{\rm SN}$ depends only weakly on the density (or density inhomogeneity) of the ambient medium and its metallicity. Considerable effort has been invested in recent years in determining the impact of clustered supernovae on this momentum boost.  It has been found that the asymptotic momentum per SN can be up to an order of magnitude greater than that delivered by isolated SNe (up to $\sim 3 \times 10^6$ $M_\odot$ km s$^{-1}$ per SN), although the results are sensitive once again on the mixing rate across the contact discontinuity between the hot and cold phases \citep{SHARMA2014,GENTRY2017,GENTRY2019,EL-BADRY2019}.

\medskip
\noindent \emph{- Cosmic Rays.}  In star-forming galaxies like the Milky Way, first-order Fermi shock accelerated particles in supernova remnants are the dominant source of cosmic rays \citep[CRs; e.g.,][]{GRENIER2015,BYKOV2018}. These cosmic rays ionize and heat the ISM, including the cold material that is optically thick to ionizing radiation, depositing $\sim$13 eV per Coulombic or hadronic interaction \citep[e.g.,][]{GRENIER2015}. 
Between 10 and 50\% of the kinetic energy from SNe ($\sim$1$-$5 $\times$ 10$^{50}$ ergs per SN) can be converted into non-thermal energy \citep{GRENIER2015,DIESING2018}, most of which is deposited in protons with energies following a power-law distribution.
In the case of a Salpeter IMF and constant star formation rate, the CR energy injection rate is thus $\dot{E}_{\rm CR} \simeq (0.7$--$3.5) \times 10^{41}~(\alpha_{\rm SN}/0.02)~(\dot{M}_*/M_\odot~{\rm yr}^{-1})$ erg s$^{-1}$ and the injected momentum rate is $\dot{p}_{\rm CR} = (0.5$--$2.5) \times 10^{33}~(\alpha_{\rm SN}/0.02)~(\dot{M}_*/M_\odot~{\rm yr}^{-1})~{\rm dyne}$. This last number does not take into account the possibility that the asymptotic injected momentum per SN may be boosted by the presence of CRs by a factor of 2-3, and perhaps as much as an order of magnitude in the denser environments. This is due to the fact that CRs act as a relativistic fluid, so they suffer less adiabatic loss than the thermal gas, and thus dominate the internal pressure at late times in the evolution of the SN remnant. Moreover, the CR energy is not radiated away during the snowplow phase, so it continues to support the expansion of the SN remnant \citep{DIESING2018}.

\medskip
\noindent \emph{- Radiation.} The stellar radiative processes most relevant to cool outflows are photoionization heating associated with the ionizing ($h\nu >$ 13.6 eV) radiation from hot stars and SNe, photo-electric heating associated with the non-ionizing radiation absorbed by dust in the warm and cold neutral media, Compton heating by hard photons from X-ray binaries,
the dissociation of H$_2$ by the Lyman-Werner (11.2--13.6 eV) radiation, and finally the radiation pressure associated with single- and multiple-scattering processes. Most of these processes provide excellent diagnostic tools to study cool outflows so they are discussed in more details in Section \ref{sec:techniques}. Radiation pressure is potentially an important driving mechanism for cool outflows so it is discussed in more detail in Section \ref{sec:radiation}. For starbursts, the luminosity is typically $L \simeq 10^{10}~\dot{M}_*~L_\odot$, where the units of the star formation rate $\dot{M}_*$ are $M_\odot~{\rm yr}^{-1}$. 

\medskip
\noindent \emph{- Type Ia Supernovae.} Stars with $M <$ 8 M$_\odot$ are the progenitors of Type Ia supernovae that go off after $\sim$100 Myr, each injecting an energy $\sim$ 10$^{51}$ erg and a mass $\sim$1.4 M$_\odot$ into the surrounding environment. The energy deposition rate from a galaxy with stellar mass $M_{*,11}$ in units of 10$^{11}$ $M_\odot$ and stellar population age $t_{9.7}$ in units of 5 $\times$ 10$^9$ yrs is $\dot{E}_{Ia} \simeq 14 \times 10^{40} M_{*,11} t_{9.7}^{-1.1}$ erg s$^{-1}$ \citep{CONROY2015}. This source of energy may play a role in preventing star formation in low-mass early-type galaxies \citep[e.g.][]{LI2018}, but is probably not an important contributor to the energetics of the cool outflows reported here in star-forming and active galaxies.

\medskip
\noindent \emph{- Stellar Winds from Old Stars.}  The thermalization of winds from asymptotic giant branch (AGB) stars, red giants, and planetary nebula phases injects energy at a rate $\dot{E}_{AGB} \simeq 5 \times 10^{40}$ $M_{*,11} \sigma_{*,300}^2~t_{9.7}^{-1.25}$ erg s$^{-1}$ \citep{CONROY2015}, where $\sigma_{*,300}$ is the stellar velocity dispersion of the host galaxy normalized to 300 km~s$^{-1}$. This energy will combine with that injected from Type Ia SNe (discussed above), but does not play a role in driving the cool outflows in starburst and active galaxies.

\subsubsection{AGN} \label{sec:agn}

The primary energy source behind AGN is accretion onto the central supermassive black hole, where the gravitational potential energy lost by the accreted material is converted into heat and partly radiated away at the rate of
\begin{eqnarray}
L = \epsilon_r \dot{M}_{\rm acc} c^2 \simeq 5.7 \times 10^{45}~(\epsilon_r/0.1) (\dot{M}_{\rm acc}/M_\odot~{\rm yr}^{-1})~{\rm erg~s}^{-1},
\end{eqnarray}
where $\epsilon_r$ is the radiative efficiency. This efficiency is a strong function of the accretion rate. At low accretion rates ($\dot{m}_{\rm  acc}$ $\equiv$ $\dot{M}_{\rm acc}/\dot{M}_{\rm acc}^{\rm Edd}$ $\equiv$ $L/L_{\rm Edd}$ $\lesssim$ 10$^{-3}$, where $L_{\rm Edd} = 4 \pi G M c / \kappa_{es} = 1.25~(M/M_\odot) \times 10^{38}$ erg s$^{-1}$ is the Eddington luminosity due to electron scattering), the accreting material forms a hot, optically thin, geometrically thick disk which is radiatively inefficient with $\epsilon_r < 10^{-3}$ because of long cooling time scales \citep[e.g.,][]{YUAN2014}. At moderate accretion rates ($\dot{m}_{\rm acc}$ $\simeq$ 0.01--0.25), the accretion flow forms a cold (relative to the virial temperature), optically thick, but geometrically thin disk that is radiatively efficient with $\epsilon_r (a_{\rm BH}) \simeq 0.05$--$0.2$, increasing with increasing black hole spin $a_{\rm BH}$ \citep[e.g.,][]{NOVIKOV1973,KORATKAR1999}. Above this accretion rate, long radiative diffusion time scales produce an optically and geometrically thick ``slim'' disk that is radiatively inefficient with $\epsilon_r \lesssim 10^{-3}$ \citep[e.g.,][]{SADOWSKI2013,SADOWSKI2014,SADOWSKI2015a,JIANG2014,MCKINNEY2014,MCKINNEY2015}. 
The relative importance of the dominant launching mechanisms for outflows in AGN -- radiation, thermal pressure, non-thermal pressure (from cosmic rays), and magnetic forces -- also varies with the mass accretion rate \citep[e.g.,][]{GIUSTINI2019}.

\medskip
\noindent\emph{- Radiative Mode of AGN Feedback.} For the radiatively efficient AGN, the mechanical energy injection rate in the so-called ``radiative'' or ``quasar'' mode of AGN feedback may be written
\begin{eqnarray}
\dot{E}_{r} = \epsilon_{f,r} \epsilon_r \dot{M}_{\rm acc} c^2 \simeq 5.7 \times 10^{45}~\epsilon_{f,r} (\epsilon_r/0.1) (\dot{M}_{\rm acc}/M_\odot~{\rm yr}^{-1})~{\rm erg~s}^{-1},
\label{eq:Er}
\end{eqnarray}
where $\epsilon_{f,r}$ is the efficiency of the radiative energy to couple with the local environment and drive the outflow. Note that this expression implies that the total amount of injected energy over the lifetime of the black hole is
\begin{eqnarray}
E_{r} = \epsilon_{f,r} \epsilon_r M_{\rm BH} c^2 \simeq 2 \times 10^{61}~\epsilon_{f,r}~ (\epsilon_r/0.1) (M_{\rm BH}/10^8~M_\odot)~{\rm erg}. 
\label{eq:Er2}
\end{eqnarray}
This number is two orders of magnitude larger than the binding energy of the bulge hosting the black hole
\begin{eqnarray}
E_{\rm bulge} \simeq M_{\rm bulge} \sigma_*^2 \simeq 2 \times 10^{59}~(M_{\rm bulge}/10^{11}~M_\odot)~(\sigma_*/300~{\rm km~s}^{-1})^2~{\rm erg}.
\label{eq:Ebulge}
\end{eqnarray}
This fact alone implies $<\epsilon_{f,r}>\;\lesssim$ 1\%, averaged over the active life cycle of the black hole. This number is not well constrained from numerical simulations, although values of 0.5\%--1\% seem to be required if AGN feedback is to suppress star formation in the host \citep[e.g.,][]{HOPKINS2010,HOPKINS2016}.

\medskip
\noindent\emph{- Kinetic Mode of AGN Feedback.} For the radiatively inefficient AGN, the mechanical energy injection rate in the ``kinetic'' or ``radio'' mode of AGN feedback may be expressed as
\begin{eqnarray}
\dot{E}_{k} = \epsilon_{f,k} \dot{M}_{\rm acc} c^2 \simeq 5.7 \times 10^{46}~\epsilon_{f,k} (\dot{M}_{\rm acc}/M_\odot~{\rm yr}^{-1})~{\rm erg~s}^{-1}, 
\label{eq:Edotk}
\end{eqnarray}
where $\epsilon_{f,k}$ is the efficiency of the kinetic energy injected by the AGN to couple to the local environment and drive the outflow. This kinetic energy may come from both light relativistic jets and slower but more massive wide-angle winds.

Recent simulations of hot accretion flows give $\epsilon_{f,k} \sim 10^{-3}$ \citep[e.g.,][]{YUAN2015}. This number includes the contribution from relativistic jets accelerated by processes associated with the magnetized accretion disk \citep[e.g.,][]{BLANDFORD1982,LYNDEN-BELL2003,LYNDEN-BELL2006,HAWLEY2006}, but may be larger in cases where the kinetic energy is delivered by jets produced by the Blandford-Znajek mechanism \citep{BLANDFORD1977}. In these objects, the jet power is extracted from the black hole spin and $L_{\rm jet} \propto B^2_{\rm pol} M_{\rm BH}^2 a_{\rm BH}^2$, where $B_{\rm pol}$ is the poloidal magnetic field at the black hole horizon and $a_{\rm BH}$ is the black hole spin \citep[$a_{\rm BH} = 0.998$ for maximally spinning black holes;][]{THORNE1974,YUAN2014}. 
Recent slim disk simulations with non-spinning black holes have shown that outflows 
are inevitable, resulting in $\epsilon_{f, k}$ $\simeq$ 3\%, and as high as $\sim$8\% with spinning black holes with $a_{\rm BH} = 0.7$ \citep[strickly speaking, these numbers include both the radiative and wind energy outputs so they correspond to $\epsilon_f \epsilon_r + \epsilon_{f,k}$ in our nomenclature; e.g.,][]{SADOWSKI2016a,SADOWSKI2016b}. 

\medskip
\noindent\emph{- Cosmic Rays.} In analogy with supernova-driven shocks, the interaction of the aforementioned wide-angle sub-relativistic winds and collimated relativistic jets with the ambient ISM may be an important source of CRs. Protons with energies $>$ 8, 21, and 140 MeV penetrate column densities $N_H$ of 2 $\times$ 10$^{22}$ cm$^{-2}$, 1.2 $\times$ 10$^{23}$ cm$^{-2}$, and 4 $\times$ 10$^{24}$ cm$^{-2}$, respectively, and therefore can contribute to heating and ionizing some of the material in cool outflows \citep{GONZALEZ-ALFONSO2018}. Depending on the details of CR propagation through the ISM, these CRs may also provide a significant source of pressure that will help drive the large-scale outflows \citep[e.g.,][Sec. \ref{sec:cosmic_rays}]{CHAN2019,YUSEF-ZADEH2019}. The most extreme ultra-high-energy CRs (UHECRs) with energies $> 10^9$ GeV collide with nuclei in the ISM and produce pions that decay into GeV-TeV $\gamma$-rays as well as PeV muonic neutrinos \citep[e.g.,][]{WANG2016,WANG2017,LIU2018}. 

\medskip
\noindent\emph{- Source Variability.} The radiative and non-radiative energy output of AGN may vary wildly on time scales that are much shorter than the dynamical timescale of large-scale cool outflows. One should therefore use caution when comparing the properties of the AGN with the dynamical properties of cool outflows. We recommend using methods that are insensitive to short-term AGN variability when estimating the AGN luminosity. For instance, methods based on reprocessed radiation (e.g., mid-infrared emission) are preferable over methods based on the hard X-ray luminosity to mitigate the effects associated with short-term ($<$10$^3$--10$^4$ yr) AGN variability.

\subsection{Driving Mechanisms} \label{sec:driving_mechanisms}

\begin{figure}[htb]
\begin{center}
\includegraphics[width=1.0\textwidth]{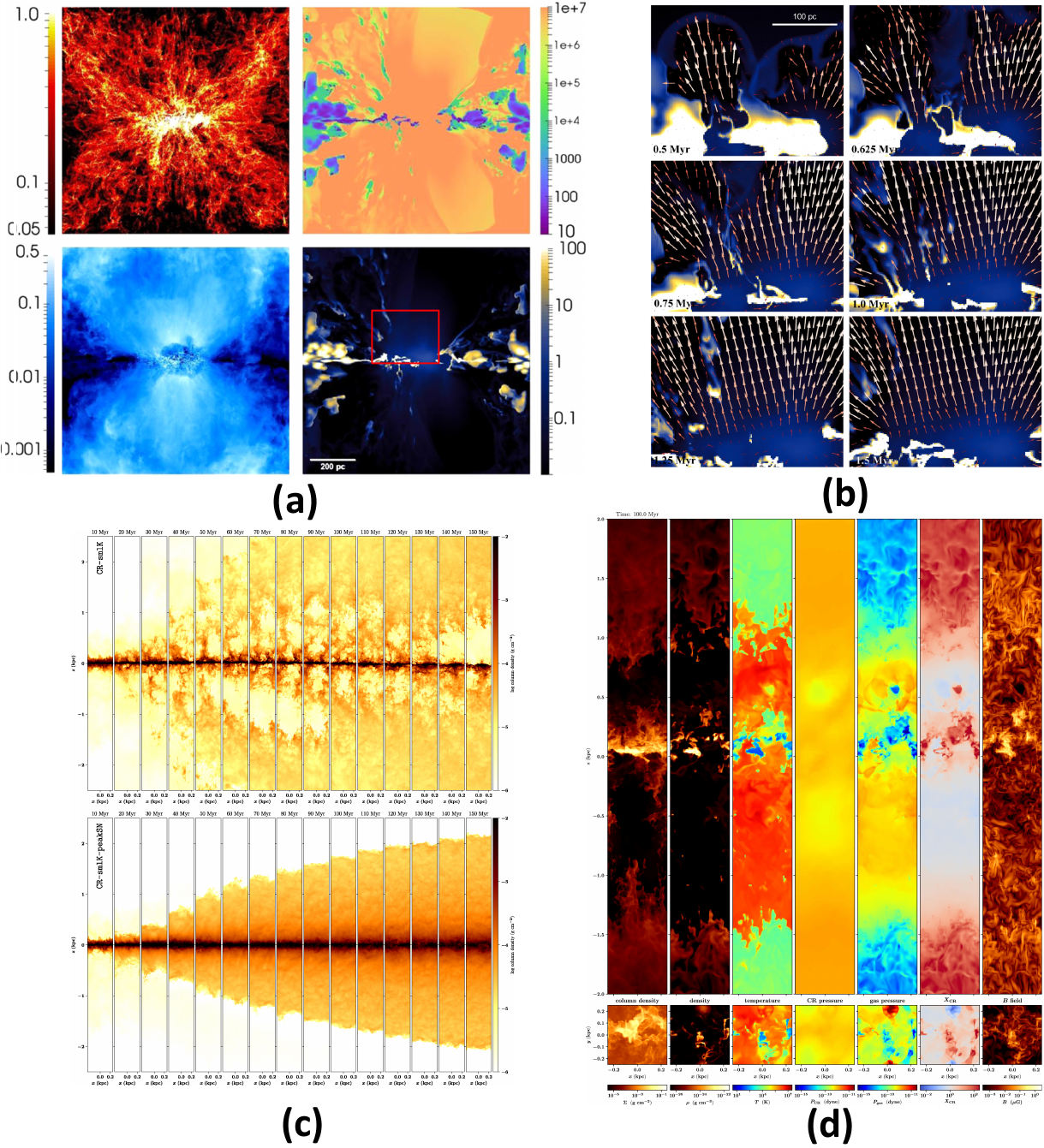}
\end{center}
\caption{Simulations of starburst-driven cool outflows: (a) Slice in $yz$ plane of a simulation at 1.5 Myr. Clockwise from top left: H$\alpha$ emission (log erg s$^{-1}$ cm$^{-2}$), temperature (K), density (cm$^{-3}$), and soft X-ray emission (log erg s$^{-1}$ cm$^{-2}$); red box indicates zoomed-in region in panel (b). (b) Close-up of the filament-forming region at six different epochs. (c) Time evolution of the edge-on column density for simulations dominated by SN thermal energy (top panel) and CR pressure (bottom panel) in steps of 10 Myr. (d) Vertical structure of the column density, density, temperature, CR pressure, thermal pressure, ratio of CR to thermal pressure, and magnetic field strength in a CR-dominated simulation at $t$ = 100 Myr. Images reproduced with permission from (a, b) \citet{TANNER2016}, copyright by AAS; and (c, d) \citet{GIRICHIDIS2018}, copyright by the authors.}
\label{fig:sims_sb_outflows}
\end{figure}

The equation of motion of an outflowing shell fragment of mass $M_{sh}$ that subtends a solid angle $\Omega_{sh}$ is
\begin{eqnarray}
\frac{d}{dt}~\left[M_{sh}(r)~\dot{r}\right] = \Omega_{sh}~r^2~(P_{th} + P_{CR} + P_{jet}) + \left(\frac{\Omega_{sh}}{4 \pi}\right)~\left(\frac{\tilde{\tau} L_{\rm bol}}{c}\right) - \frac{G M(r) M_{sh}(r)}{r^2},
\label{eq:eom}
\end{eqnarray}
where $M(r)$ is the galaxy mass enclosed within a radius $r$. The terms on the right are the forces due to the thermal, cosmic ray, and jet ram pressures, the radiation pressure, and gravity, respectively. Below, we discuss each of the pressure terms separately.  In reality, these pressure forces act together to drive the outflows. Examples of state-of-the-art outflow simulations that incorporate many of these force terms are shown in Figures \ref{fig:sims_sb_outflows} and \ref{fig:sims_agn_outflows}. 

Fundamentally, cool outflows may be ``energy-driven'' or ``momentum-driven'', depending on whether radiative losses are negligible or not, respectively. In the first case, the outflow is adiabatic and the energy injected in the ambient gas is transformed into bulk motion at a rate $\dot{E}$, where $\dot{E} \propto \dot{M} v_{\rm out}^2$. For starburst-driven winds, $\dot{E} \propto \dot{M_*}$ so the mass-loading factor $\eta \equiv \dot{M}/\dot{M_*} \propto v_{\rm out}^{-2}$ \cite[e.g.,][]{MURRAY2005}. Energy-driven AGN winds where the energy injection rate scales with the black hole mass naturally lead to $M_{\rm BH} \propto \sigma_*^5$, where $\sigma_*$ is the stellar velocity dispersion of the host galaxy \citep[e.g.,][Sect.~\ref{sec:thermal_energy}]{SILK1998,MURRAY2005}. The corresponding relations for momentum-driven outflows are  $\dot{M} \propto \dot{E} v_{\rm out}^{-1}$, $\eta \propto v_{\rm out}^{-1}$, and $M_{\rm BH} \propto \sigma_*^4$ \citep[e.g.,][Sect.~\ref{sec:radiation}]{MURRAY2005}. In that case, $v_{\rm out}$ is also expected to linearly scale with the circular velocity $v_{\rm circ}$ and, in starburst-driven outflows,  $v_{\rm out} \propto v_{\rm circ}~\dot{M_*}^{0.25-0.50}$, where the exact dependence on the star formation rate depends on whether the area of the momentum injection zone scales linearly or not with the star formation rate.

\subsubsection{Thermal Energy} \label{sec:thermal_energy}

In this scenario, the shell fragment is pushed outward by ram pressure produced by the over-pressured hot gas in the central regions of the galaxy. The spherically symmetric case where radiative losses are negligible (i.e. energy-driven outflow) and the gravitational forces are also assumed to be negligible was developed analytically by \citet[][CCC85]{CHEVALIER1985} to explain the ionized wind of M~82 in the context of a supernova-driven outflow. The central pressure in the hot cavity in this case is $P_0/k \sim 3 \times 10^5~(\dot{M}_*/M_\odot~{\rm yr}^{-1})(R_*/{\rm kpc})^{-2}$ K cm$^{-3}$, where $R_*$ is the radius of the injection zone \citep[CC85; also][]{VEILLEUX2005}. The CC85 wind solution at large radii for a monoatomic gas with adiabatic index $\gamma = 5/3$ has a constant asymptotic outflow velocity $v_{\rm hot}^\infty \simeq$ $\sqrt{5 k T_{\rm hot}/\mu} \simeq~1500~{\rm km~s}^{-1} (\xi/\beta)^{1/2}$, a gas density $\propto$ $r^{-2}$, and a gas pressure $\propto$ $r^{-10/3}$, where $T_{\rm hot} \simeq 1.5 \times 10^7~{\rm K}~(\xi/\beta)$ is the temperature of the hot gas in the center, $\mu = 1.36$ is the mean mass per particle, $\xi$ is the thermalization efficiency i.e. the fraction of the kinetic energy by the massive stars that is not lost to radiative cooling (Sec. \ref{sec:stellar_winds_supernovae}), and $\beta$ is the ratio of the mass injection rate to the star formation rate. These results agree remarkably well with the X-ray observations of M82 with $\xi \sim 1$ and $\beta \sim 0.3$ \citep[][Sec. \ref{sec:M82}]{STRICKLAND2009}. The momentum injection rate of this hot wind is $\dot{p}_{\rm hot} = \dot{M} v_X \simeq (2 \dot{E}\dot{M})^{1/2} \simeq 5~(\xi\beta)^{1/2}~L/c$, where for the last approximate equality we used $L \simeq 10^{10}~(\dot{M}_*/M_\odot~{\rm yr}^{-1})~L_\odot$ and the relation between $\dot{E}_{\rm SN}$ and $\dot{M}_*$ for $\alpha_{SN} = 0.01$ mentioned in Section \ref{sec:stellar_winds_supernovae}. This expression for $\dot{p}_{\rm hot}$ is similar to that of a radiation-pressure-driven wind in the optically thin limit (Sec. \ref{sec:radiation}).

The first energy-driven wind model for AGN was developed  by \citet{SILK1998}. In this model, the mechanical power of the wind is taken to be a constant fraction of the Eddington luminosity (thus $\epsilon_{f,k} \propto M_{\rm BH}$ in Eq.~(\ref{eq:Edotk})) and results in a dependence on the host galaxy's velocity dispersion $\sigma_*$ of the form $M_{\rm BH} \propto \sigma_*^5$. This is somewhat steeper than the observed relation between the black hole mass and spheroid velocity dispersion \citep[e.g.,][]{GULTEKIN2009,KORMENDY2013}.  A similar energy-conserving hot-bubble scenario has been proposed to explain the powerful molecular and atomic outflows seen in some ultraluminous infrared galaxies \citep{FAUCHER-GIGUERE2012a,DEBUHR2012,ZUBOVAS2012,ZUBOVAS2014,COSTA2014,NIMS2015}. In this ``blast-wave scenario'', the violent interaction of a fast inner AGN wind with the ISM of the host results in shocked wind gas at a temperature of $2.0 \times 10^{10}~(v_{\rm wind}/0.1 c)^2$ K that does not efficiently cool, but instead expands adiabatically as a hot bubble. The adiabatically expanding shocked wind sweeps up gas and drives an outer shock into the host ISM, raising the temperature of the ISM to $2.3 \times 10^7~(v_{\rm shock}/1000~{\rm km~s}^{-1})^2$ K \citep{NIMS2015}. This outflowing gas cools radiatively, and most of it ``freezes out'' into clumps of cold molecular material \citep[Fig.~\ref{fig:sims_agn_outflows}a;][]{RICHINGS2018a,RICHINGS2018b}. A variant on this scenario is that preexisting molecular clouds from the host ISM are entrained in the adiabatically expanding shocked wind, accelerated to the observed velocity without being destroyed by the many erosive forces and instabilities (the pros and cons of each of these scenarios are discussed in Section \ref{sec:origin}). 

In both scenarios, we have by energy conservation $\frac{1}{2} \dot{M}_{\rm outflow} v_{\rm outflow}^2$ = $\frac{1}{2} f \dot{M}_{\rm wind} v_{\rm wind}^2$, where the quantities with subscript ``outflow'' refer to the outer molecular/atomic outflow, while those with subscript ``wind'' refer to the inner X-ray or UV AGN wind. The efficiency $f$ is defined as the fraction of the kinetic energy of the inner wind that goes into bulk motion of the swept-up molecular/atomic material. Rearranging the terms, we get 
\begin{eqnarray}
\dot{p}_{\rm outflow} & = & f~(v_{\rm wind} / v_{\rm outflow})~\dot{p}_{\rm wind} \\
                 & \simeq & f~(v_{\rm wind} / v_{\rm outflow})~(L_{\rm  AGN}/c), 
\label{eq:pdot_outflow}
\end{eqnarray}
 where the last equality is valid only if the inner wind is radiatively accelerated, i.e.\ $\dot{p}_{\rm wind} \simeq L_{\rm AGN} / c$, which is often the case for X-ray or UV winds in AGN.

As long as the outflow remains small compared with the scale height or scale length of the ambient medium, the large-scale outflow will evolve through the same classic phases of evolution as those of individual SN remnants (Sec. \ref{sec:stellar_winds_supernovae}). The energy-driven outflows discussed above will eventually transition into non-adiabatic pressure-driven snowplows when $T \lesssim 10^6$ K and radiative losses ($\propto n^2$) are no longer negligible. Later on, once the thermal pressure of the hot gas equals that of the ambient pressure, the outflow will enter the momentum-conserving (momentum-driven) snowplow phase and rapidly slow down at it sweeps up more and more ISM gas.  The terminal outflow velocity of a cloud accelerated outward  by the ram pressure of a wind that carries momentum at a rate $\dot{p}$ into a solid angle $\Omega$ is \citep{HECKMAN2011}
\begin{eqnarray}
v_{\rm ram} = 570~\dot{p}_{34}^{1/2}(\Omega/4\pi)^{-1/2}r_{0,100}^{-1/2}N_{21}^{-1/2}~{\rm km~s}^{-1}, 
\label{eq:v_ram}
\end{eqnarray}
where $\dot{p}_{34}$ is the rate of momentum carried by the wind normalized to 10$^{34}$ dynes, $r_{0,100}$ is the initial launch radius of the cloud with respect to the central source normalized to 100 pc, and $N_{21}$ is the column density of the cloud normalized to 10$^{21}$ cm$^{-2}$.  In cases where the injected energy and momentum from stellar and AGN processes are sufficient to drive an outflow on scales comparable to the host galaxy, the forces associated with pressure gradients in the underlying ambient ISM need to be taken into account and may help drive the outflow along the paths of least resistance (e.g., minor axis of a gaseous rotating disk).

\subsubsection{Radiation} \label{sec:radiation}

 Direct UV radiation from massive young clusters absorbed by dusty gas can impart velocities to molecular clouds that are of order their escape velocities from the gravity of the cluster \citep{MURRAY2011}. Thus radiation can, in principle, drive cold galactic winds in starbursts with sufficiently high density of star formation activity if its pressure is efficiently coupled. The necessary condition is that the Eddington ratio of the radiation force to gravitational force is greater than unity. This is determined by the opacity per unit gas mass, $\kappa$, which is proportional to the dust-to-gas ratio, and the luminosity-to-mass ratio of the stellar population \citep{MURRAY2010,SKINNER2015}. In the single-scattering optically thin limit, the photon is absorbed by the gas or dust but is not re-emitted, so the momentum flux imparted onto the dust and gas is simply $L_{\rm absorbed}/c$ $\simeq$ (1--e$^{-\tau_{\rm UV}}$) $L_{\rm BOL}/c$ and the condition to drive an outflow is $L > L_{\rm Edd} \simeq 4 \pi G M c / \kappa_{\rm UV}$.  This is somewhat relaxed in a turbulent medium, where under-dense lines of sight can be radiatively accelerated even when $L<L_{\rm Edd}$ in the bulk of the material \citep{THOMPSON2016b}. Clouds that start as radiatively-driven experience gentler accelerations, and although they may later be advected in a hot flow and ultimately destroyed, they can potentially reach large distances and high velocities \citep{THOMPSON2015}. 
In this discussion, we have ignored the fact that heating of cold clouds irradiated by an energy source will create a pressure imbalance in the cloud and cause it to lose mass in the direction of the energy source. By the rocket effect, the cloud will experience an acceleration in the opposite direction \citep[similar to what occurs in photoevaporative flows; e.g.,][]{OORT1955,BERTOLDI1990}. The results from recent numerical simulations of irradiated cold clouds embedded in flows of hot and fast material suggest that the impact of this extra kick in velocity on the accelerated clouds is not significant, except perhaps in the initial stage of acceleration \citep[e.g.,][]{BRUGGEN2016}. 

Multiple scatterings, occurring when the optical depth to the re-radiated infrared $\tau_{IR}$ is large, could magnify the radiation force. 
 A convenient expression that combines both single- and multiple-scattering processes is \citep{HOPKINS2014,HOPKINS2018a}
\begin{eqnarray}
\dot{p}_{\rm rad} = \tilde{\tau} L_{\rm bol} /c~~{\rm where}~\tilde{\tau} = (1 - e^{-\tau_{\rm single}})(1 + \tau_{\rm eff,IR}).
\label{eq:pdot_rad}
\end{eqnarray}
The value of $\tilde{\tau}$ ranges from $\sim\tau_{\rm single} = \tau_{\rm UV/optical} << 1$ in the optically thin case to $\sim$ (1 + $\tau_{\rm eff, IR}$) $\gtrsim$ 1 in the infrared optically-thick limit. The effective infrared optical depth, $\tau_{\rm eff, IR} = \kappa_{\rm eff, IR} \Sigma_{\rm gas}$, 
is the ``boost factor''and is sometimes expressed as $\tau_{\rm eff, IR} = \tau_{\rm IR} \beta_{\rm IR}$, the product of the integrated infrared optical depth, $\tau_{\rm IR}$, and a factor $\beta_{\rm IR}$ that includes all of the uncertainties in calculating the effective optical depth.  To first order, the condition to drive a radiation-pressure wind under the optical-thick limit then becomes $L > L_{\rm Edd} \simeq 4 \pi G M C / \kappa_{\rm eff, IR}$ (note that the dust opacity $\kappa_{\rm eff, IR} \propto \nu^2$ at the relevant IR wavelengths, so the radiation-gas coupling becomes less efficient as the radiation field diffuses outward of the dusty cocoon). In practice, the predicted importance of radiation pressure in driving cool outflows depends strongly on the method used to solve the frequency-dependent radiative transfer equation and derive this boost factor \citep[see][for a recent summary]{KRUMHOLZ2019}.  Simulations based on the flux-limited diffusion (FLD) approximation 
suggest that such magnification is limited, as radiative Rayleigh-Taylor instabilities (RRTI) set in creating dense filaments of matter aligned with the radiation field that limit the momentum transfer and prevent gas ejection \citep{KRUMHOLZ2013}. Similarly, simulations using the M1 closure relation approximation find an anticorrelation between radiative flux and gas density that limits the radiation force magnification \citep{ROSDAHL2015,SKINNER2015}.  Other simulations, using the more reliable variable Eddington tensor algorithm \citep[e.g.,][Fig.\ \ref{fig:sims_clouds}c]{ZHANG2017}  or implicit Monte Carlo radiation transfer scheme \citep{TSANG2015}  are more successful at generating outflows after the onset of the Rayleigh-Taylor instabilities. \citet{ZHANG2017}  have argued that $\beta_{\rm IR} \sim 0.5$--$0.9$, decreasing with increasing optical depth \citep[see also][]{HUANG2019}.

The above discussion equally applies to starburst- and AGN-dominated systems,  although the star formation rate densities required to radiatively drive a wind may be very high \citep{CROCKER2018}. Radiation pressure is likely the dominant driving force behind accretion-disk scale fast X-ray winds in AGN and UV broad absorption line (BAL) outflows in quasars, and may also play a crucial role in driving cool outflows. Momentum-driven outflows involving radiation pressure forces naturally produce $M_{\rm BH} \propto \sigma_*^4$ \citep[e.g.,][]{FABIAN1999,KING2003,MURRAY2005,ISHIBASHI2018a,ISHIBASHI2018b,COSTA2018b,COSTA2018a}. 

\subsubsection{Cosmic Rays} \label{sec:cosmic_rays}

Cosmic rays have the potential to efficiently accelerate cold gas, creating a neutral gas component in the galactic halo that could be an important mass reservoir in galaxies \citep{EVERETT2008,SOCRATES2008,UHLIG2012}. A gradient in cosmic ray pressure due to cosmic ray streaming and scattering in self-excited Alfv\'en waves can transport gas from the dense regions in the mid-plane to heights of kiloparsecs. The mass flow is overwhelmingly neutral and warm ($10^4$~K), but the velocities are slower than those obtained with supernova driving \citep{GIRICHIDIS2016}.
The efficiency of the process is partially related to the global structure of the magnetic field, as cosmic rays need to stream away from the dense galactic regions along partially ordered magnetic field lines, but mass-loading factors of order unity appear feasible in a Milky Way-like galaxy \citep{RUSZKOWSKI2017}. The so-called bottleneck effect, in which cosmic ray streaming through a cold cloud embedded in hot gas causes a cosmic ray pressure buildup in the neutral/molecular material and a resulting pressure gradient, can lead to cold cloud acceleration by cosmic rays \citep{WIENER2017}. This same effect can over-pressurize cold clouds and lead them to expand and have similar densities to the hot material \citep{WERK2014}.

The ability of CRs to drive galactic outflows is related to their overall scattering optical depth $\tau_{\rm CR} \sim R/\lambda$, where $R \sim$ kpc is the distance the CRs have to travel to escape out of the galaxy and $\lambda$ is the uncertain mean free path of CRs between scattering. The effective momentum injection rate from CRs  is then $\dot{p}_{\rm CR} \simeq \tau_{\rm CR} \dot{E}_{\rm CR}/c \simeq 10^{-3} \tau_{\rm CR} (L/c)$, where the last approximate inequality comes from the relations between $\dot{E}_{\rm CR}$, $\dot{M}_*$, and $L$ (Section \ref{sec:stellar_winds_supernovae} and \citet{HECKMAN2017}). The CR momentum injection rate may thus be comparable to the radiation pressure from massive stars if $\lambda \sim$ pc and $\tau_{\rm CR} \sim 1000$. CR-driven winds are generally denser, colder, and smoother than the thermal supernova-driven outflows \citep[Fig.~\ref{fig:sims_sb_outflows}c; e.g.,][]{PAKMOR2016,RUSZKOWSKI2017,PFROMMER2017,GRONKE2018b,GIRICHIDIS2018,BUSTARD2019,VIJAYAN2019}.

\subsubsection{Radio Jets} \label{sec:radio_jets}

\begin{figure}[htb]
\begin{center}
\includegraphics[width=1.0\textwidth]{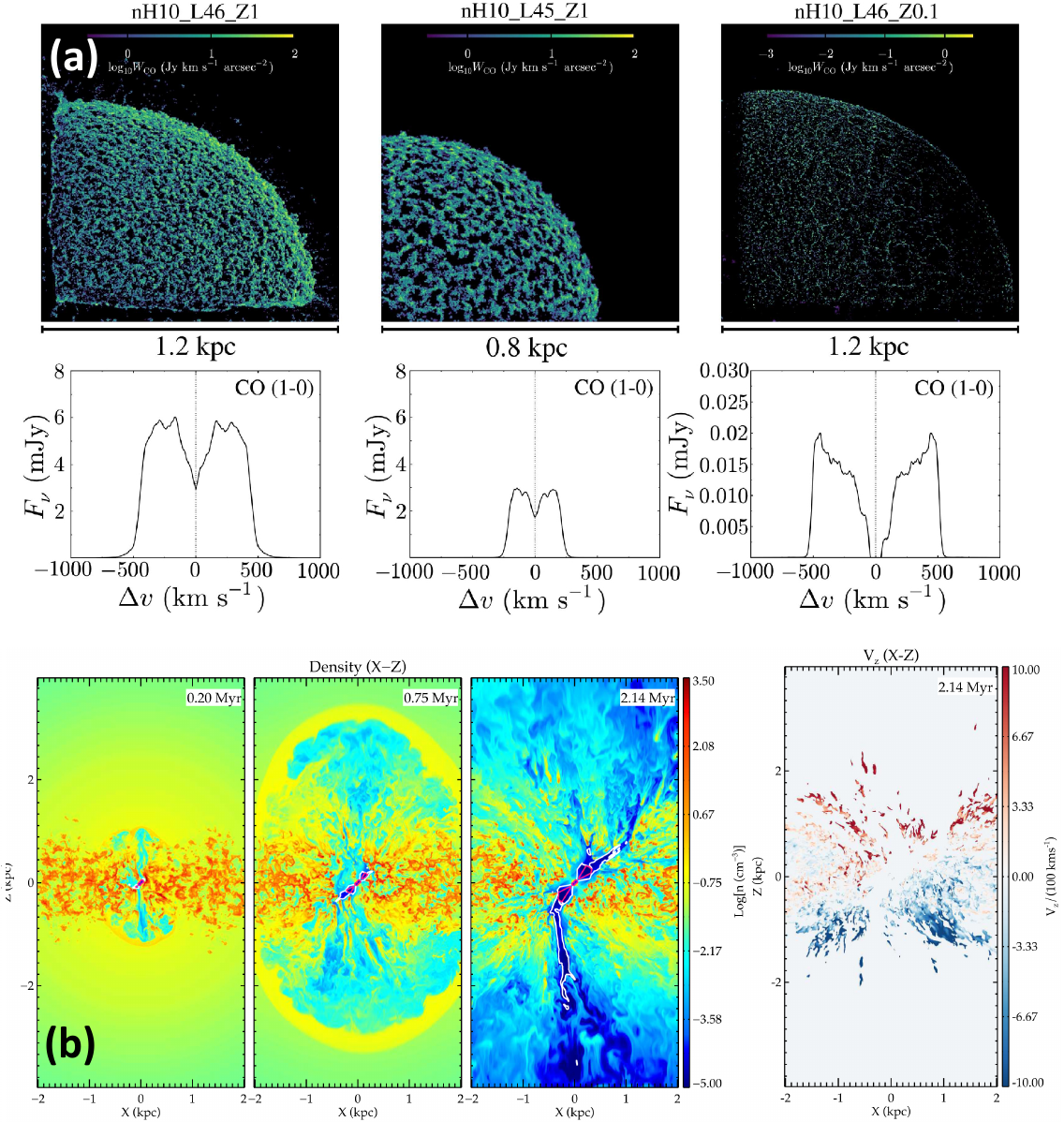}
\end{center}
\caption{Simulations of AGN-driven cool outflows: (a) Velocity-integrated maps of CO (1$-$0) emission (top row) and CO (1$-$0) spectra (bottom row) for three different simulations. (b) Left three panels: time-evolution (0.20, 0.75, and 2.14 Myr) of the density in a jetted outflow simulation, where the jet is tilted by 45$^\circ$ from the plane of the disk. The right panel shows the vertical velocity field in units of 100 km s$^{-1}$ at 2.14 Myr. Images reproduced with permission from (a) \citet{RICHINGS2018b} and (b) \citet{MUKHERJEE2018b}, copyright by the authors.}
\label{fig:sims_agn_outflows}
\end{figure}

This section considers the ram-pressure term in Eq.~(\ref{eq:eom}) generated by collimated relativistic low-density jets from AGN. In this case, the jet injects an energy $E_{\rm jet} = \frac{1}{2} \rho_j 2 \pi r_{\rm in}^2 (1\,-\,{\rm cos}~\theta_J)v_J^3 t_J$, where $\rho_J$, $v_J$, $\theta_J$, and $t_J$ are the jet mass density, velocity, half-opening angle, and duration, respectively, and $r_{\rm in}$ is the inner radius at the base of the jet where the jet energy is injected into the medium \citep[][]{BAMBIC2019}.
The impact of this jet on the ambient medium not only depends on $E_{\rm jet}$, but also on the level of interaction of the jet with the medium. For instance, a powerful, highly collimated jet will simply ``drill'' through the host galaxy's ISM without imparting much energy and momentum to it \citep[e.g.,][]{SCHEUER1974,MUKHERJEE2016}. The other extreme case where all of the jet kinetic energy is deposited on small scale to create a hot over-pressured bubble has already been discussed in Section \ref{sec:stellar_winds_supernovae} in the context of thermally-driven outflows in starbursts \citep[see also][]{MUKHERJEE2016}. Fine-tuning of the jet power, kinetic feedback efficiency, $\epsilon_{f,k}$, and ISM cloud structure is therefore needed for efficient energy \emph{and} momentum transfer to disperse a significant fraction of the gas in galaxies \citep{WAGNER2012,MUKHERJEE2016,MUKHERJEE2018a,MUKHERJEE2018b}.  Typical jet powers needed to create a jet-driven energy bubble with the right properties are $\dot{E}_{\rm jet}/L_{\rm Edd} \gtrsim 10^{-4}$. Material accelerated at the base of the jets may reach sub-relativistic velocities that are consistent with ultra-fast outflows detected in radio-loud AGN \citep{WAGNER2013}. While these radio jets launch strong outflows, virtually none of the entrained mass reaches escape velocity so it rains back onto the galaxy as a galactic fountain. In the special case of jets directed through the plane of a galaxy disk, the jet may stir the kinematics of the multi-phase ISM and affect the efficiency of star formation \citep[Fig.~\ref{fig:sims_agn_outflows}b,][]{MUKHERJEE2018b}. We discuss this issue in more details in Sect.~\ref{sec:jetted_agn}, \ref{sec:obscured} and \ref{sec:lowz_molecular} below.

\subsection{Physical Origin of the Outflowing Cool Gas} \label{sec:origin}

Mass loading of outflows with cool gas can arise from two fundamental causes: mechanical drag leading to the entrainment of cool gas in a hot, fast-moving fluid, and direct acceleration of the cool gas through radiation or cosmic ray pressure \citep[e.g.,][]{KIM2018}. A third possibility is to create the cool gas starting from the hot wind material, through cooling and condensation. Part of the complexity of understanding the driving of galactic outflows is due to the fact that several of these processes may be taking place simultaneously. Hot outflows from the blowout of over-pressured regions surrounding a starburst will be concomitant with large radiation fields and the associated radiation pressure, for example, making it difficult (and to some extent artificial) to separate them. It has been suggested that, in order to drive cool outflows efficiently in the context of galaxy evolution, several of these mechanisms need to operate simultaneously \citep{HOPKINS2012}. Below we discuss in more detail entrainment and \textit{in situ} formation. Direct acceleration by radiation and CR pressure has already been discussed in Sections \ref{sec:radiation} and \ref{sec:cosmic_rays}, respectively, so it is not repeated here.

\subsubsection{Cool Gas Entrainment} \label{sec:entrainment}

The acceleration of cold clouds by a warm, fast wind has proven a difficult problem to solve. The density contrast between the cool material and the hot supersonic wind is $\chi\simeq10^3-10^4$ \citep[e.g.,][]{SCANNAPIECO2015}, which already suggests that imparting momentum to a dense ``cloudlet'' is going to require a substantial flow of the lighter outflow warm fluid. In the standard acceleration calculation, a dense cloudlet is placed in the fast hot flow. When the flow encounters the cloud, it develops a shock that compresses it and starts to ablate and shred it on a ``cloud crushing'' timescale, $t_{cc}$, equal to the ratio of the ratio of the time for the hot flow to travel the radius of the cloud multiplied by $\chi^{1/2}$ \citep{KLEIN1994}. Accelerating the cloud to the speed of the wind requires that the mass in the flow that impacts the cloud is similar to the mass of the cloud itself. This in turn requires a length of time that is $t_{acc}\sim \chi^{1/2}\,t_{cc}\gg t_{cc}$. Therefore, in the absence of a stabilization mechanism, it would appear that clouds are destroyed through Rayleigh-Taylor fragmentation and shear-driven Kelvin-Helmholtz instabilities induced by the flow faster that they can be accelerated by it, a result also obtained by detailed hydrodynamic simulations \citep{COOPER2009,SCANNAPIECO2015,SCHNEIDER2017}. Clouds with realistic density distributions are destroyed faster than spherical clouds of uniform density \citep{SCHNEIDER2015}. The material from the destroyed cloud joins the flow, mass-loading the hot phase.

Possible stabilization mechanisms for cold material advected in a fast flow are radiative cooling and magnetic fields. As the cloud experiences the initial crushing shock its density increases. If the material can  efficiently cool radiatively, the denser clumps formed will likely survive for a longer period of time \citep{COOPER2009}. In fact, if the cooling is fast enough compared to the cloud crushing time, a cloudlet may grow in mass by acting as a condensation seed that accretes newly cooled gas \citep{MARINACCI2010,ARMILLOTTA2016,ARMILLOTTA2017,GRONKE2018a,GRONKE2019}. In supersonic flows Kelvin-Helmholtz instabilities are suppressed, slowing down cloud disruption, and as the cloud is compressed it develops a pressure gradient along the flow direction that causes a filamentary structure to develop. The lower cross section to the flow means the drag forces are reduced, resulting in a slower acceleration despite the increased lifetimes \citep{SCANNAPIECO2015}. Clouds with efficient thermal conduction experience ablation of their outer layers, in contact with the hot flow, which results in an additional acceleration due to the rocket effect but also hastens their destruction \citep{MARCOLINI2005}. The lifetime of the cloud in these circumstances depends on the balance between radiation, conduction, and the disrupting instabilities \citep{ORLANDO2005}, but it appears that on balance longer lifetimes do not translate into reaching faster flow velocities before destruction \citep{BRUGGEN2016}. Magnetic fields can suppress conduction \citep{ORLANDO2008}, and tangled magnetic fields threading a cloud may help stabilize it against disruption \citep{MCCOURT2015}. In strongly magnetized winds, simulations suggest that magnetized clouds experience additional drag forces that result in much faster accelerations \citep{MCCOURT2015,BANDA-BARRAGAN2016,BANDA-BARRAGAN2018,BANDA-BARRAGAN2019,GRONNOW2018}. There may be some recent observational support for this magnetized outflow scenario in one object \citep{LEAMAN2019}. The effects of the different cloud destruction mechanisms are a function of parameters such as the cloud column density and the temperature of the medium, with different physics dominating in different regions of parameter space \citep{LI2019}.

\begin{figure}[htbp]
\begin{center}
\includegraphics[width=0.95\textwidth]{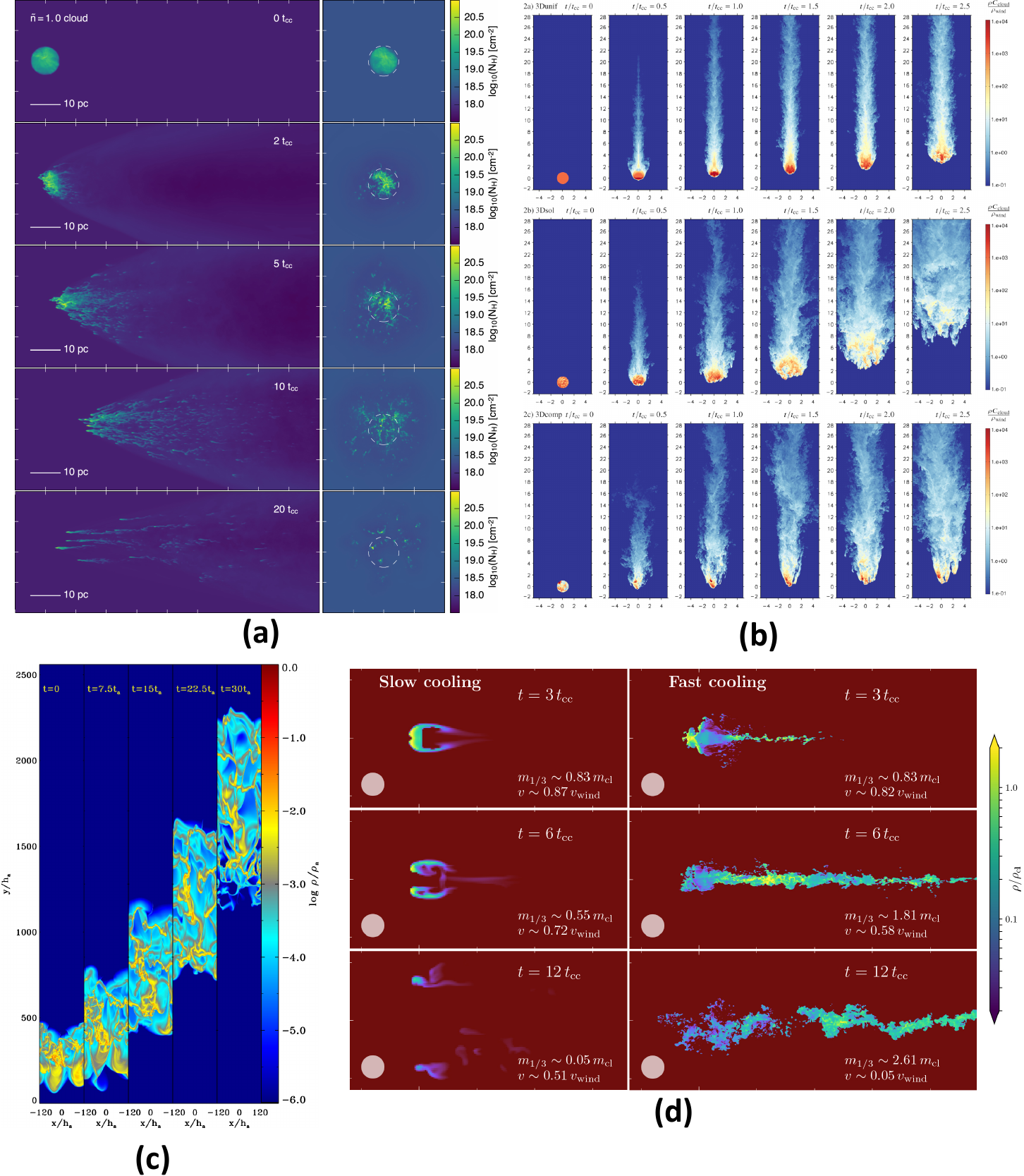}
\end{center}
\caption{Simulations of accelerated cool clouds: (a) Time series evolution of a turbulent cloud embedded in a wind. The panels on the left show the surface density projected along the y-axis, with the wind entering the box on the left, while the right column shows the same quantity in the direction of the wind velocity. (b) Same as (a) but for different internal density distributions. The wind enters from the bottom in this case. (c) Density distribution at five different epochs, from two-dimensional radiation hydrodynamical simulations of dusty winds. (d) Density distributions at three different times when cooling is slow (left panels) and fast (right panels). Images reproduced with permission from (a) \citet{SCHNEIDER2017}, (c) \citet{ZHANG2017}, copyright by AAS; and (b) \citet{BANDA-BARRAGAN2019},  (d) \citet{GRONKE2018a}, copyright by the authors.}
\label{fig:sims_clouds}
\end{figure}

\subsubsection{In-Situ Formation from the Hot Wind Material} \label{sec:in-situ}

It is possible for a cold phase to form \textit{in situ} via thermal instabilities and condensation from hot gas with a cooling time shorter than its dynamical time \citep{EFSTATHIOU2000,SILICH2003}. Because adiabatic cooling due to expansion acts to lower the gas temperature into a regime where runaway radiative cooling takes over, this appears to be a general expectation. The efficiency of this process is strongly dependent on the mass-loading of the hot phase, thus increased mass-loading due to destruction of entrained cold clouds in the inner regions of the flow may play an important role in causing thermal instabilities to develop \citep{THOMPSON2016}. Also, systems with larger surface densities of star formation rate are expected to have shorter cooling times. In strongly mass-loaded systems with high star-formation rates, which may correspond to outflows from ULIRGs, the result from the cooling instability may be material that has enough density to turn molecular and enough column density to self-shield from the dissociating metagalactic radiation field \citep{THOMPSON2016}. In this picture X-ray emission arises from recombination, as the outflow progressively cools down, which may explain the strong spatial coincidence between H$\alpha$ and soft X-ray emission observed in galaxy winds. Because of the expected strong correlations between mass loading, velocity, and temperature, this scenario predicts a correlation between the observed wind velocities and the local escape velocity, as well as faster velocities for the lower column densities \citep{MARTIN2005,RUPKE2005c}. The ultimate fate of the wind would be to populate halos with cool gas, as the expanding wind cools adiabatically first to a point where runaway cooling takes over, or shocks the ambient circumgalactic medium which then cools in less than a Hubble time \citep{THOMPSON2016}.

Can cooling and condensation explain the observations of molecular material in winds? For the material to become molecular, hydrogen atoms need to combine to produce molecular hydrogen. Because of the symmetry of the system and the lack of a permanent dipole, gas phase reactions between neutral atoms for formation of H$_2$ proceed extremely slowly \citep{GOULD1963}. The dominant gas channel for H$_2$ formation in the gas phase starts with the formation of a H$^-$ ion through radiative association, followed by formation of H$_2$ via associative detachment. But this process is also slow, and likely dominant only in the very early universe \citep{GLOVER2006}. To explain its abundance in the dense ISM, it has been long accepted that H$_2$ formation is catalyzed by dust grains, where the dust surface provides a substrate for H atoms to find each other efficiently and combine \citep{GOULD1963,HOLLENBACH1971}. If the dust grains immersed in low density, high temperature gas are warm, this may impact their ability to form molecular hydrogen (note that the equilibrium temperature of the grains is determined by the radiation field at these densities). The precise mechanism through which the dust catalysis of H$_2$ formation occurs, and hence its temperature dependence, are matters of debate, and hence it is unclear how efficiently H$_2$ formation would proceed on warm dust-grain surfaces \citep{LEBOURLOT2012}. Moreover, dust grains immersed in hot gas experience destructive sputtering due to grain charging and gas-grain collisions. Dust grains of size $\sim0.1$\,$\mu$m immersed in a $T>10^6$\,K hot plasma of density $\sim10^{-2}$\,cm$^{-3}$ have a lifetime of only a few 10$^6$ years \citep[][see eq.\ \ref{eq:t_sput} in Sec. \ref{sec:dust_cycle} below]{DRAINE1979b}. This implies that even when neglecting the effects of shocks \citep[which are also very destructive,][]{JONES1994,SLAVIN2015}, once grains are immersed in the hot phase they only last for a few million years before being destroyed. Taken together, this suggests that reforming a molecular phase from hot material through cooling and condensation could be very difficult.

Independently, it seems clear that many of the observations of molecular gas in winds cannot be explained by material that is condensed from the hot phase. The mm-wave observation of complex molecules (HCN, HNC, HCO$^+$) in some of the molecular outflows strongly suggests that the gas started as molecular, dense, and chemically rich \citep{AALTO2012,WALTER2017}. Some of these molecular transitions appear enhanced in outflows, but this is likely to be due to either shock chemistry or excitation effects, not incorporation of material condensed from the hot phase \citep{AALTO2012}. Some observations show that the dynamical age of the outflowing features that contain these molecules is short, likely too short for the molecules to have formed in the flow. Time-dependent calculations of chemistry and molecular cloud formation in colliding flows find that, even in relatively dense and benign conditions in the neutral galactic ISM, it takes a few Myr for the material to become predominantly molecular, and several Myr for it to emit brightly in CO \citep{CLARK2012}. By contrast, the dynamical time scale of some CO-emitting features in outflows is $\sim1$\,Myr  \citep[for example in NGC~253,][]{WALTER2017}, strongly pointing to an entrainment origin for the molecular gas. 

\subsection{Minimum Requirement for the Outflowing Material to Escape the Host Galaxy} \label{sec:escape}

An important question is whether the outflowing material escapes the host galaxy altogether or remains bound to it and thus is deposited in the CGM. A necessary condition to escape the gravitational potential of a galaxy is for the material to have a radial outflow velocity that exceeds the local escape velocity (it is not a sufficient condition since drag forces may be important). Assuming a spherically symmetric galaxy with gravitational potential $\Phi(r)$, the local escape velocity is given by $v_{\rm esc}^2(r) = 2 [\Phi(\infty) - \Phi(r)]$. For a singular isothermal sphere, where the mass $M(< r)$ diverges at large $r$ but the circular velocity $v_{\rm circ}$ is a constant, $v_{\rm esc}^2(r) = 2~v_{\rm circ}^2~{\rm ln}(r_{\rm halo}/r)$, so it slowly diverges for large $r_{\rm halo}/r$ \citep{BINNEY2008}. 
To avoid this divergence, one often truncates the singular isothermal sphere at a radius $r_{\rm halo} = r_{\rm max}$. In that case, $v_{\rm esc}^2(r) = 2 [\Phi(\infty) - \Phi(r_{\rm max}) + \Phi(r_{\rm max}) - \Phi(r)] = 2 v_{\rm circ}^2 [1 + {\rm ln}(r_{\rm max}/r)]$, so
\begin{eqnarray}
v_{\rm esc}(r) = v_{\rm circ} \sqrt{2 \left[1 + {\rm ln}\left(\frac{r_{\rm max}}{r}\right)\right]}. 
\label{eq:v_esc}
\end{eqnarray}
In more realistic galaxies where the gas is supported by a combination of rotational and random motions, the circular velocity may be replaced by $v_{\rm circ} = \sqrt{2} S$, where the kinematic parameter $S$ is defined as $S^2 \equiv 0.5 v_{\rm rot}^2 + \sigma^2_{\rm gas}$ ($v_{\rm rot}$ and $\sigma_{\rm gas}$ are the observed rotation velocity and velocity dispersion of the ISM in the host galaxy) and scales with the stellar mass following log~$S$ = 0.29 log~$M_*$--0.93, where $S$ is in km s$^{-1}$ and $M_*$ is in $M_\odot$ \citep{WEINER2006,KASSIN2007}, in low-$z$ star-forming galaxies \citep{HECKMAN2015,SIMONS2015}. The value of $v_{\rm esc}$ is not sensitive to $r_{\rm max}/r$ (e.g., for $r_{\rm max}/r = 10-100$, $v_{\rm esc}/v_{\rm circ} = 2.6-3.3$). A rather conservative simplification when studying galaxy-scale outflows is to choose $v_{\rm esc}(r_{\rm outflow}) \simeq 3 v_{\rm circ}$, which implicitly assumes $r_{\rm max}/r_{\rm outflow} \simeq 33$. When studying CGM-scale flows, the radius $r_{\rm max}$ may be substituted with the virial radius $r_{\rm vir} \simeq v_{\rm circ}/10 H(z)$ (in Mpc), where $H(z)$ is the Hubble parameter at redshift $z$ \citep[e.g.,][]{SCHROETTER2016}. In that case, $v_{\rm esc}(r_{\rm CGM}) \simeq 2 v_{\rm circ}$ generally is a reasonable approximation ($r_{\rm vir}/r_{\rm CGM} \simeq e$). The issue of the fate of the cool outflowing material is discussed in Sections \ref{sec:lowz_fate} and \ref{sec:highz_fate}.

\section{Observational Techniques and their Limitations} \label{sec:techniques}

This section is a critical review of the methods used to detect and characterize the properties of cool outflows in galaxies, near and far. First, we briefly discuss the most commonly used criteria to distinguish the outflowing gas from the quiescent material in the host galaxy, and describe the basic formulae used to derive the key dynamical quantities of the outflow. The remainder of this section describes from a practical point of view the various techniques and diagnostics that have been used to detect and characterize the outflowing material, starting with the neutral-atomic gas, followed by the molecular gas, and concluding with the dust in these outflows.

\subsection{Kinematic Definition of an Outflow} \label{sec:kin_def}

All of the properties of the outflows discussed in this review rely on a robust decomposition of the outflowing material from the dominant non-outflowing material.  This is a non-trivial task that adds uncertainties to the outflow properties.

\subsubsection{Spatially Unresolved Outflows} \label{sec:spatially_unresolved}

When the outflow is spatially unresolved, one has to rely on the detection of faint wings in the velocity profile of the spatially integrated absorption and/or emission line to determine the portion of the gas that is outflowing out of the galaxy. Emission line features have the disadvantage of being ambiguous in terms of flow direction, when the line-emitting gas is optically thin and unaffected by dust or other absorbing material within the outflow structure. Velocity thresholds have traditionally been used to identify the outflowing gas, but these thresholds vary wildly in the literature. In principle, these kinematic thresholds should be adjusted on an object-by-object basis to account for the range of velocities associated with gravitational motions (rotation, random motion, inflow), based on the mass of the host galaxy. Parametric decomposition of the observed profiles into a ``narrow'' and ``broad'' velocity components is often carried out to quantify the fraction of the emission / absorption that lies in the wing(s) of the profiles. The dominant narrow component, responsible for the core of the line profile, is associated with gas in the gravitational potential of the host galaxy while the broad component is associated \emph{in part} with the outflow.  The width of the narrow component is a proxy for the mass of the galaxy and therefore can be used to define the kinematic threshold above which the outflowing material becomes significant in the broad component \citep[e.g.,][]{LUTZ2019}  

The use of the formulae in Section \ref{sec:energetics} requires a characteristic outflow velocity. This velocity may be derived from the above parametric decomposition procedure or using non-parametric methods such as interpercentile range measurements \citep[e.g.,][]{WHITTLE1985,VEILLEUX1991}. Typically, one selects $v_{\rm 50}$ or $v_{\rm 84}$. The maximum outflow velocity (e.g., 98-percentile velocity, $v_{\rm 98}$, of the integrated or broad-component profile) should not be used in these formulae since (1) it overestimates the characteristic velocity and thus the energetics of the outflow, and (2) it is much more sensitive to noise than other more appropriate measurements of the maximum outflow velocity such as $\vert v_{84}\vert$ or $\vert v_{50}\vert + 2 \sigma$, where $\sigma$ is the velocity dispersion of the broad component \citep[note that $\vert v_{98}\vert = \vert v_{50}\vert + 2 \sigma$ for a Gaussian velocity distribution; e.g.,][]{RUPKE2013b}.

\subsubsection{Spatially Resolved Outflows} \label{sec:spatially_resolved}

The identification of the outflowing gas in cases where the outflow is spatially resolved usually relies on comparisons between the observed velocity field and line width distribution and the predictions from that of a differentially rotating disk of material with some level of random motion normal to the disk, taking into account the finite spectral resolution of the instrument as well as the effects of beam smearing of the velocity field associated with the finite width of the instrumental beam size. Beam smearing will effectively flatten velocity gradients and broaden the widths of the observed line profiles where the gradients are steeper \citep[e.g., $^{\rm 3D}$BAROLO;][]{DITEODORO2015}. This method works well when the outflow is fast and powerful but fails when the signatures of the outflow in the line profiles are more subtle. The next level of analysis consists in trying to capture these more subtle signs of the outflow in the line profiles (shoulders, wings) by parametrizing the line profiles at each position in the outflow and galaxy as a sum of 2-3 components and assign the component that shows signatures of rotation (e.g., clear velocity gradient in the same direction as the stars) to the gravitationally dominated component in the host galaxy and the other(s) to the outflow \citep[e.g.,][]{RUPKE2013b}. The definitions discussed in Sec.\ 
\ref{sec:spatially_unresolved} for the characteristic velocity and flux assigned to the outflow material can then be applied for the individual profiles at each position. Constraints or initial guesses on the parameters of the individual components based on the results of fits in the neighboring pixels may be used to guard against unphysical discontinuities on sub-beam scale. An iterative method may be used to remove contamination from a bright central point source \citep[e.g., quasar;][]{RUPKE2017}. 

The above method will fail if the line-emitting material is not well resolved by the beam. Assuming the source is strongly detected, one solution is to look for the residuals in the integrated profiles after fitting and removing a Gaussian source model at each velocity channel \citep[e.g.,][]{VEILLEUX2017}. First, a two-dimensional Gaussian is fit to the image for each channel in the region where a source is detected. These results are then used to make a smooth source model with linearly changing position as a function of velocity (to account for a possible velocity gradient), Gaussian-changing intensity, but constant size and orientation. This smooth source model is then removed from each velocity slice to arrive at a ``residuals'' cube. The high S/N of the detection allows to centroid the source in each channel with very good sub-beam precision. The ``high-velocity'' emission, which cannot be accounted for by the gas in pure rotation, is identified as the outflowing material. In cases where the data cubes are based on interferometric data, the results from the analysis on the imaging data may also be checked independently by deriving the sizes and fluxes of the wing emission from a fit of the data directly in the \emph{uv} plane \citep[e.g., using CASA \emph{uv}-Plane Model Fitting routines \emph{uvmodelfit} and \emph{uvmultifit} from the library of][]{MARTI-VIDAL2014}.

\subsection{Derivation of Outflow Energetics: Basic Formulae} \label{sec:energetics}

In cases where the outflow is not spatially resolved, one has to assume an outflow model. The simple case of an optically thin,  spherically symmetric mass-conserving free wind, with an instantaneous mass outflow rate and velocity that are independent of radius within the wind and zero outside, is discussed in \citet{RUPKE2005c}. The mass outflow rate at a radius r within the wind is given by
\begin{eqnarray}
\dot{M} = \Omega \mu m_p n(r) v r^2,
\end{eqnarray}
where $m_p$ is the proton mass, $\mu \simeq 1.36$ is the mass per H nucleus taking into account the relative He abundance, $v$ is the outflow velocity, and $\Omega$ is the solid angle subtended by the wind as seen from its origin (i.e. the wind's global covering factor). This last term includes a large-scale covering factor (related to the wind's opening angle), given by $C_\Omega$, and a small-scale covering factor (related to the wind's clumpiness), $C_f$. Thus, $\Omega$/4 $\pi$ = $C_\Omega C_f$ \citep{RUPKE2005b}.  The hydrogen number density $n(r)$ is related to the measured hydrogen column density $N = \int^{r_2}_{r_1} n(r)~dr$ for a wind extending from inner radius $r_1$ to outer radius $r_2$. Assuming a density profile $n(r) \propto r^{-2}$ (e.g., an isothermal sphere), the instantaneous (local) mass outflow rate across any radius $r$ within the wind is
\begin{eqnarray}
\dot{M}^{inst}_{thick} = \Omega \mu m_p N v \frac{r_1 r_2}{\Delta r}, 
\label{eq:mdot_inst_thick}
\end{eqnarray}
where $\Delta r = r_2 - r_1$, and the corresponding mass of the wind is
\begin{eqnarray}
M_{thick} = \Omega \mu m_p N r_1 r_2.
\label{eq:m_thick}
\end{eqnarray}
The mass outflow rate averaged over the wind lifetime $\sim r_2/v$ is
\begin{eqnarray}
\dot{M}^{avg}_{thick} = \Omega \mu m_p N v r_1, 
\label{eq:mdot_avg}
\end{eqnarray}
If the wind is a thin shell of radius $r \sim r_1 \sim r_2$, the corresponding equations are
\begin{eqnarray}
\label{eq:mdot_inst_thin}
\dot{M}^{inst}_{thin} & = & \Omega \mu m_p N v \frac{r^2}{\Delta r}, \\
\label{eq:mdot_avg_thin}
\dot{M}^{avg}_{thin} & = & \Omega \mu m_p N v r, \\
\label{eq:m_thin}
M_{thin} & = & \Omega \mu m_p N r^2. 
\end{eqnarray}
Note that, in both the thick- and thin-shell cases, the time-averaged mass outflow rates are a factor of $\Delta r / r$ lower than the local instantaneous rates.

The momenta and kinetic energies and their outflow rates are
\begin{eqnarray}
\label{eq:p}
p & = & M~v,\\
\label{eq:pdot}
\dot{p} & = & \dot{M}~v,\\
\label{eq:E}
E & = & M~(v^2/2 + 3 \sigma^2/2), \\
\label{eq:Edot}
\dot{E} & = & \dot{M}~(v^2/2 + 3 \sigma^2/2), 
\end{eqnarray}
where the energy includes both the ``bulk'' kinetic energy due to the outflowing gas and ``turbulent'' kinetic energy (where we assume the same velocity dispersion $\sigma$ in each dimension).\footnote{Note that there is a typographical error in the numerical formulae of \citet{RUPKE2005c}:  in their equations 13-18, the normalization factor of the column density $N$ should be 10$^{20}$ cm$^{-2}$ rather than 10$^{21}$ cm$^{-2}$. The outflow energetics published in \citet{RUPKE2005c} are based on the correct formulae and not affected by this error.}  

\begin{figure}[htb]
\begin{center}
\includegraphics[width=1.0\textwidth]{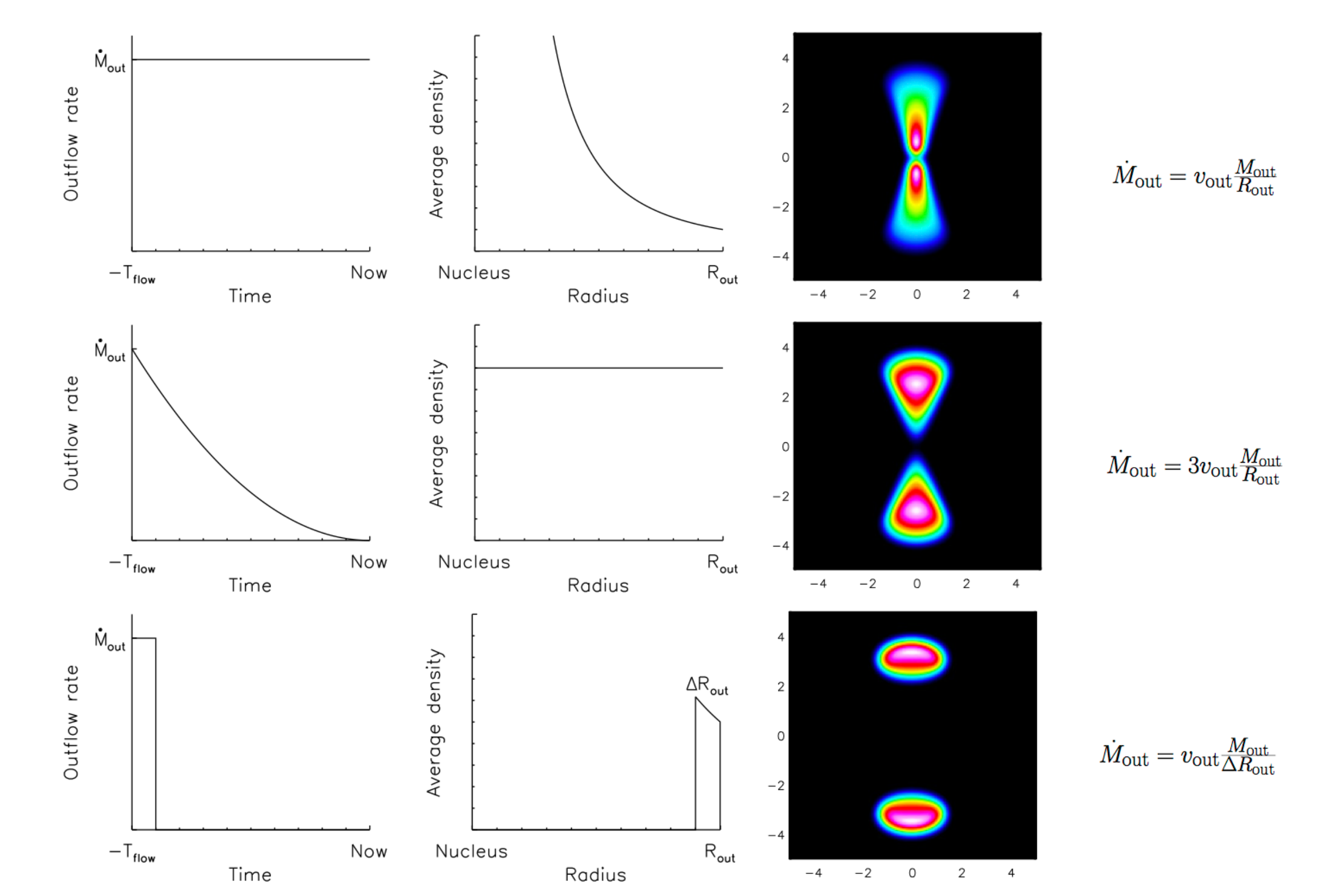}
\end{center}
\caption{Three commonly adopted outflow histories and their effects on the radial density profile and conversion from outflow mass to outflow mass rate. From left to right, the panels show the outflow history, the average radial density profile (local cloud conditions may differ from this average), the simulated moment 0 map (for a modest 20$^\circ$ half opening angle flow seen with flow axis inclined by 60$^\circ$ with respect to the line of sight), and the equation for conversion from outflow mass to outflow mass rate. From top to bottom, the three cases are constant outflow history, constant average volume density in cone (requiring a decaying outflow history), and thin outflowing shell. Images reproduced with permission from \citet{LUTZ2019}, copyright by the authors.}
\label{fig:outflow_histories}
\end{figure}

Recall that $n(r) \propto r^{-2}$ was assumed. All of the above quantities need to be multiplied by a factor of 3 if the spherical (or multiconical) volume is instead filled with uniform density.  
Note, however, that the quick drop-off in the radial intensity profile generally seen in outflows seems inconsistent with this picture (Fig.~\ref{fig:outflow_histories}). Thus we favor the above equations for the calculations of the energetics.

In cases where the wind is spatially resolved, the integrated mass of outflowing material is simply
$M = \mu m_p \int N dA$, 
where the integral of the two-dimensional hydrogen column density distribution of the outflowing material is carried out over the projected area of the outflow. If the observed outflow velocity varies with position, then one needs to modify the above equations for the outflow mass, momentum, and energy rates by letting each position on the sky (corresponding to a unique angular coordinate with respect to the wind's center) have its own velocity, mass, momentum, and energy on the thin shell \citep[see][]{SHIH2010}.  The outflow rates are derived by dividing these quantities by the dynamical time $r/v$ at each position in the wind.  Note that all of the above formulae are expressed in terms of the column density, and are in this form more appropriate for absorbing material. In cases of line-emitting material, then one needs to compute the gas mass at each location on the sky. 

Projection effects add uncertainties to the results of these computations. In nearby systems where the outflows are well resolved, attempts have been made to account for these projection effects by fitting the observed velocity field with the predictions from simple three-dimensional kinematic models \citep[e.g., bipolar bubbles, biconical outflows, filled-in geometries instead of shells; e.g.,][]{VEILLEUX1994,RUPKE2013b}. This often cannot be done at higher redshifts.

\subsection{Diagnostic Tools of the Neutral-Atomic Gas Phase} \label{sec:diagnostics_neutral_atomic}

\begin{figure}[htb]
\begin{center}
\includegraphics[width=1.0\textwidth]{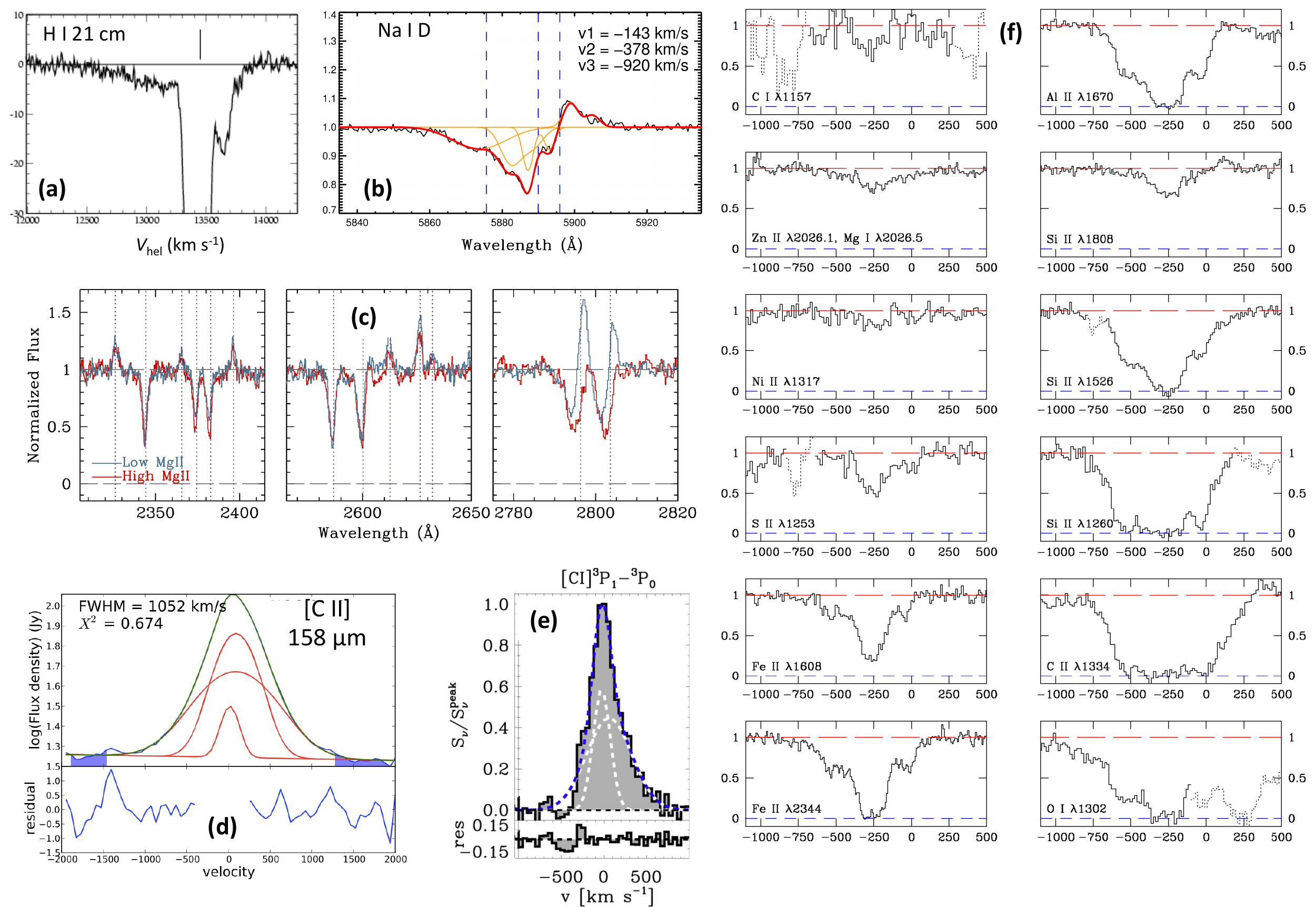}
\end{center}
\caption{Some of the diagnostic tools used to study neutral-atomic gas outflows. (a) H~I 21 cm from, (b) optical Na I D, (c) NUV Mg II and Fe II, (d) FIR [C~II] 158 $\mu$m, (e) mm-wave [C~I], and (f) FUV lines from neutral and low-ionization elements. Images reproduced with permission from (a) \citet{MORGANTI2005}, copyright by ESO; and (b) \citet{RUPKE2017}, (c) \citet{ERB2012}, (d) \citet{JANSSEN2016}, (e), \citet{CICONE2018a}, (f) \citet{PETTINI2002}, copyright by AAS.}
\label{fig:diagnostics_neutral_atomic}
\end{figure}

A broad diversity of techniques and diagnostic tools have been used to study the neutral atomic component of galactic outflows (Fig.\ \ref{fig:diagnostics_neutral_atomic}). These tools and techniques can be divided into three broad categories based on the use of the strongest transitions from: (1) hydrogen itself,
(2) the most abundant neutral chemical elements, 
or (3) the most abundant singly-ionized chemical elements with first ionization potential below 13.6 eV. 
Below we discuss in turn each of these diagnostic tools. In the following discussion, a ``resonant transition'' corresponds to the absorption of a photon from the ground state and re-emission to the same lowest level of the ground state, while a ``non-resonant transition'' takes place when the photon is re-emitted to an excited level of a ground state that has multiple levels due to fine structure splitting (non-resonant transitions are traditionally labeled with a $^*$ such as Fe II$^*$). Below, we adopt the usual nomenclature for absorption systems based on their hydrogen column density $N_{\rm H}$: damped Ly$\alpha$ absorbers  or ``DLAs'' have $N_{\rm H} > 2 \times 10^{20}$ cm$^{-2}$ and Lyman Limit Systems or LLS have $N_{\rm H} > 10^{16.2}$ cm$^{-2}$.

\subsubsection{H~I 21 cm} \label{sec:21cm}

Most of the hydrogen in space far from the sources of ionization is neutral and in the ground state, but this ground state exhibits hyperfine splitting. The ``spin-flip'' H~I 21-cm (1.420-GHz) emission line is produced when the slightly excited electron in neutral hydrogen flips its spin to enter a lower energy state, going from parallel to anti-parallel with the spin of the proton. The so-called spin temperature, $T_s$, sets the relative populations of these two energy levels. H~I 21 cm has been detected in several cool outflows, either in emission or in absorption against the radio continuum. This technique has been reviewed by \citet{MORGANTI2018} in a general extragalactic context, so here we only focus on the essentials relevant to cool outflows. The strength of H I 21 cm is generally a direct measure of the H~I column density, although corrections apply at very low densities \citep[e.g.,][]{LISZT2001a}. When H~I 21 cm is in emission,
\begin{eqnarray}
N_{\rm H~I}^* = 1.823 \times 10^{18} {\rm cm^{-2}} \int T_b(v) {\rm [K]}~dv~[{\rm km~s}^{-1}],
\label{eq:N_HI*}
\end{eqnarray}
where $T_b$(v) is the observed profile of the H~I emission line as a function of velocity, so the H~I equivalent width EW(H I) = $\int T_b(v)~dv$. This expression underestimates the actual value of $N_{\rm H I}$ if the H~I emission is not optically thin. The ratio $N_{\rm H I}/N_{\rm H I}^* = \tau_{\rm H I} / [1 - {\rm exp}(-\tau_{\rm H I}$)], where $\tau_{\rm H I}$ is given by the following equation \citep{FUKUI2015}, 
\begin{eqnarray}
\tau_{\rm H~I} = \frac{N_{\rm H~I}~[{\rm K~km~s}^{-1}]}{1.823 \times 10^{18}} \times \frac{1}{T_s[K]} \times \frac{1}{\Delta v_{\rm H~I}~[{\rm km~s}^{-1}]}.
\label{eq:tau_HI}
\end{eqnarray}

When H~I is seen in absorption against the continuum, 
\begin{eqnarray}
N_{\rm H I} = 1.823 \times 10^{18} {\rm cm^{-2}}~T_s[K] \int \tau(v)~dv~[{\rm km~s}^{-1}].
\label{eq:N_HI}
\end{eqnarray}
In this expression, the optical depth is defined as $\tau(v)$ = $-$ln[1 $+$ $\Delta T (v) / c_f T_c]$, where $\Delta T (v)$ is the observed absorption spectrum, expressed in terms of the brightness temperature relative to $T_c$, the brightness temperature in the continuum, and $c_f$ is the covering factor, i.e. the fraction of the continuum source that is covered by the H~I absorbers. 
If we make the (generally reasonable) assumption that $\tau$(v) $<<$ 1, this expression becomes
\begin{eqnarray}
N_{\rm H I} = 1.823 \times 10^{18} {\rm cm^{-2}}~\frac{T_s}{c_f T_c} \int \vert \Delta T(v) \vert~dv~[{\rm km~s}^{-1}],
\label{eq:N_HI2}
\end{eqnarray}
The H~I column density derived from the absorption feature thus depends on the spin temperature and optical depth, which itself depends on the covering factor. These dependencies make the H~I column densities derived from H~I 21 cm absorption line more uncertain than those derived from the H~I 21 cm emission line. $T_s$ is typically $\sim$100 K unless the absorbing gas is affected by the radiation field of a powerful nearby continuum source (e.g., AGN or starburst). In that case, $T_s$ will depend on the density of the absorbing gas and the location of this gas relative to the source(s) of radiation \citep[e.g.,][]{BAHCALL1969}. 

\subsubsection{Hydrogen Ly$\alpha$} \label{sec:lya}

In most astrophysical environments, the electron of neutral hydrogen is in its ground electronic state ($n = 1$) so the resonance electric-dipole transition between the ground state and the first energy level ($n = 2$), H~I Ly$\alpha$ $\lambda$1216, is the strongest line in the spectrum. The detection of this line in absorption is a powerful tracer of the column density of neutral hydrogen along the line of sight from the background continuum source to the observer. In the linear regime, we simply have $N$(H I) = $W$(Ly$\alpha$)$_\lambda$ /($\lambda f$), where $f$, $\lambda$, and $W$(Ly$\alpha$)$_\lambda$ are the oscillator strength, wavelength (in vacuum, in the rest frame), and equivalent width of Ly$\alpha$. Unfortunately, this simple equation only applies when the optical depth is $<<$ 1, corresponding to $N(HI) \lesssim 10^{14}$ cm$^{-2}$. 
Above these values, the two quantities follow a curve-of-growth relation, with a flat portion where EW $\propto \sqrt{{\rm ln}(N)}$ for $N$(H I) $\sim$ 10$^{14}$--$10^{18}$ cm$^{-2}$ and a square-root portion EW $\propto \sqrt{N}$ above 10$^{18}$ cm$^{-2}$ \citep[e.g.,][]{HUMMELS2017}. In most circumstances, Ly$\alpha$ is not in the linear regime and detailed fitting of the damping wings is required to derive accurate $N$(H I). An alternate way of deriving $N(H~I)$ is to avoid Ly$\alpha$ altogether and use the fainter lines of the Lyman series from $\sim$926 down to $\sim$918 \AA, ignoring the higher wavelength transitions due to blending \citep[e.g.,][]{ZECH2008}. A standard curve-of-growth analysis can then be used to derive $N$(H~I).

We have so far ignored sources (e.g., H~II regions, AGN) or sinks (e.g., dust, H$_2$) of Ly$\alpha$ photons, but these need to be taken into account when evaluating the hydrogen column densities. Ly$\alpha$ photons may be produced through a number of processes: (1) recombination radiation following hydrogen photoionization by a nearby AGN or star-forming galaxy, (2) collisional excitation, (3) collisional ionization followed by recombination radiation, and (4) Ly$\alpha$ scattering (sometimes called ``fluorescence'' in the literature) of the photons produced in the AGN or within the ISM of a star-forming galaxy.
In the first case of recombination radiation, the strength of Ly$\alpha$ is a measure of the amount of \emph{ionized} material rather than neutral material:
 \begin{eqnarray}
M_{\rm ionized} & = & C_e \frac{m_p L(Ly\alpha)}{n_e \alpha_{Ly\alpha}^{\rm eff} h\nu_{Ly\alpha}},
\label{eq:M_ionized}
\end{eqnarray}
where $\alpha_{Ly\alpha} \sim 2 \times 10^{-13} T_4^{-1.26}$ cm$^3$ s$^{-1}$ is the effective Case B Ly$\alpha$ recombination coefficient \citep{CANTALUPO2008}, and $C_e \equiv \langle n_e^2 \rangle / \langle n_e \rangle^2$ is the electron density clumping factor, which can be assumed to be of order unity on a cloud-by-cloud basis (i.e. each cloud has uniform density). Note, however, that for the reasons discussed below, H$\alpha$ or strong collisionally excited metal lines, such as [O~III] $\lambda$5007, are more reliable indicators of the ionized mass of outflows than Ly$\alpha$. So, for completeness, we list the relations between the luminosities of these lines and the ionized masses\footnote{Both eq.\ (\ref{eq:M_ionized2}) and (\ref{eq:M_ionized3}) assume $T \sim 10^4$ K, while eq.\ (\ref{eq:M_ionized4}) further assumes that $n_e \lesssim 7 \times 10^5$ cm,$^{-3}$ the critical density associated with the [O~III] 5007 transition above which collisional de-excitation becomes significant.}:
 \begin{eqnarray}
\label{eq:M_ionized2}
M_{\rm ionized} & = & 3.3 \times 10^8~\frac{C_e L_{44}(H\alpha)}{n_{e,3}} M_\odot,\\
\label{eq:M_ionized3}
M_{\rm ionized} & = & 5.3 \times 10^7~\frac{C_e L_{44}([O~III~5007])}{n_{e,3} 10^{[O/H]}} M_\odot,
\label{eq:M_ionized4}
\end{eqnarray}
where $L_{44}$(H$\alpha$) and $L_{44}$([O III] 5007) are respectively the luminosities of H$\alpha$ and [O III] $\lambda$5007, normalized to 10$^{44}$ erg s$^{-1}$, $n_{e,3}$ is the average electron density, normalized to 10$^3$ cm$^{-3}$, $C_e$ is the electron density clumping factor defined earlier, and 10$^{\rm [O/H]}$ is the oxygen-to-hydrogen abundance ratio relative to the solar value.

Ly$\alpha$ emission via collisional excitation may be more important than recombination emission if $T$ $\simeq$ 2--5 $\times$ 10$^4$ K and the medium is partially ionized \citep[i.e. both neutral hydrogen and free electrons are present;][]{CANTALUPO2008}. For temperatures above $\sim$10$^5$ K, collisional ionization dominates over collisional excitation. An external source of heating (e.g., AGN, starburst, shocks) is needed to balance radiative losses due to collisional excited Ly$\alpha$, otherwise the electron temperature quickly drops, effectively stopping the production of Ly$\alpha$ emission. If collision excitation dominates, one can estimate the neutral gas mass:
 \begin{eqnarray}
M_{\rm neutral}  =  C_{\rm e,HI}~\frac{m_p L({\rm Ly}\alpha)}{n_e q_{{\rm Ly}\alpha}^{\rm eff} h\nu_{{\rm Ly}\alpha}},
\label{eq:M_neutral}
\end{eqnarray}
where $q_{Ly\alpha}^{\rm eff}$ is the effective collisional excitation coefficient for Ly$\alpha$ emission and is a steep function of $T$ \citep[e.g.,][]{GIOVANARDI1987}, and $C_{\rm e,HI} \equiv \langle n_e n_{\rm H I} \rangle / \langle n_e \rangle \langle n_{\rm H~I}\rangle$ is the clumping factor of the partially ionized gas, which can be assumed to be of order unity on a cloud-by-cloud basis (i.e. each cloud has uniform H~I and electron densities). 

The scattering scenario is only important when the gas is not (photo-)ionized i.e. the density of neutral hydrogen is high enough for the Ly$\alpha$ depth at line center,
 \begin{eqnarray}
\tau_0 & = & 3.3 \times 10^{-14} T_4^{-1/2} N(HI), 
\label{eq:tau_0}
\end{eqnarray}
to be significant. The source of radiation must therefore produce copious amounts of Ly$\alpha$ and continuum photons slightly blueward of Ly$\alpha$, but not below 912 \AA\ to avoid photoionization. During scattering, the Ly$\alpha$ photons will perform a random walk both in space and frequency, and will only be able to escape from the outflow when scattered in the wings of the line profile.

Dust may absorb Ly$\alpha$ photons as they perform their random walk.  The optical depth due to dust $\tau_d = N_H \sigma_d \simeq 0.72~\times$~10$^{-21}$ $N_H$ at the wavelength of Ly$\alpha$ for a LMC-type dust \citep{LAURSEN2009}. The energy absorbed will increase the temperature of the dust grain, which will cause the grain to re-radiate in the infrared. The impact of dust increases with increasing $N$(H~I)
because Ly$\alpha$ scattering becomes more important. This effectively lengthens by a factor of $\tau_0$ the path of the Ly$\alpha$ photons through the dusty medium and thus increases the chance that Ly$\alpha$ is absorbed by the dust. The dust grain can also scatter the Ly$\alpha$ photon with a probability given by its albedo, $A_s = 0.32$ at the Ly$\alpha$ wavelength for dust made mainly of graphite and silicates \citep{LI2001}. Finally, in cases where the source of Ly$\alpha$ is surrounded by molecular gas, Ly$\alpha$ may pump lines of the H$_2$ Lyman bands that are $<$100 km s$^{-1}$ redward of Ly$\alpha$, and thus may be destroyed \citep[e.g.,][]{NEUFELD1990}.

Overall, the calculation of the emergent Ly$\alpha$ emission is complex, since it depends on the details of the gas structure, ionization, and kinematics as well as the relative distribution of the sources and sinks of Ly$\alpha$ photons. A quantitative interpretation of the observed Ly$\alpha$ profiles generally requires three-dimensional radiative transfer calculations where all of the relevant spatial and kinematic information is provided as input. Remarkably, simple expanding spherical shell models \citep[e.g.,][]{VERHAMME2008,DIJKSTRA2017,GRONKE2017} provide a relatively good match to the observed integrated Ly$\alpha$ profiles of outflowing gas in high-z galaxies, 
though not always \citep{DIJKSTRA2017}. More recent simulations now consider well resolved turbulent cloud structures \citep{KIMM2019} and the impact of cosmic rays on the Ly$\alpha$ spectrum \citep[][]{GRONKE2018b}.

\subsubsection{Na I~D $\lambda\lambda$5890, 5896} \label{sec:NaID}

The Fraunhofer D$_1$ and D$_2$ lines of neutral sodium at 5896 and 5890 \AA, first identified in the solar spectrum, are by far the strongest lines from this element in the visible range. Since the first ionization potential of sodium is only 5.14 eV ($\sim$2400 \AA), Na~I~D is an excellent tracer of neutral hydrogen, but dust is generally required to provide shielding for Na I against non-ionizing UV sources. Na~I~D is often detected in absorption against the continuum of galaxies, and indeed a correlation is seen between the strength of the Na I D absorption feature and dust extinction traced by the color excess, $E$(B--V) \citep{VEILLEUX2005}. Na~I~D has also been observed as redshifted emission in a few objects \citep{PHILLIPS1993,RUPKE2015,ROBERTS-BORSANI2019}, in which case it can be used in combination with the blueshifted absorption feature to further constrain the geometry of the cool outflow \citep{PROCHASKA2011}. 

The general expression for the intensity of a single absorption line, assuming a continuum level of unity, is
\begin{eqnarray}
I(\lambda) = 1 - C_f(\lambda) + C_f(\lambda)e^{-\tau(\lambda)}, 
\label{eq:I_lambda}
\end{eqnarray}
where $C_f$ is the line-of-sight covering fraction or the fraction of the background source producing the continuum that is covered by the absorbing gas. The optical depth $\tau(\lambda)$ can be expressed as a Gaussian  under the curve-of-growth assumption (i.e. a Maxwellian velocity distribution)
\begin{eqnarray}
\tau(\lambda) = \tau_0e^{-(\lambda - \lambda_0)^2/(\lambda_0b/c)^2}, 
\label{eq:tau_lambda}
\end{eqnarray}
where $\tau_0$, $\lambda_0$, and $b$ are the central optical depth, central wavelength, and Doppler width ($b = \sqrt{2} \sigma$ = FWHM/$[2 \sqrt{{\rm ln 2}}]$), respectively.  Note that Gaussians in optical depth translate into observable non-Gaussian intensity profiles for optical depths greater than unity. $C_f$ is generally unity in the case for quasar, stellar or GRB sightlines through the ISM, CGM, or IGM, since the background light source in each of these cases can be approximated as a point source. However, when the background light source subtends a significant angle on the sky relative to the absorbing gas (as in the case that concerns us where the background continuum source is the host galaxy itself), $C_f$ may depend on wavelength (velocity) and position (if the absorbing gas is spatially resolved). In these cases, a degeneracy arises when solving for optical depth and covering fraction for a single line. This degeneracy is typically broken by observing two or more transitions in the same ion, with the same lower level, whose relative optical depths can be computed exactly from atomic physics. This is the origin of the doublet ratio method where it can be shown that the ratio of the equivalent widths of the two lines in the doublet is a monotonic function of the central optical depth of either line \citep[see equation 2-39 and Table 2.1 in][]{SPITZER1968} as long as $C_f$ is independent of velocity and the velocity distribution is Maxwellian. With knowledge of the optical depth, the covering fraction then follows from the residual intensity of the doublet lines.

Unfortunately, this method cannot be applied to Na I~D in cool outflows where the lines in the doublet are usually blended. \citet{RUPKE2005b} discussed the more general case where the lines in the doublet are blended together and are produced by two distinct velocity components that either completely, partially, or do not, overlap with each other. For partially overlapping velocity components, Eq.~(\ref{eq:I_lambda}) becomes
\begin{eqnarray}
I(\lambda) = [1 - C_{f,1}(\lambda) + C_{f,1}(\lambda)e^{-\tau_1(\lambda)}] \times [1 - C_{f,2}(\lambda) + C_{f,2}(\lambda)e^{-\tau_2(\lambda)}] = I_1(\lambda) I_2(\lambda)
\label{eq:I_lambda2}
\end{eqnarray}
and the total covering fraction of the continuum source at a given wavelength is
\begin{eqnarray}
C_{f,{\rm total}} = C_{f,1} + C_{f,2} - C_{f,1} C_{f,2}. 
\label{eq:C_ftotal}
\end{eqnarray}

Once the optical depth profile $\tau_\lambda(\lambda)$ is known, one can derive the column density producing the absorption line. The general expression for the column density of an ionic species is \citep{SPITZER1978}
\begin{eqnarray}
\label{eq:Nspitzer1}
N & = &  \frac{m_e c}{\pi e^2 f \lambda_0} \int \tau(\nu)~d\nu \\
\label{eq:Nspitzer2}
  & = & \frac{3.768 \times 10^{14}~{\rm cm^{-2}}}{f \lambda_0} \int \tau(\nu)~d\nu,
\end{eqnarray}
where $f$ and $\lambda_0$ are the oscillator strength and wavelength (in vacuum, in the rest frame) of the transition. For neutral sodium, this expression becomes 
\begin{eqnarray}
N({\rm Na~I}) = 3.56 \times 10^{11}~(\tau_{1,c}~b)~{\rm cm}^{-2},
\label{eq:N_NaI}
\end{eqnarray}
where $\tau_{1,c}$ is the central optical depth of the Na I D$_1$ $\lambda$5896 line and $b$ is the Doppler parameter in km s$^{-1}$. 

Stellar absorption from late-type stars in the underlying galaxy spectrum may contribute to the observed profile of Na I D. The expected stellar contribution to Na I D has been estimated in the past by scaling the equivalent width of the Mg I b triplet at 5167, 5173, and 5185 \AA. This is possible because Na and Mg are created in stars by similar mechanisms and they have similar first ionization potentials (5.14 eV for Na, 7.65 eV for Mg). There is thus a correlation between the stellar equivalent widths of Mg I b and Na I D in galaxy spectra: $W^*$(Na I D) = $\alpha_{\rm Na}$ $W^*$(Mg I b) + $\beta_{\rm Na}$  where $(\alpha_{\rm Na},\beta_{\rm Na})$ = (0.5, 0) in U/LIRGs \citep{RUPKE2005c} and (0.685, 0.8) in emission-line galaxies \citep{ALATALO2016}. Another weaker stellar feature may also be present at a rest wavelength of 5857.5 \AA, 33 \AA\ (1700 km s$^{-1}$) blueward of the D$_2$ line. A more important contaminant near this region is the He I $\lambda$5876 recombination emission line from ionized gas; this line is generally fit with one or two Gaussians in intensity.

Once $N$(Na I) has been derived, the column density of hydrogen, $N$(H), can be estimated from $N$(Na I) given a metallicity and ionization fraction \citep{RUPKE2005b}:
\begin{eqnarray}
N({\rm H}) = N({\rm Na~I})~X_{\rm Na^0}^{-1} 10^{-(A_{\rm Na}+B_{\rm Na})},
\label{eq:N_H_N_NaI}
\end{eqnarray}
where $X_{\rm Na^0}$ $\equiv$ $N$(Na~I) / $N$(Na) is the fraction of Na that is neutral Na~I, $A_{\rm Na}$ $\equiv$ log[$N$(Na)/$N$(H)]$_{\rm gas}$ is the gas-phase Na abundance relative to H, and $B_{\rm Na}$ $\equiv$ log[$N$(Na)/$N$(H)]$_{\rm gas}$--log[$N$(Na)/$N$(H)]$_{\rm total}$ is the level of depletion of Na onto dust. Typically, Milky-Way like metal abundance of sodium \citep[$A_{\rm Na}$ = $-$5.69;][]{SAVAGE1996b} and depletion factor of sodium onto dust \citep[$B_{\rm Na}$ = $-$0.95;][]{SAVAGE1996b} are adopted in this expression. The Na neutral fraction $X_{\rm Na^0}$ is uncertain. A value $X_{\rm Na^0} = 0.1$ has been measured toward Galactic stars \citep{STOKES1978} and also a cold HI cloud in NGC 3067 \citep{STOCKE1991}, but this number could be substantially smaller in the more extreme environment of an AGN or star-forming galaxy, therefore resulting in a larger $N$(H). 

Direct measurements of the H~I columns from 21-cm radio data may provide an independent check on the Na-based $N$(H).  For instance, \citet{TENG2013} detected two H~I absorption components at the same velocities as the Na~I outflow components in a ULIRG \citep{RUPKE2015} and derived $N$(H) = 10$^{22}$ cm$^{-2}$, the same column inferred from Na I D in the dustiest regions of this galaxy.  Other more indirect measures of $N$(H) in cool outflows, based on reddening measurements of the underlying host galaxy light, have brought additional support for these correction factors \citep{RUPKE2013b}, although they likely remain uncertain to within factors of $\pm$ 2-3. 

Note finally that Na~I~D has also been seen in emission in a few cool outflows (see Sec. \ref{sec:lowz_neutral_atomic}). Since the Na~I~D emission lines are resonant transitions like Ly$\alpha$, they are subject to scattering and ``fluorescence'' by the cool gas and thus can also be used to trace this component of the outflows. Simple models of metal-line absorption and emission from cool gas outflows that track the scattered and fluorescent photons in a spherical homogenously expanding wind have been discussed in \citet{PROCHASKA2011}, and these are directly applicable to the observed Na~I~D P-Cygni like profiles \citep[e.g.,][]{RUPKE2015}.

\subsubsection{UV Lines from Neutral Elements} \label{sec:neutral_elements}

UV transitions from abundant neutral chemical elements have been used to probe cool outflows.  The strongest features between 1000 \AA\ and 3000 \AA, most commonly used to study the cool ISM/CGM in the Milky Way and external galaxies, include O~I
1302, N~I 1134, 1200, C~I 1277, 1280, 1329, 1657, and Mg~I 2853.
Outside ISM environments, these features are generally weak enough that the column densities of these elements may be derived using a simple curve-of-growth analysis of the fitted absorption line profiles against the continuum of the background UV-bright source. These column densities can then be compared with H~I column density measurements, derived for instance from the H~I 21-cm line emission, to determine the metallicity of the absorbing material, modulo an ionization correction to take into account the ionized portion of the gas. The O~I/H~I ratio is particularly well suited for this measurement because it is only moderately depleted onto dust grains \citep{JENKINS2009} and, since oxygen and hydrogen have virtually identical first ionization potentials, strong charge-exchange reactions tie the two species together \citep{FIELD1971,DOMGORGEN1994}.
This method has for instance been used in HVCs associated with the nuclear outflow in the Milky Way \cite[e.g.,][see Sect.~\ref{sec:MW} below]{KEENEY2006,ZECH2008,BORDOLOI2017a}. Since the HI 21-cm observations have a finite beam and the UV observations are effectively infinitesimal, beam-smearing corrections must be applied to take into account potential small-scale structure in the H~I beam \cite[e.g.,][]{WALTER2017,JENKINS2009,BORDOLOI2017a}.
At sufficiently high spectral resolution ($R \gtrsim 30,000$), it is possible to distinguish between the three fine-structure levels of the ground electronic state of C~I, and use the C~I / C~I$^*$ / C~I$^{**}$ ratios to derive the thermal pressure in the absorbing gas \citep{JENKINS2001,SAVAGE2017}.

\subsubsection{UV lines from Singly Ionized Elements} \label{sec:lowion_elements}

Many of the most abundant elements have first ionization potentials that fall below 13.6 eV, and therefore are also good tracers of neutral hydrogen. Key elements include C$^+$ (11.26 eV), Na$^+$ (5.14 eV), Mg$^+$ (7.65 eV), Al$^+$ (5.99 eV), Si$^+$ (8.15 eV), S$^+$ (10.36 eV), Ca$^+$ (6.11 eV) (and also Ca$^{++}$, 11.87 eV), and Fe$^+$ (7.87 eV). The strongest absorption lines between 1000 \AA\ and 4000 \AA, most commonly used in the study of the cool ISM/CGM in the Milky Way and external galaxies, include Ca~II 3935 (K) and 3970 (H), Mg~II 2796, 2804, Fe II
2344 (UV3), 2374 + 2383 (UV2), 2587 + 2600 (UV1),   Si II 1190, 1193, 1260, 1304, 1527,  
C II 1334,
and Al II 1671. Weaker absorption lines from slightly less abundant but volatile elements that are less subject to dust depletion [e.g., S~II 1251, 1254, 1260 from S$^+$ (10.36 eV), Zn~II 2026, 2063 from Zn$^+$ (9.39 EV)] can be used in conjunction with the lines from refractory elements \citep[e.g,][]{SAVAGE2017} to estimate the dust depletion factor in the neutral gas phase of cool outflows.

Several resonant transitions (e.g., Mg II 2796, 2804) and non-resonant optically-thin transitions (e.g., Fe II* 2626, 2612, 2396, 2365, C II* 1335, Si II* 1265, 1309, 1533) are also commonly ($\sim$ 15-30\%) observed in emission from cool outflows, particularly in $z \sim$ 1 galaxies with blue UV slopes and small stellar masses \citep{ERB2012,KORNEI2013}. This line emission can be used in combination with the absorption line to further constrain the properties of the cool outflows \citep{PROCHASKA2011}. On the other hand, if the spectral resolution is poor, this line emission ``infills'' the absorption lines of the resonant transitions (e.g., Mg II 2796, 2804), particularly at velocities near the systemic velocity of the galaxy, and thus complicates the analysis of these lines \citep{ZHU2015}. Contamination from stellar absorption features at systemic velocity also needs to be taken into account for the analysis of Mg~II and Fe~II. Fits with two separate velocity components for each line are generally used to mitigate these effects \citep{RUBIN2010,COIL2011}. Typically, Mg~II (IP$_1$ = 7.65 eV, IP$_2$ = 15.0 eV) is assumed to be the dominant ionization state in photoionized $T \sim 10^4$ K winds \citep{MURRAY2007}, so an ionization correction factor $X_{{\rm Mg}^+} \simeq 1$ in Eq.~(\ref{eq:N_H_N_NaI}) is used to derive the total Mg abundance in the gas phase, and a dust depletion corrections $B_{\rm Mg}$ = $-$0.5 dex is used in galaxies with near-solar abundances \citep{RUBIN2014}. For Fe II, $X_{{\rm Fe}^+} = [0.5 - 1.0]$ and $B_{Fe} = [-0.69, -1.0]$ have been used in the literature \citep{MARTIN2012,RUBIN2014}.

Absorption lines originating from the (collisionally) excited metastable state of these ions (e.g., Fe II$^*$, Si II$^*$) have been detected in several objects, particularly in LOw-ionization Broad Absorption Line (LoBAL) QSOs, and have been used to constrain the electron density in the absorbing material. This information can be used to derive the distance $r$ of the absorber from the central ionizing source via the dimensionless ionization parameter:
\begin{equation}
U_H \equiv \frac{n(h\nu > 13 {\rm eV})}{n_H} = \frac{Q_H}{4 \pi r^2 n_H c}, 
\label{eq:U_H}
\end{equation}
where $n_e \simeq 1.2 n_H$ for highly ionized plasma, $c$ is the speed of light, $n(h\nu > 13 eV)$ and $Q_H$ are the density and rate of ionizing ($>$13.6 eV) photons hitting the absorber, respectively. $Q_H$ is derived by integrating the UV-soft X-ray portion of the AGN spectral energy distribution that best fits the broad-band data on the object. Dust between the source and the absorber needs to be taken into account in the calculations of $Q_H$. $U_H$ is derived independently by comparing the measured column densities with the predictions of photoionization models under the assumption of ionization equilibrium \citep[e.g.][]{MOE2009, DUNN2010}.
This method has more recently been extended to the broad absorption lines from more highly ionized species (e.g., Si IV$^*$) in the more common High-ionization Broad Absorption line (HiBAL) QSOs \citep[e.g.][]{ARAV2013,ARAV2018,MILLER2018,XU2018},
and for all BAL QSOs in general \citep{HAMANN2019}. 
The timescale of variability of these absorption lines has also been used to constrain the density of the absorbers, hence their distance, assuming it is due to changes in the ionizing radiation incident on the absorbing gas rather than tangential movement of the absorbing gas across the line of sight \citep{FILIZAK2012,WANG2015}. In this case, the variability time scale sets a constraint on the ionization or recombination timescale, which depends solely on the incident ionizing continuum and gas density \citep[e.g.][]{HE2019}.

\subsubsection{Far-Infrared [C~I] and [C~II] Emission Lines} \label{sec:CI_CII}

While FIR [C~I] and [C~II] lines are formally tracers of the neutral-atomic component of cool outflows, they are also excellent tracers of the molecular component so they are instead discussed in Section \ref{sec:m_and_mdot} below.

\subsection{Diagnostic Tools of the Molecular Gas Phase} \label{sec:diagnostics_molecular}

Millimeter and submillimeter-wave techniques can provide excellent direct estimates of the luminosity of CO and other molecules, as well as their kinematics, and morphology (the latter through interferometric imaging). This information can be converted into estimates for column density, volume density, and mass, momentum, energy, and their fluxes for the molecular phase. The fact that the millimeter/submillimeter-wave window is very rich in molecular transitions means observations provide also a unique view of the chemistry and excitation in the outflowing gas. 

\begin{figure}[htb]
\begin{center}
\includegraphics[width=1.0\textwidth]{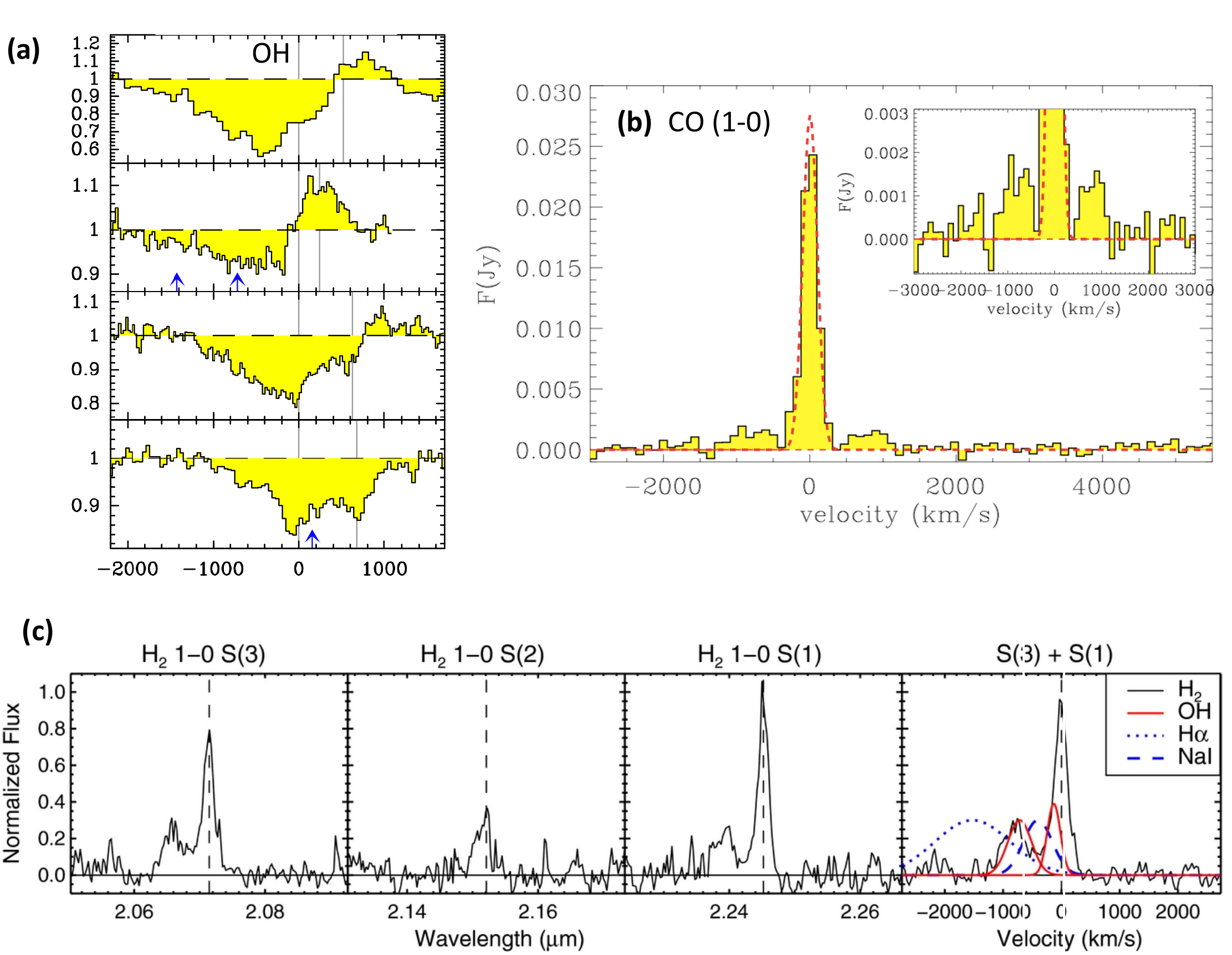}
\end{center}
\caption{Some of the diagnostic tools used to study molecular gas outflows. (a) OH absorption and emission line profile, (b) CO emission line profile, (c) near-infrared ro-vibrational H$_2$ emission lines. Images reproduced with permission from (a) \citet{GONZALEZ-ALFONSO2017b}, (c) \citet{RUPKE2013a}, copyright by AAS; and (b) \citet{CICONE2014}, copyright by ESO.}
\label{fig:diagnostics_molecular.pdf}
\end{figure}

Measurements obtained at single-dish radio telescopes are usually expressed in Rayleigh-Jeans brightness temperature units of Kelvins, corresponding to $T_B=I_\nu\,c^2/2k\nu^2$, which are meaningful for the angular resolution (beam size) of the observation. In interferometers, the measurements are usually expressed in flux density units. The conversion between temperature surface brightness units and flux density is

\begin{equation}
    T_B=1360~{\rm K}~\frac{\lambda^2}{\theta^2} S_\nu
\label{eq:TbSnu}
\end{equation}

\noindent where $T_B$ is the Rayleigh-Jeans brightness temperature in K, $\lambda$ is the wavelength of the observation in cm, $\theta$ is the full-width at half-maximum of the Gaussian beam of the observation in arcseconds, and $S_\nu$ is the flux density in Jy\,beam$^{-1}$. The Rayleigh Jeans brightness temperature can be related to the excitation temperature of the transition through

\begin{equation}
    T_B=\frac{h\nu}{k}\left(1-e^{-\tau_J}\right)\left(\frac{1}{e^{\frac{h\nu}{k T_{ex,J}}}-1}-\frac{1}{e^{\frac{h\nu}{k T_{cmb}}}-1}\right),
\end{equation}

\noindent where the Cosmic Microwave Background temperature $T_{cmb}$ is the background against which the emission is measured. The last term is usually a small correction for $z\sim0$, and it is frequently neglected.

Luminosities are frequently expressed in units of K\,km\,s$^{-1}$\,pc$^2$, and when using these units they may be noted as $L^\prime$ (strictly speaking this is not the usual meaning of luminosity, since K are units of brightness temperature rather than energy flux per unit time, and km\,s$^{-1}$ are units of velocity rather than units of frequency). In general 

\begin{equation}
    L^\prime=3.25\times10^7\,S_\nu\,\Delta v\,\nu_{obs}^{-2}\,(1+z)^{-3}\,D_L^2
    \label{eq:Lprime}
\end{equation}

\noindent where $S_\nu$ is the transition flux density in Jy, $\Delta v$ is the line width in km\,s$^{-1}$, $\nu_{obs}$ is the observed frequency of the transition in GHz, $z$ is the source redshift, and $D_L$ is its luminosity distance in Mpc \citep{SOLOMON2005}.

\subsubsection{Probes of the Volume Density} \label{sec:volume_density}
Determination of volume densities using millimeter-wave data mostly rely on access to rotational transitions sampling a range of critical densities. In the following discussion we restrict ourselves to transitions between rotational energy levels characterized by a quantum number $J$, although most of it can easily be generalized to other types of transitions.  The critical density is the density of the collisional partner at which the chance of spontaneous de-excitation and that of collisional de-excitation are equal, so that $n_{cr}=A_J/\gamma_J$ where $A_J$ is the Einstein $A$ coefficient to spontaneously transition from $J\rightarrow J-1$, and $\gamma_J$ is the collisional coefficient for the same transition (note sometimes this is defined using the sum of the coefficients for all possible downward collisional transitions in the denominator). At densities much higher than $n_{cr}$ the populations of the levels are in collisional equilibrium, and follow the distribution corresponding to the Boltzmann equation with an excitation temperature $T_{ex}$ equal to the kinetic temperature of the gas, $T_K$. At densities much lower than $n_{cr}$, on the other hand, molecules are preferentially de-excited by spontaneous emission, and their excitation temperature is lower than $T_K$. 

Over a very wide range of conditions, the brightest transitions from molecular gas at these wavelengths are the rotational transitions of the main isotopologue of carbon monoxide, $^{12}$C$^{16}$O. These transitions of carbon monoxide are excited mostly through collisions with H$_2$ molecules. Radiative trapping will play an important multiplicative role in the excitation if the transition is optically thick, which is easy to achieve since CO is an abundant molecule that is frequently the largest reservoir of carbon in the gas phase. The critical density for the ground rotational transition of CO (at $\nu=115.27$~GHz) to collisions with molecular gas is $n_{cr}({\rm H}_2)\approx2000\,T_{100}^{-0.019}$~cm$^{-3}$ \citep{YANG2010}. The excitation is such that bright emission usually requires densities that are similar to or larger than $n_{cr}$ \citep[e.g.,][]{LEROY2017}. In species like CO that are abundant and optically thick, radiative trapping acts to reduce the effective critical density by a factor of approximately $(1-e^{-\tau_J})/\tau_J$, which is a significant correction when the optical depth $\tau_J$ of the $J\rightarrow J-1$ transition is large.

Critical densities scale approximately as $n_{cr}\propto\mu^2$, where $\mu$ is the electronic dipole moment of the molecular species, so that molecules with higher $\mu$ require higher volume densities to be excited collisionally. For a given molecular species $n_{cr}\propto J^3$, so that higher transitions have higher critical densities. Among the abundant molecules with bright millimeter-wave transitions, HCN, HCO$^+$, and CS have high dipole moments and are usually employed as tracers of high density gas. For example, the ground rotational transition of HCN at $\nu=88.63$~GHz has $n_{cr}({\rm H}_2)\approx2.67\times10^6\,T_{100}^{+0.349}$~cm$^{-3}$ \citep{DUMOUCHEL2010}. Therefore, the HCN/CO ratio is, to first order, an indicator of density\footnote{Just as for CO, high opacities can impact the critical density for e.g. HCN which may be very abundant in warm regions. In addition, the critical density can be strongly reduced (by factors 4-6) in regions of high temperature. This requires a multi-level treatment of the critical density.}. Comparison across molecular species, however, can be affected by their relative abundances determined by the chemistry. Moreover, some molecules can be excited by more than one collisional partner. For example, HCN has a cross-section to collisions with electrons that is $10^5$ times larger than for collisions with H$_2$ molecules, which may affect its use as a density tracer in regions where electron abundance is high \citep{GOLDSMITH2017}. 
Note also that molecules such as HCN may be affected by infrared radiative excitation and pumping, which can impact the interpretation of the molecular line emission. HCN (and also other molecules) has degenerate bending modes in the infrared. The molecule absorbs infrared photons to the bending mode (its first vibrational state) and then it can decay back to the ground state via its P- or R-branch. In this way, a vibrational excitation may produce a change in the rotational state in the ground level and can be treated (effectively) as a collisional excitation in the statistical equations \citep[e.g.][]{CARROLL1981,AALTO2007}

The presence of bright emission from one of the high dipole moment species usually is taken to suggest that the volume density is similar to or higher than the critical density of the transition. A full treatment of the gas density requires solving the detailed balance equations to establish the populations of each rotational level. As we discuss below, this can provide simultaneous constraints on volume density, column density, and temperature.  

\subsubsection{Probes of the Column Density and Optical Depth} \label{sec:column_density}
The optical depth of a rotational transition relates to the column density in its upper level, $N_J$, and velocity width, $\Delta v$, through

\begin{equation}
\tau_J = \frac{8\pi^3}{3h}\mu^2\frac{J}{2J+1}\left({e^{h\nu_J/kT_{ex,J}}-1}\right)\frac{N_J}{\Delta\,v},
\label{eq:tauco}
\end{equation}

\noindent where $\mu$ is the electric dipole moment, and $h$ and $k$ are Planck's and Boltzmann's constants respectively. 
In Local Thermodynamic Equilibrium (LTE, implying $T_{ex,J}=T_{ex}\approx T_K$), the population of a level, $N_J$, and the total column density, $N$, are related through the partition function $Q(T_{ex})$ and the corresponding Boltzmann factor,

\begin{equation}
    N=\frac{Q(T_{ex})}{2J+1}\,N_J\,e^{\frac{h\nu}{kT_{ex}}},
\end{equation}

\noindent where $Q(T_{ex})=2kT_{ex}/h\nu$ is a good approximation for $T_{ex}\gg h\nu/k$. There are a number of different methodologies for measuring optical depths. For a much more extensive discussion on using molecular measurements to estimate column density see \citet{MANGUM2015}. 
 
A good way to estimate the optical depth for an emission line, and consequently have a good handle on the column density of the gas, is to obtain measurements in an optically thin isotopologue of known abundance. If the ratio of column densities of the optically thick to the optically thin isotopologue is $R$, the intensity ratio $r$ of the same transition observed in an optically thick ($R\tau\gg 1$) to an optically thin ($\tau\ll 1$) isotopologue will be

\begin{equation}
    r\approx\frac{(1-e^{-R\tau})T_{thick}}{(1-e^{-\tau})T_{thin}}\approx\frac{1-e^{-R\tau}}{1-e^{-\tau}}\sim\frac{1-e^{-R\tau}}{\tau}.
    \label{eq:isotopic}
\end{equation}

\noindent For example, the observed ratio of $^{12}$CO/$^{13}$CO intensities is frequently observed to be $r\sim10$, while the abundance ratio is closer to $R\approx50$ at the Solar circle \citep{WILSON1994}. Selective photodissociation effects, chemical fractionation, and changes in the $^{13}$C/$^{12}$C ratio all possibly impact the calculation, as well as the implicit assumption that both lines emerge from the same parcel of gas.
This suggests that the typical optical depth of the $^{12}$CO emission is $R\tau\sim5$. Note that this assumes the excitation temperature is similar for both isotopologues, to cancel out its ratio. This is not necessarily correct, particularly in the presence of temperature gradients: the relevant excitation temperature for the optically thick isotopologue is that at the $R\tau\sim1$ surface, while for an optically thin isotopologue what matters is the temperature throughout the medium. For an internally heated cloud, for example, where the temperature increases toward the center, the optically thick isotopologue may reach optical depth of unity in a colder region far from the center of the cloud. As a consequence the ratio of excitation temperatures will be $T_{thick}/T_{thin}<1$, and neglecting this will result in  too large an inferred optical depth.

Sometimes it is not possible to detect an optically thin isotopologue in a reasonable amount of time, because the emission is likely to be very weak. Using Eq.~(\ref{eq:tauco}) it is possible to show that in LTE the ratio of Rayleigh-Jeans brightness temperature of two consecutive rotational transitions $r_J=T_J/T_{J-1}$ that fill the same area of the beam will be

\begin{equation}
    r_J=\left(\frac{J}{J-1}\right)^2\,e^{-\frac{h\nu}{kT_{ex}}}
\label{eq:Rj}    
\end{equation}

\noindent for optically thin emission, which for gas with $T_K\approx T_{ex}> h\nu_J/k$ is $r_J\approx J^2/(J-1)^2$. For optically thick emission, under the same conditions $r_J\approx 1$. This opens the possibility of using measurements of bright line ratios to estimate emission optical depth.

\subsubsection{Probes of the Temperature} \label{sec:temperature}
Under the frequent assumption of optically thick emission, the surface brightness of the CO emission becomes a probe of the excitation temperature of the transition, $T_{ex,J}$, which at densities higher than the effective critical density approaches the kinetic temperature of the gas, $T_K$. Obtaining a good estimate requires resolving the CO emission, which is not always possible, otherwise the surface brightness measurement represents the true brightness times an emission dilution factor (the ratio of solid angle $\Omega$ occupied by the emission to that of the telescope beam) that can be very small and uncertain. Thus the observed brightness temperature always provides a lower limit to the gas temperature (unless very unusual excitation and/or radiative transfer conditions occur).

Because it is the combined effects of volume density, column density, and temperature that affect the excitation,  the most comprehensive way to constrain these parameters requires modeling the spectral line energy distribution (SLED) across several rotational transitions in perhaps more than one molecular species. Because rotational transitions are equally spaced in frequency, the energy of the upper level $J$ of a rotational transition is $E_J=h\nu_{1}\,J(J+1)/2$, and the SLED will be particularly sensitive to the temperature of the gas as long as the highest transition probed has $E_J/k$ larger than the gas temperature \citep[see for example,][]{WEISS2007,ISRAEL2015}. The modeling is usually accomplished by solving the detailed balance equations assuming a large velocity gradient. An example of a widely available code to do this is {\tt RADEX} \citep{VANDERTAK2007}.

In general accurate temperature measurements may be derived from emission from symmetric top molecules, such as NH$_3$ \citep[e.g.,][]{MANGUM2013}. The relative intensity of the rotational ground-state inversion transitions for different $K$-ladders provides a direct measurement of the kinetic temperature. This is because the exchange of population between the $K$-ladders occurs via collisional processes. Asymmetric rotor molecules, such as H$_2$CO, can also be useful temperature probes \citep[e.g.,][]{MANGUM1993}. However, to detect emission from such rarer molecules in outflows, requires very deep and sensitive observations.

\subsubsection{Estimating the Masses and Mass-Outflow Rates} \label{sec:m_and_mdot}

The determination of mass usually relies on using an appropriate mass-to-light ratio for a bright line (although it is also possible to use the dust continuum assuming a dust-to-gas ratio). The resulting equation is 

\begin{equation}
    M_{mol}=\alpha_{X} L_{X}^\prime, \label{eq:alpha}
    \label{eq:M_mol}
\end{equation}

\noindent where $\alpha_{X}$ is the mass-to-light ratio for species \emph{X}, and $L_X^\prime$ is its luminosity in the units of Eq.~(\ref{eq:Lprime}). Note that, frequently, the molecular mass M$_{mol}$ is not just that of H$_2$ but it includes a correction for the cosmic abundance of He by mass ($\sim1.36$), although this is not entirely consistent across authors. 

\medskip
\noindent \emph{- Using CO Emission to Estimate Outflow Mass.} The ground transition of $^{12}$C$^{16}$O is frequently used for this purpose, even though it is optically thick. For a full discussion on the foundations and the caveats of this approach, see \citet{BOLATTO2013a}. Here we will summarize a few relevant equations.

For optically thick emission from a self-gravitating cloud (or an ensemble of clouds), where the velocity dispersion is a reflection of self-gravity, the mass-to-light ratio of the ground rotational transition has been calibrated to be $\alpha_{\rm CO}\approx4.4$ M$_\odot{\rm (K\,km \,s^{-1}\,pc^2)}^{-1}$ for large (M$_{mol}>10^5$ M$_\odot$) Giant Molecular Clouds in the disk of the Milky Way. This value depends on the average temperature of the molecular gas, the mass of the cloud, and the virial parameter of the object, becoming smaller for hotter clouds, smaller clouds, and less bound clouds. Given that, where we can resolve it, molecular emission in outflows is frequently highly turbulent and it probably arises from clouds that may have experienced a shock or are perhaps heated by conduction by the neighboring hot fluid, it is likely that the ``Galactic'' $\alpha_{\rm CO}$ is an overestimate of the true mass-to-light ratio of the gas in a galactic outflow. The lowest likely mass-to-light ratio (and consequently the lowest mass estimate) comes from assuming optically thin emission, for which $\alpha_{\rm CO}\approx0.34\,(T_{ex}/30\,{\rm K})$ assuming CO/H$_2$=10$^{-4}$ and $T_{ex}\gtrsim30$~K \citep[see][]{BOLATTO2013b}. From these considerations it appears that assuming a typical $\alpha_{\rm CO} \approx 1$ M$_\odot~{\rm (K\,km\,s^{-1}\,pc^2)}^{-1}$ yields reasonably bracketed estimates of molecular masses in outflows from CO observations. 

Since $\alpha_{\rm CO}$ is defined for the ground transition, when using higher transitions of CO to determine masses it is necessary to correct their intensities by a brightness temperature line ratio, $r_{J1}=T_B(J\rightarrow J-1)/T_B(1\rightarrow0$). The applicable line ratio depends on the excitation conditions in the medium: hotter and denser media will lead in general to higher line ratios, and as we saw in Eq.~(\ref{eq:Rj}) optically thin emission may also lead to higher line ratios. If the emission is optically thick, however, the ratios are $r_{J1}\le1$, and $r_{J1}\sim1$ usually requires temperatures hot enough to populate the $J$ level and densities higher than the effective critical density. In galaxies, typical observed line ratios for bright emission are $r_{21}\sim0.8$ and $r_{31}\sim0.67$ \citep[e.g.,][]{ROSENBERG2015,KAMENETZKY2016}. As the $J$ level increases and the required excitation conditions become more stringent, the ratios become increasingly variable from source to source \citep[e.g.,][]{KAMENETZKY2018}. 

Note that line ratios and SLEDs depend on the units used for their luminosities. When line intensities are expressed in Jy\,km\,s$^{-1}$ or equivalent units, there will be an additional factor of $\nu^2$ (see Eq.~(\ref{eq:TbSnu})) in the integrated intensities, resulting in line ratios $r_{J1}\,J^2$. When line intensities are expressed in erg\,cm$^{-2}$\,s$^{-1}$ or equivalent units, there will be another factor of $\nu$ associated with the conversion from velocity to frequency, resulting in line ratios $r_{J1}\,J^3$.

\medskip
\noindent\emph{- Using [CI] Emission to Estimate Outflow Mass.} The fine structure lines of atomic carbon, [C~I] (1-0) $^3$P$_1$--$^3$P$_0$ and $^3$P$_2$--$^3$P$_1$ transitions at 609 $\mu$m (492 GHz) and 370 $\mu$m (809 GHz), have received increased attention in recent years as probes of the cool gas in galaxies. 
How [C~I] compares to CO in terms of producing reliable molecular masses is still unclear. 
\citet{PAPADOPOULOS2004} argue on theoretical grounds that the (generally) optically thin [C~I] (1-0) line may be a better tracer of the molecular gas than $^{12}$CO under typical ISM conditions, especially for diffuse ($\sim 10^2-10^3$ cm$^{-3}$) molecular gas in metal-rich environments far from intense UV sources. Similar conclusions are reached by \citet{OFFNER2014} and \citet{GLOVER2015} based on simulations of resolved molecular clouds in conditions similar to those at the Solar circle. \citet{ISRAEL2015}, on the other hand, analyze the suitability of CO and [C~I] to measure molecular masses using an extensive dataset of 76 ULIRGs and starburst galaxy centers. They conclude that the scatter in molecular mass estimation using [C~I] in these galaxies is not better than using CO (and arguably, based on their Figure 9, it appears worse). Because of the competing effects of photoionization by far-UV to produce [C~II], and combination with oxygen to produce CO, [C~I] is usually not a large reservoir of carbon in the gas, and so its abundance may be subjected to large system-to-system variations depending on the physical conditions.

Adopting an excitation temperature $T_{\rm ex}$ = 30 K and a neutral carbon abundance $X_{\rm C}$ = (3.0 $\pm$ 1.5) $\times$ 10$^{-5}$, appropriate for (U)LIRGs \citep{WEISS2003,WEISS2005b,PAPADOPOULOS2004,WALTER2011,JIAO2017}, \citet{CICONE2018a} suggests
$\alpha_{\rm [C I]}$ = 9.43 $M_\odot$ (K km s$^{-1}$ pc$^2$)$^{-1}$ as the [C I](1-0)-to-H$_2$ mass-to-light ratio. \citet{CROCKER2019}, on the other hand, report $\alpha_{\rm [C I]}$ = 7.3 $M_\odot$ (K km s$^{-1}$ pc$^2$)$^{-1}$ in a sample of local star-forming galaxy disks targeted by the Herschel Beyond the Peak program. 

\medskip
\noindent\emph{- Using [C II] Emission to Estimate Outflow Mass.} The [C~II] 157.737 $\mu$m fine-structure transition is the major cooling transition for the cold atomic medium, at densities $n\sim 1-10^3$ cm$^{-3}$ and temperatures T$\gtrsim50$~K. With an ionization potential of 11.3~eV, smaller than that of hydrogen, ionization of neutral carbon into C$^+$ by far-ultraviolet photons is pervasive. In fact, C$^+$ is the major form of carbon in regions illuminated by stars with $A_V<1-2$. 
In molecular regions, far-ultraviolet photons that penetrate the surfaces of interstellar molecular clouds at $A_V < 1-3$ mag create the so-called photon-dominated or photodissociation regions \citep[PDRs, e.g.,][]{HOLLENBACH1999,KAUFMAN1999,WOLFIRE1995,WOLFIRE2003}, where photoelectric heating effect on dust grains heats the gas and [C~II] emission cools it. Some fraction of [C~II] is also generated in neutral atomic and in ionized gas, for example in H~II regions themselves, where collisions with electrons excite the emission. 

Far-infrared [C~II] emission is excited by collisions. The collisional partners can be H, H$_2$, or electrons depending on their abundance in the type of gas the [C~II] ion is embedded in --- atomic, molecular, or ionized. Because the energy carried away by the 157.7 $\mu$m photons is significant and the excitation requirements for the transition are not particularly stringent, wherever [C~II] is abundant it tends to dominate the cooling. The critical densities for [C~II] collisions with H, H$_2$, and electrons are respectively n$_{cr}$=3000, 6100, and 44 cm$^{-3}$ taking as typical temperatures 100~K for atomic clouds and PDRs and 8,000~K for ionized regions \citep{GOLDSMITH2012}. [CII] emission is usually not significantly optically thick, although in photodissociation regions $\tau$ can be moderate \citep{BOREIKO1996,STACEY1991}, so optical depth effects are unlikely to have a significant impact on these critical densities. 

In the case of optically thin emission and neglibible background, the intensity of the [C~II] emission (I$_{\rm [CII]}$, in erg\,s$^{-1}$\,cm$^{-2}$\,sr$^{-1}$) relates to the temperature ($T$), density ($n$), and column density ($N$) of the gas through \citep{CRAWFORD1985,GOLDSMITH2012}

\begin{equation}
{\rm I_{[C~II]}=2.38\times10^{-21}\left[\frac{2\,e^{-91.25/T}}{1+2\,e^{-91.25/T}+n_{cr}/n}\right]\,N({\rm C^+})}.
\label{eq:I_CII}
\end{equation}

\noindent This can be converted to a relation between luminosity and mass using \citep[see also][]{HAILEY-DUNSHEATH2010}

\begin{equation}
    M_{\rm H}=0.68 M_\odot \left(\frac{1.6\times10^{-4}}{X({\rm C^+})}\right)\,\left[\frac{1+2\,e^{-91.25/T}+n_{cr}/n}{2\,e^{-91.25/T}}\right]\, \frac{L_{{\rm [CII]}}}{L_\odot},
\label{eq:M_H_CII}
\end{equation}

\noindent where $X({\rm C^+})$ is the abundance of C$^+$, and $1.6\times10^{-4}$ represents an approximate gas-phase carbon abundance near the Solar circle in the Milky Way \citep{SOFIA2004}, although the precise value is somewhat uncertain \citep[e.g.,][]{SOFIA2011}. In the limit of $T \gg 91$\,K and $n \gg n_{cr}$ (i.e., the high excitation limit) this becomes

\begin{equation}
    \frac{M_{\rm H}}{M_\odot} \approx \left(\frac{1.6\times10^{-4}}{X({\rm C^+})}\right)\,\frac{L_{\rm [CII]}}{L_\odot}.
    \label{eq:M_H_CII2}
\end{equation}

Because we are assuming maximal excitation, this is effectively a lower limit to the amount of mass traced by the [CII] emission. However, if the main contributor to the emission is C$^+$ embedded in ionized gas due to HII regions, where dust may not be an important reservoir of carbon, all carbon will be in the gas-phase and its abundance may be a factor of 3 larger \citep{SOFIA2011}. Conversely, if the radiation field is such that most carbon is in a more highly ionized state, the $X({\rm C^+})$ would be lower and the [CII] emissivity per unit gas mass will be suppressed \citep[e.g.,][]{LANGER2015}. Using the relation between integrated Rayleigh-Jeans brightness temperature and intensity for this transition, $\int T_B dv=1.43\times10^5 I_{\rm [CII]}$ \citep[e.g.,][]{GOLDSMITH2012}, we can express this in a form analogous to Eq.~(\ref{eq:alpha}), keeping in mind that the left hand side represents all hydrogen (not just molecular gas), and that this does not include the 1.36 correction factor by mass due to the contribution from He and other heavy elements

\begin{equation}
    M_{\rm H}=0.15\,{\rm \frac{M_\odot}{K\,km\,s^{-1}\,pc^2}} \left(\frac{1.6\times10^{-4}}{X({\rm C^+})}\right)\, L^\prime_{\rm [CII]}.
    \label{eq:M_H_CII3}
\end{equation}

\noindent Note that this is a lower limit not just because of the assumed excitation and the contribution from moderate optical depth regions, but also because extinction limits the penetration of carbon-ionizing photons, sharply limiting the ionization of carbon inside molecular clouds. For $A_V>2-3$ most carbon will be neutral or molecular, and therefore that gas will not emit in [CII] \citep[e.g.,][]{NARAYANAN2017}.

\medskip
\noindent\emph{- Using OH Absorption to Estimate Outflow Mass and Spatial Extent.}  The OH electronic ground-state $^2\Pi$ is split due to spin--orbit coupling where spin angular momentum can be oriented either parallel or anti-parallel to the orbital angular momentum. Thus, the rotational states are split into two ladders: $^2\Pi_{3/2}$ and $^2\Pi_{1/2}$ \citep{STOREY1981}, where the total angular momentum is given by $J = N \pm 1/2$ ($N$ is the rotational quantum number). Additionally, the orbital angular momentum can have two orientations with respect to the molecular axis, each having a slightly different energy. This $\Lambda$-doubling further splits each $J$ level into two levels of opposite parity. OH and its isotopologues $^{18}$OH and $^{17}$OH have been detected in absorption in several infrared-bright galaxies. With many transitions lying at far-IR wavelengths, where the bulk of the luminosity of these galaxies is emitted, OH is mainly excited through absorption of far-IR photons and selectively traces a region close to the central source of strong far-IR radiation density.
OH doublets with a broad range of lower-level energies, from $E_{\rm low}$ = 0 K (OH 119, 79, 53.3, and 35 $\mu$m originating from the $\Lambda$-doublet state of the ground $^2\Pi_{3/2}~J = 3/2$ rotational level) to $E_{\rm low}$ = 620 K (OH 56 originating from $^2\Pi_{1/2}~J = 7/2$), have been detected in (U)LIRGs, often showing clear signatures of outflowing gas (prominent blue wings or P-Cygni profiles). The detection of OH 79, OH 53.3, and OH 35 implies that OH 119, with a much higher opacity, is optically thick i.e. the continuum optical depth at 100 $\mu$m, $\tau_{100} \gtrsim$ 1, or $N_H \gtrsim$ 10$^{24}$ cm$^{-2}$. However, the OH 119 absorption feature is not black, so the OH 119 doublet at a given velocity only covers a fraction of the total 119 um continuum. When P-Cygni profiles are observed, the relative strength of the redshifted emission (which is produced in the receding cocoons) and blueshifted absorpion (from the approaching parts along the line of sight to the continuum source) can be used to constrain the importance of extinction of the line-emitting photons arising from the back side of the far-IR source and/or significant departure from spherical symmetry \citep{GONZALEZ-ALFONSO2014}.

Since the OH energy levels are radiatively pumped in the outflows, transitions at different wavelengths and energy levels, in combination with continuum component fits, yield crucial information about the radial location where the lines are formed. This information enables the estimation of the outflow physical parameters using the equations in Sect.~\ref{sec:energetics}. In practice, the observed OH profiles are carefully compared with the predictions from radiative-transfer models. The code calculates the line excitation due to the dust emission and collisions with H$_2$, and takes into account opacity effects (i.e. radiative trapping), non-local effects, velocity gradients, extinction by dust, and line overlaps. For simplicity, the outflow is usually modeled as a series of concentric expanding shells around a nuclear continuum source, allowing for each source one to three components with different velocity gradients and distances to the central source.  The density profile for each velocity component is determined through mass conservation assuming a constant mass outflow rate. Free parameters are the inner and outer radii, the velocity field of each velocity component, the OH density at the inner radius, the covering factor of the continuum FIR source, and the solid angle of the outflow.  For a given model, the code first calculates the statistical equilibrium populations in all shells that make up the source, and then the emerging line shapes are computed, convolved with the instrumental profile, and compared directly with the observations.

The derived outflow mass directly depends on the assumed value of the OH abundance relative to H nuclei, $X_{\rm OH} \equiv N_{OH} / N_H$, which is uncertain. A value of 2.5 $\times$ 10$^{-6}$ has been used in the models of \citet{GONZALEZ-ALFONSO2017b}, consistent to within a factor of $\sim$3 with the value inferred from multi-transition observations of OH in the Galactic Sgr B2 and Orion KL outflow \citep{GOICOECHEA2002,GOICOECHEA2006} and in buried galaxy nuclei \citep{GONZALEZ-ALFONSO2012,FISCHER2014,FALSTAD2015}, and with chemical models of dense photodissociation regions \citep[the peak value in][]{STERNBERG1995} and of cosmic-ray- and X-ray-dominated regions \citep{MEIJERINK2011}. New constraints on the value of $X_{\rm OH}$ in the outflows have recently been derived using the strength of the optically thin cross-ladder $^2\Pi_{3/2} \rightarrow ^2\Pi_{1/2}~J = 3/2 \rightarrow 5/2$ $\Lambda$-doublet transitions at 34.60 and 34.63 $\mu$m as a powerful OH column density estimator \citep{STONE2018}. The OH column densities in a number of OH outflows were compared with the hydrogen column density for a typical optical depth at 35 $\mu$m of $\sim$0.5 and Galactic gas-to-dust ratio ($\sim$125; see Sect.~\ref{sec:lowz_dust} below) to yield an OH-to-H abundance ratio of $X_{\rm OH} = (1.01 \pm 0.15) \times 10^{-6}$. This abundance ratio is formally a lower limit since (1) some fraction of the OH molecules are in excited levels, so $N_{\rm OH}$ derived from OH 35 is a lower limit to the column density of all OH states, and (2) the covering fraction of the 35 $\mu$m continuum source by the obscuring material may not be 100\%, as assumed. 

\subsubsection{Importance of Shocks, X-rays, and Cosmic Rays} 
\label{sec:shocks}

Chemistry is an important tracer of the conditions in the outflow and hence of its power source and evolution. Chemistry can for example probe the impact of shocks and irradiation by e.g. cosmic rays (CR) and X-rays. Important shock tracers include SiO, which traces fast shocks and the destruction of grains \citep{BACHILLER1991,GARCIA-BURILLO2001}. Slower shocks that evaporate the ices (without disrupting the grain core) can be probed by e.g. H$_2$O, CH$_3$OH, HNCO, OH and possibly also HCN  \citep[e.g.][]{RUIZ1992,CERNICHARO2006,RODRIGUEZ-FERNANDEZ2010, AALTO2012,MEIER2012, PELLEGRINI2013,GARCIA-BURILLO2014,GOICOECHEA2015,SAITO2017}.  The high HCN luminosity of the Mrk~231 outflow (Sect.~\ref{sec:Mrk231}) is likely due both to a large filling factor of dense ($n>10^4$ cm$^{-3}$) molecular gas, as well as an elevated HCN abundance \citep[$X{\rm (HCN)}=10^{-8}-10^{6}$,][]{AALTO2012,AALTO2015c,LINDBERG2016}. Chemical differentiation in the outflow between HCN, HNC and HCO$^{+}$ can be partially caused by shock chemistry \citep{LINDBERG2016}.  Suspected shock-enhanced HCN, SiO and CH$_3$OH is found in the outflow regions of the merger NGC~3256  \citep[Sect.~\ref{sec:obscured};][]{HARADA2018a}.  The CO line-to-continuum ratio \citep{MEIJERINK2013} is also a robust diagnostic of shock excitation, as is elevated [C~I] line emission \citep{PELLEGRINI2013}.  Shocks are also traced by near-IR [Fe~II] 1.257 and 1.644 $\mu$m lines and H$_2$ line ratios (see Sec~\ref{sec:H2}).

Note that high H$_2$O, HCN, and CH$_3$OH abundances may also reflect a hot-core like chemistry where ice mantles are evaporating in high-column density gas \citep[e.g.][]{CERNICHARO2006,GONZALEZ-ALFONSO2012,QIN2015}.  \citeauthor{GONZALEZ-ALFONSO2012} suggest that such an ``undepleted'' chemistry may be traced by, for example, studying the HCN/H$_2$O and HCN/NH$_3$ column density ratios. Methanol (CH$_3$OH) primarily forms on cold ($T < 24$ K) dust grains \citep[e.g.][]{SOMA2015} and can therefore be used as an important ``chemical clock''. Once depleted from the dust grains it cannot reform as long as the dust grains remain warm.

Radiative impact on the gas can be traced by radicals such as CN that thrives in the interface regions between photon dominated region (PDRs), or X-ray dominated regions (XDRs), and dense molecular clouds.  Far-UV photons in post-shock gas may also impact the chemistry. CN has been detected in the double outflows of the merger NGC~3256 \citep{SAKAMOTO2014} and more studies are required to understand its origin. Very luminous CN emission is also detected in the quasar-driven molecular outflow of Mrk~231 \citep[][]{CICONE2019} where X(CN)$>$X(HCN), suggesting strong radiative impact on the outflowing dense molecular gas. 
C$_2$H has been found in the starburst-driven outflow of Maffei~2 \citep{MEIER2012} and also in the interface region between the molecular disk and ionized gas outflow of NGC~1068 \citep{GARCIA-BURILLO2017}. The high fractional C$_2$H abundances in this interaction region can be reached at very early times ($t$ of about $10^{2-3}$ yr) in models of UV or X-ray irradiated dense gas.
 
Molecular ions are important tools to probe the ionization rates of the interstellar medium in XDRs/PDRs or in CR-dominated regions (CRDRs).  \citet{GONZALEZ-ALFONSO2018} suggest that high OH$^+$/H$_2$O$^+$ and OH$^+$/H$_3$O$^+$ ratios imply high ionization rates per unit density.  X-rays appear to be unable to explain the ionization rates inferred by these elevated abundance ratios, and thus they suggest that low-energy (10--400 MeV) CRs are primarily responsible for the ionization (Sect.~\ref{sec:Mrk231}).  Ionization fraction can also be probed by the HOC$^{+}$/HCO$^{+}$ ratio that can be studied at mm and submm wavelengths  \citep[e.g.][]{GOICOECHEA2009,MARTIN2009}. One of the most important probes of ionization rate in diffuse molecular clouds is the H$_3^{+}$ molecule that can be studied at infrared wavelengths. \citep[e.g.][]{OKA2005,INDRIOLO2007}. 

The CH$^+$ molecule is a very useful tracer of turbulence in interstellar gas \citep[e.g.][]{FALGARONEANDPUGET1995,MYERS2015}. Only few extragalactic studies of CH$^+$ have been carried out \citep[e.g.][]{WELTY2014,MULLER2017}. One concerns the discovery of large turbulent reservoirs of molecular gas surrounding some high-redshift galaxies through the detection of the CH$^+$(J =1-0) transition, both in emission and absorption \citep{FALGARONE2017}. Given the high critical density of this transition and the very large width of the emission line (FWZI $>$ 2,500 km s$^{-1}$) in these objects, it is likely tracing very dense gas ($>$ 10$^5$~cm$^{-3}$) ejected by outflows.

It is important to be aware, however, that to safely apply chemistry as a diagnostic tool usually requires a combination of lines to address the dichotomy between PDR, XDRs, CRDRs -- and the chemistry of shocks and mechanical heating \citep[e.g.][]{KAZANDJIAN2012,AALTO2015b,VITI2014,VITI2016,HARADA2018b}.

Molecular chemistry and abundances are usually probed at the mm/submm and far-infrared wavelengths. However, X-ray absorption lines and edges may also be used to identify molecules \citep{GATUZZ2015}. Molecules may also be probed at longer cm wavelengths--for example maser lines of OH and H$_2$O that can be seen in dense dusty outflows \citep{TURNER1985,BAAN1989}. Also thermal absorption/emission lines of e.g. OH, H$_2$CO, NH$_3$ can be detected \citep[e.g.][]{MANGUM1993,SALTER2008, MANGUM2013,FALSTAD2015}, also in outflows \citep{FALSTAD2017}.

\subsubsection{Compact Obscured Nuclei: CONs}
 \label{sec:obscured}

There is a population of luminous infrared galaxies that have highly obscured nuclei with visible extinction $A_{\rm V}>$1000 (corresponding to $N$(H$_2$)$>10^{24}$ cm$^{-2}$).  At these high levels of extinction, mid-infrared diagnostic methods suffer from opacity effects and one has to resort to longer wavelength lines and continuum to probe the enshrouded activity.  The obscuring columns appear to be warm with temperatures in the range of 100--300 K \citep{SAKAMOTO2008,COSTAGLIOLA2010,SAKAMOTO2013,GONZALEZ-ALFONSO2012,AALTO2015a,FALSTAD2015,SCOVILLE2017,SAKAMOTO2017,BARCOS-MUNOZ2018,AALTO2019}. These enshrouded nuclei may be previously undetected accreting SMBHs and/or compact and luminous young stellar clusters (possibly also with top-heavy stellar initial mass functions \citep[IMFs;][]{AALTO2019}). Radiatively excited molecular emission of e.g. H$_2$O and OH \citep{VANDERWERF2011,GONZALEZ-ALFONSO2012,VEILLEUX2013a,GONZALEZ-ALFONSO2014,FALSTAD2015,FALSTAD2017} are good probes of the nuclei as long as column densities stay below $N$(H$_2$)$=10^{25}$ cm$^{-2}$. However, several enshrouded nuclei have $N$(H$_2$)$>10^{25}$ cm$^{-2}$ and in some cases column densities may reach extreme values of $N$(H$_2$)$>10^{26}$ cm$^{-2}$ \citep{SCOVILLE2017,SAKAMOTO2017,BARCOS-MUNOZ2018,AALTO2019}.  These Compact Obscured Nuclei \citep[CONs][]{AALTO2015b} are found in the nuclei with radii $r<10-50$ pc and high temperatures $>$200 K. In the most extreme cases there may also be issues of photon trapping and self-heating of the gas and dust \citep[e.g.][]{KAUFMAN1999, GONZALEZ-ALFONSO2019b}.

To probe the most deeply enshrouded CONs one needs to go to mm and even cm wavelengths, for example using vibrationally excited lines of HCN \citep{SALTER2008,SAKAMOTO2010,AALTO2015a,MARTIN2016,IMANISHI2016,AALTO2019}.  The HCN molecule has degenerate bending modes in the IR and may absorb IR-photons to its first vibrational state. There are rotational transitions within the vibrationally excited state that can be observed with e.g. ALMA. The energy above ground of the $\nu_2$=1 state is $T_{E/k}$=1050 K requiring high surface-brightness mid-IR ($T_{\rm B}$(14$\mu$m)$\approx$100 K) emission to be excited. 
CONs are also very rich sources of molecular emission that can be used to diagnose the nuclear activity \citep{MARTIN2012,GONZALEZ-ALFONSO2012,COSTAGLIOLA2016}.

\subsubsection{Direct H$_2$ Probes} \label{sec:H2}

Molecular hydrogen is the most abundant molecule in the Universe, but direct observations of H$_2$ are difficult. The molecule exists in two, almost independent isometric forms, named ortho-H$_2$ (spin of H nuclei parallel) and para-H$_2$ (spins antiparallel). The ortho/para ratio is a complex function of the environment (e.g., Bron et al. 2016). In LTE, the ortho/para ratio is close to 3 under warm ($T > 200$ K) conditions, but is generally smaller otherwise. Given its symmetry, H$_2$ has no permanent dipole moment and all its ro-vibrational and rotational transitions within the electronic ground state are forbidden electric quadrupolar and therefore weak.  The ro-vibrational lines lie in the near-IR, and thus can be studied from the ground at low redshifts, while the rotational lines occur in the mid-IR and generally require space-borne facilities. The far-UV electronic transitions from nearby systems can only be detected above the Earth's atmosphere and are strongly affected by extinction. Indeed the very presence of H$_2$ requires $A_V \gtrsim$ 0.01-0.1 mag.\ so that it is shielded from the ultraviolet photons responsible for its photo-dissociation.

\medskip
\noindent\emph{- Electronic Transitions.}  Photons with an energy in the range of 11.2 to 13.6 eV (912 \AA\ to 1126 \AA) that coincides with
one of the lines in the Lyman band (between the ground and first electronic states; $B - X$) or Werner band (between the ground and second electronic states; $C - X$) will be absorbed by H$_2$ and place the molecule in an excited electronic state, followed by rapid radiative decay that may lead to its dissociation. This destruction process is balanced by formation of H$_2$ on the surface of dust grains (Sect.~\ref{sec:dust_cycle}).  The Lyman and Werner electronic absorption bands in the far-UV provide a very sensitive tool to detect even very diffuse H$_2$, down to column densities as low as a few 10$^{12}$ cm$^{-2}$, provided a UV-bright background source is available \citep[e.g.][]{SPITZER1973}. A curve-of-growth analysis that relies on the measured equivalent widths of all H$_2$ lines is possible when $N$(H$_2$) $\lesssim$ 10$^{19}$ cm$^{-2}$, while a fit to the damping wings of the low-J H$_2$ lines is needed to derive larger column densities. H$_2$ column densities in low rotational state ($J$ = 0 and 1) may be used to derive rotational and/or kinetic temperatures of the diffuse gas. The higher $J$ abundances can constrain models of the UV radiation fields and gas densities \citep[e.g.][]{SHULL2000,WAKELAM2017}. 

\medskip
\noindent\emph{- Ro-Vibrational Transitions.} In the ground-state electronic state, the $\nu$ = 1 vibration level is $\sim$ 6000 K above the ground state, so that ro-vibrational excitation requires kinetic temperatures $T >$ 1000 K or fluorescent (``FUV pumping'') excitation.  The method most commonly used to differentiate collisional excitation from fluorescence consists in using flux ratios of various ro-vibrational H$_2$ lines visible in the K band, particularly H$_2$ $\nu$ = 1-0 S(1) 2.121 $\mu$m and the weaker $\nu$ = 1-0 S(0) 2.223 $\mu$m and $\nu$ = 2-1 S(1) 2.247 $\mu$m transitions. 
FUV pumping dominates if $G_0/n$ is large ($G_0$ is the strength of the UV radiation field and $n$ is the hydrogen density in the PDR).
In the alternative scenario of collisional excitation, three mechanisms may provide the heating: (1) UV radiation from a starburst or AGN, (2) X-rays from a starburst, AGN, or wind, or (3) shocks induced by the outflow. A strong constraint on the importance of UV heating can be derived from the ratios of H$_2$ 2.12 $\mu$m to H$\alpha$ \citep[or, equivalently, Br$\gamma$; e.g.,][]{PUXLEY1990,DOYON1994}, while the relative H$_2$ 1-0 S(1) and X-ray fluxes provide an excellent way to test the mechanism of X-ray heating. The third and final scenario may be tested using shock diagnostics (e.g. [Fe~II] 1.644 $\mu$m, mm-wave SiO; see Sec. \ref{sec:shocks}).

\medskip
\noindent\emph{- Pure Rotational Transitions.}  The H$_2$ pure-rotational emission lines in the mid-IR trace warm molecular gas at $T = 100 -
1000$ K. Due to the low Einstein coefficients
of these transitions, optical depth effects are usually unimportant and thus these mid-IR lines provide an accurate measure of the warm gas mass. At densities $n_H \gtrsim$ 10$^4$ cm$^{-3}$ of PDRs, collisions maintain the lowest rotational levels of H$_2$ ($\nu$ = 0, $J \lesssim$ 5) in thermal equilibrium. The rotational transitions between these levels therefore provide a thermometer for the bulk of the gas above $\sim$80 K.  The excitation temperatures measured from these line ratios should be considered an upper limit to the kinetic temperature since UV pumping may contribute to the excitation of H$_2$ even for low $J$. Shocks may also contribute to exciting the H$_2$ molecules without dissociating them. The excitation temperature is a steep function of shock velocity and therefore can be used to constrain it if shocks are dominant \citep[e.g.,][]{GUILLARD2012,BEIRAO2015}.

\subsection{Diagnostic Tools of the Dust} \label{sec:diagnostics_dust}

There is extensive literature on this topic including the excellent review of \citet{GALLIANO2018} and the classic monograph by \citet{DRAINE2011}.  The many diagnosic tools that have been used to derive the dust properties in galaxies may also be applied to study the dust within cool outflows. These tools can largely be divided into five main categories based on the physical process at work: (1) extinction and reddening, (2) scattering, (3) emission, (4) polarization, and (5) depletion. The derived properties include the dust composition (the exact mixture of carbonaceous and silicate grains), the grain structure (e.g., amorphous, crystalline, porous), the size distribution of the dust grains (typically, sizes $a$ range from 0.3 $n$m to 0.3 $\mu$m with a power-law distribution $\propto a^{-3.5}$ to first order), the abundance of the dust grains relative to the gas \citep[i.e.\ the dust-to-gas ratio by mass, $D/G$, of order $\sim$ 1\% in the Milky Way, and a decreasing function of metallicity;][]{ROMAN-DUVAL2014}, and the fraction of dust present in the ionized gas \citep[$f_{\rm ion}$ ranging from  $\sim$ 10$^{-4}$ in the hot intracluster medium of galaxy clusters to $\sim$ 1 in some luminous infrared galaxies;][]{ LAURSEN2009}.

\begin{figure}[htb]
\begin{center}
\includegraphics[width=1.0\textwidth]{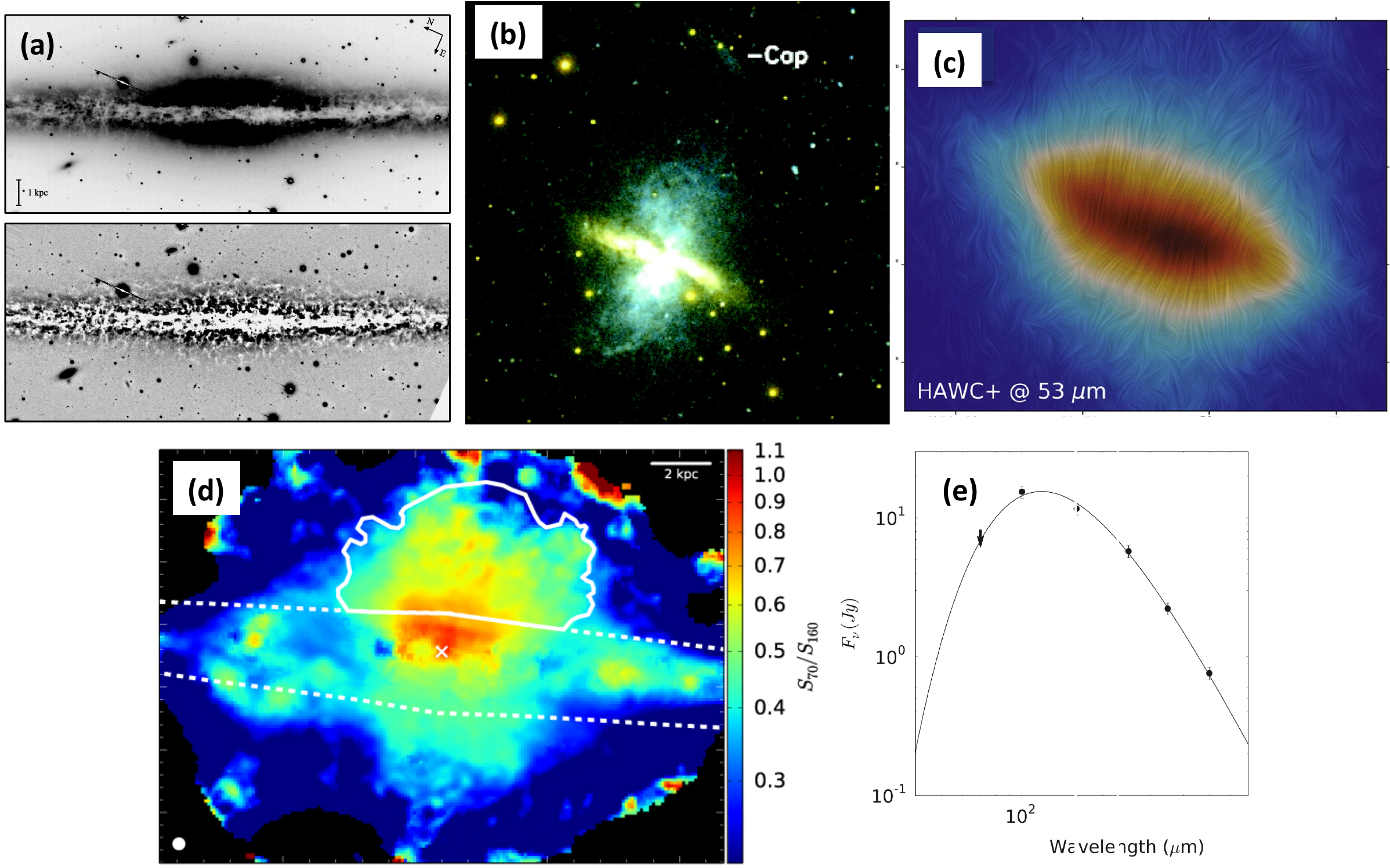}
\end{center}
\caption{Some of the diagnostic tools used to study dust in cool outflows. (a) Extinction in the dusty extraplanar material of NGC~891, (b) UV scattering in the dusty wind of M~82, (c) polarization in the disk and wind of M~82, (d)--(e) thermal infrared emission from the extraplanar material in NGC~4631. Images reproduced with permission from (a) \citet{HOWK2000}, (b) \citet{HOOPES2005}, (c) \citet{JONES2019a}, and (d, e) \citet{MELENDEZ2015}, copyright by AAS.}
\label{fig:diagnostics_dust}
\end{figure}

\subsubsection{Extinction and Reddening} \label{sec:ext_red}

The optical depth due to dust is $\tau_d = \int \sigma_d n_H ds \simeq \sigma_d N_H$, where $\sigma_d$ is defined as the effective cross section per hydrogen atom. 
This term includes contributions from both absorption and scattering: $\sigma_d = \sigma_a + \sigma_s$. A related quantity is the dust opacity, $\kappa_d = \kappa_a + \kappa_s$, which is the effective cross section per dust mass so that $\rho \kappa_d = n_H \sigma_d$. The observed extinction $A_\lambda$ in magnitudes is directly related to $\sigma_d$ and scales linearly with $N_H$: 
\begin{eqnarray}
A_\lambda = \frac{2.5}{\rm ln~10} \tau_d = 1.086~\sigma_d N_H.
\label{eq:A_lambda}
\end{eqnarray}
Both absorption and scattering increase at shorter wavelengths causing reddening of background objects. The slope of the extinction curve in the optical region is often measured by the dimensionless quantity
\begin{eqnarray}
R_V \equiv A_V / (A_B - A_v) = A_V / E(B - V),
\label{eq:R_V}
\end{eqnarray}
where B and V refer to the B and V bands and $E(B - V)$ is the color excess. This quantity ranges from $R_V \simeq 1.2$, for very small grains (grain size $a << \lambda$) where Rayleigh scattering ($A_\lambda \propto \lambda^{-4}$) is dominant, up to $R_V \rightarrow \infty$ for grey extinction from very large grains ($a >> \lambda$).  The ``classical'' value, $R_V = 3.1$, applies to the diffuse ISM in the Milky Way \citep[in that case, $A_V \simeq 0.53~(N_{\rm H}/10^{21}~{\rm cm}^{-2})$ mag;][]{CARDELLI1989}, but in practice $R_V$ shows significant scatter ($R_V \approx 2$--$6$) within the Milky Way and in external galaxies, which underscores the environmental dependence of extinction.

The overall 0.1--30 $\mu$m extinction curve is characterized by a number of features \citep[see Fig.\ 6 in][]{GALLIANO2018}: (a) a steep rise in the FUV (sometimes as steep as $\lambda^{-1}$), primarily attributed to absorption by small dust grains, (b) a bump around 2175 \AA\ bump, produced by small carbon grains such as PAHs, graphite, or amorphous carbon, (c) a knee around $\sim$ 4000 \AA\ due to scattering by large grains, and (d) a power-law optical-NIR continuum (roughly $\propto \lambda^{-1}$). In addition, bright infrared sources with $A_{\rm 18~\mu m}~\gtrsim~1$ mag.\ ($A_V$ $\gtrsim$ 12 mag.) often present two prominent absorption features at 9.7 and 18 $\mu$m, caused mainly by amorphous silicate although the distinct absorption features of unprocessed crystaline silicate have been detected in some ULIRGs \citep{SPOON2006}.
The infrared spectra of the most obscured sources also present H$_2$O, CO, and CO$_2$ ice absorption features at 3.1, 4.7, and 15.2 $\mu$m, produced by icy grain mantles deeply shielded from heating sources.

\subsubsection{Scattering} \label{sec:scattering}

The albedo, $A_{s,\lambda} \equiv \sigma_s/\sigma_d \le 1$, is a measure of the importance of scattering on the overall extinction. For instance, in the NUV, $A_{s,\lambda} \simeq 0.5$ \citep{DRAINE2003}. For a Galactic dust-to-gas ratio, $D/G \approx 0.01$, the dust scattering cross-section in the UV-optical is $>$ 1-2 orders of magnitude larger than the Thompson scattering cross-section of electrons ($\sigma_T = 6.65 \times 10^{-25}$ cm$^{-2}$).  The dust scattering cross-section and its phase function (angular dependence) may be decomposed into two functions \citep{WEINGARTNER2001,DRAINE2003}:
\begin{eqnarray}
\left(\frac{d\sigma_s}{d\Omega}\right) = C_s(\lambda) \times p(\theta),
\label{eq:dsigma_domega}
\end{eqnarray}
where $\theta$ is the scattering angle between the observer's line of sight and the initial direction of propagation of the photon from the light source to the scattering event.  For simplicity, $C_s$ may be approximated as a power-law with wavelength $C_s(\lambda) \propto \lambda^{-\alpha}$,
where the index $\alpha$ depends on $x \equiv (2 \pi a)/\lambda$, ranging from $\alpha \rightarrow 4$ in the Rayleigh scattering limit ($x << 1$) to $\alpha \rightarrow 0$ in the geometric optics limit ($x >> 1$). If all dust grains had the same size, one would therefore expect a break in the wavelength dependence of the scattered light at a critical wavelength $\lambda_c \simeq a$. In reality, the grain sizes show a broad distribution which smears the spectral break. 

In the weak scattering approximation (i.e. each photon scatters at most once), and assuming the energy source is a point source of luminosity $L_\lambda$, the radiative transfer equation of photons through a scattering screen is
\begin{eqnarray}
\frac{dI_{s,\lambda}}{ds} \simeq \left(\frac{\Phi_\lambda}{4 \pi}\right) \left(\frac{L_\lambda}{4 \pi r^2}\right)n_H \sigma_{s,\lambda},
\label{eq:dI_ds}
\end{eqnarray}
where $s$ is the path length, $I_{s,\lambda}$ is the measured scattered intensity, $L_\lambda/(4 \pi r^2)$ is the incident intensity at the scattering location, and $\Phi_\lambda$ is a wavelength-dependent factor that captures the anisotropy of dust scattering. In the idealized case of a spherically symmetric scattering screen surrounding a point source (e.g., dusty halo around an AGN or compact starburst), the integrated scattered luminosity is simply 
\begin{eqnarray}
L_{s,\lambda} \simeq L_\lambda(1 - e^{-\tau_{s,\lambda}}) \simeq L_\lambda \tau_{s,\lambda},
\label{eq:L_s_lambda}
\end{eqnarray} 
where a scattering optical depth $\tau_{s,\lambda} \equiv (D/G)N_{\rm H}\sigma_{s,\lambda} << 1$ was assumed. While this expression can be used in principle to estimate the dust column densities $(D/G)N_{\rm H}$ of the scattering screen, a more accurate measurement requires using a radiative transfer scattering model where a geometric model for the halo dust and disk emission is adopted (Sect.~\ref{sec:lowz_dust}). 

\subsubsection{Emission} \label{sec:emission}

Dust grains are heated by the absorbed photons and this energy is re-emitted at longer wavelengths as infrared radiation. The dust mass may be estimated to first order by fitting a simple single-temperature modified black body (MBB) to the infrared spectral energy distribution using 
\begin{eqnarray}
F_\nu = \frac{M_d \kappa_\nu B_\nu(T_d)}{d^2}.
\label{eq:F_nu}
\end{eqnarray}
Here $M_d$ is the dust mass (free parameter), $B_\nu$ is the Planck function, $T_d$ is the dust temperature (free parameter), $d$ is the distance to the galaxy, and $\kappa_\nu$ is the dust emissivity, $\kappa_\nu = \kappa_0 (\nu/\nu_0)^\beta$, where $\kappa_0$ is the dust opacity at 350 $\mu$m. A commonly used value for $\kappa_0$ is 0.192 m$^2$ kg$^{-1}$; this value of the dust opacity is based on the best fit of the average FIR dust emissivity for the Milky Way model presented in \citet{DRAINE2003}, which yields a best-fit spectral index value of $\beta$ = 2.0. Caution must be taken as the normalized dust model opacity cross-section, $\kappa_0$, is associated with a dust model with $\beta$ = 2.0 and thus, discrepancies may arise between the results from a single-temperature MBB fit with a fixed emissivity and with an emissivity as a free parameter \citep[see][for a review]{BIANCHI2013}. Note also that more recent fits to the infrared SED of the Milky Way, and other Local Group galaxies, suggest a lower value for $\beta$ \citep[= 1.62; e.g.,][]{PLANCKCOLLABORATION2014a}. 
These fits may be refined by adding additional MBB components with different dust temperatures to better capture the full infrared SED. These simple MBB fits are usually carried out using the infrared SED above $\sim$30 $\mu$m, which is largely featureless and dominates the dust mass determinations. The far-infrared and submm-wave emissivity increases with the porosity of dust grains \citep{JONES1988}.

The underlying assumption of this fitting method is that the dust is in thermal equilibrium wih the ambient radiation field. This is true for large dust grains with large enthalpy, but that is not the case for small grains ($a \lesssim 0.02$ $\mu$m) where the absorption of a single photon will cause a significant temperature spike (``stochastic'' heating). These temperature fluctuations complicate the small-grain dust mass estimates, although constraints from {\em Spitzer} spectra have helped refine the absorption properties of these small grains and thereby their contributions to the overall dust masses \citep[e.g.,][]{DRAINE2007,CHASTENET2019}. The small grains that emit in the mid-infrared have characteristic broad emission features with Drude-like profiles centered at 3.3, 6.2, 7.7, 8.6, 11.2, 12.7, and 17 $\mu$m. These features are most likely due to optically active vibrational modes of C-C, C-C-C, and C-H bonds in polycyclic aromatic hydrocarbons (PAHs) with $\lesssim$10$^3$ carbon atoms. The relative intensity of these features may be used to diagnose two key properties of the PAH population: brighter short-wavelength PAH features relative to the long-wavelength features indicates an excess of small PAHs in the PAH population, while bright 6--9 $\mu$m features indicate more abundant PAH$^+$ ions relative to PAH$^0$ molecules \citep[e.g.,][]{DRAINE2007b,BEIRAO2015}. 

Two broad silicate features from warm ($\sim$ 200 K) dust have been detected in emission at 10 $\mu$m and 18 $\mu$m in several type 1 and 2 quasars and AGN, including a low-luminosity LINER \citep[e.g.,][]{SIEBENMORGEN2005,HAO2005,STURM2005,STURM2006,TEPLITZ2006,SCHWEITZER2008}. The strength of the 10 $\mu$m bump relative to that of the 18 $\mu$m bump is a diagnostic of the dust temperature and can thus be used to estimate the location of the warm dust with respect to the energy source as long as the incident heating spectrum and intensity are known. For instance, \citet{SCHWEITZER2008}  have argued that the bulk of the silicate emission in type 1 PG QSOs arises from dust in the innermost portion of the NLR ($\sim$ 10$-$300 pc), well outside of the dust sublimation radius (Sec. \ref{sec:dust_cycle}). 

Finally, dust continuum emission has also been detected below $\sim$ 10 $\mu$m in the Milky Way and external galaxies \citep[e.g., the ``extended near-infrared emission'' at $\sim$ 3--7 $\mu$m in][]{SELLGREN1983,BARVAINIS1987,STURM2000,XIE2018}.  In general, grains heated by single photons of energy $h\nu$ radiate most of their energy at $T_{\rm peak}$ given by $T_{\rm peak} \propto a^{-3/4}(h\nu)^{1/4}.$ The dust emission below 10 $\mu$m therefore indicates the presence of very small grains (VSGs) and nanometer-size particles at $T = 300$--$1500$ K.

\subsubsection{Polarization} \label{sec:polarization}

Dust may induce polarization through scattering, dichroic extinction, and dichroic emission. The level of linear polarization due to dust scattering reflects the dust composition, grain size, and spatial distribution with respect to the light. It peaks in the FUV  where $\sigma_s$ is largest and strongly forward directed \citep[e.g.,][]{ZUBKO2000}.
Dichroic extinction is produced when photons propagate through a screen of dust where the elongated grains are aligned along the magnetic field. The polarization is perpendicular to the magnetic field and peaks in the optical according to $P(\lambda) = P_{\rm max}~{\rm exp}[-0.92~{\rm ln}^2(\lambda_{\rm max}/\lambda)]$, where $\lambda_{\rm max} = 0.55~\mu$m \citep{SERKOWSKI1975,DRAINE2006}, followed by a power-law extension from 1.4 $\mu$m up to 4 $\mu$m with an index of $-$1.7 \citep{MARTIN1992,DRAINE2009b}. The warm dust in this screen will also produce polarized thermal emission, but in this case the polarization is parallel to the magnetic field.
Spectropolarimetry in the UV-optical has also been used to measure the velocity of the dusty scatterers relative to that of the line-emitting gas \cite[e.g.,][]{YOSHIDA2011,YOSHIDA2019}.

\subsubsection{Depletion} \label{sec:depletion}

Dust primarily consists of silicates \citep[Mg, Fe, Si, O, Na, and Al in various combination including Mg$_2$SiO$_4$, MgFeSiO$_4$, MgSiO$_3$, Mg$_{0.5}$Fe$_{0.5}$SiO$_3$, Na$_{0.5}$Al$_{0.5}$SiO$_3$, SiO$_2$, and SiC; e.g.,][]{BOUWMAN2001,MIN2007}
and carbonaceous material (C in graphite and organics).  Refractory elements with high condensation temperatures (e.g., many of the iron peak elements primarily produced in Type Ia SNe) are over-represented in dust grains compared with volatile elements (e.g., elements arising from $\alpha$-capture processes), resulting in differential elemental depletion in the gas phase when dust is present. This depletion pattern has been used to constrain the dust content of a wide variety of environments in the Milky Way and elsewhere \cite[e.g.,][]{SAVAGE1996b,JENKINS2009}. Depletion is seen to scale with the mean density of the medium \citep[e.g.,][]{SAVAGE1979,CRINKLAW1994}. Following \citet{JENKINS2009} and \citet{GALLIANO2018}, the
depletion of an element X is defined as
\begin{eqnarray}
\left[\frac{X_{\rm gas}}{H}\right] \equiv {\rm log}\left(\frac{X}{H}\right)_{\rm gas}-~{\rm log}\left(\frac{X}{H}\right)_{\rm ref} \simeq \left[\frac{X_{\rm gas}}{H}\right]_0 + A_X \times F_*
\label{eq:depletion}
\end{eqnarray}
where the term with subscript ``ref'' is the reference value of the abundance when no dust is present. The depletion is approximated as a first-order polynomial with a fixed term, the minimum depletion, which corresponds to the core of the grains, and a varying environmental factor attributed to accretion of mantles in denser environments ($F_*$ is called the depletion strength). While most depletion measurements in the neutral and ionized phases of our Milky Way and external galaxies have been made using UV-optical absorption lines, X-ray spectroscopy may be used to determine the depletion pattern in hot plasmas by measuring the depth, structure, and energy of the photoelectric absorption edges of C (K, 284 eV), O (K, 538 eV), Fe (L$_{2,3}$, 708, 721 eV) Mg (K, 1311 eV), Si (K, 1846 eV), and Fe  (K, 7123 eV) \citep[e.g.,][]{SCHULZ2002,DRAINE2003,GATUZZ2015,GATUZZ2016}.

\subsubsection{Dust Cycle: Formation and Destruction} \label{sec:dust_cycle}

Dust is created in the cooling ejecta of asymptotic giant branch (AGB) stars with initial masses of 0.8--8 $M_\odot$ and in Type II SNe from 8--40 M$_\odot$ stellar progenitors.\footnote{\citet{SARANGI2019} have recently suggested that nuclear AGN winds may also be sites of dust formation.} The detection of dust at $z > 6$ implies a significant  contribution from supernovae in the early universe \citep{MAIOLINO2004}. The dust grains produced by AGBs and SNe act as seeds that grow, via accretion, in the ISM, on a time scale \citep{ASANO2013} 
\begin{eqnarray}
t_{\rm acc} \simeq 2.2 \times 10^4~\left(\frac{a}{0.1~\mu {\rm m}}\right)\left(\frac{n_H}{10^4~{\rm cm}^{-3}}\right)^{-1} \left(\frac{T}{10^4~K}\right)^{-1/2}\left(\frac{Z}{0.0129}\right)^{-1}~{\rm yr}.
\label{eq:t_acc}
\end{eqnarray}
Several processes may destroy this dust including collisions with other grains, sputtering due to collisions with ions, sublimation or evaporation, explosion due to ultraviolet radiation, and alteration of grain material  by cosmic rays and X-rays.  Many of these processes are associated with SN shocks. In this context, a distinction is made between thermal sputtering, where the sputtering rate only depends on the local gas properties (namely $n_H$ and $T$), and non-thermal (inertial) sputtering, where the dust-gas relative velocity is a crucial parameter.
The thermal sputtering timescale is
\begin{eqnarray}
t_{\rm sput} \simeq 3.3 \times 10^3~\left(\frac{a}{0.1~\mu{\rm m}}\right)~\left(\frac{n_H}{10~{\rm cm}^{-3}}\right)^{-1}~\left(\frac{Y_{\rm tot}}{10^{-6}~\mu{\rm m}~{\rm yr}^{-1}~{\rm cm}^3}\right)^{-1}~{\rm yr},
\label{eq:t_sput}
\end{eqnarray}
where $Y_{\rm tot}$ is the sputtering yield. For both silicate and carbonaceous dust, the sputtering yield hovers around $\sim$10$^{-6}$ for $10^6$ K $\lesssim$ $T$ $\lesssim$ 10$^9$ K, and drops precipitously below $\sim$10$^6$ K \citep{NOZAWA2006,HU2019}. 

This timescale is shortened by factors of $\sim$2-3 in SN environments where non-thermal sputtering is important \citep[e.g.,][]{HU2019}. The threshold temperature below which sputtering becomes inefficient corresponds to a shock velocity $v_s \simeq 300~{\rm km~s}^{-1}~(T_s/10^6~{\rm K})^{1/2}.$
It is not clear at present to what extent dust is able to survive faster shocks \cite[e.g.,][]{ARENDT2010,LAKICEVIC2015,TEMIM2015,DOPITA2016}, although the presence of dust behind the reverse shock in some SN remnants \citep[e.g.,][]{KOCHANEK2011,MATSUURA2019} and in galaxies in general \citep{GALL2018} suggests that dust is either more resilient than originally predicted \citep[e.g.,][]{SILVIA2010,BISCARO2016}  or it reforms rapidly behind the shocks \citep{HUMPHREYS2012,SEALE2012,GALL2014} despite models predicting the contrary \citep[e.g.,][]{BISCARO2014}.

In a galaxy with a powerful nuclear source of UV radiation like an AGN, dust sublimation will also be a factor. Dust will sublimate if located within
\begin{eqnarray}
r_{\rm subl} = 1.3 \left(\frac{L_{\rm UV}}{10^{46}~{\rm
    erg~s}^{-1}}\right)^{1/2}~\left(\frac{T}{1500~K}\right)^{-2.8}~{\rm pc}
\label{eq:r_subl}
\end{eqnarray}
from the AGN \citep[][]{BARVAINIS1987,NETZER2013}. This expression assumes graphite grains with $a$ = 0.05 $\mu$m and an evaporation temperature of $\sim$1500 K. Silicate grains have a lower evaporation temperature of $\sim$800-1500 K, depending on the grain size and exact composition, resulting in a larger sublimation radius \citep{KIMURA2002}.

\section{Best-Studied Cases of Cool Outflows} \label{sec:best_cases}

\subsection{The Milky Way and the Magellanic Clouds} \label{sec:MW_MC}

\subsubsection{The Milky Way} \label{sec:MW}

The Milky Way is not only the nearest laboratory at our disposal to study cool outflows, it is also an excellent test case for quenched galaxies since it lies squarely in the ``green valley'' \citep{MUTCH2011,BLAND-HAWTHORN2016} and has not undergone any significant galaxy mergers in the past $\sim$8$-$11 Gyr \citep[e.g.,][]{BELOKUROV2018} that would have erased the signatures of fossil outflows or revived the star formation activity by bringing new fuel for the next generation of stars.  

\begin{figure}[htb]
\begin{center}
\includegraphics[width=1.0\textwidth]{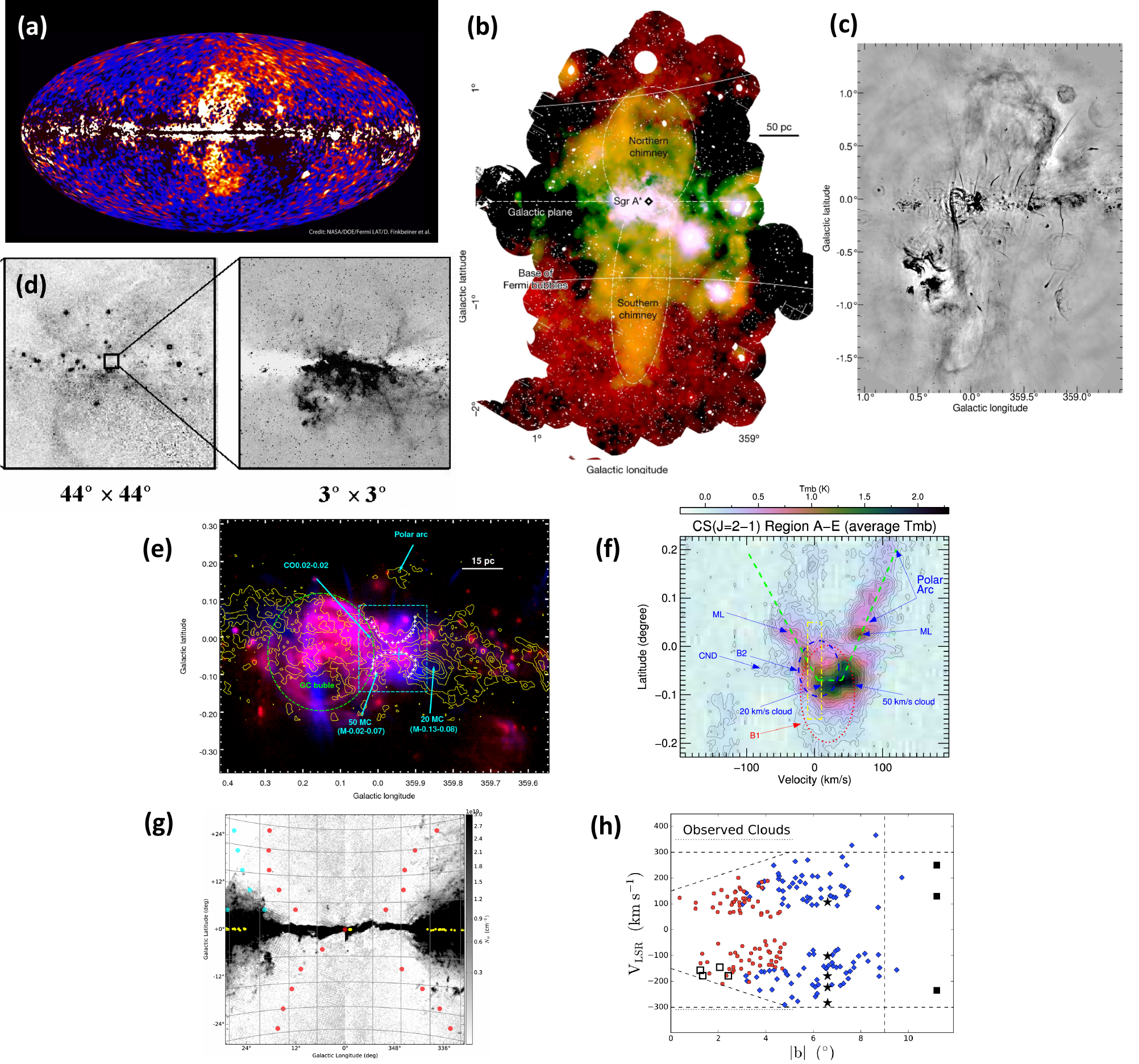}
\end{center}
\caption{The multi-phase outflow of the Milky Way. (a) All-sky gamma-ray emission map showing the $\pm$ 55$^\circ$ Fermi Bubbles. (b) X-ray lobes extending $^{+1.0}_{-1.5}$ degrees from the Galactic plane. (c) Radio structure approximately on the same scale as in panel (b). (d) X-shaped 1.5 keV emission (left; 44$^\circ$ $\times$ 44$^\circ$ centered on the GC) and inner bipolar dusty structure at 8.3 $\mu$m (right; 3$^\circ$ $\times$ 3$^\circ$). (e) CS(2-1) polar arc and hourglass structure in the central 0.8$^\circ$ $\times$ 0.6$^\circ$. (f) Vertical ($\pm$ 0.2$^\circ$) velocity structure of the CS(2-1) polar arc, averaged over Galactic longitudes $l$ = [$-$0.18, 0.07]. (g) Deficiency in H~I 21-cm emitting gas in the central 50$^\circ$ $\times$ 50$^\circ$. (h) Evidence of acceleration among the high-velocity H~I clouds in the central 20$^\circ$ $\times$ 20$^\circ$ around the GC. Images reproduced with permission from (a) \citet{SU2010}, (d) \citet{BLAND-HAWTHORN2003}, (e, f) \citet{HSIEH2016}, (g) \citet{LOCKMAN2016}, (h) \citet{LOCKMAN2019}, copyright by AAS; (b) \citet{PONTI2019}, (c) \citet{HEYWOOD2019}, copyright by the author(s).}
\label{fig:mw}
\end{figure}

The existence of a large-scale outflow centered on the Galaxy center \citep[GC; $d$ = $R_0$ = 8.18 kpc, 1 pc = 25.2\arcsec;][]{GRAVITYCOLLABORATION2019} has been suspected for many years. Long before the Fermi discovery of two giant $\sim$ 12-kpc 1-100 GeV $\gamma$-ray emitting bubbles extending $\sim55^\circ$ on either side of the Galactic center \citep[Fig.~\ref{fig:mw}a;][]{SU2010,DOBLER2010,ACKERMANN2014} and the earlier discovery of a microwave ``haze'' in the data from the Wilkinson Microwave Anisotropy Probe \citep[\emph{WMAP};][]{FINKBEINER2004,DOBLER2008,PLANCKCOLLABORATION2013}, 
it has been argued that the large radio-emitting shells and filaments seen in the direction of the GC \citep[particularly the $\Omega$-shaped Galactic Center Lobe reported by][]{SOFUE1984} and their X-ray counterparts in the 1.5-keV ROSAT all-sky maps \citep[Fig.~\ref{fig:mw}d;][]{BLAND-HAWTHORN2003} are indicative of a large ($\pm$ 3 kpc) bipolar outflow of warm-hot plasma centered on the GC. 
Recently, it has been shown that the Galactic Center Lobe is part of a larger bipolar radio structure spanning 430 pc centered on the GC \citep[Fig. \ref{fig:mw}c;][]{HEYWOOD2019}. The X-ray emission in the central $\pm2^\circ$ ($\pm$250 pc) has also recently been mapped in exquisite details with XMM-Newton and Chandra \citep[Fig.~\ref{fig:mw}b;][]{PONTI2019} \citep[see also][]{NAKASHIMA2019}. 
Two bipolar lobes are observed, one reaching the base of the northern Fermi Bubble $\sim$160 pc from the GC and the other extending beyond the base of the southern Fermi bubble out to $\sim$250 pc (Fig.~\ref{fig:mw}b).  
The thermal energy of these X-ray structures is modest, $\sim$4 $\times$ 10$^{52}$ ergs, corresponding to a power of $\sim$4 $\times$ 10$^{39}$ erg s$^{-1}$ for a dynamical timescale of $\sim$3 $\times$ 10$^4$ yr \citep{PONTI2019}. Similar numbers are derived from the radio structures \citep{HEYWOOD2019}. This power can in principle be provided by supernovae in the central star cluster \citep{LU2013} or a minor accretion or stellar tidal disruption event \citep{REES1988} onto the SMBH in the GC \citep[$M_{\rm BH}$ = 4.15 $\times$ 10$^6$ M$_\odot$;][]{GRAVITYCOLLABORATION2019}.

In contrast, the power needed to create the Fermi Bubbles is several orders of magnitude larger, of order 10$^{42 - 44}$ erg s$^{-1}$ (or 10$^{56}$--$10^{57}$ ergs in total), depending on the exact duration of the activity and how long ago this energy was injected in the GC. These numbers
favor a Seyfert-like SMBH accretion event in the recent past (a few Myr ago) rather than a starburst event as the energy source for the Fermi Bubbles and related structures \citep[e.g.][]{ZUBOVAS2011,GUO2012,YANG2012,BLAND-HAWTHORN2013,MOU2014,RUSZKOWSKI2014,MILLER2016,BLAND-HAWTHORN2019}.\footnote{ There is also evidence that the AGN in the GC  was $\sim$10$^5$ more active in the recent past ($\sim$10$^{2 -3}$ yrs), although still greatly sub-Eddington ($L/L_{\rm Edd} \lesssim 10^{-5}$), based on the detection of strong fluorescent Fe~K$\alpha$ line emission off of molecular clouds near the GC \citep{SUNYAEV1993,KOYAMA1996,PONTI2010}.} The scenario in which slow but sustained ($\gtrsim 10^8 - 10^9$ yr) energy injection from protracted star formation activity in the GC is responsible for the Fermi Bubbles \citep[e.g.,][]{CROCKER2011,CROCKER2015} appears to be inconsistent with the energy injection rate inferred from the X-ray properties of the MW hot halo \citep[$\sim$2 $\times$ 10$^{42}$ erg s$^{-1}$,][]{MILLER2016} and the spatially resolved stellar population within the inner $\sim$500 pc of the GC \citep[time-averaged star formation rate $\dot{M_*}$ $\lesssim$ 0.1 $M_\odot$ yr$^{-1}$ over the last 10$^6 - 10^{10}$ yr, implying an energy injection rate $\lesssim$ 10$^{40}$ erg s$^{-1}$; e.g.,][]{YUSEF-ZADEH2009,BLAND-HAWTHORN2013,BARNES2017}. While the exact value of this average star formation rate is still a matter of debate,\footnote{ For instance, \citet{KRUMHOLZ2017b} argue that the average SFR is closer to $\sim$ 1 $M_\odot$ yr$^{-1}$, more in line with the estimated mass supply rate to the CMZ \citep[$\gtrsim$ 1 $M_\odot$ yr$^{-1}$,][]{SORMANI2019}.} all estimates fall short by at least an order of magnitude to explain the energy injection rate inferred from the MW X-ray data.

The first evidence for a cooler component associated with this large-scale outflow was reported by \citet{BLAND-HAWTHORN2003},  who 
showed the existence of a limb-brightened bipolar structure of cool ($\sim$20-30 K) dust extending $\pm$1$^\circ$ $\simeq$ $\pm$140 pc from the mid-plane of the Galaxy, coincident with the radio-emitting Galactic Center Lobe structure (Fig.~\ref{fig:mw}d).  The total mass in the shells is $\sim$ 5 $\times$ 10$^6$ M$_\odot$, using a Galactic dust-to-gas ratio, and the kinetic energy is $\sim$ 1 $\times$ 10$^{56}$ ($v_{\rm sh}/100$)$^2$ erg, assuming an uncertain shell expansion velocity of 100 km s$^{-1}$.
The morphology, kinematics, and physical conditions of the molecular gas in the central 250 pc of the Milky Way (the so-called Central Molecular Zone or CMZ) have since been refined considerably using a large number of molecular gas tracers \citep[e.g.,][]{KRUIJSSEN2015,TANAKA2018}. While most of the gas kinematics in the CMZ is dominated by rotation on closed elliptical orbits and radial inflow along the stellar bar \citep{SORMANI2015,KRUMHOLZ2017b,ARMILLOTTA2019}, or open streams and spiral arms \citep{HENSHAW2016}, it is clear that a number of features in this region have anomalous kinematics \citep{HENSHAW2016,OKA2019,YUSEF-ZADEH2019}.

One of these features is the extraplanar ``polar arc'' detected by \citet{HSIEH2016} in the CS(2-1) and CS(1-0) lines (Fig.~\ref{fig:mw}e-f). This feature extends up to $\sim$ 30 pc above the GC and shows a radial velocity gradient of $\sim$5.5 km s$^{-1}$ pc$^{-1}$ (from 0 to 165 km s$^{-1}$), suggestive of gas acceleration. This feature merges at the base of the Galactic disk with a ``molecular loop'' with opposite (blueshifted) velocities, and has elevated CS (2-1)/CS (1-0) ratios that are clearly distinct from that of the rotationally-dominated central nuclear disk (CND). Another, fainter bubble is seen $\sim$7 pc below the GC with apparent expansion velocity of $\sim$50 km s$^{-1}$.  The scale and velocities of these anomalous features point to an expanding outflow with a timescale of a few $\sim$ 10$^5$ yr, slightly shorter than the estimate of \citet{BLAND-HAWTHORN2003}. The CS(2-1) material at systemic velocity also shows a remarkable hourglass geometry suggestive of a biconical ($\pm$6 pc) cavity evacuated by a recent outflow. In contrast, the molecular gas on smaller scales shows surprisingly little evidence of outflowing kinematics \citep{GOICOECHEA2018}. 
Similarly, there is little to no evidence for outflowing motion in the material making up the warm ionized gas streamers traced for instance by [Ne II] 12.8 $\mu$m \citep{LACY1991,ZHAO2009,IRONS2012}. Clearly, the event that took place a few 10$^5$--$10^6$ yrs ago is not active anymore. 

In parallel to this effort to map the large-scale molecular outflow in the GC, a number of studies have searched for, and successfully detected, the neutral and warm-ionized gas phases entrained in this outflow. The inner 3 kpc of the Milky Way disk have long been known to be deficient in diffuse HI 21-cm emitting gas \citep{LOCKMAN1984},  and recent work by \citet{LOCKMAN2016} has revealed an anti-correlation between HI and $\gamma$-ray emission suggesting a physical connection between the Fermi Bubbles and the HI central void (Fig.~\ref{fig:mw}g). A population of $\sim$200 compact high-velocity HI clouds has been detected within $\pm$1.5 kpc of the Galactic plane, centered on the GC \citep{MCCLURE-GRIFFITHS2013,DITEODORO2018,LOCKMAN2019}. \citet{DITEODORO2018} have argued that the kinematics of these clouds do not follow Galactic rotation but are instead consistent with a simple GC-centered biconical volume-filling outflow model with a constant radial velocity of $\sim$330 km s$^{-1}$ and an opening angle $>$ 140$^\circ$. These clouds show a broad range of properties ($r$ = 10--40 pc, $N_{\rm HI}$ = 10$^{18.2}$--10$^{19.7}$ cm$^{-2}$, $n_{\rm HI}$ = 0.3--25 cm$^{-3}$, M$_{\rm HI}$ = 10--10$^5$ M$_\odot$, assuming that they are at the distance of the GC), adding up to a total mass of $\sim$10$^6$ M$_\odot$ and a kinetic energy of $\sim$1.6 $\times$ 10$^{54}$ erg. The best-fitting model implies constant neutral-gas mass outflow rate of $\sim$ 0.1 M$_\odot$ yr$^{-1}$ and kinetic energy injection rate of 5 $\times$ 10$^{39}$ erg s$^{-1}$ over the past $\sim$10 Myr. While these numbers may be off by factors of a few, given new evidence of acceleration of these HI clouds \citep[Fig.~\ref{fig:mw}h;][]{LOCKMAN2019}, it is clear that this H~I outflow is considerably less energetic than that associated with the Fermi Bubbles, and may in principle be driven by a past starburst rather than a recent AGN episode. 

Using UV-bright background AGN and high Galactic-latitude stars projected behind, near, or in front of, the Fermi Bubbles, FUSE and HST have provided a complementary view of the large-scale outflow in the Milky Way over $R$ $\sim$ 1--6.5 kpc \citep{KEENEY2006,BOWEN2008,ZECH2008,FANG2014,FOX2015,BORDOLOI2017b,SAVAGE2017,KARIM2018}. Blueshifted and redshifted HVCs with maximum absolute LSR velocities that decrease with increasing Galactic latitude $\vert b  \vert$ have been detected in absorption in both low- and high-ionization species in the direction of the Fermi bubbles. These data are consistent with a decelerating multi-phase outflow with initial (launch) velocity of $\sim$ 1000--1300 km s$^{-1}$, where the cool and warm clouds are entrained \emph{within} the flow \citep{BORDOLOI2017b}. While the implied launch velocities of the UV-detected clouds exceed the velocities of the HI 21-cm clouds in the HI model of \citet{DITEODORO2018}, the dynamical time scale, or age, of the UV-detected outflow is similar ($\sim$6-9 Myr) to that of the H~I outflow, pointing to a common outflow event. \citet{BORDOLOI2017b} estimate an average mass outflow rate of $\gtrsim$ 0.2 $M_\odot$ yr$^{-1}$ and a total mass of cool gas entrained in both Fermi Bubbles of $\gtrsim$ 2 $\times$ 10$^6$ M$_\odot$, assuming [Si/H] $\approx$ [O/H] $\approx$ $-$0.5, no correction for depletion onto dust grains, and mirror symmetry between the bubbles \citep{KARIM2018}.  Using an outflow velocity of 1300 km s$^{-1}$, the total kinetic energy associated with this outflow is $\sim$ 6 $\times$ 10$^{55}$ erg, and the inferred energy injection rate is $\gtrsim$ 2 $\times$ 10$^{41}$ erg~s$^{-1}$. The energy source (AGN or starburst) of this outflow is again ambiguous, e.g. a star formation episode of SFR $\simeq$ 1 $M_\odot$ yr$^{-1}$ sustained over $\lesssim$ 5 Myr \citep[e.g.,][]{KRUMHOLZ2017b} may be capable of driving this outflow. Moreover, these numbers remain uncertain since the small number (8) of sightlines clearly passing through the Fermi bubbles prevents a detailed 3D modeling of the outflowing gas kinematics. 

Overall, most of the neutral and molecular gas entrained in the outflow will not escape the Milky Way. Instead, this material will participate in a large-scale galactic fountain \citep{BREGMAN1980,ARMILLOTTA2019}, feeding the Milky Way CGM \citep{BLAND-HAWTHORN2016,HODGES-KLUCK2016b,ZHENG2019,WERK2019,BISH2019,FOX2019}, before falling back onto the outer disk of the Galaxy, and providing new fuel for star formation. This may be a common occurrence in quenched galaxies, but the very subtle signatures of the cool, warm, hot, and relativistic fluids in this tenuous outflow will generally be challenging to detect in external galaxies.

\subsubsection{The Magellanic Clouds} \label{sec:MC}

\begin{figure}[htb]
\begin{center}
\includegraphics[width=1.0\textwidth]{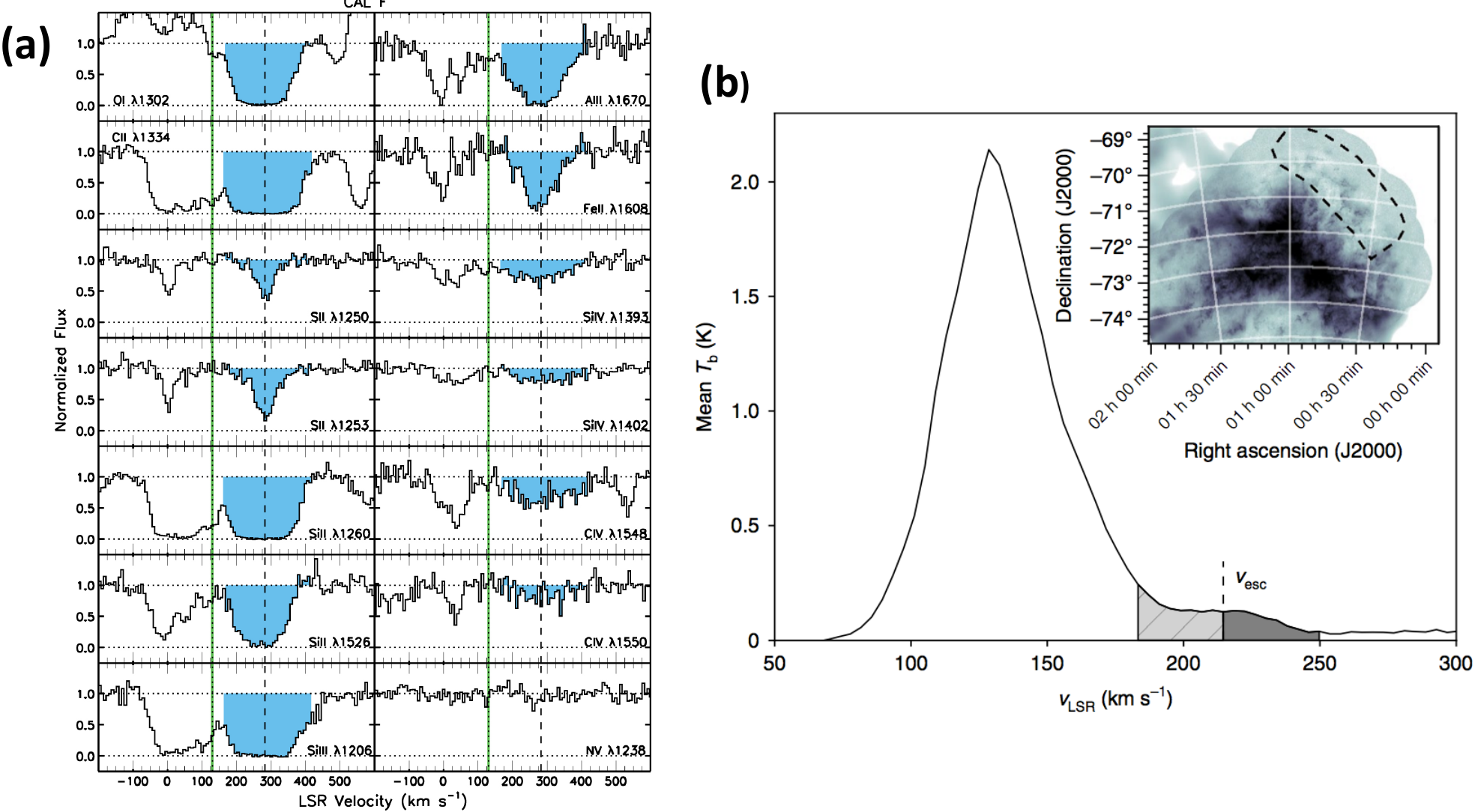}
\end{center}
\caption{Cool outflows in (a) LMC and (b) SMC. The blue shaded region in (a) marks the kinematic extent ($\sim$ $\pm$100 km~s$^{-1}$) of the multi-phase outflow in LMC, traced in both low- and high-ionization absorption lines. In (b), the inset shows the peak H~I intensity image of the SMC and the dashed box indicates the region used to generate this spectrum. The dark grey shaded region in the spectrum indicates the H~I gas with velocity in excess of the escape velocity. Images reproduced with permission from (a) \citet{BARGER2016}, copyright by AAS; and (b) \citet{MCCLURE-GRIFFITHS2018}, copyright by the authors.}
\label{fig:lmc_smc}
\end{figure}

The Large Magellanic Cloud (LMC; $d \simeq$ 50 kpc) is a cauldron of star formation activity likely triggered by the tidal interactions with the Small Magellanic Cloud (SMC; $M_{\rm SMC} \simeq 1/10~M_{\rm LMC}$) and the Milky Way ($M_{\rm MW}$ $\simeq$ 100 $M_{\rm LMC}$). Fast (up to $\sim$200 km s$^{-1}$) ionized outflow events have long been known to take place across the disk of the LMC, particularly around the giant H~II region 30 Doradus powered by the mini-starburst centered on super star cluster R136 \citep[e.g.,][]{CHU1993,REDMAN2003}. The spectacular filamentary and bubbly appearance of the H~I gas distribution in the LMC \citep[e.g.,][]{KIM1999,KIM2003}, and the connection between the H~I shell velocities and the presence of H~II regions and OB associations, clearly point to small-scale injection of energy from supernovae and stellar associations into the ISM of the LMC, but the question of whether these separate outflow events combine to drive a large-scale outflow has been a matter of debate over the years.  For instance, \citet{STAVELEY-SMITH2003} detected a complex of HI 21-cm line emitting HVCs at $\sim$ $-$100 km s$^{-1}$ from  systemic velocity that project onto the positions of giant HI voids. Given the nearly face-on orientation of the disk of the LMC \citep[$i = 22^\circ$;][]{KIM1998},  this is strong evidence that this material has been ejected from the disk of the LMC, but it is not clear that these HVCs are part of a large-scale outflow rather than separate events associated with individual bursts of star formation. UV studies with HST and FUSE of stars embedded in the disk of the LMC \citep{HOWK2002,LEHNER2007,LEHNER2009,PATHAK2011} have successfully answered this question.  The high detection rate of blueshifted F/UV absorption lines found in these studies implies that the high-velocity material covers most of the galaxy, including both quiescent and active regions of star formation. These data also indicate that the high-velocity material has a multi-phase structure with neutral (O I, Fe II), weakly ionized (Fe II, N II), and highly ionized (O VI) components, and shows evidence of dust but no molecules. A recent analysis by \citet{BARGER2016} of the spectrum of an AGN \emph{behind} the LMC has shown that the outflow is kinematically symmetric with respect to the disk of the LMC with velocities reaching $\sim$110 km s$^{-1}$ (after deprojection), perhaps large enough for some of the gas to escape the LMC (Fig.~\ref{fig:lmc_smc}a; from Eq. (\ref{eq:v_esc}), $v_{\rm esc} \simeq 1.4-3~ v_{\rm circ} \simeq 90$--$200$ km s$^{-1}$, where $v_{\rm circ} = 65$ km s$^{-1}$; Kim et al. 1998) and become part of the Milky Way halo. \citet{BARGER2016} estimate a total (neutral + ionized) outflowing gas mass of $\sim$1.3 $\times$ 10$^6$ M$_\odot$ and a mass outflow rate of 0.04 M$_\odot$ yr$^{-1}$, assuming a typical outflow velocity of 50 km s$^{-1}$. This mass outflow rate is comparable to the star formation rate of the LMC, $\lesssim 0.2$ M$_\odot$ yr$^{-1}$ \citep{HARRIS2009}.

The latest addition to the list of MW companions with cool outflows is the Small Magellanic Cloud (SMC; $d$ $\simeq$ 60 kpc). \citet{MCCLURE-GRIFFITHS2018} recently reported the detection of a complex of H~I 21-cm comet-shaped head-tail clouds, looping filaments, and compact ($\sim$10-30 pc) high-velocity clouds, extending $\gtrsim$ 2 kpc from the star-forming bar of SMC, and with measured velocities that deviate by up to 100 km s$^{-1}$ with respect to the portion of the galaxy nearest to these features (Fig.~\ref{fig:lmc_smc}b). The total gas mass of this complex adds up to at least $\sim$1.3 $\times$ 10$^7$ M$_\odot$ or about $\sim$ 3\% of the total atomic gas in this galaxy. The implied mass outflow rate, calculated without correcting for possible projection effects, is $\sim$ 0.2-1.0 M$_\odot$ yr$^{-1}$, up to an order of magnitude larger than the star formation rate of the SMC. Given the inclination of the SMC bar \citep[$i \sim$ 40$^\circ$;][]{STANIMIROVIC2004} and the escape velocity of the SMC \citep[$v_{\rm esc} \simeq 1.4-3~v_{\rm circ} \simeq 85-180$ km s$^{-1}$, where $v_{\rm circ}$ = 60 km s$^{-1}$;][]{STANIMIROVIC2004}, \citet{MCCLURE-GRIFFITHS2018} have argued that nearly 20-40\% of the outflowing material may be escaping out of the SMC. This may be considered a lower limit since tidal stripping by the LMC and Milky Way, and ram-pressure forces associated with the fast \citep[$\sim$300 km s$^{-1}$;][]{KALLIVAYALIL2006} motion of the SMC through the Milky Way halo, will promote the removal of any material that is outside of the host's main body. The implied dynamical timescale of the outflow ($\sim$30-60 Myr) coincides roughly with the age of the recent burst of star formation in the portion of the bar nearest of the HI features, so it may very well have been at the origin of this outflow event. The detection of diffuse H$\alpha$ and X-ray emission \citep{WINKLER2015,STURM2014}, as well as strong and broad O~VI absorption features \citep{HOOPES2002} throughout this region, indicates that the cool outflow is also accompanied by a warm-hot ionized gas phase. The recent detection by \citet{DITEODORO2019} of molecular gas in two of the outflowing H~I clouds indicates that a cold component is also associated with this outflow.

\subsection{Starburst galaxies: M 82 and NGC 253} 
\label{sec:M82_N253}

\begin{figure}[htb]
\begin{center}
\includegraphics[angle=270,width=0.8\textwidth]{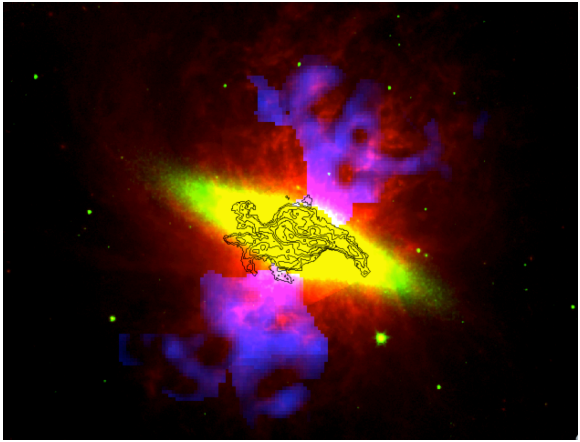}
\end{center}
\caption{Cool outflows in M~82. The red, green, and blue planes correspond to IRAC 8.0 $\mu$m (PAHs), IRAC 3.6 $\mu$m (stars), and H~I 21~cm line emission masked along the minor axis to show the velocities corresponding to the outflow. Observations are from \citet{ENGELBRACHT2006} and \citet{LEROY2015a}, respectively. The contours are the extended CO emission from the interferometric observations by \citet{WALTER2002}.}
\label{fig:m82_n253}
\end{figure}

The nearest starburst galaxies, M~82 and NGC~253, are both hosts of multi-phase outflows. In both cases the driver of the outflow is star formation activity: neither M~82 nor NGC~253 host energetically important AGN. The proximity of both galaxies enables highly detailed views of the outflowing gas (Figs. \ref{fig:intro_outflows} and \ref{fig:m82_n253}).

\subsubsection{M~82} \label{sec:M82}

M~82, a small galaxy part of the M~81 group, is the archetypal example of a starburst galaxy. The group is strongly interacting with H~I gas tails connecting M~81 to the nearby M~82, NGC~3077, and NGC~2976 satellite galaxies \citep{YUN1994,DEBLOK2018}. The interaction has likely triggered the starburst activity in M~82. Modelling of near and mid-infrared observations indicates M~82 has a complex history \citep{FORSTERSCHREIBER2003}. It likely underwent a powerful starburst in its central 500~pc 8--15~Myr ago peaking at $160\,M_\odot$\,yr$^{-1}$, followed by a second burst in a circumnuclear ring (likely caused by a bar driving gas into the core) 4--6~Myr ago peaking at 40~M$_\odot$\,yr$^{-1}$. The average recent star formation rate is $13$--$33\,M_\odot$\,yr$^{-1}$, depending on the contribution from low mass stars \citep{FORSTERSCHREIBER2003}. 

The M~82 outflow is prominent in X-rays. Soft X-ray observations reveal an extended structure perpendicular to the major axis of the galaxy, extending several arcminutes north and south of the nucleus \citep[e.g.,][]{STRICKLAND1997,STRICKLAND2009}. Modeling of these data result in estimates for the mass loss rate of the hot wind fluid of $1.4$--$3.6\,M_\odot$\,yr$^{-1}$, and a terminal velocity of 1400 to 2200 km\,s$^{-1}$, much larger than the escape velocity. The inferred hot plasma temperatures are 3 to $8\times10^7$~K, much hotter than a soft X-ray emitting plasma despite the cold mass loading in the central regions.

The wind is bright in warm, ionized, H$\alpha$ emitting gas. This includes several filaments associated with the disk and lower wind \citep{SHOPBELL1998}, and a large nebular shell-like feature only weakly connected to the galaxy called the ``H$\alpha$-cap'', located at 12~kpc over the midplane of M~82 and coincident with soft X-ray emission \citep{DEVINE1999,LEHNERT1999}. This feature is likely caused by a  $40$--$80$~km\,s$^{-1}$ shock driven by the wind \citep{MATSUBAYASHI2012}. In any case, the wind appears to move fast enough to escape, and it suggests the starburst has been active for $\gtrsim10$ Myr and possibly much longer \citep{STRICKLAND2009}.

The M~82 outflow is also prominent in far-infrared indicators of cold, dusty phases. The wind has been imaged in polycyclic aromatic hydrocarbon emission out to 6~kpc away from the midplane with varying ratios between PAH features, suggesting some processing of the PAH carriers (perhaps grain shattering) takes place in the wind \citep[PAH;][]{ENGELBRACHT2006,YAMAGISHI2012,BEIRAO2015}. Modeling of the dust properties across the near-UV, optical, and infrared finds that the UV/optical emission in the wind cone is caused by scattering from dust grains entrained in the flow \citep{HUTTON2014}, and it also points to grain processing. The dependence of the scattering on wavelength is best fit by either a steep power-law of grain size, or by a model with a flatter power-law and a maximum size for the grains, suggesting that large grains are missing from the entrained material. Polarization in the far-infrared dust emission in both the disk and the outflow is observed to be a few percent at 53 $\mu$m to a few tenths of a percent at 154 $\mu$m \citep[where the central region is much more depolarized,][]{JONES2019a}. The polarization traces an ordered component of the magnetic field that is mostly oriented along the axis of the outflow, perpendicular to the minor axis of M~82. Spectropolarimetric measurements in the optical emission lines suggest that the dust is moving substantially slower than the ionized gas in the wind of M~82  \citep[][]{YOSHIDA2011,YOSHIDA2019}.

Near infrared H$_2$ ro-vibrational transitions 
with $E/k$ of a few thousand degrees Kelvin have been observed in the outflow, likely collisionally excited by a combination of UV-heating from the central starburst and shocks \citep{VEILLEUX2009a}. Lower energy mid-infrared H$_2$ rotational quadrupole transitions are also observed, with ratios that suggest they are excited by relatively slow velocity shocks with $v_{shock}\sim40$~km\,s$^{-1}$ \citep{BEIRAO2015}. The entrained warm molecular gas mass observed in the outflow is estimated to be $0.5$ to $1.7\times10^7$~M$_\odot$. 

The colder phases of the M~82 wind are also bright in CO and H~I emission. \citet{WALTER2002} imaged the inner regions of the starburst and outflow, revealing a complex geometry of molecular streamers including faint emission along the outflow. Modeling of the CO excitation suggests densities $n\sim10^3$~cm$^{-3}$ and warm temperatures of up to 100~K \citep{WEISS2005a}. The complexity of the cold neutral wind of M~82 is also reflected in the multi-component Na~D absorption line profiles \citep{SCHWARTZ2004}.
\citet{LEROY2015a} find CO emission out to 3~kpc away from the starburst, and significant HI emission associated with the outflow out to 5~kpc in the north and 10~kpc in the south. The kinematics show line splitting, likely indicative of a tilted bi-conical geometry. The speed of the molecular outflow is critically dependent on the inclination assumed, but likely $v\sim300-450$~km\,s$^{-1}$. \citet{LEROY2015a} find a decreasing mass loss rate in the outflow as a function of distance to the mid-plane, rapidly decreasing for H$_2$ traced by CO, and more slowly decreasing for H~I and for total gas as inferred from the dust continuum. The total mass outflow rate in the colder phases seems to be considerably higher than that in the hot wind fluid, as modeled by \citet{STRICKLAND2009}. \citet{MARTINI2018} point out that the H~I kinematics are not compatible with an outflow launched only from the nuclear region, but instead require launching over a $\sim1$~kpc region that matches better the extent of the starburst. The kinematics are also not compatible with H~I formation through cooling from the much faster X-ray emitting hot phase. Moreover, \citeauthor{MARTINI2018} find that the H~I is decelerating along the outflow: this is particularly well constrained on the more extended south side emission. The deceleration experienced by the H~I is too large to be due to gravity alone, suggesting that the ejected material is experiencing other drag forces, perhaps from pre-existing material associated with tidal debris. This casts further doubt on what fraction, if any, of the cold material may escape the system \citep{MARTINI2018}. 

\subsubsection{NGC 253} \label{sec:N253}

NGC~253, the Sculptor galaxy, is another prototypical example of a nearby starburst \citep[$d\sim3.5$ Mpc;][]{REKOLA2005}. It is a barred galaxy and its starburst is circumnuclear, meaning it is highly concentrated in its central 200--300~pc. The star formation rate in this region is $\sim2$--$3\,M_\odot$\,yr$^{-1}$ \citep{OTT2005,LEROY2015b,BENDO2015}, and there is no evidence for an energetically important AGN. High resolution submillimeter-wave imaging reveals a complex of massive clusters in formation in the central 200~pc \citep{LEROY2018}, which encompass a large fraction of the total nuclear star formation activity. A bi-conical wind emerges out of this region, apparent in X-rays \citep{STRICKLAND2000,STRICKLAND2002}, H$\alpha$ emission and other ionized lines \citep{SHARP2010,WESTMOQUETTE2011}, neutral sodium emission \citep{HECKMAN2000}, polycyclic aromatic hydrocarbon emission \citep{TACCONI-GARMAN2005}, OH absorption \citep{TURNER1985,STURM2011}, and spatially resolved molecular emission \citep[Fig.~\ref{fig:intro_outflows}e;][]{BOLATTO2013b,WALTER2017,ZSCHAECHNER2018,KRIEGER2019}. The geometry of the wind is well characterized through H$\alpha$ observations, emerging approximately perpendicular to the plane of the galaxy and almost in the plane of the sky (inclination of $12^\circ$) with an opening angle of $\sim60^\circ$. NGC\,253 has prominent lobes associated with its outflow, extending to distances of $\sim10$ kpc away from the central region along the galaxy minor axis.  The extended emission is observed in X-rays, far-UV, H$\alpha$, and radio-synchrotron \citep{BAUER2007,KAPINSKA2017}.

The molecular outflow in NGC\,253 has a CO-emitting mass of $\sim3-4\times10^7$\,M$_\odot$, and molecular mass loss rates estimated to be $\sim20$\,M$_\odot$\,yr$^{-1}$ \citep{ZSCHAECHNER2018,KRIEGER2019}. By comparison, the mass of the H$\alpha$-emitting phase is $\lesssim10^7$\,M$_\odot$ \citep{WESTMOQUETTE2011}, and the inferred mass loss rate is $\sim4$\,M$_\odot$\,yr$^{-1}$ \citep{KRIEGER2019}. The mass-loss rate in the hot, X-ray emitting fluid is estimated to be $<2.2$\,M$_\odot$\,yr$^{-1}$ for reasonable flow parameters \citep{STRICKLAND2000}. As a consequence the cold, slower phase carries appears to dominate the mass loss from the central starburst region. 

Both the H$\alpha$ and the CO-emitting material appear to be located in the walls of the hot, X-ray emitting outflow, at least in the inner 0.5 kpc of the flow. \citet{WESTMOQUETTE2011} report a physical outflow speed for the H$\alpha$ emitting material of $\sim200-300$\,km\,s$^{-1}$. They measure a velocity gradient that they interpret as acceleration, with a magnitude of $\sim0.6$\,km\,s$^{-1}$ per parsec along the axis of the outflow cone. \citet{WALTER2017} measure a very similar (within de-projection uncertainties) velocity gradient of $\sim1$\,km\,s$^{-1}$ per parsec for the molecular material along their brightest filament. It remains unclear what fraction of this material reaches escape speed, if any. The existence of the extended lobes suggests that at least part of the flow can reach very far distances from the starburst regions. 

\citet{WALTER2017} identify several molecules in the molecular outflow, including species usually associated with dense gas such as HCN. They also point out that the line ratios, to the extent they can measure, are similar in the filamentary ejecta and in the central starburst. This suggests the molecular material may be entrained from the central regions, rather than being formed \textit{in situ} through condensation which would require very fast chemistry to recreate the implied molecular abundances. \citet{WALTER2017} also calculate the expected acceleration due to radiation pressure likely experienced by the material in the brightest filament. Their conclusion is that radiation pressure cannot be the main source of momentum to explain the observed ejection velocity, as it would require an extremely favorable and unlikely geometry.

\subsection{Seyfert Galaxy: NGC 1068}

\label{sec:N1068}

NGC~1068 is a nearby ($d$ = 14 Mpc, 1\arcsec\ = 70 pc) galaxy with a nuclear bar and a massive pseudo-bulge \citep{KORMENDY2013} that is often regarded as the prototypical Seyfert 2 galaxy.  It has a prominent starburst ring of $\sim1$--$1.5$ kpc radius and a 200 pc molecular circumnuclear disk surrounding the AGN \citep[e.g.,][]{SCHINNERER2000}.
Non-circular motions as traced by CO and also near-IR H$_2$ in the inner regions can be interpreted as due to in- and outflow motions \citep{GALLIANO2002,DAVIES2008,GARCIA-BURILLO2010, KRIPS2011}.  The nuclear gas is dense ($n>10^{5-6}$ cm$^{-3}$) and can be probed by polar molecules such as HCN and high-$J$ CO emission. Studies suggest that the chemistry and physical conditions of the gas is strongly impacted by the AGN,  either by X-rays or shocks \citep[e.g][]{TACCONI1994,STERNBERG1994,USERO2004,AALTO2011,VITI2014,GARCIA-BURILLO2017}.
The nucleus is launching a radio jet that is (on small scales) perpendicular to the nuclear water maser disk \citep{GALLIMORE1997} and creates a bow shock on 100 pc scales.

While the AGN in NGC 1068 is not as powerful as quasars (which are thought to drive outflows energetic enough to quench star formation in their massive hosts), it is much closer than any quasar (e.g. the closest quasar, Mrk~231 discussed in the next section, is more than ten times more distant) and therefore it can be investigated in far greater detail. Moreover, it offers the possibility of investigating both the outflow associated with the radio jet and the radiatively driven component of the outflow.

\begin{figure}[htbp]
\begin{center}
\includegraphics[width=0.90\textwidth]{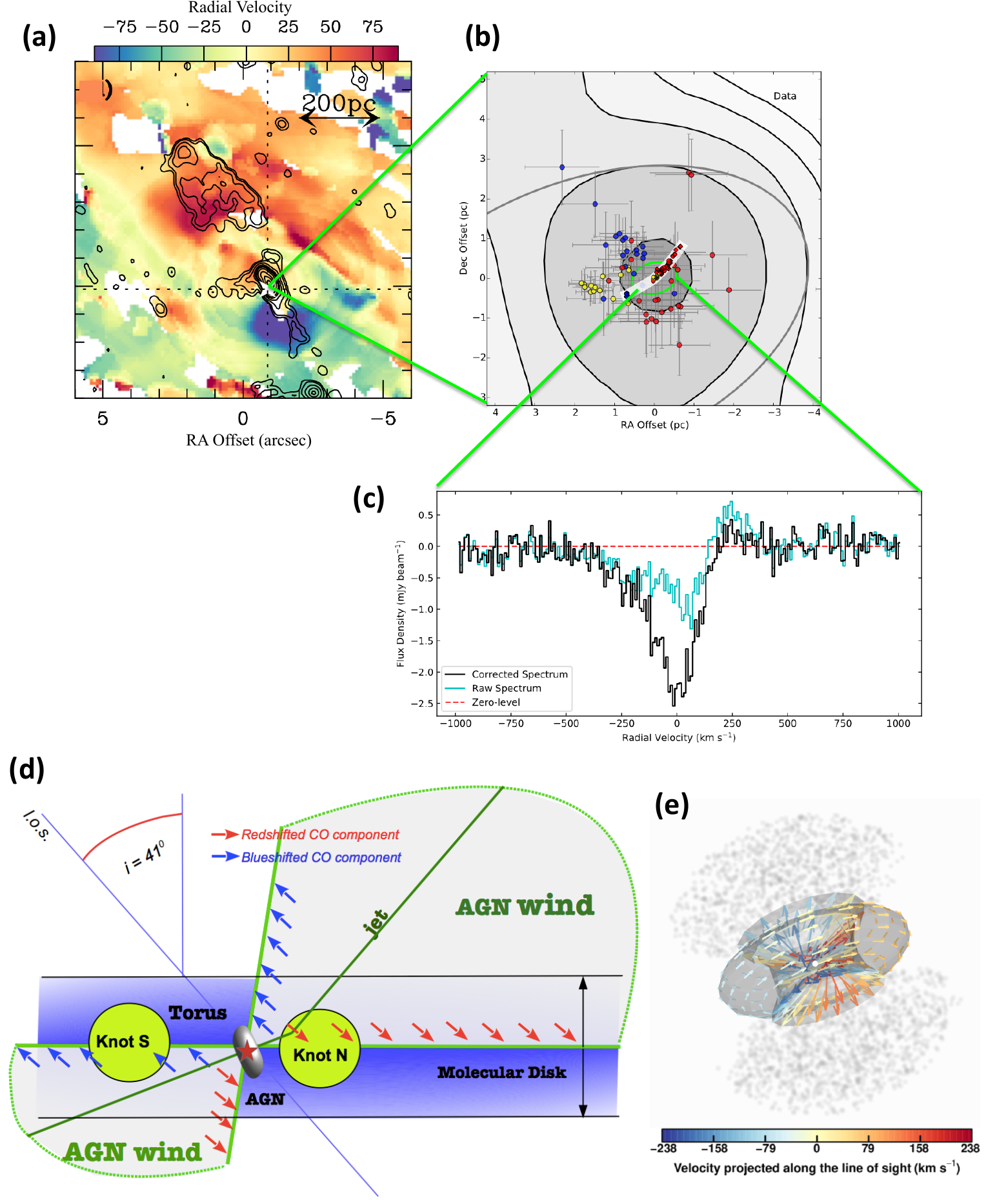}
\end{center}
\caption{Cold outflows in NGC~1068.  
(a) Residual velocity field (after subtracting the rotation field) traced by the CO(3-2) line, with the 22 GHz radio map overlayed with contours. Adapted from \cite{GARCIA-BURILLO2014}. (b) Location of different kinematics components in the central few pc of NGC~1068; blue $v$ $<$ $-$70~km s$^{-1}$, red $v$ $>$ 70~km~$^{-1}$, yellow $-$70~km~s$^{-1}$ $\le v \le$ 70~km~s$^{-1}$; circles for CO(6-5), diamonds for the nuclear water maser disk. Note that the CO(6-5) outflow has inverted velocity profile with respect to the large-scale molecular outflow. Adapted from \cite{GALLIMORE2016}. (c) HCN(3-2) profile extracted from the central 1.2 pc, at the location of the radio core, showing a clear P-Cygni profile. Adapted from \cite{IMPELLIZZERI2019}. (d) Sketch of the central region of NGC~1068 illustrating the outflow on large (100 pc) scales, driven by the expansion of the bow-shock generated by the radio-jet.   \citep[from][]{GARCIA-BURILLO2019}. (e) Sketch of the kinematic model of the molecular torus described in \citet{GARCIA-BURILLO2019}. Images reproduced with permission from (a) \citet{GARCIA-BURILLO2014} and (d, e) \citet{GARCIA-BURILLO2019}, copyright by ESO; and (b) \citet{GALLIMORE2016} and (c) \citet{IMPELLIZZERI2019}, copyright by AAS.}
\label{fig:n1068_outflow}
\end{figure}

The presence of an ionized outflow has been clearly detected by various authors, through optical, UV and near-IR nebular tracers, in the shape of a biconical or hourglass structure co-aligned with the radio-jet (NE-SW), with velocities of several hundred km~s$^{-1}$, exceeding 1,000 km~s$^{-1}$ in some locations \citep[e.g.][]{POGGE1988,CECIL1990,CRENSHAW2000,TECZA2001,CECIL2002,DAS2006,ZHENG2008,BARBOSA2014,VERMOT2019}. However, one peculiarity is that the high ionization lines and lower ionization lines, tracing denser gas, have opposite velocities. For instance in the NE cone (which is the one on the near side of the disk), [O~III] 5007 \AA\ is blueshifted while [Fe~II] 1.64 $\mu$m is redshifted. This has been interpreted (Fig.~\ref{fig:n1068_outflow}d) as evidence that the bow shock generated by the radio jet accelerates low-density, more highly ionized gas (being directly
exposed to the AGN ionizing radition) out of the galactic plane in our direction, while the bow shock on the far side ploughs into the galactic disk and accelerates denser and less-ionized gas in the opposite direction.

Recent high-resolution ALMA observations of NGC~1068 reveal molecular outflows on what may be several scales. \cite{GARCIA-BURILLO2014} mapped various high transitions of CO, HCN, HCO$^+$ and CS showing motions of dense gas, on scales from 50 to 400 pc, that deviate from rotation (Fig.~\ref{fig:n1068_outflow}a). This emission is consistent with the outflow previously traced by the low ionization tracers, i.e. in the opposite direction of the high ionization transitions (e.g. [O~III] 5007 \AA). One interpretation is that this apparently outflowing molecular gas is also part of the dense gas in the galactic disc that is being accelerated by the oblique bow shock (Fig. \ref{fig:n1068_outflow}d). This molecular outflow is estimated to have $\dot{M} \sim 63$ \msun\ yr$^{-1}$,  which is an order of magnitude higher than the star formation rate at these radii. Nuclear depletion times are short (less than 1 Myr). The inner gas reservoir may be rapidly replenished by inflows from the outer disk, especially given the streaming gas resulting from the torques exerted by the stellar bar. A fast ($\sim$ 400 km s$^{-1}$) molecular outflow has been detected on smaller scales, a few parsecs from the nucleus \citep[Fig.~\ref{fig:n1068_outflow}b;][]{GALLIMORE2016}, but this nuclear molecular outflow has velocities in the opposite direction as that of the molecular outflow on large scales, i.e.\ going out of the disk. This nuclear molecular outflow appears to be disconnected from the molecular outflow on larger, 100-pc scales, and may have a different origin. While the large-scale molecular outflow is associated with gas in the galactic disk accelerated by the bow shock entering into the disc, the nuclear molecular outflow is likely associated with nuclear clouds being directly accelerated by the AGN radiation pressure or ram pressure from the nuclear disk-wind \citep[as in the disk-wind scenario proposed by][]{ELITZUR2006}.

\citet{GARCIA-BURILLO2019} have recently presented a model of the nuclear wind (the outflowing torus) as well as an update of the 2014 jet expansion model (Fig.~\ref{fig:n1068_outflow}d). A vertical component, in addition to the co-planar flow, is now included, that can explain the CO line-splitting and large line widths. The model can also account for the velocity structure of the highly ionized wind. This wind encounters less resistance against expansion above and below the disk mid-plane resulting in opposite velocities with respect to CO. \citeauthor{GARCIA-BURILLO2019} also find that half of the mass of the torus is outflowing (Fig.~\ref{fig:n1068_outflow}e), which means that this AGN feedback on very small scales is also regulating the fueling and
thus likely setting the AGN flickering timescale. 

Interestingly, the pc-scale nuclear wind has also been directly detected as a P-Cygni profile of the HCN(3-2) transition against the nuclear radio core (Fig\ \ref{fig:n1068_outflow}c), therefore implying that the nuclear outflowing gas must be very dense \citep[$\sim 10^7$ cm$^{-3}$;][]{IMPELLIZZERI2019}. The fact that such dense molecular clouds are already outflowing at high velocities ($\sim$ 450 km~s$^{-1}$) so close to the nucleus, where the dynamical timescales (a few thousand years) are far too short for the formation of molecules, favor the scenario of outflowing molecular clouds resulting from acceleration of circumnuclear clouds rather than formation of molecular clouds from gas cooling inside the hot outflow, at least on such small scales. Clearly, in NGC~1068 jet/bow-shock driven molecular outflow and radiation-driven molecular outflow are both present at the same time and both of them are relevant for the evolution of the nuclear region.

\subsection{Quasar: Mrk 231} \label{sec:Mrk231}

Mrk~231 is the nearest quasar known ($z$ = 0.0422; $d$ $\simeq$ 180 Mpc; 1\arcsec\ = 0.867 kpc) and arguably the best example of a ``wind-dominated'' fast-accreting ($L/L_{\rm Edd} \simeq 1$) quasar where the signatures of the AGN-driven outflow are detected on virtually all spatial scales, from the AGN accretion disk to the CGM. As such, Mrk~231 offers a unique laboratory to study a phenomenon that might be more common in the early universe during the epoch of formation and fast growth of SMBHs. Mrk~231 is the end-product of a (locally) rare infrared luminous gas-rich galaxy merger which provided the right conditions to trigger both a luminous AGN and a powerful circumnuclear starburst \citep[$\dot{M}_* \sim 140$ M$_\odot$ yr$^{-1}$; this starburst accounts for about 30\% of the bolometric luminosity of the system; e.g.,][]{VEILLEUX2009b}.

\begin{figure}[htbp]
\begin{center}
\includegraphics[width=0.95\textwidth]{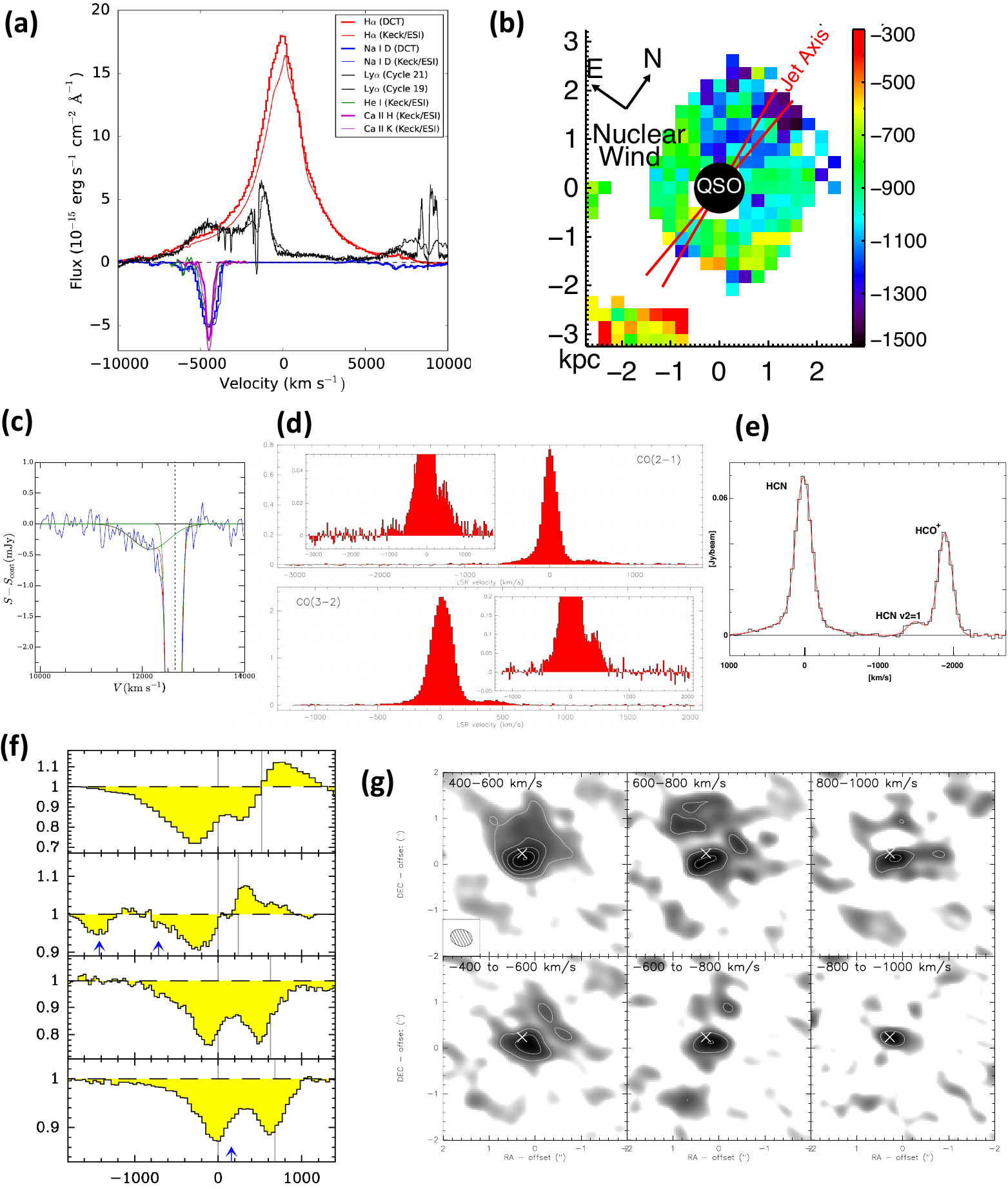}
\end{center}
\caption{The nuclear and large-scale cool outflows of Mrk~231. (a) Broad low-ionization absorption features in the central AGN. (b) Spatially resolved velocity field of the Na I D absorption line. (c) Blue wing in H~I 21-cm absorption line. (d) Blue and red wings in the CO (1-0) and (2-1) line profiles. (e) HCN signatures of the outflow. (f) OH signatures of the outflow. (g) Spatially resolved blue- and red-shifted CO line emission. Images reproduced with permission from (a) \citet{VEILLEUX2016}, (b) \citet{RUPKE2011}, (f) \citet{GONZALEZ-ALFONSO2017b}, copyright by AAS; and (c) \citet{MORGANTI2016}, (d, g) \citet{FERUGLIO2015}, (e) \citet{AALTO2015c}, copyright by ESO.}
\label{fig:mrk231}
\end{figure}

As shown in Figure \ref{fig:mrk231}a, broad spatially unresolved blueshifted absorption line features are detected in the nucleus across the range $\sim$ $-$3500 to $-$6000 km~s$^{-1}$ in several transitions including many low-ionization species \citep[e.g. Na I D 5890, 5896, He I$^*$ 3889 and 10830, Ca II H and K, Mg II 2796, 2803, and Mg I 2853, as well as several NUV Fe II features, hence fitting the rare FeLoBAL category; e.g.,][]{BOKSENBERG1977,SMITH1995,GALLAGHER2002,GALLAGHER2005,RUPKE2002,VEILLEUX2013b,VEILLEUX2016,LEIGHLY2014}. Dust within the outflowing clouds is required to provide shielding for Na I D 
and also explain the blueshifted Ly$\alpha$ and C~IV line emission. The presence of excited-state Fe II UV60, 61, 62, 63, and 78 and He I$^*$ 3889 absorption lines leads to a distance of $\lesssim$2-20 pc between the absorber and the ionizing source and implies a mass outflow rate and kinematic luminosity of the BAL outflow of $\lesssim$10-100 M$_\odot$ yr$^{-1}$ and $\lesssim$10$^{44-45}$ erg s$^{-1}$ \citep{VEILLEUX2016}. Other indicators of outflows in the nucleus include Blazar-like radio flares at 20 GHz, associated with highly relativistic ejecta on pc scale \citep{REYNOLDS2009,REYNOLDS2013}, and the possible (3.5-$\sigma$) detection of a P-Cygni Fe K$\alpha$ 6.7 keV profile, indicative of an ultra-fast ($\sim$ 20,000 km s$^{-1}$) outflow on smaller scales \citep[0.001--0.01 pc;][]{FERUGLIO2015}.

A slower ($\lesssim$1500 km s$^{-1}$), galactic-scale ($\sim$ 0.1--3 kpc) 
cool-gas outflow has been detected in Mrk~231 using multiple techniques (Fig.~\ref{fig:mrk231}b-g): neutral-gas absorption line tracers \citep[Na I~D:][in HI]{RUPKE2002,RUPKE2011,RUPKE2017,MORGANTI2016}, molecular-gas FIR absorption line tracers \citep{FISCHER2010,STURM2011,GONZALEZ-ALFONSO2014,GONZALEZ-ALFONSO2017b}, and molecular-gas mm-wave emission line tracers \citep{FERUGLIO2010,FERUGLIO2015,AALTO2012,AALTO2015c,CICONE2012,ALATALO2015a,LINDBERG2016}.
The inferred cool-gas mass outflow rate ranges from $\sim$  100--200 M$_\odot$ yr$^{-1}$ \citep[Na~I~D tracer;][]{RUPKE2017} to 500--1000 M$_\odot$ yr$^{-1}$ \citep[OH and CO tracers;][]{GONZALEZ-ALFONSO2017b,FERUGLIO2015}, and the corresponding kinetic energy rate of the cool outflow is $\sim$ 0.2\% (Na~I) and $\sim$ 1--3\% of the AGN bolometric luminosity.  Remarkably, the warm and hot ionized components of the large-scale outflow are energetically insignificant in comparison to the cool-gas component, barely detected in H$\alpha$ \citep[e.g.,][]{RUPKE2017} 
and the X-rays \citep{VEILLEUX2014}, respectively. Overall, the energetics of the large-scale outflow strongly suggest that the AGN in Mrk~231 plays a dominant role in driving this powerful outflow event. 

The velocities, masses, and energetics of the neutral and molecular components of the large-scale outflow in this object significantly exceed those discussed thus far in this section. This is not surprising given the extremely powerful source of energy at the center, but this is also due to the remarkably dense and compact merger-induced dust-obscured environment encompassing this energy source \citep[$N$(H$_2$) $\sim$ 1 $\times$ 10$^{24}$ cm$^{-2}$ in the inner $\sim$100 pc;][]{AALTO2015c}, which helps capture an unusually large fraction of the radiative and mechanical energy produced by the quasar. These special conditions may also help explain the other distinctive characteristics of this outflow, including the virtual lack of an ionized component and the fact that the cool outflowing gas clouds are embedded in the flow itself rather than distributed along edge-brightened biconical structures like those in M82 and NGC~253. The cooler molecular portion of the outflow in Mrk~231 is also denser on average (up to $\sim$ 5 $\times$ 10$^5$ cm$^{-3}$) than those in these two objects, and shows velocity-dependent HCN/HNC/HCO$^+$ ratios, signs of chemical differentiation possibly due to shocks \citep[Sect.~\ref{sec:shocks};][]{LINDBERG2016}. Comparisons of the mm-wave spectra with the predictions of RADEX non-LTE radiative transport models \citep{VANDERTAK2007} suggests that the molecular clouds making up this portion of the outflow can be represented as self-gravitating dense clouds, if a normal, Galactic HCN-to-H$_2$ abundance is assumed, but this result can be relaxed if HCN abundances are enhanced by shocks or X-rays from the central quasar  \citep[Sect.~\ref{sec:shocks};][]{AALTO2015c}. A recent study reveal very bright CN emission from the Mrk~231 outflow. This may be due to UV photons originating from star formation in the outflow itself, or from the quasar \citep[][]{CICONE2019}.
\citet{GONZALEZ-ALFONSO2018} have recently argued that the high OH$^+$ abundances relative to OH, H$_2$O, and H$_3$O$^+$ measured from the Herschel spectra imply a high ionization rate that cannot be explained by the unattenuatd hard X-ray flux measured in this object \citep{TENG2014}. They suggest instead that low-energy (10-400 Mev) cosmic-rays accelerated in the forward shock associated with the molecular outflow are responsible for the ionization. 

The current best estimate for the momentum injection rate of the large-scale outflow in Mrk~231 exceeds by a factor of 5--10 the \emph{total} radiation pressure $L/c$ from the central energy source \citep[quasar + starburst;][]{FERUGLIO2015,GONZALEZ-ALFONSO2017b,FLUETSCH2019}, and thus favors the energy-driven scenario discussed in Sect.~\ref{sec:thermal_energy}. A comparison of the kinetic energy of the X-ray nuclear wind with that of the cool gas outflow (using Eq.~(\ref{eq:pdot_outflow}) from Sect.~\ref{sec:thermal_energy}) implies that most of the kinetic energy in the X-ray wind goes into bulk motion of the swept-up molecular material \citep{FERUGLIO2015}, although it is important to repeat that the detection of the X-ray wind in this system is only at the 3.5-$\sigma$ level. 

\subsection{Jetted AGN: Centaurus A, IC 5063, NGC~1266, and NGC~1377} \label{sec:jetted_agn}

\subsubsection{Centaurus A} \label{sec:CenA}

At a distance of only $\sim$3.8 Mpc \citep[][1\arcsec = 19 pc, similar to that of M82 and NGC~253]{HARRIS2010}, Centaurus A or Cen A for short is the nearest radio galaxy and a classic example of jetted energy making its way through the ISM of the host galaxy, which in this case is an elliptical galaxy with a prominent dust lane and a complex of HI-rich tidal debris left over from the merger or stripping interaction of a small gas-rich galaxy 250-750 Myr ago \citep[e.g.,][]{SCHIMINOVICH1994,STRUVE2010,MORGANTI2018}.
Cen A is a low-power Fanaroff type I radio source so it is perhaps not surprising to find very little evidence of outflowing cool gas in the central inner kpc of this object. The dynamics of both the neutral material (traced by H~I 21 cm line emission and absorption) and molecular gas (traced by CO, HCO$^+$, HCN, and HNC absorption features) are dominated by a rotating circumnuclear disk and intervening material likely associated with the dust lane \citep[e.g.,][]{LISZT2001a,MORGANTI2008,ESPADA2010}. The only cool-gas  kinematic features that may not fit this picture are the moderately broad (FWHM $\sim$ 50 km s$^{-1}$) H~I and HCO$^+$ absorption lines associated with the base of the nuclear jet, located $\sim$ 10 pc from the AGN, where molecular hydrogen appears to be efficiently dissociated by X-rays from the AGN \citep{ESPADA2010}. A shell-like, bipolar structure 500 pc to the north and south of the nucleus has also been detected by \citet{QUILLEN2006} in {\em Spitzer} dust maps at 8.0 and 24 $\mu$m, suggestive of a coherent expanding structure similar to that seen in the MW (Sect. \ref{sec:MW}), but this dusty outflow has not yet been confirmed kinematically.

\begin{figure}[htbp]
\begin{center}
\includegraphics[width=0.90\textwidth]{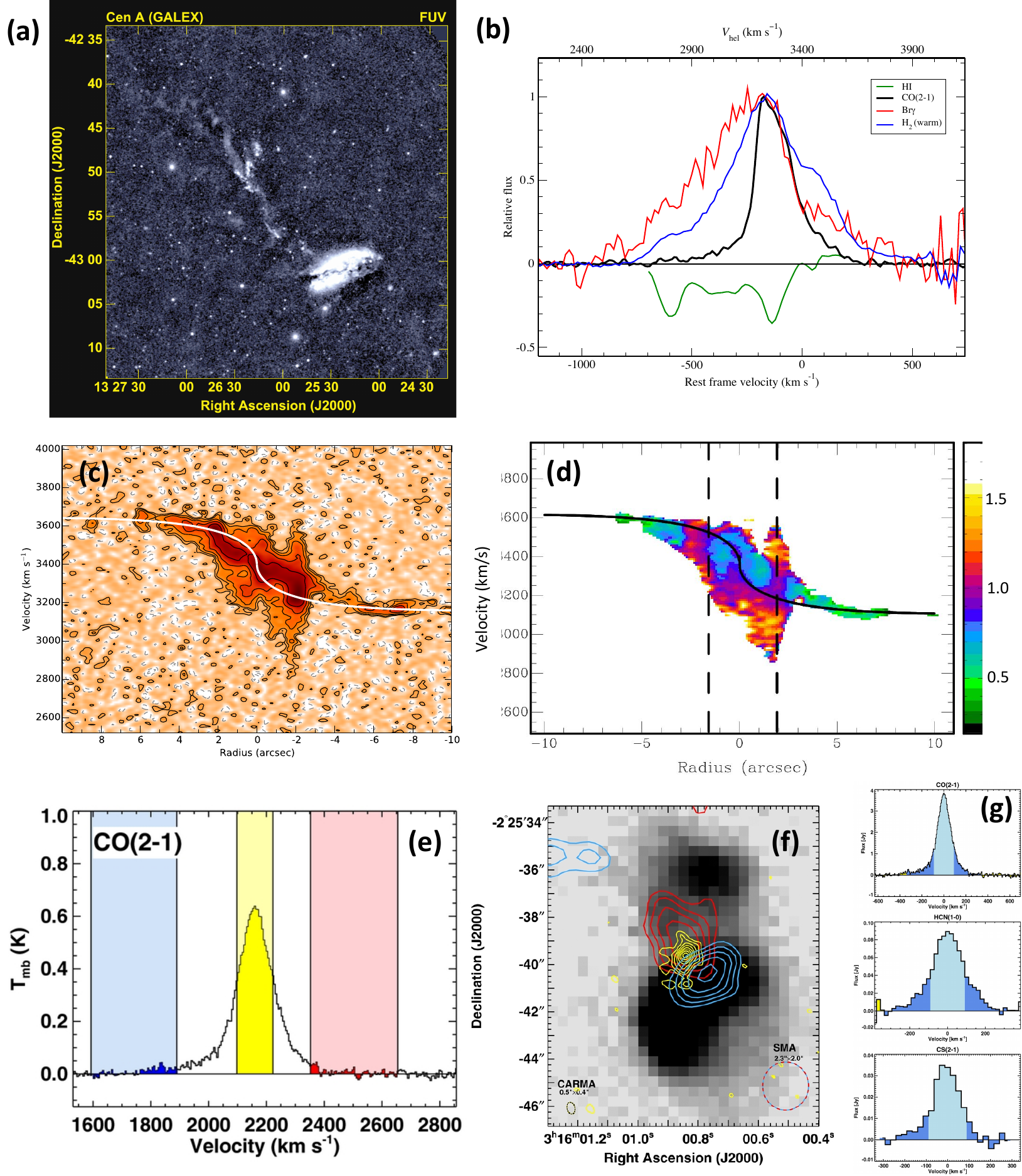}
\end{center}
\caption{Jetted cool outflows. (a) Evidence for jet-induced star formation in filaments along the radio jets of Centaurus A. (b) Multi-phase outflow of IC~5063 detected in Br$\gamma$, H~I 21 cm, H$_2$ 2.12 $\mu$m, and CO (2-1), at the location of the north-west radio lobe. (c) Position-velocity diagram of CO(2-1) brightness temperatures along the radio jet axis of IC 5063. (d) Same as (c) but for  the ratio of the HCO$^+$(4-3) and CO(2-1) brightness temperatures,  (e)  CO(2-1) spectrum of NGC~1266. The blue and red shaded areas represent the outflow blue and red wing emission, respectively. (f) Map of the CO (2-1) blue and red wing emission defined in panel (e). (g) Profiles of CO(2-1) (top), HCN(1-0) (middle), and CS(2-1) (bottom) in NGC~1266. Dark blue shading indicates the outflow emission.  Images reproduced with permission from (b) \citet{MORGANTI2015}, (c, d) \citet{OOSTERLOO2017}, copyright by ESO; and (a) \citet{NEFF2015}, (e, f) \citet{ALATALO2011}, (g) \citet{ALATALO2015b},  copyright by AAS.}
\label{fig:jetted_outflows}
\end{figure}

The influence of the radio jet on the ISM is more apparent on larger scales. Deep UV, infrared, and optical images have revealed two linear dusty line-emitting filaments located $\sim$8 and $\sim$15 kpc north of the central AGN, coincident with young stars with ages $\lesssim$10 Myr \citep[e.g.,][]{MOULD2000,CROCKETT2012} and well aligned with the jet radio structures on the same scale \citep[Fig.~\ref{fig:jetted_outflows}a;][]{NEFF2015,HAMER2015,SALOME2016,MCKINLEY2018}. FIR emission has also been detected in a southern filament along the direction of the southern jet \citep{SALOME2016}. The disturbed kinematics of the neutral and ionized gas in the northern filaments suggest an on-going interaction with the radio jet \citep{OOSTERLOO2005,SANTORO2015a,SANTORO2015b,SANTORO2016}. These filaments have been interpreted as evidence for jet-induced star formation at the level of a few 10$^{-3}$ M$_\odot$ yr$^{-1}$ \citep[e.g.,][]{DEYOUNG1981}, although this is probably not the whole story. The high level of ionization in the optical filaments strongly argues for additional ionizing radiation originating from the central AGN \citep[e.g.,][]{SHARP2010,SANTORO2015a,SANTORO2015b,MCKINLEY2018} or shocks induced by the interaction with the radio jet \citep{SANTORO2015b,SALOME2019} or with a wide-angle starburst/AGN-driven wind that coexists with, and extends on similar scales as, the jetted outflow \citep{NEFF2015,MCKINLEY2018}.

Taken at face value, the star formation rate in the northern filaments of Cen A is 2-3 orders of magnitude less than the inferred star formation rate in the central galaxy \citep{SALOME2016}, and thus the current level of jet-induced star formation will not have a significant impact on the overall evolution of the host galaxy. Interestingly, the star formation efficiency (or $1/t_{\rm depl} = \dot{M}_*/M_{\rm mol}$) in the filaments, measured from the ratio of the star formation rate to the molecular gas mass derived using a CO-to-H$_2$ that is appropriate for the low metallicities of the filaments ($Z = 0.4 - 0.8~Z_\odot$), is also lower by about two orders of magnitude relative to that in the host galaxy. So it shows that, while the jet-gas interaction may be needed to compress and cool the gas to form stars, the local turbulent energy associated by this interaction \citep[as evidenced by the broad CO line widths;][]{SALOME2016}  prevents the gas from forming new stars efficiently.

\subsubsection{IC~5063 and NGC 1266} \label{sec:IC5063_N1266}

IC~5063 and NGC~1266 are two other well-studied low-power radio sources hosted by early-type hosts with prominent dust lanes and filaments. However, contrary to Cen~A, both of these objects harbour clear neutral and molecular outflows within a kpc of their respective AGN. The case of the multi-phase jet-driven outflow in IC~5063 is well documented and is summarized in Figure \ref{fig:jetted_outflows}b-d: large bulk and turbulent motion of the ionized, neutral, and molecular gas phases is present at the position of the bright north-western radio lobe (1.8\arcsec\ $\sim$ 520 pc from the nucleus), based on the detection at this location of a HI 21-cm absorption feature with a blue wing extending to about $-$700 km s$^{-1}$ \citep{MORGANTI1998,OOSTERLOO2000}  and broad emission lines with FWZI up to $\sim$1000 km s$^{-1}$ from [O~III] 5007 \citep[][]{MORGANTI2007}, [Fe~II] 1.644 $\mu$m and H$_2$ 2.128 $\mu$m \citep{TADHUNTER2014,DASYRA2015},  and CO and HCO$^+$ \citep{MORGANTI2013,DASYRA2016,OOSTERLOO2017}.  More modest bulk and turbulent velocities of $\sim$100 km s$^{-1}$ and $\sim$600 km s$^{-1}$, respectively, are detected at the position of the weaker eastern radio lobe (2.2\arcsec\ $\sim$ 450 pc from the nucleus). Ionized, neutral, and molecular mass outflow rates of $\sim$0.1, $\sim$35, and $\gtrsim$12 M$_\odot$ yr$^{-1}$ are deduced from these data. The outflowing gas is of higher density ($\gtrsim$2000 cm$^{-3}$ for the ionized gas, from the [S~II] ratio, and up to $\sim$10$^6$ cm$^{-3}$ in the molecular phase, based on the ratio of HCO$^+$/CO(3-2)) than the surrounding gas, as expected if the gas is compressed by the radio jet \citep[e.g.,][]{WAGNER2012}.   Note that the relative brightness of the $^{12}$CO lines indicate that the outflowing molecular gas is optically thin, however \citep{DASYRA2016,OOSTERLOO2017}.  \citet{MUKHERJEE2018b} have recently produced three-dimensional hydrodynamic simulations of a jet with powers 10$^{44}$--10$^{45}$ erg s$^{-1}$ propagating through the multi-phase gaseous disk of a galaxy that qualitatively reproduce many of the kinematic features observed in IC~5063, assuming that the jet turned on 0.24 Myr ago (Fig.~\ref{fig:sims_agn_outflows}b).

The kinematics of the neutral and molecular outflows in NGC~1266 are less extreme than those of IC~5063 \citep[outflow velocities typically less than 250 km s$^{-1}$; Fig.~\ref{fig:jetted_outflows};][]{ALATALO2011,DAVIS2012,NYLAND2013}, but the recent detection of broad wings in the HCN and CS spectra \citep{ALATALO2015b} indicates that the outflowing molecular gas is dense and optically thick, which boosts the inferred molecular mass outflow rate to $\sim$110 M$_\odot$ yr$^{-1}$, or $\sim$8 $\times$ larger than the optically thin value reported in earlier papers.  This large mass outflow rate and corresponding kinetic power of $\sim$ 3.4 x 10$^{45}$ erg s$^{-1}$ are surprisingly high given the relatively modest radio power of the central VLBA-detected source in NGC~1266 \citep{NYLAND2013}. 
The high brightness temperature ($\gtrsim$ 1.5 $\times$ 10$^7$ K) of this source is consistent with an AGN origin rather than a compact starburst. Indeed, star formation (total $\dot{M}_*$ $\sim$ 0.87 M$_\odot$ yr$^{-1}$) seems to be suppressed by a factor $\sim$50 given the amount of molecular gas in this object. \citet{ALATALO2015b} have argued that the energy injected by the AGN stirs the host ISM and makes it more turbulent and less efficient at making new stars. Shock excitation has indeed been proposed to play a role in producing the line emission in the outflowing molecular gas \citep{GLENN2015} and the fast (up to $\pm$900 km s$^{-1}$) ionized outflow that extends to $\pm$4\arcsec\ \citep[$\pm$ 600 pc;][]{DAVIS2012}. Shocks could also conceivably be responsible for some of the X-ray emission seen on a similar scale \citep{ALATALO2011}.

\subsubsection{NGC~1377}\label{sec:N1377}

The most collimated molecular outflow discovered to date resides in the unassuming lenticular galaxy NGC~1377.  It is nearby ($d$ = 21 Mpc) and of moderate luminosity \citep[log $L_{\rm IR}$ = 9.63;][]{ROUSSEL2006}. Deep mid-infrared silicate absorption features imply that the nucleus is enshrouded by large masses of dust \citep[e.g.][]{SPOON2007}. High resolution mm- and submm-wave studies suggest a nuclear $N$(H$_2$) of $\sim10^{24}$ cm$^{-2}$ \citep{AALTO2016,AALTO2017}, but NGC~1377 is less obscured (on similar scales) than extreme CONs such as IC~860, Arp~220 or NGC~4418 (discussed in Sect.~\ref{sec:CONs}). NGC~1377 has a post-starburst optical spectrum which is likely a relic from a vigorous star formation event about a Gyr ago, possibly as a result of an interaction.

\begin{figure}[htb]
\begin{center}
\includegraphics[width=1.0\textwidth]{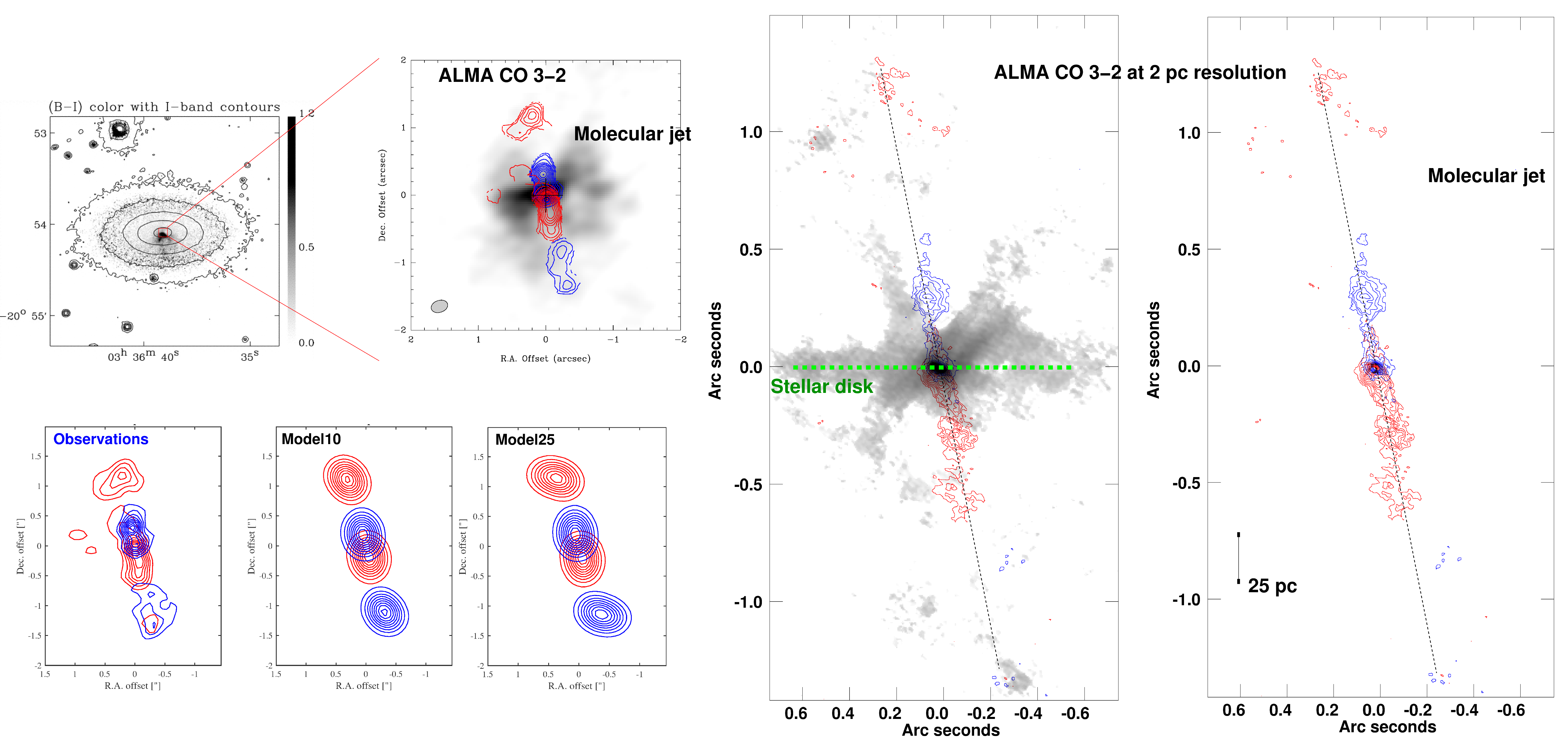}
\end{center}
\caption{Jetted molecular outflow in NGC~1377. Left: Optical (B-I) image \citep[from][]{ROUSSEL2006} showing a southern outflow dust feature. Centre left: ALMA 20 pc resolution CO (3--2) data where emission at systemic velocity (0 - 60 km s$^{-1}$) is shown in gray scale and the high velocity (projected 80 - 160 km s$^{-1}$) jet emission in red and blue contours.  Bottom left panels:  Models of the high velocity emission of a precessing jet. The left panel shows the observed data and the two right ones show models with different precession angle \citep{AALTO2016}. Right panels: High resolution (3 pc) ALMA CO (3--2) image (without the merged lower resolution data) showing the inner region and jet of NGC~1377 (Aalto et al. 2019, in prep.). Images reproduced with permission from \citet{ROUSSEL2006}, copyright by AAS; and \citet{AALTO2016}, copyright by ESO.}
\label{fig:n1377}
\end{figure}

The molecular outflow was first detected in CO (2--1) at relatively low 0."6 (60 pc) resolution \citep{AALTO2012} and appeared to be a young (1-2 Myr) biconic molecular flow. However, higher resolution (0."2) ALMA observations revealed an extremely collimated, 150 pc long jet-like structure in the CO 3-2 transition (Fig.~\ref{fig:n1377}), surrounded by a slower, wide-angle wind.  There are so far no indications of an outflow in ionized or atomic gas--although a v-shaped dust lane 
can be seen on kpc scale in optical images. Interestingly, NGC~1377 is one of the most radio-quiet objects in the sky \citep[e.g.][]{ROUSSEL2006,COSTAGLIOLA2016} and there is so far no evidence of a radio jet that could be responsible for the high collimation of the high-velocity gas. Also, in contrast to molecular gas structures being carried out by radio-loud jets, the high-velocity molecular gas is found on the spine of the collimated flow.  The CO (3--2) emission shows deviations from the symmetry axis, with velocity reversals, wiggles, and intensity variations on several size scales.  The velocity reversals led to the suggestion that the molecular jet is precessing \citep{AALTO2016}.  The velocity of the outflowing gas is difficult to determine but a lower limit is $v_{\rm out} >$ 240 km s$^{-1}$ and may be as high as 850 km s$^{-1}$ if the disk inclination is high. The total molecular mass in the collimated jet-like structure exceeds $2 \times 10^6$ \msun.  The CO emission is clumpy along the jet with a high velocity dispersion ($\sigma$ = 40-90 km s$^{-1}$) suggesting a highly turbulent medium and/or sideways ejection/expansion.  The v-shaped minor axis structures (surrounding the molecular jet) are either part of a slower wide-angle wind, or material entrained by the molecular jet. In the latter case, the small (150 pc) jet may potentially feed much of the mass into the kpc-scale bipolar structures.

The lack of star formation indicators (e.g no NIR Pa$\alpha$ or Br$\gamma$ and the extreme paucity of radio emission) have led to suggestions that NGC~1377 could be a nascent starburst \citep{ROUSSEL2006}. However, the highly collimated, and relatively powerful high-velocity outflow, indicate that the molecular gas outflow is not driven by star formation.  It may be powered by SMBH accretion -- either by inefficient, hot (``radio-mode'') accretion, or by effective accretion of cold gas (``quasar-mode'').  In the former case the driving radio-jet is invisible, and in the latter case the molecular jet would represent an unexplored form of AGN feedback, where the high velocity gas is expelled in a collimated outflow and not a wide-angle wind. 

It is also possible that the radio-quiet molecular jet of NGC~1377 represents accretion onto the nuclear disk (instead of onto the SMBH accretion disk) driving a magnetohydrodynamic (MHD) wind in a process reminiscent of that seen (on much smaller scales) in protostars.

\section{Cool Outflows in the Local ($z \lesssim 1$) Universe} \label{sec:lowz}

Only very few galaxies have been studied at the level of detail as those discussed in Section \ref{sec:best_cases}. 
Many other galactic cool outflows have been identified  using only one or a few tracers, and often with data that do not enable the extraction of detailed information, but only some basic outflows properties. The samples are also necessarily biased and incomplete. Yet, these several studies are useful to obtain statistical information of some of the basic properties of cool outflows as a function of other galactic properties, which can shed light on the driving mechanisms of cool outflows and their effects on galaxy evolution. This section summarizes the results on the local systems defined as $z \lesssim 1$. Section \ref{sec:highz} discusses the results at higher redshifts.

\subsection{Neutral Atomic Gas Component} \label{sec:lowz_neutral_atomic}

Neutral atomic outflows have been detected in several $z \lesssim$ 1 galaxies via ``down-the-barrel'' absorption-line spectroscopy of the host galaxies (including GRBs embedded in the hosts), direct emission-line imaging and 3D IFS of the cool outflows, and transverse absorption-line spectroscopy of background sources (quasars and galaxies) to probe the host CGM.
The full panoply of diagnostic tools discussed in Section \ref{sec:diagnostics_neutral_atomic} has been brought to bear on the characterization of these outflows.

Historically, fast neutral-atomic outflows were first unambiguously detected in absorption in the spectra of quasars. Quasars with LOw-ionization Broad Absorption Lines or LoBALs are characterized by broad ($\gtrsim$ 1000 km~s$^{-1}$) highly blueshifted ($\lesssim$ $-$1000 km~s$^{-1}$) absorption lines from Mg I and Mg II. This class of quasars is rare, accounting for only $\sim$1.3\% of all quasars \citep{TRUMP2006}.  LoBALs with excited states of Fe II or Fe III absorption (``FeLoBALs'' such as Mrk~231; Sect.~\ref{sec:Mrk231}) are even less common \citep[$\sim$0.3\%;][]{TRUMP2006}. The sizes of these low-ionization outflows, derived using Eq.~(\ref{eq:U_H}) in the few $z \lesssim 1$ objects where the electron density can be constrained, span a broad range from a few pc to a few kpc
\citep{MOE2009,DUNN2010,BAUTISTA2010,AOKI2011,VEILLEUX2016}. These fast nuclear winds may be the driving force behind the outflows seen on larger scales in these same objects (Sect. \ref{sec:thermal_energy} and \ref{sec:lowz_driving_mechanisms}). 

\begin{figure}[htb]
\begin{center}
\includegraphics[width=0.9\textwidth]{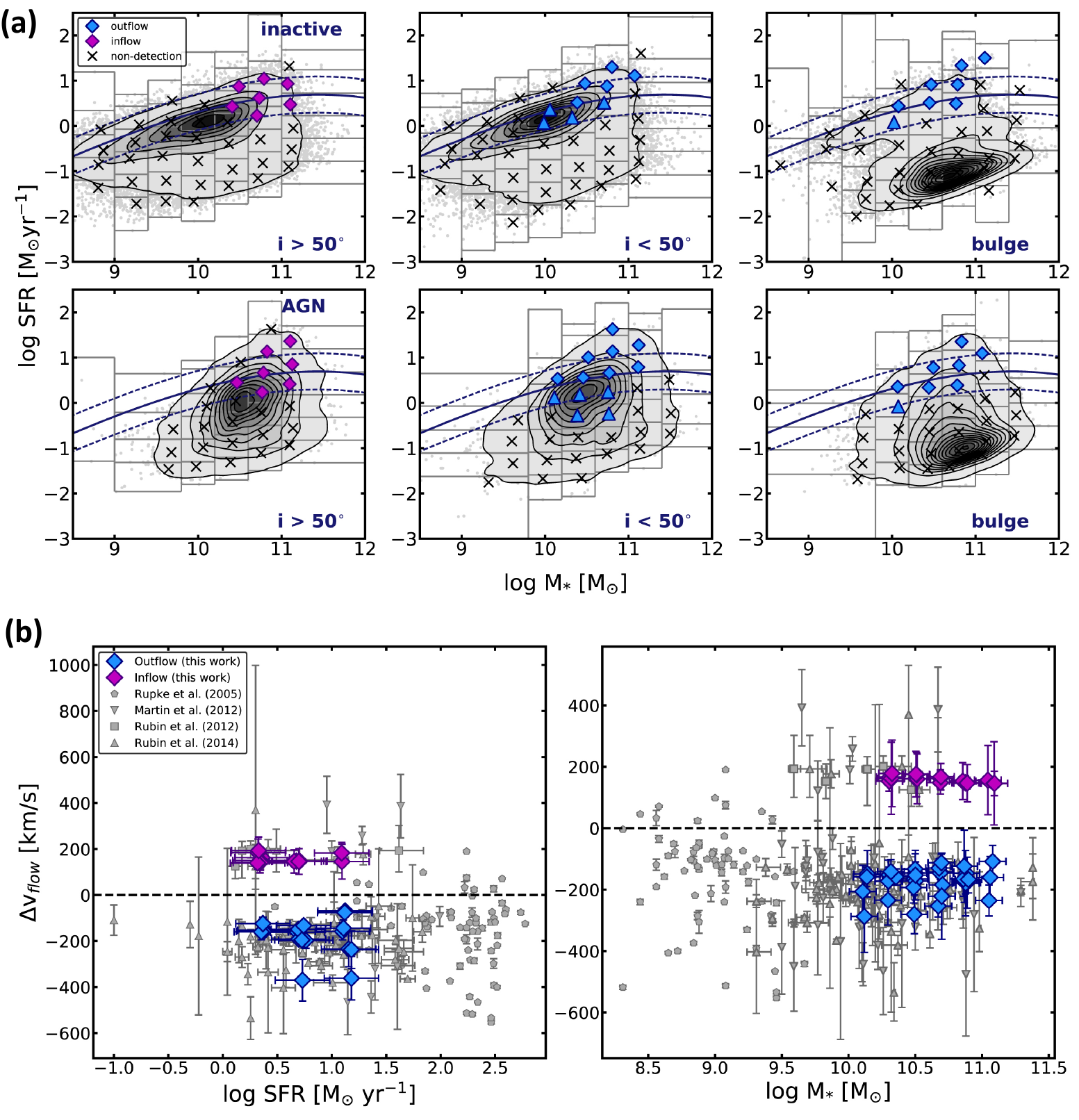}
\end{center}
\caption{Statistical properties of nearby neutral-atomic outflows based on the Na~I D absorption doublet. (a) Detections of inflows and outflows across the SFR $- M_*$ plane for disk galaxies (left and middle panels are for different inclinations) and bulge galaxies (right panels). The top row shows inactive galaxies while the bottom row shows AGN hosts. (b) Average inflow and outflow velocities as a function of SFR (left panel) and $M_*$ (right panel). Images reproduced with permission from \citet{ROBERTS-BORSANI2019}, copyright by the authors.}
\label{fig:nai_outflows}
\end{figure}

Ironically, few neutral-atomic outflows have been detected using H~I 21 cm (Sect.~\ref{sec:21cm}), until recently \citep[see review by][]{MORGANTI2018}. While this largely reflects the technical limitations of early-generation radio facilities, H~I photoionization also plays a role behind the low detection rate of H~I 21-cm absorption line in powerful AGN with ionization rates $Q_{\rm HI}$ $\gtrsim$ 3 $\times$ 10$^{56}$ s$^{-1}$
\citep[e.g.,][]{CURRAN2016,CURRAN2019,ADITYA2018}. 
In contrast, down-the-barrel ground-based observations of absorption lines from neutral-atomic tracers such as Na~I~D 5890, 5896, Mg~II 2796, 2803, and Fe II 2586, 2600 against the host galaxy light (Sect.~\ref{sec:NaID}, \ref{sec:neutral_elements}, and \ref{sec:lowion_elements}) have been extremely successful at detecting cool outflows in gas-rich $z \lesssim 1$ galaxies. Fast Na~I outflows with inferred hydrogen column densities $N_{\rm H} \simeq 10^{20} - 10^{22}$ cm$^{-2}$ (based on Eq.~(\ref{eq:N_H_N_NaI})) are ubiquitous in powerful dusty starburst and active galaxies, where dust in the hosts and within the outflowing gas shields Na~I (5.1 eV) against the $\lesssim$2432 \AA\ radiation emitted by the starbursts and AGN \citep[e.g., nearby U/LIRGs,][]{HECKMAN2000,RUPKE2002,RUPKE2005b,RUPKE2005c,RUPKE2005a,MARTIN2005,MARTIN2006,CAZZOLI2016}. The detection rate (and velocities) of Na~I outflows drops precipitously with decreasing dust content $A_V$, SFR, SFR per unit area, AGN power, and stellar mass, averaging a value $\lesssim$ 1\% among the general population of optically-selected star-forming and active galaxies in SDSS
\cite[e.g.,][]{SARZI2016,BAE2017,CONCAS2019,NEDELCHEV2019,ROBERTS-BORSANI2019}, in MaNGA-selected star-forming galaxies \citep{ROBERTS-BORSANI2020}, as well as in IR-faint quasars \citep{KRUG2010}. Na~I outflows in non-ULIRGs show a dependence on galaxy disk inclination that indicates a preference for Na~I outflows to align along the minor axis of the host galaxy disk \citep[Fig.~\ref{fig:nai_outflows}; e.g.,][]{CHEN2010,CONCAS2019,ROBERTS-BORSANI2019}. Recent Na~I IFS observations of local ULIRGs and IR-bright quasars hint at a similar alignment of the neutral outflowing gas along the minor axis of the underlying rotating gas structures on kpc scales, although these outflows clearly subtend a wider angle and are more irregular than those found in systems of lower SFRs and AGN powers \citep{RUPKE2013b,RUPKE2017}.

\begin{figure}[htb]
\begin{center}
\includegraphics[width=0.9\textwidth]{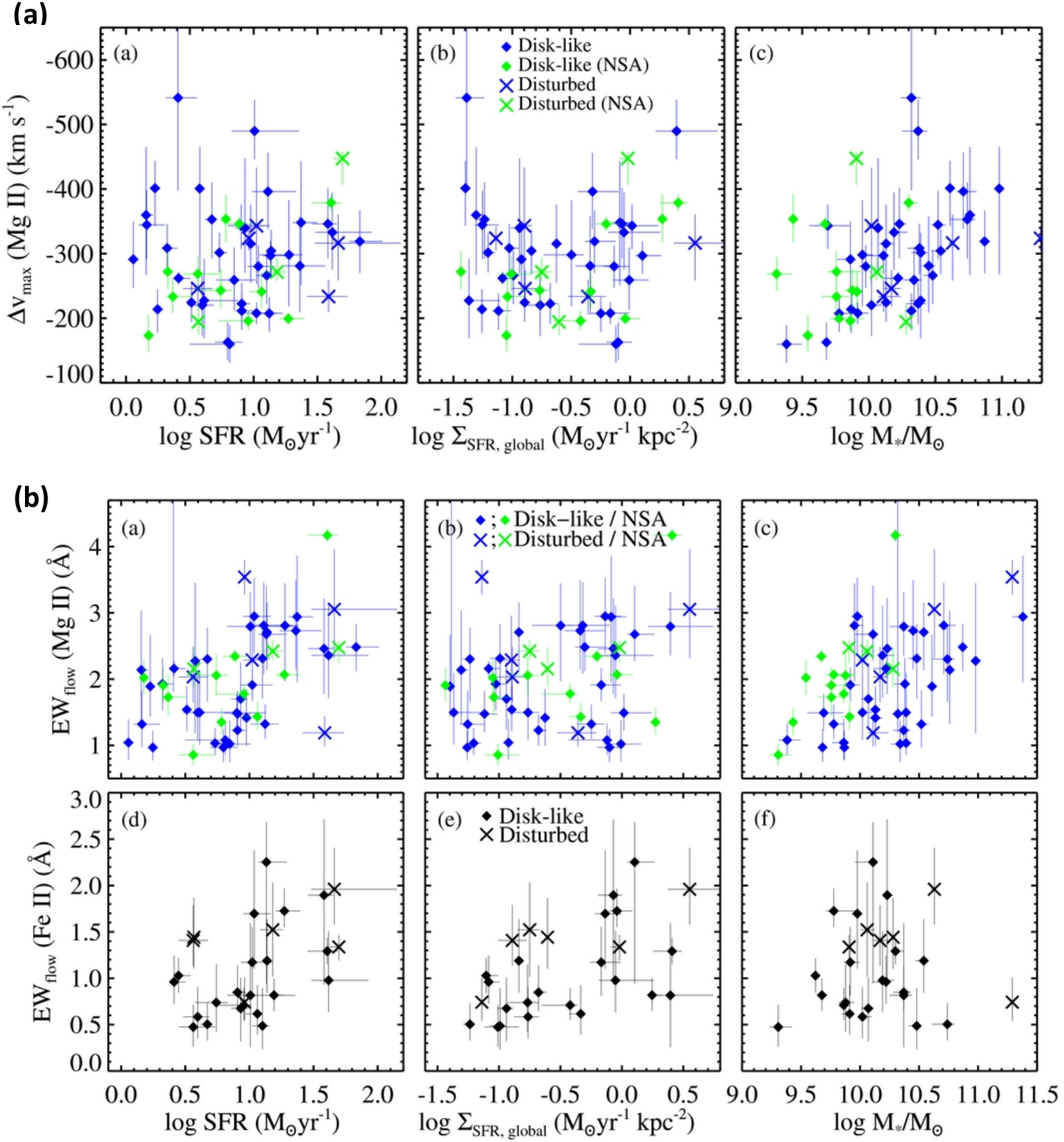}
\end{center}
\caption{Statistical properties of neutral-atomic outflows based on the NUV Mg II and Fe II absorption lines. (a) Maximum outflow velocities as a function of SFR, SFR surface density, and stellar mass. (b) Equivalent widths of Mg~II (top panels) and Fe~II (bottom panels) as a function of the same quantities as in (a). Images reproduced with permission from  \citet{RUBIN2014}, copyright by AAS.}
\label{fig:mgii_outflows}
\end{figure}

Large ground-based optical spectroscopic surveys in the past decade have extended these down-the-barrel studies of neutral-atomic outflows to higher redshifts using the Mg~II 2796, 2803, Fe II 2586, 2600, and Mg~I 2853 absorption features redshifted into the optical band \citep[e.g.,][using GRBs]{WEINER2009,RUBIN2010,COIL2011,ERB2012,KORNEI2012,KORNEI2013,MARTIN2012,BORDOLOI2014b,RUBIN2014,ZHU2015,HEINTZ2018}.
Mg~II outflows with typical inferred hydrogen column densities $N_{\rm H} \gtrsim 10^{19}$ cm$^{-2}$ (Mg II is saturated when EW$_r \gtrsim 1$ \AA\ so these columns are lower limits) are detected in most ($\gtrsim$60\%) $z \sim 0.5 - 1.5$ star-forming galaxies with SFR $\gtrsim$ 1 M$_\odot$ yr$^{-1}$, particularly face-on galaxies ($\sim$90\%), implying ubiquitous biconical outflows with opening angle $\sim$ 100$^\circ$ \citep{RUBIN2014}. The wind maximum velocities (typically $\sim 200-400$ km s$^{-1}$) and equivalent widths correlate only weakly with galaxy stellar mass and SFR \citep[Fig.~\ref{fig:mgii_outflows},][]{WEINER2009,RUBIN2014}.  ``Relic'' outflows with typical velocities $\sim$ 200 km s$^{-1}$ are often seen in post-starburst galaxies \citep{COIL2011,YESUF2017}, although in exceptional cases, possibly due to recent quasar activity or extremely compact optically obscured starbursts, these winds reach velocities in excess of $\sim$1000 km s$^{-1}$ \citep{TREMONTI2007,DIAMOND-STANIC2012,GEACH2014,SELL2014,MALTBY2019}. In general, low-luminosity AGN do not have a strong influence on the velocities and equivalent widths of these winds \citep{COIL2011,YESUF2017}.

\begin{figure}[htb]
\begin{center}
\includegraphics[width=0.9\textwidth]{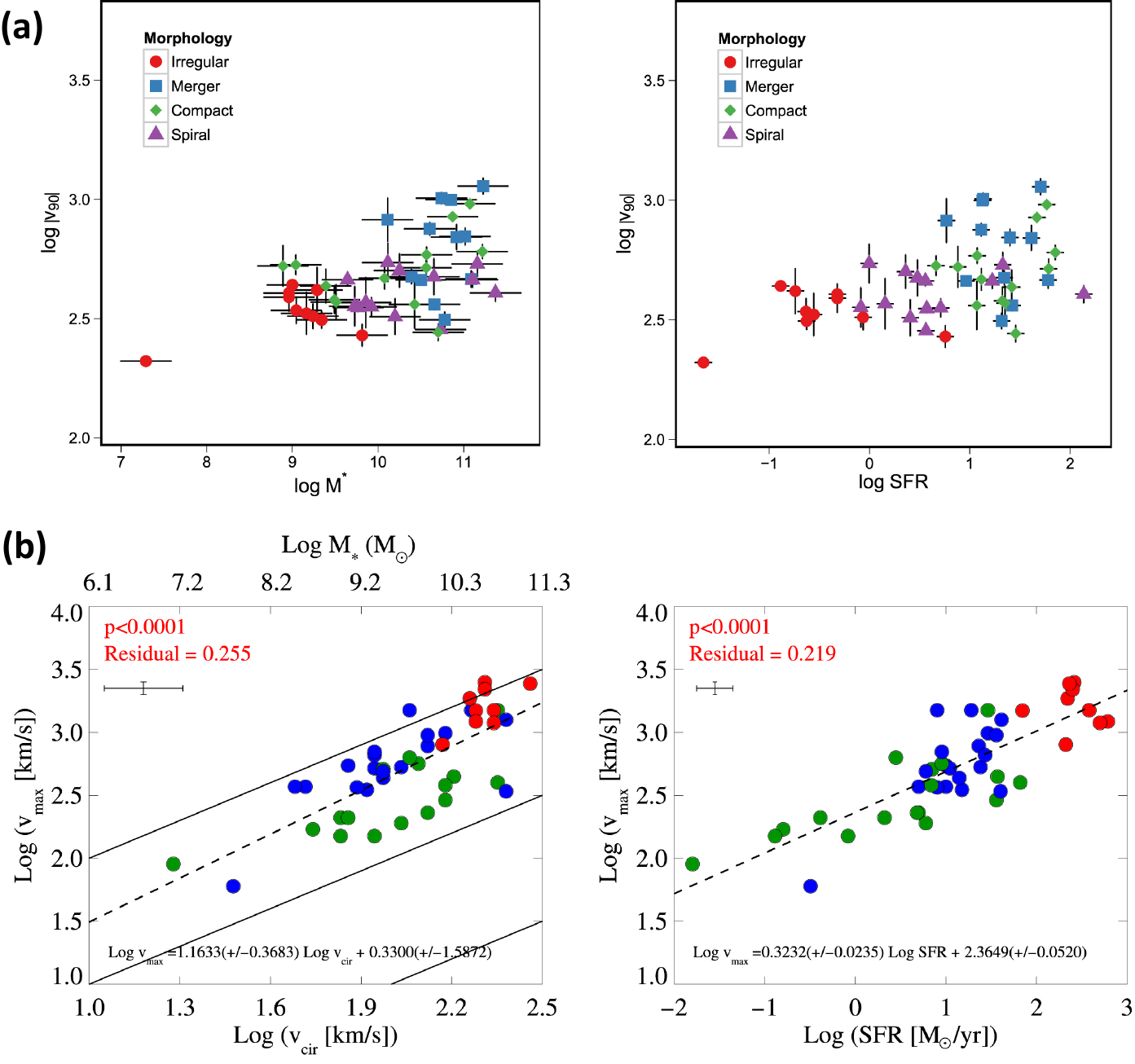}
\end{center}
\caption{Statistical properties of nearby low-ionization outflows based on HST spectra. (a) 90-percentile Si~II velocities as function of stellar mass and SFR, (b) Maximum Si~III velocities as a function of the circular velocity and SFR. Images reproduced with permission from (a) \citet{CHISHOLM2015}, (b) \citet{HECKMAN2016}, copyright by AAS.}
\label{fig:hst_lowion_outflows}
\end{figure}

The excellent FUV sensitivity of the Cosmic Origins Spectrograph (COS) on-board HST has allowed to study in detail, albeit with limited spatial information, the cool neutral-atomic and warm ionized gas phases of outflows in some of the nearest and UV-brightest star-forming galaxies \citep[e.g.,][]{HECKMAN2011,HECKMAN2015,HECKMAN2016,CHISHOLM2015,CHISHOLM2016b,CHISHOLM2016a,CHISHOLM2017,CHISHOLM2018} and even in a few infrared-bright systems \citep{LEITHERER2013,MARTIN2015,VEILLEUX2016}. \citet{CHISHOLM2015} find weak (3.0--3.5 $\sigma$) correlations between the Si~II-based outflow velocities and SFR ($v_{\rm out} \propto \dot{M_*}^{0.08-0.22}$), the stellar mass ($\propto M_*^{0.12-0.20}$), and the circular velocity ($\propto v_{\rm circ}^{0.44-0.87}$), where the index depends on whether the maximum or centroid velocity is used as the outflow velocity (Fig.~\ref{fig:hst_lowion_outflows}a). In contrast, \citet{HECKMAN2015} and \citet{HECKMAN2016}  find stronger correlations between the maximum velocity and SFR ($v_{\rm out} \propto \dot{M}_*^{0.32}$), and the star formation rate surface density $\Sigma_{\rm SFR}$ (Fig.~\ref{fig:hst_lowion_outflows}b), consistent with momentum-driven winds where the momentum comes from a combination of ram pressure from the wind, radiation pressure, and cosmic rays (Sec. \ref{sec:driving_mechanisms}).
The origin of the apparent discrepancy between these kinematic studies is not clear, but is probably due to more than just one factor.  The results of \citet{HECKMAN2015} and \citet{HECKMAN2016} are based on Si~III 1206 (IP$^+$ = 16.3 eV) rather than Si~II 1260 (IP$^0$ = 8.15 eV), so they trace the ionized component of these outflows rather than the neutral-atomic component. However, this cannot fully explain the apparent discrepancy between these studies since both \citet{HECKMAN2015} and \citet{CHISHOLM2015} find that Si~IV, Si~III, Si~II, and O~I have virtually the same profiles and are thus co-moving within a single outflowing structure. The limited number ($\lesssim$ 50) of objects examined in these studies no doubt adds noise to possible underlying relations.  Sample selection may also be critical. While there is considerable overlap between the two samples, it is also clear that the addition of the extreme starbursts from \citet{DIAMOND-STANIC2012} and \citet{SELL2014} in the study by \citet{HECKMAN2016} extends the range of SFR and $\Sigma_{\rm SFR}$ over the previous studies and therefore provides more leverage for the fits. This comparison also raises a cautionary flag that the wide-spread practice of reducing the velocity fields of these complex three-dimensional multi-phase outflows to a single parameter ($v_{50}$ or $v_{\rm max}$) in a single tracer is bound to introduce considerable scatter in the scaling relationship with the host galaxy properties \citep[see Table 1 of][for a summary]{RUPKE2018}.

\begin{figure}[htb]
\begin{center}
\includegraphics[width=0.95\textwidth]{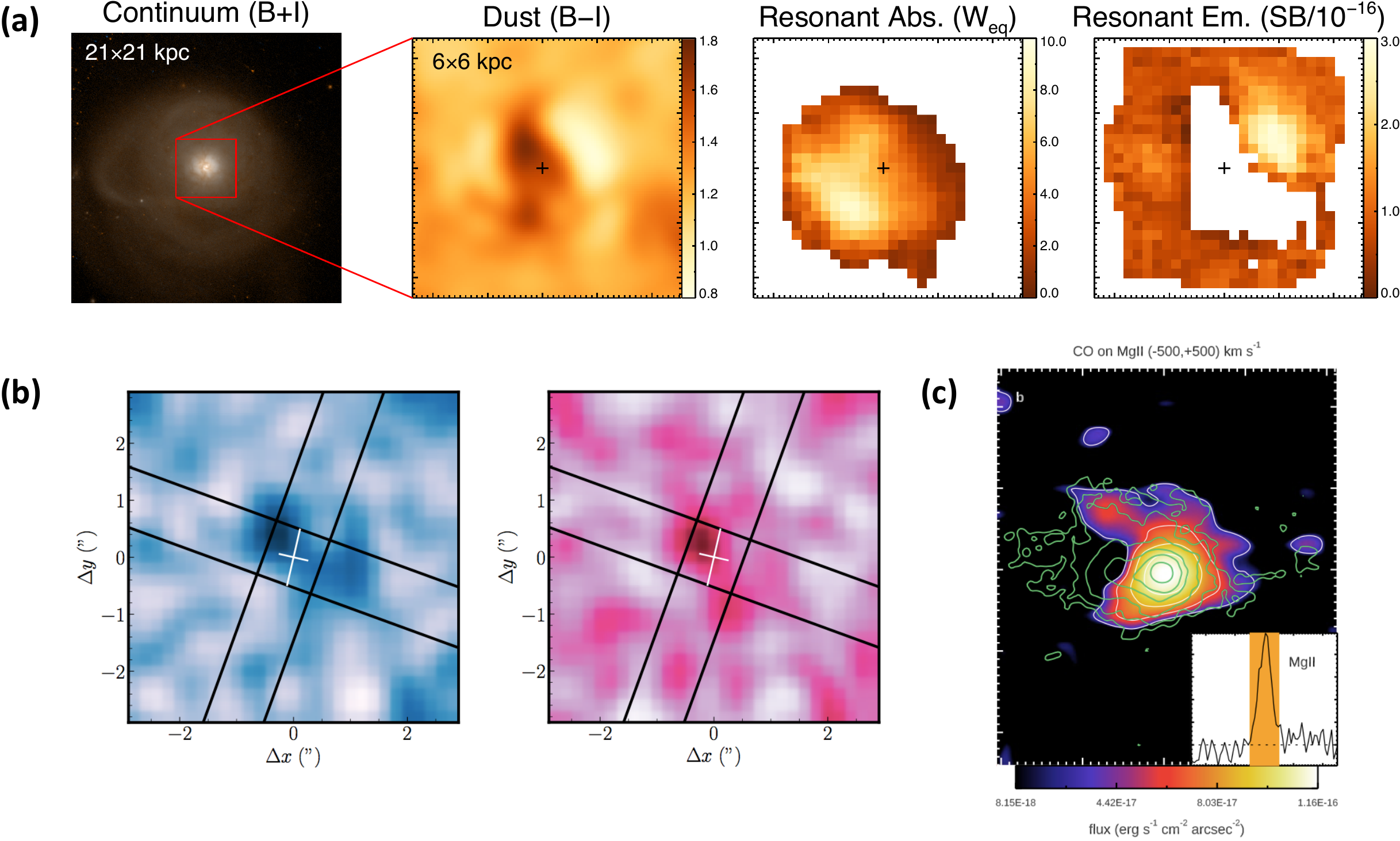}
\end{center}
\caption{Nearby neutral-atomic outflows in line emission. (a) Resonant Na I line emission (rightmost panel) from the redshifted outflow component in a nearby ULIRG, compared with the Na I absorption and dust distributions on the same scale. (b) Non-resonant Fe~II$^*$ line emission from the blueshifted (left) and redshifted (right) outflow components in a $z \simeq 1.24$ galaxy. (c) Resonant Mg~II line emission in $z = 0.46$ Makani. Images reproduced with permission from \citet{RUPKE2015}, copyright by AAS; and \citet{FINLEY2017}, copyright by ESO; and \citet{RUPKE2019}, copyright by the authors.}
\label{fig:na_mg_emission_outflows}
\end{figure}

To go beyond a discussion of the kinetics of cool neutral-atomic outflow and address their dynamical properties requires a measure of their physical extents (Sec. \ref{sec:energetics}). The sizes of the outflows studied in the above HST data ($r \lesssim$ 1 kpc) are derived indirectly from detailed modeling of the photoionization conditions in the outflowing material, carefully taking into account object-by-object variations in the metallicity and ionization corrections \citep[e.g.,][]{CHISHOLM2016b,CHISHOLM2016a}.  In contrast, direct size measurements exist for a significant fraction of the cool neutral-atomic outflows based on the Na~I absorption line; they typically extend to a few kpc, with a range from $\lesssim$ 1 kpc to $\sim$15 kpc \citep[][]{RUPKE2005c,MARTIN2006,RUPKE2017,ROBERTS-BORSANI2020}. These measurements should be considered lower limits since they are limited by the sensitivity of optical spectrographs to detect this absorption line against the faint galaxy continuum at large $r$.  This has been made clear by the detection in a few objects of redshifted resonant Na~I emission well beyond the scale where blueshifted Na~I absorption is detected \citep[Fig.~\ref{fig:na_mg_emission_outflows}a;][]{PHILLIPS1993,RUPKE2015}.  Scattered resonant and non-resonant line emission has also proven to be very useful to constrain the size of the Mg~II and Fe~II outflows, which would otherwise remain largely unconstrained from the absorption-line studies. The Mg~II and Fe~II$^*$ emission line sizes derived from stacked long-slit composite spectra range from a few kpc up to $\sim$20 kpc \citep{RUBIN2011,ERB2012,MARTIN2013}, although recent direct imaging and 3D IFS data indicate that most of the line emission appears to come from $\lesssim$4$-$10 kpc \citep[Fig.~\ref{fig:na_mg_emission_outflows}b;][]{FINLEY2017,RICKARDSVAUGHT2019}, and may be localized and strongly influenced by the nearest star-forming clump, based on IFS of a few lensed systems at $z \sim$ 1 \citep[][most of the work on lensed systems has been done at higher redshifts; see Sec. \ref{sec:highz_neutral_atomic}]{BORDOLOI2016,KARMAN2016}. 
Direct size measurements of neutral-atomic outflows also exist for a few nearby sources mapped in the [C~II] 158 $\mu$m line emission using Herschel or SOFIA \citep[][Stone et al. 2019, in prep.]{CONTURSI2013,KRECKEL2014,APPLETON2018}. All of these direct size measurements should be considered lower limits since they are severely limited by the sensitivity of the observations. The recent detection of Mg~II-line emitting material out to 20 kpc in Makani, a compact starburst-dominated galaxy at $z = 0.459$ mapped with deep KCWI IFS data \citep[Fig.\ \ref{fig:na_mg_emission_outflows};][]{RUPKE2019}, underscores the need for deeper observations.

\begin{figure}[htb]
\begin{center}
\includegraphics[width=0.95\textwidth]{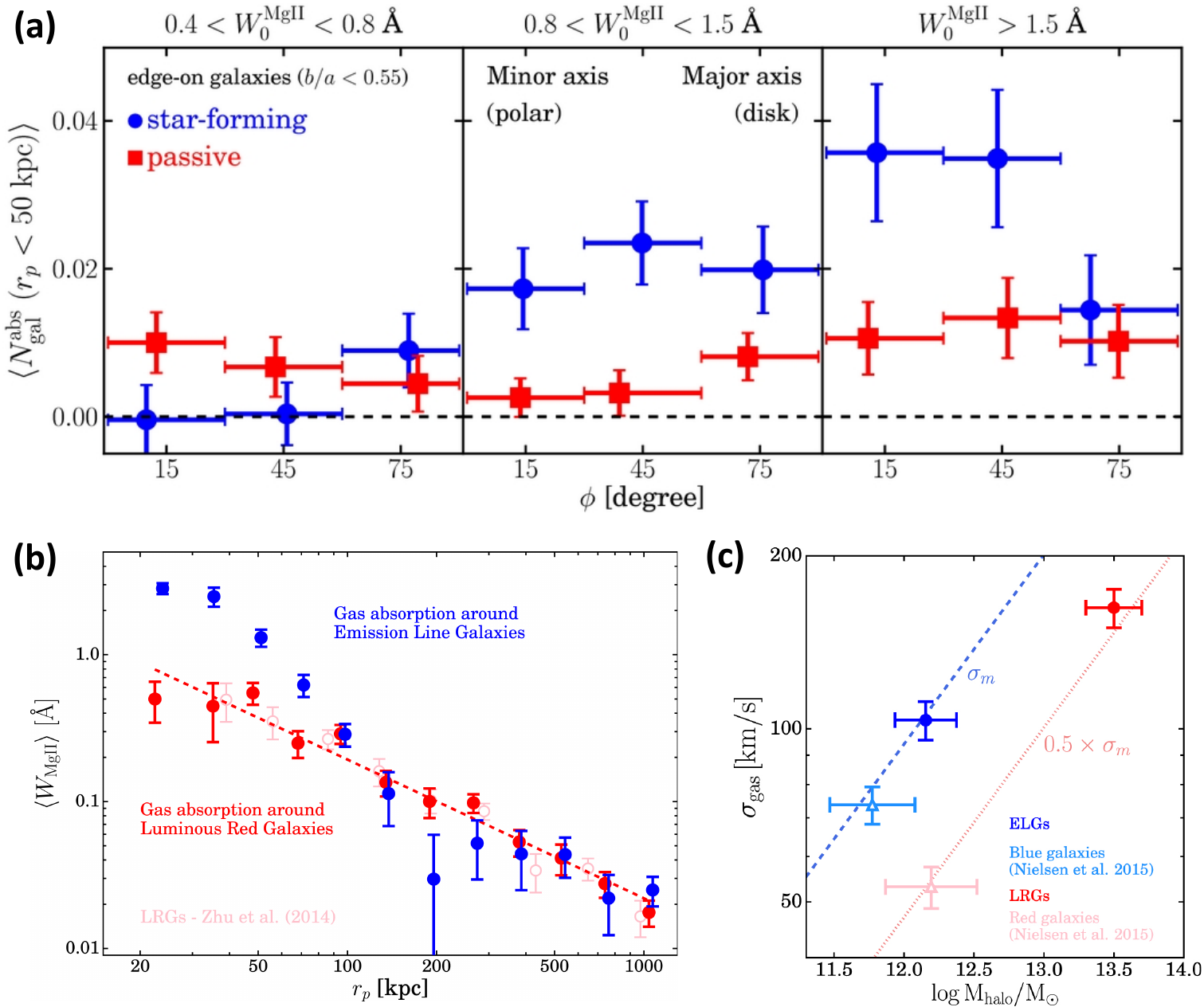}
\end{center}
\caption{Statistical properties of NUV Mg~II absorbers. (a) Frequency of occurrence as a function of position angle with respect to the disk minor axis, for star-forming (blue) and passive (red) galaxies. (b) Mg~II equivalent width as a function of the distance from emission-line galaxies (blue) and luminous red galaxies (red). (c) Line-of-sight Mg~II velocity dispersion as a function of halo mass and galaxy types (color-coded). Images reproduced with permission from (a) \citet{LAN2014} and (b, c) \citet{LAN2018}, copyright by AAS.}
\label{fig:mgII_absorbers}
\end{figure}

Transverse absorption-line studies of $z \lesssim 1$ galaxies using background quasars and galaxies as probes of the CGM have provided additional constraints on the sizes of cool neutral-atomic outflows \citep[see review by][]{TUMLINSON2017}. Studies of individual star-forming galaxies \citep{KACPRZAK2012,MARTIN2019} 
and stacked spectra \citep{BORDOLOI2011,MENARD2011,LAN2014,LAN2018,LAN2019} have shown that the strongest $EW_r \gtrsim 0.3$ \AA\ Mg~II absorbers with $N_{\rm H} \gtrsim
3 \times 10^{18}$ cm$^{-2}$
lie within $\sim$ 50-80 kpc ($\lesssim 0.25 - 0.40$ $R_{\rm vir}$) and are more prevalent along the minor axis of disk galaxies, as expected if they trace biconical cool outflows emerging from the disk (Fig.~\ref{fig:mgII_absorbers}). A similar result was found by \citet{ZHU2013} using strong Ca II H+K absorbers. The remaining strong Mg~II absorbers seem largely aligned with the major axis of the disk and in co-rotation with the disk \citep{BOUCHE2012,NIELSEN2015,MARTIN2019,ZABL2019}. 
The excess of strong Mg~II absorbers within $\sim$50--80 kpc may be even more pronounced among quasars, especially luminous ones, although Mg~II is rarely detected down-the-barrel in these quasars suggesting that the nearest Mg~II clouds have been photoionized by the central AGN \citep{FARINA2014,JOHNSON2015}. As discussed in Sect.~\ref{sec:highz_neutral_atomic}, a similar deficit is seen in quasars at higher redshifts, based on C~II absorbers \citep{PROCHASKA2014,LAU2016}.

Passive galaxies, on the other hand, show a lower--though non-zero--rate of incidence of strong Mg II absorbers within $\sim$50 kpc than star-forming galaxies \citep[Fig.~\ref{fig:mgII_absorbers}ab;][]{LAN2014,LAN2018,LAN2019}, while the rate of incidence of H~I absorption in general, and that of strong Mg II absorption on scales greater than $\sim$100 kpc, are virtually the same for both galaxy populations \citep{THOM2012,LAN2014,KEENEY2017,LAN2018}. Similarly, C~II and Si~II absorption from the CGM around local dwarfs with modest SFR is much less common than in massive star-forming galaxies \citep{PROCHASKA2014,JOHNSON2017}. This last result seems at odds with model expectations of increasing mass-loading (and metal-loading) factors with decreasing stellar masses \citep{MURATOV2015,MURATOV2017}, and the recent observations that support these predictions \citep{CHISHOLM2017},  unless the CGM of these dwarf galaxies are dominated by unexpectedly high ionization states like C~IV and O~VI \citep{BORDOLOI2014a}, contrary to the results of \citet{MATHES2014}. The difference between passive and actively star-forming galaxies is also reflected in (1) the velocity distribution of the clouds in the absorption profiles, with the blue star-forming galaxies showing a larger velocity dispersion than red galaxies of the same masses, consistent with outflows in the star-forming galaxies \citep[Fig.~\ref{fig:mgII_absorbers}c; e.g.,][]{NIELSEN2016,LAN2018}, and (2) the radial profiles of the Fe~II/Mg II ratio, which indicate a substantial contribution from Type Ia SN enrichment in the inner halos of passive galaxies, but dominant enrichment from core-collapse SNe throughout the halos of star-forming galaxies and in the outer halo of passive galaxies \citep{ZAHEDY2016}. The well-known correlation between velocity width and metallicity in DLAs with $N_{\rm H} > 10^{20.3}$ cm$^{-2}$ has been shown to extend down to sub-DLAs with $N_{\rm H} = 10^{19} - 10^{20.3}$ cm$^{-2}$ \citep{SOM2015,QUIRET2016}. This correlation has traditionally been interpreted as a mass-metallicity relation of the DLA galaxy hosts, but in the present context it may instead reflect a connection between the wind kinematics and the metallicity of the outflowing material.

Overall, down-the-barrel observations of cool outflows and transverse absorption line studies of the cool CGM paint a picture where stellar feedback is capable of driving cool neutral-atomic outflows with radial sub-virial velocities of $\sim$100$-$200 km s$^{-1}$ out to 50--80 kpc (0.25--0.40 $R_{\rm vir}$) into the halo, expelling cool gas out of the central regions of the galaxies at rates that scale loosely with $\dot{M}_*$ and $\Sigma_{\rm SFR}$. 
It is also clear that the presence of a powerful quasar provides a significant boost in velocity, momentum, and energy over and above those provided by the stellar processes, but quasars also destroy the neutral gas in its proximity, reducing the rate of incidence of line-of-sight neutral-atomic outflows unless the gas is shielded by dust in the host or embedded in the outflow itself, as it is the case in Na~I outflows. AGN of lower luminosities do not significantly impact the kinematics of the neutral-atomic gas in the host galaxies \citep[this statement also seems to apply in general to the ionized phase of these systems,][]{WYLEZALEK2019}, although heating by these AGN may still significantly reduce the ability of the cool gas to form stars (see also Sect.\ \ref{sec:lowz_molecular}).

The cool outflowing material no doubt evolves on its way out of the galaxy, eventually decelerating and becoming more diffuse on average as it reaches CGM scales. The inferred hydrogen column densities of the cool clouds cited above ($N_{\rm H} \simeq 10^{20}$--$10^{22}$ cm$^{-2}$ at the base of the outflows based on Na~I~D measurements and $N_{\rm H} \simeq 10^{18}$--$10^{20}$ cm$^{-2}$ in the strongest Mg II absorbers in the CGM) are uncertain and model-dependent. The results depend sensitively on the ionizing field (is the starburst and/or AGN contributing to the ionization in addition to the UV background radiation field?) and the cloud structure \citep[is the gas along each line of sight characterized by a single density or a range of densities; e.g.,][]{WERK2014,STERN2016}. The inferred volume densities of these clouds are equally uncertain, although it is clear that they span a broad range of values: $n_{\rm H} \simeq 10^{-3} - 0.1$ cm$^{-3}$ for the Mg~II CGM clouds \citep{PROCHASKA2011,KACPRZAK2014,WERK2014,STERN2016,THOMPSON2016,ZAHEDY2019}, while $n_{\rm H} \simeq 0.1 - 10^2$ cm$^{-3}$ in the inner Mg~II and Na~I outflows \citep{PROCHASKA2011,TANNER2016,TANNER2017}.  The inferred cloud thickness of the Mg~II absorbers in the CGM ranges from $\lesssim$10 pc to $\sim$1 kpc \citep{STERN2016,ZAHEDY2019}, while their coherence length scale across the sky, based on comparisons between results on QSO-galaxy and galaxy-galaxy pairs \citep{RUBIN2018b,RUBIN2018a}, exceed $\sim$1.9 kpc, reflecting the scale over which the number and velocity dispersion of these structures are spatially correlated. In the few cases where it is measured reliably, the metallicity of the outflow on $\lesssim$ kpc scales is roughly solar, regardless of the host stellar mass \citep{CHISHOLM2018}. The metal loading factor inversely scales with stellar mass, consistent with models where outflows help shape the galaxy mass-metallicity relation \citep[][and references therein]{MACLOW1999,MURATOV2015,MURATOV2017,EMERICK2018,FORBES2019,MAIOLINO2019}.The metallicity of Mg~II absorbers on CGM scales are typically 1/10 that of the hosts, although there is significant scatter \citep[][]{KACPRZAK2014,KACPRZAK2019}. No doubt, complex mixing between the outflowing material and the accreting and recycled gas is taking place in the CGM, and can account for some of the large point-to-point metallicity variations \citep{POINTON2019}.

\subsection{Molecular Gas Component} \label{sec:lowz_molecular}

The detection of P-Cygni profiles (or simply blueshifted absorption) of the OH transitions of 79 $\mu$m or 119 $\mu$m (as discussed in Section \ref{sec:m_and_mdot}), which identified the first cool molecular outflow \citep[in Mrk~231, Sect.~\ref{sec:Mrk231},][]{FISCHER2010}, has now been extended to $\sim$ 50 galaxies thanks to Herschel spectroscopy \citep{STURM2011,VEILLEUX2013a,SPOON2013,STONE2016}. These samples are generally mostly restricted to (U)LIRGs, given the requirement of having a strong far-IR source in order to get enough signal to noise on the continuum to properly detect these features in absorption. Studies involving the detection of multiple molecular species and multi-transition absorption features, which enable a more detailed modelling of the outflow, are still restricted to about a dozen galaxies \citep{GONZALEZ-ALFONSO2012,GONZALEZ-ALFONSO2013,GONZALEZ-ALFONSO2014,GONZALEZ-ALFONSO2017b}. In these cases detailed information on the outflow rate, energetics and size of the outflow can be inferred \citep[although the abundances of the various molecular species are still difficult to constrain properly despite recent progress;][]{STONE2018}. Interestingly, the joint analysis of multiple transitions has shown a stratification of the outflow with the inner regions (with sizes ranging from a few pc to $\sim$100 pc) having higher temperatures (up to 400~K) and high column densities (up to $\rm N_H>10^{25}~cm^{-2}$) and an outer, cooler ($<100~K$) and more rarefied ($\rm N_H \sim 10^{22}-10^{23}~cm^{-2}$) envelope, distributed on scales of several hundred parsec. One should however take into account that the far-IR absorption features tend to preferentially probe the inner, more IR-luminous regions of the galaxy (given by the requirement of having a strong background far-IR source), hence more extended component of outflows may be missed. For all other cases, for which only one or a few transitions are available, generally the molecular absorption only provides some basic information on the outflow velocity distribution.

The identification of molecular outflows through the detection and mapping of CO transitions associated with high velocity gas has now been possible for more than fifty local galaxies, especially in recent years thanks to ALMA enabling high sensitivity and high angular resolution observations \citep[see e.g. compilations in ][]{CICONE2014,FIORE2017,FLUETSCH2019,LUTZ2019}. 
Some of these observations are very detailed and reveal very clumpy structures \citep[e.g.][]{PEREIRA-SANTAELLA2016,PEREIRA-SANTAELLA2018,CICONE2019} 
and complex morphologies on different scales, often difficult to describe with simple models \citep{COMBES2014,TSAI2012,SAKAMOTO2014,GARCIA-BURILLO2014,GARCIA-BURILLO2016,GALLIMORE2016,AALTO2016,PEREIRA-SANTAELLA2016,FALSTAD2017,PEREIRA-SANTAELLA2018,HERRERA-CAMUS2019b,TREISTER2020}. However, the bulk of these observations are often still limited to the detection of high velocity wings of the CO transitions,  and only marginally resolved. This makes it difficult to properly characterize the outflow properties, such as the outflow rate, kinetic power and momentum rate, as the direct determination of the outflow extent is an important piece of information to constrain these quantities (Sec. \ref{sec:energetics}). One should also keep in mind that these studies obviously suffer from biases. In particular, weak and low velocity outflows tend to be missed, as the CO emission associated with these motions are difficult to disentangle from the CO emission coming from the galactic disk, unless detailed, high angular resolution mapping is available which enable modelling (and subtracting) the galactic rotation curve \citep[e.g.][]{ZSCHAECHNER2016,PEREIRA-SANTAELLA2016,GALLIMORE2016}.

Generally, the outflow is traced in only one CO transition. In those few cases for which multiple transitions are observed, these are typically consistent with an excitation similar to that observed in the host galaxy \citep[e.g.][]{CICONE2012,FERUGLIO2015}, although (as discussed in Sec. \ref{sec:Mrk231}) there are also cases where the CO excitation in the outflow is remarkably different than typically observed in galactic disks, and in a few cases the very high CO excitation suggests an optically thin molecular medium \citep{DASYRA2016}.

The detection of additional molecular species in the outflow is even rarer. However, it is often found that dense gas molecular tracers (such as HCN, HCO$^+$, CN) are enhanced in galactic outflows \citep{AALTO2012,GARCIA-BURILLO2014,AALTO2015c,WALTER2017,FALSTAD2018,BARCOS-MUNOZ2018,HARADA2018a,MICHIYAMA2018,IMPELLIZZERI2019}. Evidence that the outflowing gas is systematically characterized by higher densities than in the host galaxy is also found for the warm ionized component, as highlighted by various studies of the density-sensitive nebular diagnostics, both in AGN and SF-driven winds \citep[][Gallagher et al.\ 2019, in prep., Fluetsch et al.\ 2019, in prep.]{PERNA2017,MINGOZZI2019,HINKLE2019}. These results suggest that gas compression makes the dense phase of the molecular gas more prevalent in outflows than in galactic disks \citep[as suggested by some models, ][]{ZUBOVAS2014,RICHINGS2018b}, or that the more diffuse molecular phase is more easily evaporated and destroyed \citep{SCANNAPIECO2017,DECATALDO2017} or both phenomena are at work. Furthermore, recent models have suggested that massive molecular clouds in the outflows are self-gravitating and may evolve into dense cores \citep{DECATALDO2019}.

The range of CO excitation properties and the range of dense-to-diffuse molecular gas ratio suggests that the CO-to-H$_2$ conversion factor in the outflow is also subject to variations from case to case, which adds uncertainty to the determination of the outflow properties. Those few outflows with multiple transitions available and also observed in other species, suggest a broad range of conversion factors, spanning about an order of magnitude, ranging from $\alpha _{CO} \sim 2~M_\odot~{\rm (K~km~{s}^{-1}pc^2)}^{-1}$ to  $\alpha _{CO} \sim 0.3~M_\odot~{\rm (K~km~{s}^{-1}pc^2)}^{-1}$ \citep{WEISS2001,DASYRA2016}. The vast majority of studies cannot really constrain $\alpha_{CO}$, hence a fixed $\alpha_{CO}$ is generally assumed. In \citet{AALTO2015b}, the impact of self-gravitating versus non-self gravitating dense gas on the mass estimate is discussed.

Within this context, as already mentioned in Section \ref{sec:m_and_mdot}, molecular outflows have also started to be traced by exploiting sensitive observations of [C~I] \citep[e.g.][]{CICONE2018a}, which (although locally these transitions are in a frequency range more difficult to observe) is arguably a better tracer of the molecular gas. In particular, the [C~I]-to-H$_2$ conversion is considered less affected by environmental conditions and scales only linearly with metallicity (as opposite to $\alpha _{CO}$ which has a debated and not fully understood super-linear dependence on metallicity). As already mentioned, ALMA [C~I] observations of the study case NGC~6240 have confirmed the presence of a very massive fast molecular outflows extending on scales of several kpc scale \citep{CICONE2018a}. The same observations have suggested that the CO-to-H$_2$ conversion factor in the outflow is as high as $\alpha _{CO} \sim 2~M_\odot~{\rm (K~km~{s}^{-1}pc^2)}^{-1}$. 

Despite all of the biases and uncertainties discussed above, statistical studies of molecular outflows in local galaxies have provided interesting information on the scaling relations between the properties of molecular outflows and those of their host galaxies.

Far-IR OH P-cygni profiles have provided the clearest identifications of outflows and clear statistics on the outflow velocities, generally avoiding ambiguity with other types of non-virial motions in studies based on transitions in emission. However, geometrical effects and orientation have certainly prevented the detection of some molecular outflows, specifically those outflows not oriented along our line of sight (the case of NGC~1068, which has a nuclear molecular outflow detected in multiple transitions, but no outflow signatures in its OH spectrum, is a clear example; Sec. \ref{sec:N1068}).

\begin{figure}[htb]
\begin{center}
\includegraphics[width=0.95\textwidth]{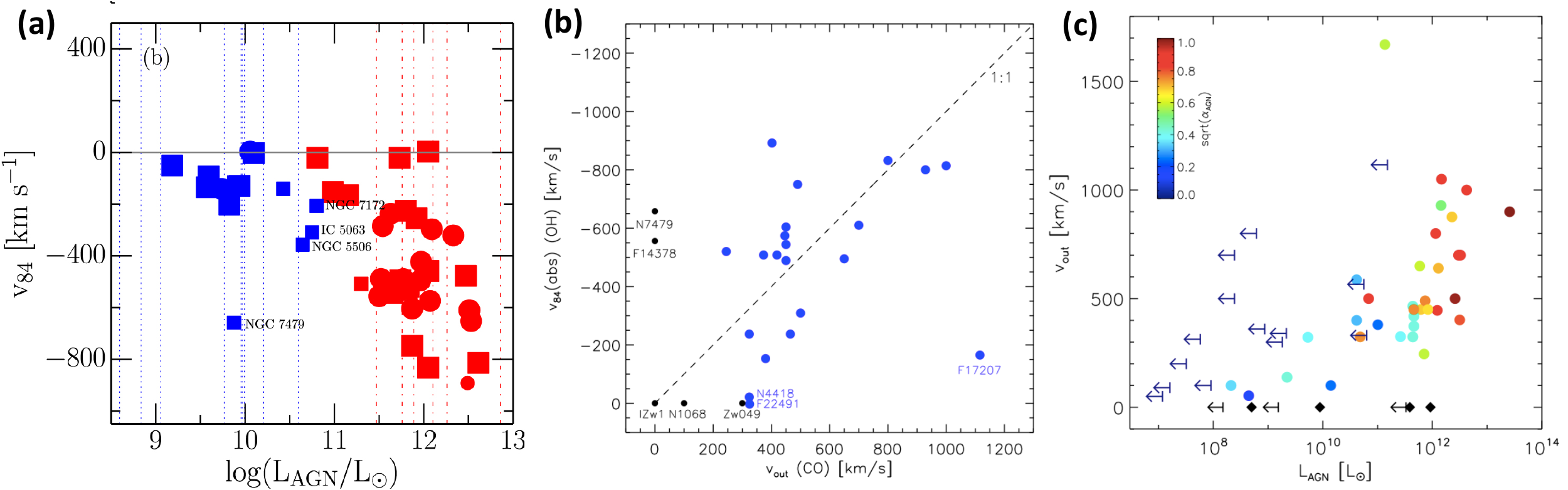}
\end{center}
\caption{Velocities of nearby molecular outflows. (a) OH-based 84-percentile outflow velocities in U/LIRGs (red) and BAT AGN (blue), (b) Comparisons of CO and OH 84-percentile outflow velocities in nearby gas-rich galaxies. (c) CO outflow velocities as a function of AGN luminosities, color-coded by AGN fraction. Images reproduced with permission from (a) \citet{STONE2016}, copyright by AAS; and (b, c) \citet{LUTZ2019}, copyright by the authors.}
\label{fig:v_molecular}
\end{figure}

The OH absorption studies reveal a broad range of velocities $v_{84}$ from a few 100 km~s$^{-1}$ up to a few 1000 km~s$^{-1}$ \citep[Fig.~\ref{fig:v_molecular};][]{VEILLEUX2013a,SPOON2013}. Interestingly, the outflow velocity does not correlate with the SFR, while it shows a clear correlation with the AGN luminosity. The fastest outflows are systematically seen to be associated with very luminous (quasar-like) AGNs, clearly revealing that luminous AGNs play a primary role in driving fast molecular outflows. In particular, in some cases, the outflow velocity exceeds 1000 km~s$^{-1}$. Such fast outflows are difficult to explain with models of starburst-driven outflows. Interestingly, ionized outflows generally have higher velocities, suggesting a different origin of the two phases. However, there are also outflows in which molecular and ionized phases have nearly identical velocities; in these few cases the scenario of molecular gas forming out of cooling from the hot phase may be a viable explanation (see Sect.~\ref{sec:thermal_energy} and \ref{sec:in-situ} for more details). We return to this point in Sect.~\ref{sec:lowz_fate}.

\begin{figure}[htbp]
\begin{center}
\includegraphics[width=0.95\textwidth]{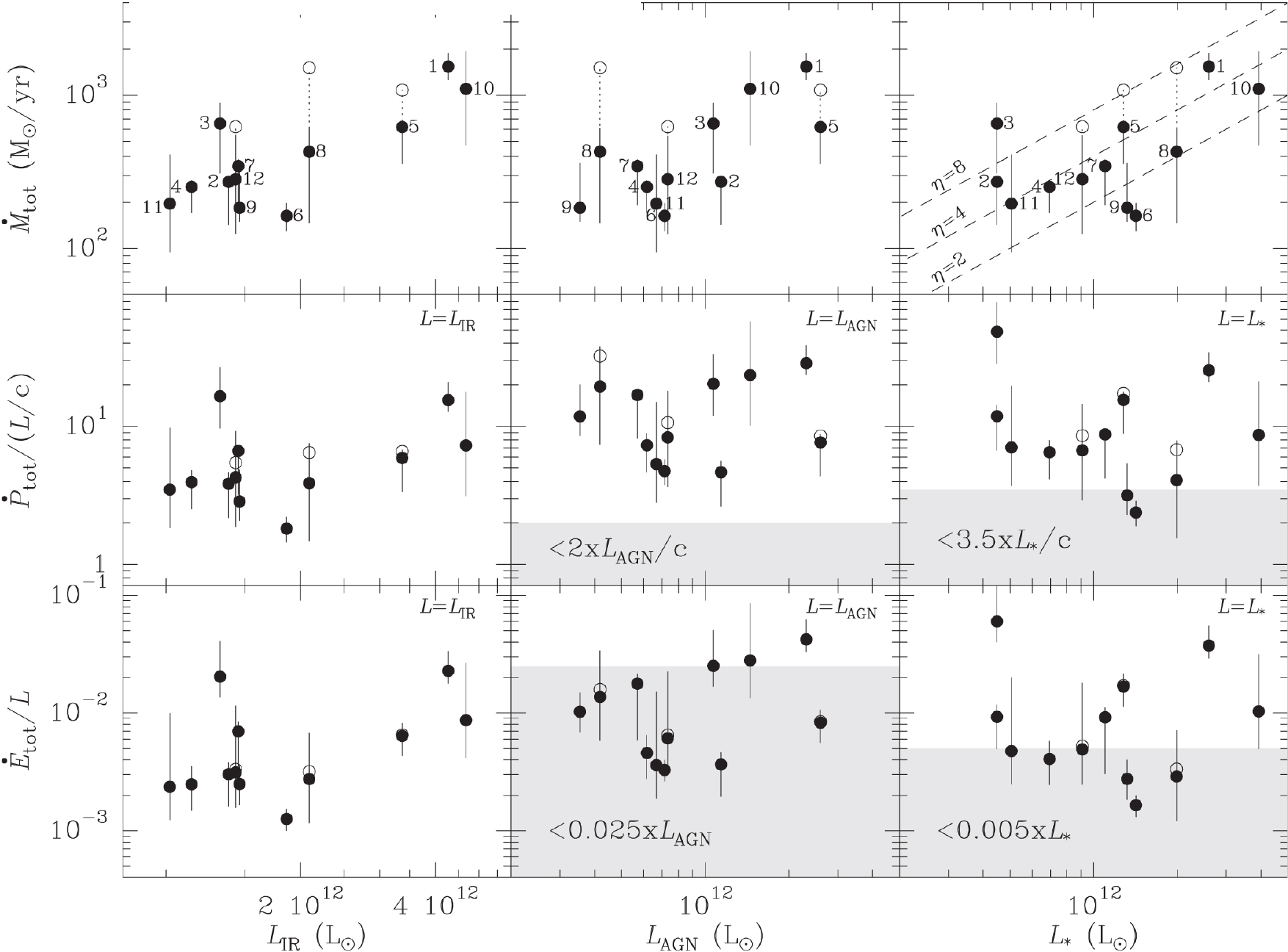}
\end{center}
\caption{Energetics of nearby OH-based molecular outflows. The mass outflow rates (top row), momentum rates normalized by the radiation pressure, $L/c$ (middle row), and kinetic energy rates normalized by the luminosity, $L$, are plotted as function of the total infrared luminosity (left column), AGN luminosity (middle), and starburst luminosity (right). Filled circles show the values obtained by ignoring the more uncertain low-velocity components, while open circles include them. Dashed lines in the upper right panel trace different mass-loading factors $\eta \equiv \dot{M}/\dot{M_*}$ = 2, 4, and 8. The grey-shaded rectangles mark the momentum and energy rates that can be supplied by an AGN and a starburst according to various models described in \citet{GONZALEZ-ALFONSO2017b}.   Images reproduced with permission from \citet{GONZALEZ-ALFONSO2017b}, copyright by AAS.}
\label{fig:oh}
\end{figure}

\begin{figure}[htbp]
\begin{center}
\includegraphics[width=0.95\textwidth]{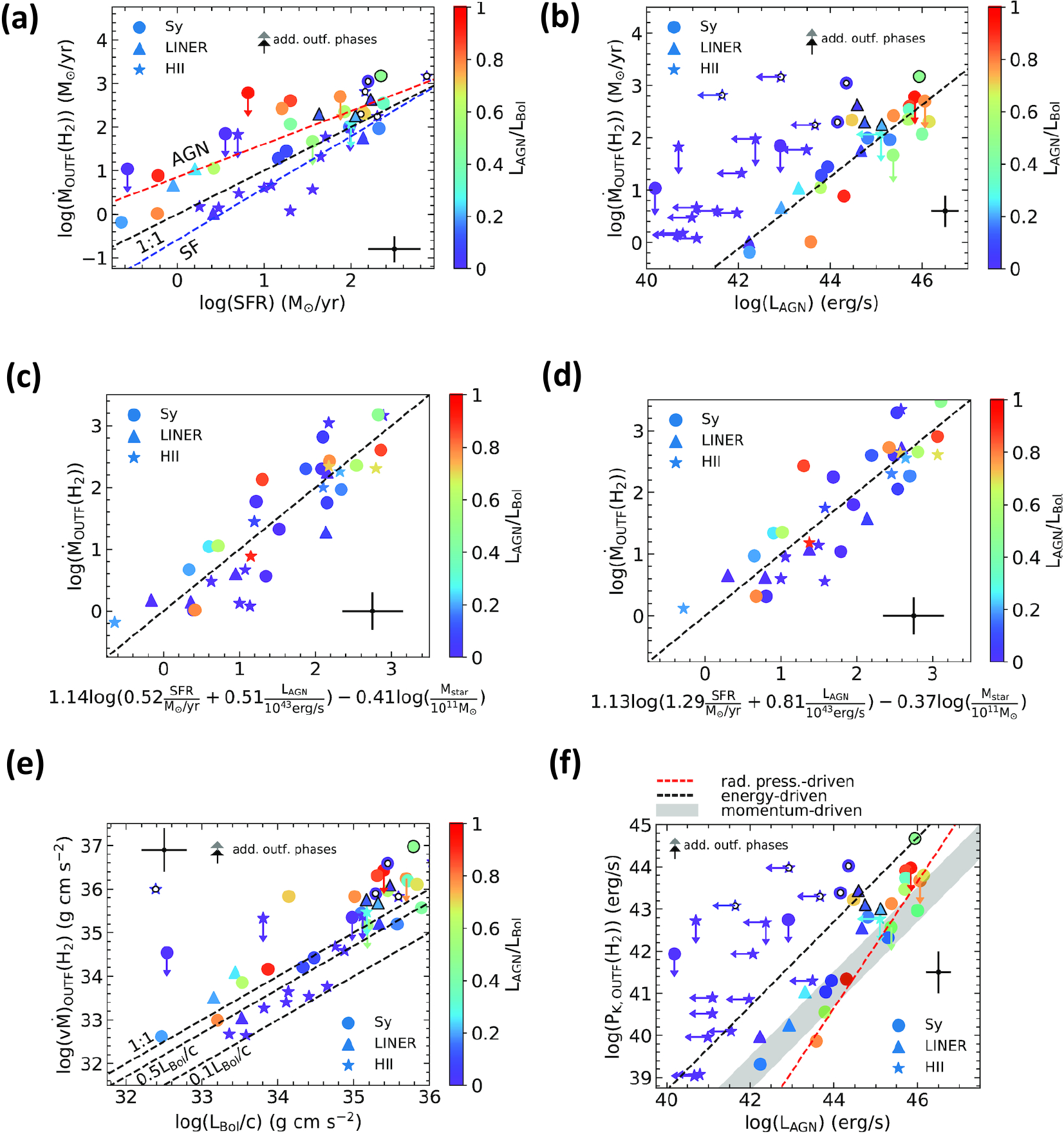}
\end{center}
\caption{Energetics of nearby CO-based molecular outflows. (a) Total (ionized + neutral + molecular) mass outflow rates are plotted as a function of SFR. The black dashed line indicates a mass-loading factor of unity, while the red and blue dashed lines are the best fits for the AGN and star-forming galaxies, respectively. (b) Same as (a) but plotted as a function of AGN luminosity.  The black dashed line indicates the best fit to the AGN host galaxies. (c) Molecular mass outflow rates as a function of a combination of SFR, AGN luminosity, and stellar mass. The black dashed line represents the best fit to the data. (d) Same as panel (c) but for the total mass outflow rate. (e) Momentum outflow rate as a function of the total radiation pressure, $L/C$. The dashed lines show different ratios of momentum rate to radiation pressure. (f) Kinetic power as a function of the AGN luminosity. The grey band indicates the predictions of momentum-driven outflows, while the black and red dashed lines indicate the predictions for energy-driven and radiation pressure driven outflows. In all panels, the symbols are color-coded according to the AGN fraction, and the shapes are used to differentiate between Seyferts, LINERs, and star-forming galaxies. Images reproduced with permission from \citet{FLUETSCH2019}, copyright by the authors.}
\label{fig:co}
\end{figure}

In star-forming galaxies, the outflow rate is roughly proportional to the star formation rate (upper right panel of Fig. \ref{fig:oh} and Fig.~\ref{fig:co}a). The molecular outflow rate to SFR rate is typically lower than unity. However, when taking into account the contribution from the ionized and atomic neutral phases, the total outflow rate to SFR ratio, i.e. the so-called ``outflow mass loading factor'', is around unity, as expected and required by models of galaxy evolution in order to regulate star formation \citep[e.g.][]{FINLATOR2008,DAVE2011,SOMERVILLE2015}. The same correlation extends to powerful starbursts, although there are indications that it might saturate in the most extreme objects in the early Universe (see Sec. \ref{sec:highz_molecular}).

The presence of an AGN can boost the outflow rate, and therefore the mass loading factor, by one or even two orders of magnitude (top-middle panel in Fig.\ref{fig:oh} and Fig.\ref{fig:co}a-b), providing further, unambiguous evidence that AGN feedback plays a key role in regulating star formation in galaxies. However, the relation between outflow rate and AGN luminosity has a large scatter. \cite{FLUETSCH2019} clarify that part of this scatter is due to the fact that one has to simultaneously take into account the contribution of star formation and AGN when trying to identify scaling relations for outflows, and suggest that a diagram involving a near-linear dependence on both SFR and AGN luminosity greatly decreases the scatter in the outflow scaling relations. The scatter decreases even further if one includes also the dependence on the galaxy stellar mass (Fig.\ref{fig:co}c-d), which is expected to play an additional important role as the deeper gravitational potential well of massive galaxies suppress the development of massive outflows (Eq.~(\ref{eq:eom})). Interestingly, once the dependence on SFR and AGN luminosity are simultaneously taken into account, the dependence on galaxy stellar mass follows the relation $\rm \dot{M}_{H2} \propto M_*^{\alpha}$ with $\alpha = -0.41\pm 0.25$, i.e. consistent with the theoretical value $\alpha _{th}=-0.5$ expected by many past models of outflows driven by star formation \citep{MITRA2015,SOMERVILLE2015,CHISHOLM2017}. While more recent models and numerical simulations have obtained a more complex dependence on stellar mass \citep{NELSON2019,TOLLET2019}, they are however still broadly consistent with the observations given the current large uncertainties and scatter. Often there is also some level of misunderstanding between theoretical/simulated quantities and observations on what is the definition of ``outflow rate'', as observationally it is often identified as the rate at which gas leaves the active region of the galaxy, while often in theoretical models the 
outflow rate refers to the fraction of the gas permanently leaving the galaxy or the halo.

The residual dispersion in these relations is certainly partly due to uncertainties in the measurements, but also likely associated with the flickering nature of AGNs (Sect.~\ref{sec:agn}). Indeed, black hole accretion can vary on timescales as short as a few years, while the outflow dynamical scales are of the order of at least a few million years \citep[e.g.][]{SCHAWINSKI2015,GILLI2000}. Therefore, AGN-driven outflows can easily outlast the black hole active phase (AGN-quasar) and one would expect a significant population of ``fossil'' outflows. This phenomenon has been modelled in detail by \cite{ZUBOVAS2018}. The observational identification of such ``fossil'' outflows is not easy, by definition, as the tendency is always to associate the outflow with the currently observed property of the host galaxy. However, recent studies have identified massive and energetic molecular outflows which are extremely difficult to explain in terms of star formation or (currently observed) AGN luminosity \citep[][]{SAKAMOTO2014,FLUETSCH2019,LUTZ2019}; the most plausible explanation in these cases is that the outflow is the result of a past ejection event produced by a (more active) AGN/quasar phase, that has recently faded.

Various authors have attempted to infer the effect of the observed  molecular outflows, especially the AGN-driven ones, on galaxy evolution. Given that molecular gas is the main fuel for star formation, these massive molecular outflows are expected to affect significantly the evolution of star formation in galaxies. In some of the AGN-driven outflows the ejection rate is so high (significantly higher than the star formation rate) that, if maintained, it can potentially clean the galaxy of its molecular gas content within only a few tens million years, hence potentially be effective in totally quenching star formation on short timescales \citep{STURM2011,CICONE2014,FIORE2017,GOWARDHAN2018,FLUETSCH2019}. However, these depletion times assume that there is no significant accretion on similar timescales (assumption which may be appropriate locally but not necessarily in the distant universe) and that the observed outflow is not an isolated ejection event but a continuous process (which is not really the case in the blast wave scenario and also in the case of AGN flickering). Moreover, the depletion times are often calculated by using the molecular gas content. If the total gas content is used, i.e. including the atomic H~I component, then the resulting depletion timescales are much longer. This suggests that most of these outflows may be effective in cleaning (and quenching) the nuclear region of galaxies, but not globally, across the entire disk. This is indeed expected by detailed, zoom-in 3D numerical simulations, illustrating that even for the most energetic outflows the bulk of the outflow escapes along the path of least resistance (e.g. perpendicular to the galaxy disk in the case of spiral galaxies) leaving the bulk of dense gas in the host galaxy unaffected \citep{COSTA2014,COSTA2015,GABOR2014,ROOS2015,HARTWIG2018,RICHINGS2018b,RICHINGS2018a,NELSON2019,KOUDMANI2019}.

The additional concern about the ejective mode in really being an effective mechanism in quenching galaxies is that generally only a small fraction of the outflowing gas reaches the escape velocity from the galaxy. This will be discussed more extensively in Section \ref{sec:lowz_fate}.

Most of the studies reported above refer to wide-angle outflows driven by the thermal or radation pressure from AGN, supernovae or young stellar populations. However, there is growing evidence that radio jets, despite being highly collimated, can also uplift molecular gas and produce wide angle outflows in the ISM of galaxies. In Section \ref{sec:jetted_agn},
we discussed in detail four specific cases (Cen~A, IC~5063, NGC~1266, and NGC~1377), but evidence is found for other galaxies with radio jets \citep[e.g.][]{DASYRA2012,GUILLARD2012,MORGANTI2018,MURTHY2019,FOTOPOULOU2019,ZOVARO2019,FERNANDEZ-ONTIVEROS2019}. It is generally found that the jet-driven outflowing gas in the molecular phase is as massive as the neutral-atomic phase, or even more massive, and largely exceeding the mass in the ionized phase. The morphology, excitation and general physical properties of jet-driven molecular outflows are consistent with models and numerical simulations in which relativistic jets, while collimated, can generate wide angle molecular outflows by depositing energy into the surrounding medium while piercing through the ISM of the galaxy \citep{WAGNER2012,MUKHERJEE2018b,MUKHERJEE2018a,BOURNE2017}. In the cases of clearly jet-driven outflows the mass loading outflow rates are generally significantly lower than outflows driven by thermal/radiation pressure. However, a systemic study of a large sample of jet-driven outflows is still missing, so it is not really possible to assess the impact of this phenomenon on the evolution of galaxies.  

In this context, it is interesting to remind the readers of the recent discovery of molecular outflows that are highly collimated, which also have indication of precession \citep[the best such case, NGC~1377, was discussed in  Sect.~\ref{sec:jetted_agn}; see also Sect.~\ref{sec:CONs} below;][]{SAKAMOTO2014,AALTO2016}. Such highly collimated outflows may be more common than the few cases identified so far, as their detection requires mapping with angular resolution (and sensitivity) high enough to fully resolve the structure of the molecular outflows, which has become possible only recently. The nature and driving mechanism of these highly collimated molecular outflows is not yet clear. Faint (undetected) or faded radio jets may have been responsible for driving these molecular outflows. Alternatively, these collimated molecular outflows may be driven by a disk-wind similar to those seen in protostars, but scaled up by a large factor.

Still in connection with radio jets, but on much larger scales, the last few years has seen the ubiquitous detection of molecular filaments in the vicinity of the bright central galaxies (BCG) of galaxy clusters, which are generally radio galaxies. These molecular structure extend from a few kpc to several tens kpc, have molecular gas masses in the range of $\rm 10^8-10^{10}~M_{\odot}$, and generally are seen to trail the X-ray hot cavity inflated by the radio jets \citep{RUSSELL2014,RUSSELL2017a,RUSSELL2017b,RUSSELL2019,TREMBLAY2018,VANTYGHEM2018,VANTYGHEM2019,OLIVARES2019}. The velocity gradients along the filaments are smooth and shallow, and generally inconsistent with free-fall. The origin of these large-scale molecular structures is not yet clear. At least in a few cases the mass of the molecular gas in the filament is so large that it is unlikely to result from direct uplifting of molecular clouds by the radio-jet or by the radio-bubbles \citep[e.g.][]{RUSSELL2017a}. Generally, the scenario favoured by most authors is that the molecular filaments result from low-entropy (X-ray emitting) gas uplifted by the radio bubbles and which becomes thermally unstable and therefore cools rapidly forming molecules \textit{in situ}. As we shall see in the next sections, the gas seems to cool to the point of becoming gravitationally unstable and result in prominent star formation. However, in some cases even the gas cooling from low entropy gas is unable to explain the large masses of gas observed in filaments in the intracluster medium and in these cases sloshing of molecular gas induced by galaxy merging appears to be a reasonable alternative \citep{VANTYGHEM2019,OLIVARES2019}. Finally, it is important to note that in all cases the velocity of these molecular filaments is significantly lower than the escape velocities, implying that this molecular gas will rain back onto the galaxy to fuel additional star formation and black hole accretion, in line with scenarios of chaotic cold accretion and precipitation \citep[e.g.][]{GASPARI2017}

\subsection{Dust Component} \label{sec:lowz_dust}

The evidence for dust in outflows and the halos of galaxies is wide-spread, varied, and quickly growing. Extinction and reddening measurements (Sect.~\ref{sec:ext_red}) have a proven track record of successfully detecting dust using background light from the host galaxy itself. Dark dust ``worms'' and ``filaments'' seen against the bright optical stellar continuum of nearby edge-on disk galaxies with modest SFRs are the telltale signs that stellar feedback is efficient at transporting dust at least $\sim$ 2-3 kpc from the sites of star formation \citep[Fig.~\ref{fig:diagnostics_dust};][]{HOWK1997,HOWK1999,THOMPSON2004}. At the other extreme of star formation activity, reddening measurements based on galaxy colors and H$\alpha$/H$\beta$ line ratios have long been known to be a good predictor of deep Na~I absorption features and Na~I outflows in U/LIRGs \citep{VEILLEUX1995,HECKMAN2000,RUPKE2005c}.  More recently, spatially resolved reddening measurements across the host galaxies of Na~I outflows have shown a good spatial correspondence between $E$(B-V) and $N$(Na~I), proving without a doubt that dust is entrained with the neutral gas in these cool outflows \citep{RUPKE2013b}. 

\begin{figure}[htb]
\begin{center}
\includegraphics[width=0.95\textwidth]{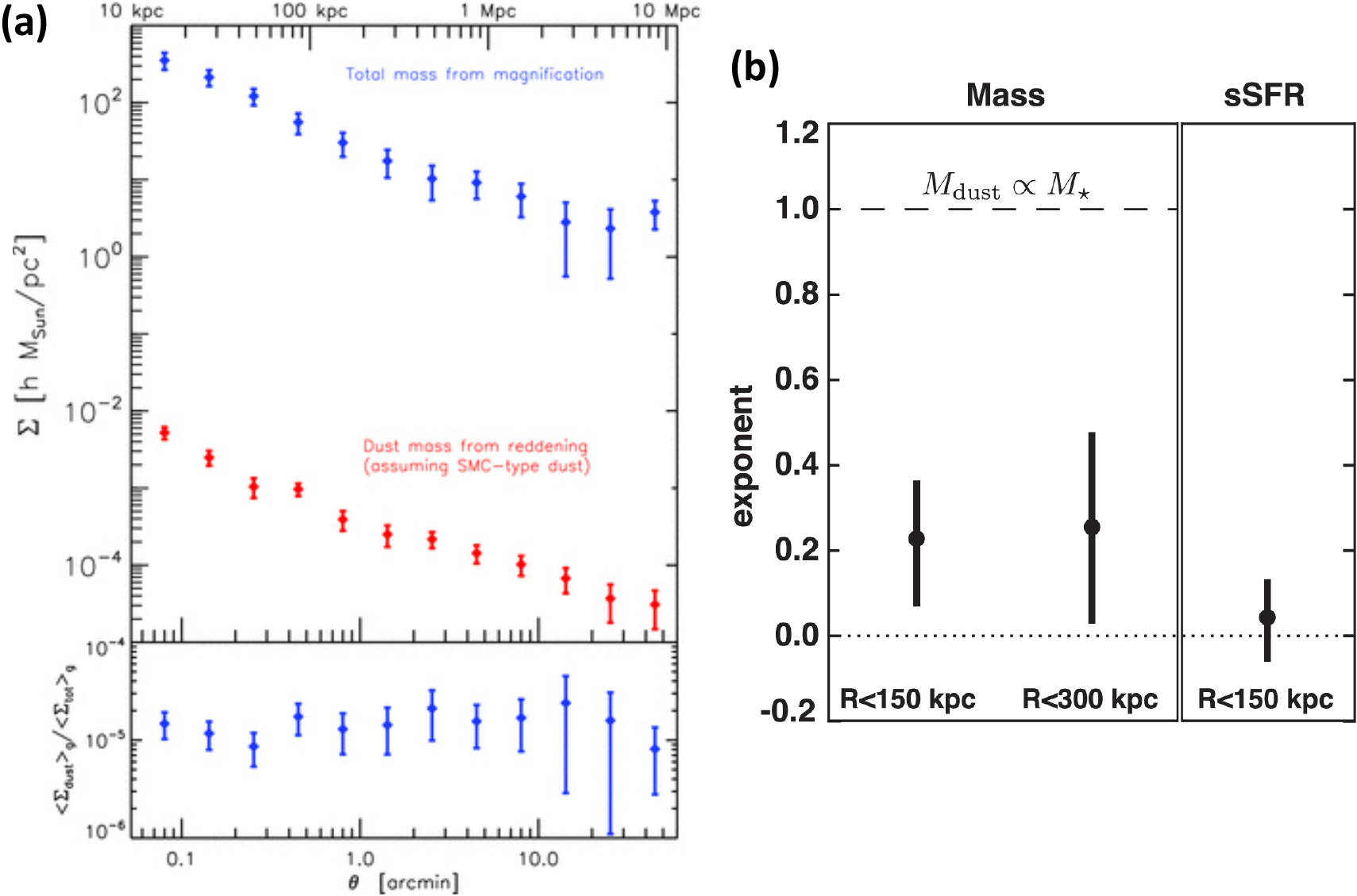}
\end{center}
\caption{Circumgalactic and intergalactic dust derived from reddening in nearby galaxies. (a) Mean surface mass density (blue), dust mass density (red), and the ratio of dust to total mass (bottom in blue) as a function of the angular distance (lower scale) and effective projected distance (upper scale) from $z \sim 0.3$ galaxies. (b) Power-law index of the dependence of the circumgalactic reddening on the stellar mass and sSFR in $z \sim 0.05$ galaxies. Images reproduced with permission from (a) \citet{MENARD2010}, copyright by the authors; and (b) \citet{PEEK2015}, copyright by AAS.}
\label{fig:dust_reddening}
\end{figure}

In recent years, extinction and reddening of background probes by dusty foreground galaxies has become arguably one of the most powerful techniques to search for dust in the CGM. Following the pioneering work of \citet{ZARITSKY1994}, studies of the reddening measurements of background quasars \citep{MENARD2010} and passively evolving galaxies \citep[``standard crayons'';][]{PEEK2015} by foreground $z \lesssim$ 1 galaxy halos have detected dust from 20 kpc to a few Mpc from galaxies, i.e. extending all the way to the IGM (Fig.~\ref{fig:dust_reddening}a). The inferred amount of dust outside of galaxies is comparable to that within galaxies, and the sum of the two adds up to the amount of dust produced in stellar evolution over the entire history of the universe \citep{FUKUGITA2011}. This remarkable result implies that most of the intergalactic dust survives over a cosmic time ($t_{\rm sput} \simeq 10^{10}$ yrs if $n_{\rm H} \simeq 10^{-5}$ cm$^{-3}$ from Eq.~(\ref{eq:t_sput})).
The circumgalactic reddening shows surprisingly no dependence on sSFR and only a weak correlation with the stellar mass ($M_{\rm dust} \propto M_*^{0.2}$; Fig.~\ref{fig:dust_reddening}b). The reddening curve derived from these data is SMC-like (steeply rising in the UV) and inconsistent with graphite (no strong 2175 \AA\ bump). The data favor silicate with $a \sim 0.03$ $\mu$m or amorphous carbon with $a \sim 0.01 - 0.03$ $\mu$m, although if the main dust composition in galaxy halos is silicate there is a tension in the grain size between the constraint from the reddening and that from the evolution of extinction with redshifts \citep{HIRASHITA2018}.  Regardless, these results indicate that large dust grains in the CGM are ruled out, perhaps a sign that the dust grains in the CGM experience erosion as they travel to large distances from the host galaxies. However, these conclusions should be tempered by the fact 
that some of the circumgalactic dust may be associated with extended gaseous disks rather than cool winds \citep{SMITH2016}. A study that takes into account the orientation of the sight lines with respect to the minor axis of foreground galaxies should be able to weigh in on the question of the origin of the circumgalactic dust.

\begin{figure}[htbp]
\begin{center}
\includegraphics[width=0.95\textwidth]{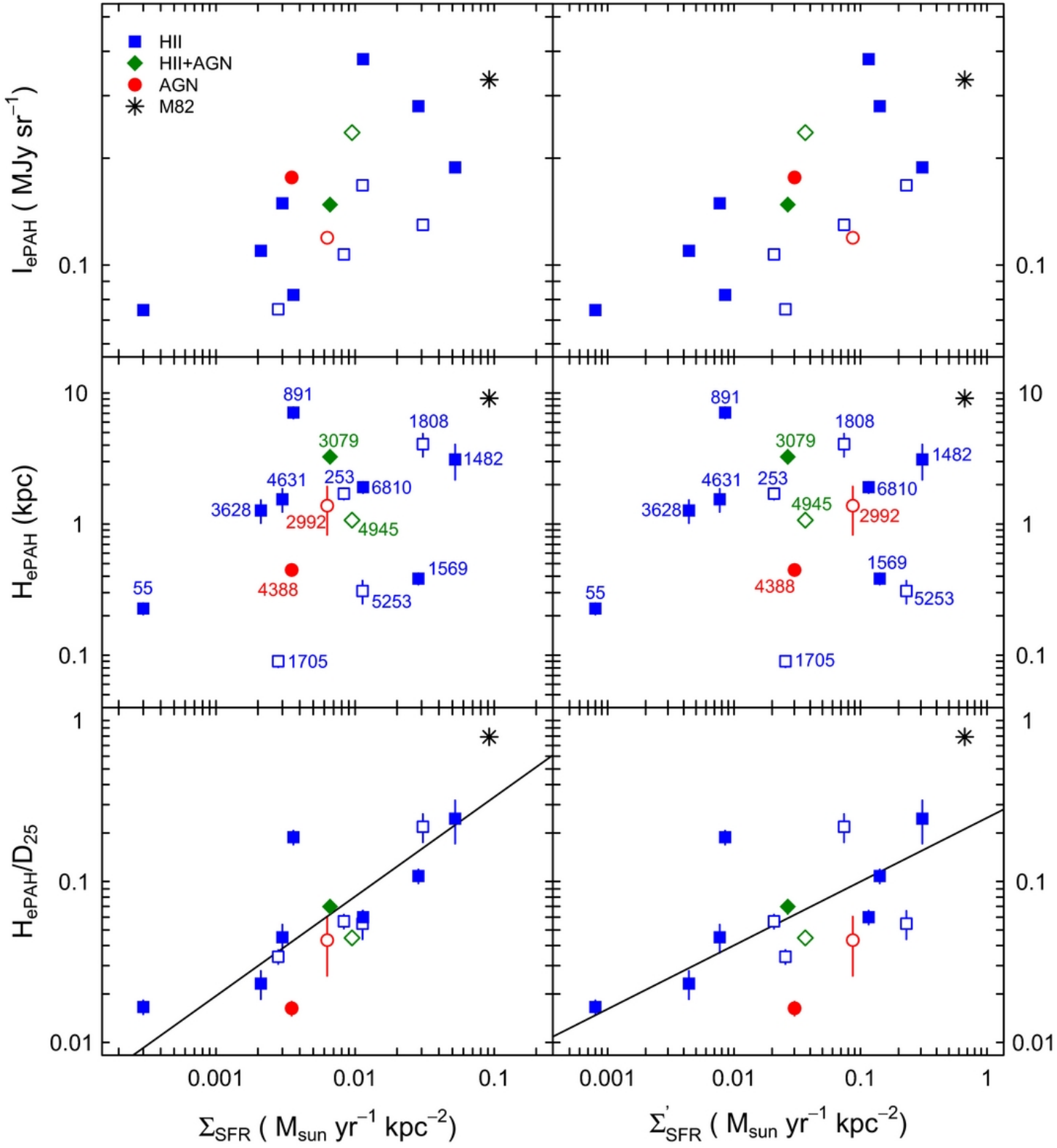}
\end{center}
\caption{Extraplanar PAH emission in six nearby disk galaxies. Characteristic extraplanar PAH surface brigthtness (top row), characteristic extraplanar emission height (middle row), and characteristic extraplanar emission height normalized by the stellar disk diameter (bottom row) as a function of the galaxy's SFR surface density (left column) and characteristic SFR surface density (right column). Images reproduced with permission from \citet{MCCORMICK2013}, copyright by AAS.}
\label{fig:dust_pah}
\end{figure}

\begin{figure}[htb]
\begin{center}
\includegraphics[width=0.95\textwidth]{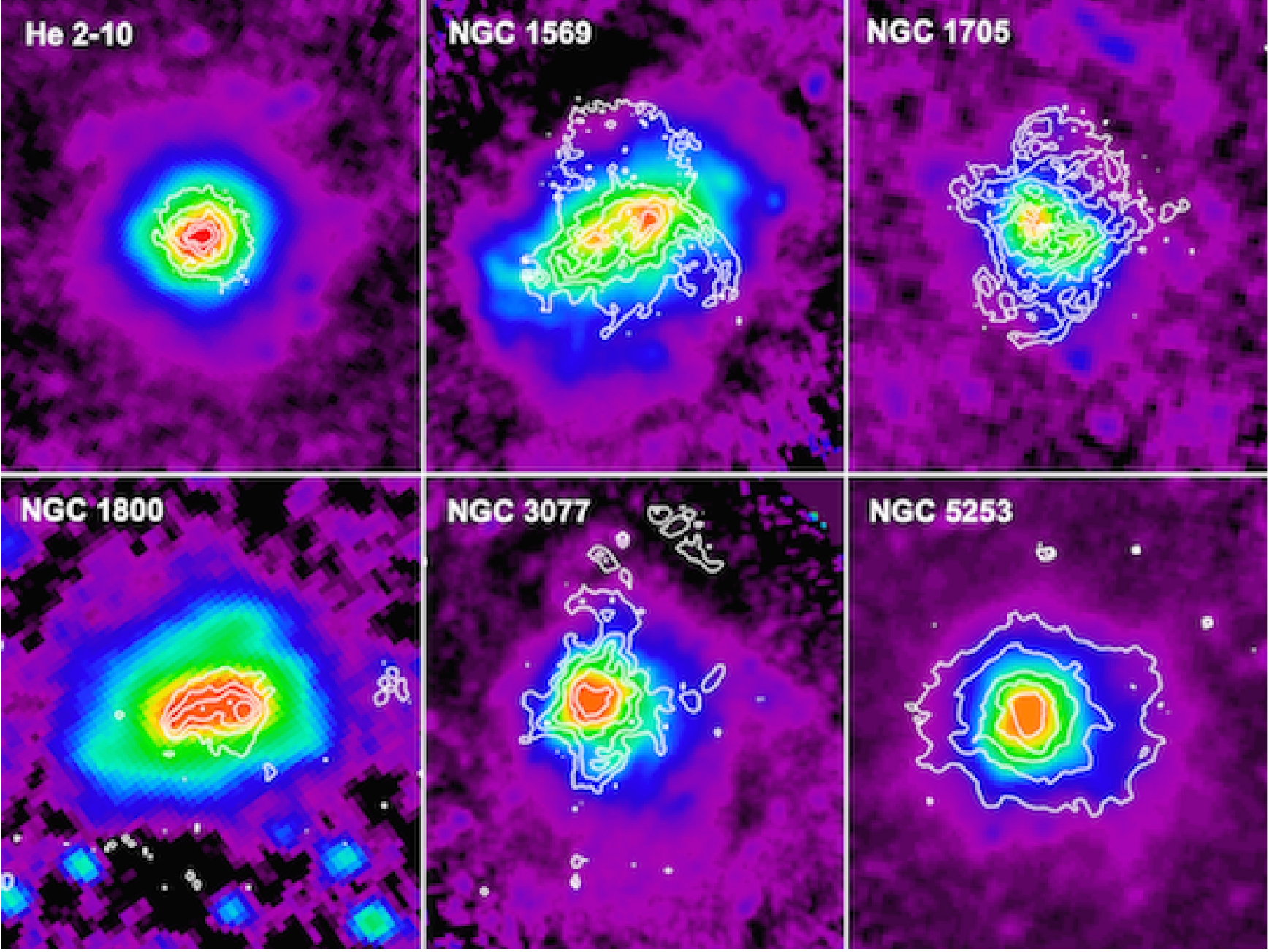}
\end{center}
\caption{Circumgalactic infrared dust emission in nearby dwarf galaxies. Herschel PACS 160 $\mu$m maps are overlaid with H$\alpha$ contours to compare the distribution of the cold dust with that of the warm ionized material. Images reproduced with permission from \citet{MCCORMICK2018}, copyright by the authors.}
\label{fig:dust_ir_dwarfs}
\end{figure}

Most of our knowledge on the distribution and properties of the circumgalactic dust in nearby galaxies has come from using \emph{direct} methods of detection. In the past $\sim$15 years, \emph{ISO}, \emph{Spitzer} and Herschel have been able to directly map the thermal emission from dust in the halos of about two dozen galaxies (Sect.~\ref{sec:emission}). 
A linear correlation between the amount of extraplanar PAH emission and SFR has been reported by \citet{MCCORMICK2013}. Their results also indicate a correlation between the height of the extraplanar PAH emission and the star formation rate surface density, $\Sigma_{\rm SFR}$ (Fig.~\ref{fig:dust_pah}), although objects with extensive PAH emission (up to $\sim$10 kpc above the mid-plane) and low SFR and $\Sigma_{\rm SFR}$ do exist \citep[e.g., NGC~5907 and NGC 5529:][respectively]{IRWIN2006,IRWIN2007}. 
NGC~4631 is another object which does not fit these simple relations. A rich complex of cool-dust filaments and chimney-like features extends to projected distance of 6 kpc above the plane of this galaxy despite a modest SFR and $\Sigma_{\rm SFR}$ \citep[Fig.~\ref{fig:diagnostics_dust}d;][]{MELENDEZ2015}. 
Star formation in the disk of this galaxy seems energetically insufficient to lift the implied $\sim$10$^8$ $M_\odot$ material out of the disk, unless it was more active in the past or the dust-to-gas ratio in the superbubble region is higher than the assumed Galactic value. Another important factor is the galaxy mass, which sets not only the escape velocity but also the gas pressure in the ISM working against that of the hot bubble created by the SNe (Sect. \ref{sec:thermal_energy}). Indeed, the fraction of the dust mass that resides outside of the stellar disks of the few star-forming dwarf galaxies studied so far is higher on average than in larger galaxies, typically $\sim$ 10$-$20\% instead of $\lesssim$5\% in the more massive galaxies.  This fraction is even higher (perhaps $\gtrsim$ 50\%) in NGC~1569, host of a well-known galactic wind \citep[][Fig.~\ref{fig:dust_ir_dwarfs}]{MCCORMICK2018}. Evidence for dust grain processing by shocks and other processes has been reported in a few objects. A pixel-by-pixel analysis of the Herschel data on NGC~4631 shows that dust coincident with an 3-kpc X-ray superbubble has a higher temperature and/or an emissivity with a steeper spectral index ($\beta > 2$) than the dust in the disk, possibly the result of the harsher environment in the superbubble. \citet{BOCCHIO2016}  have also found that the abundance of small grains relative to large grains $\sim$2 kpc from the midplane of NGC~891 is almost halved compared to levels in the midplane. 

\begin{figure}[htb]
\begin{center}
\includegraphics[width=1.0\textwidth]{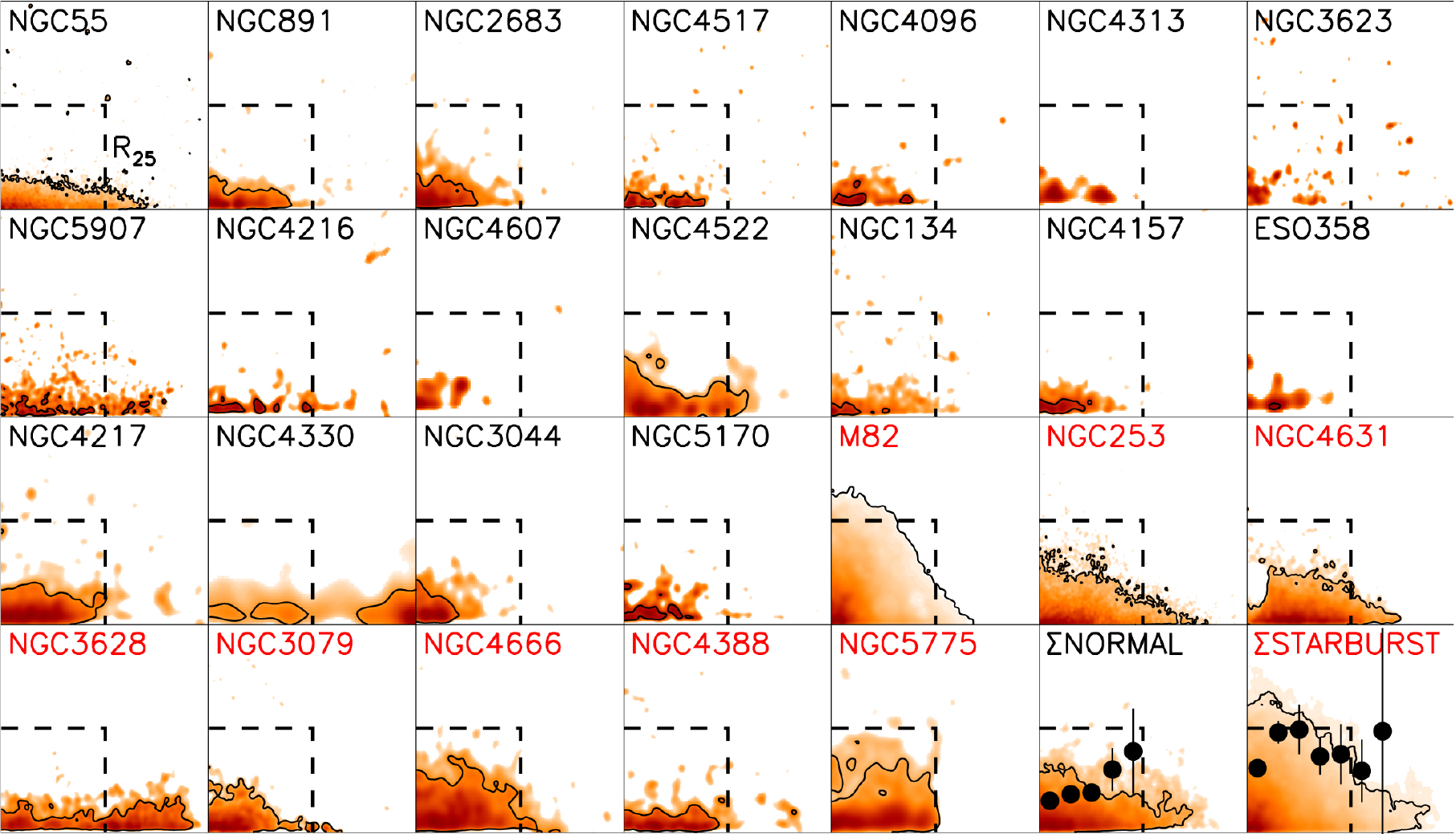}
\end{center}
\caption{Dust scattering in nearby edge-on star-forming disk galaxies. Combined UV image stacked in the first quadrant are shown relative to the optical radius, $R_{25}$ (indicated by the dashed lines). Starburst galaxies are labeled with red names. The final two panels show the combined normal and starburst galaxies, respectively, with scale heights measured as a function of galactocentric radius shown as points. Image reproduced with permission from \citep{HODGES-KLUCK2016a}, copyright by AAS.}
\label{fig:dust_scattering}
\end{figure}

As discussed in Section \ref{sec:scattering}, galaxy light scattered off of dust (``reflection nebulae'') may also be used to detect dust in galaxy halos and constrain its properties (since the scattering cross-section $\sigma_{s,\lambda}$ depends on dust composition and grain size distribution; Sec. \ref{sec:scattering}). 
This method has been particularly fruitful in the UV, where dust scattering is efficient and the sky background in space is dark. Systematic searches for vertically extended scattered UV starlight in $\sim$ 50 nearby edge-on galaxies with \emph{GALEX} and \emph{Swift} have revealed UV halos extending up to $\sim$ 20 kpc from the galaxy midplanes \citep[e.g.,][]{HODGES-KLUCK2014,SEON2014,HODGES-KLUCK2016a,SHINN2018,JO2018}. A strong linear correlation between the integrated luminosities of UV halos and the de-reddened luminosities (and thus SFR) of the host galaxies is observed, fitting both starburst and less actively star-forming galaxies. However, starburst galaxies with known ionized galactic winds have qualitatively different halos that are more extensive and have complex filamentary structures that often match H$\alpha$ and cool-dust infrared filaments \citep[Fig.~\ref{fig:dust_scattering};][]{HODGES-KLUCK2016a,JO2018}.
Dust masses of a few $\times$ 10$^{6}$ $M_\odot$ (or gas masses of a few 10$^8$ $M_\odot$ for a MW-like dust-to-gas ratio) within $2-10$ kpc of the disk are implied, based on Monte Carlo radiative transfer (MCRT) scattering models and relatively simple assumptions about the geometry of both the scattering halos and galaxy disk luminosity distribution \citep{HODGES-KLUCK2016a,BAES2016}.  A good match is found between the dust masses derived from the UV data and those derived from FIR emission for the few objects where this comparison is possible \citep{HODGES-KLUCK2016a}.

In recent years, differential elemental depletion onto dust grains (Sect.~\ref{sec:depletion}) has independently confirmed the existence of dust in the halos of $z \lesssim 1$ galaxies inferred from reddening, infrared, and UV measurements. But, for all practical purposes, this method has so far been limited to the stronger metal-line absorbers with $N_{\rm H} \gtrsim 10^{18}$ cm$^{-2}$, where the UV absorption lines from ions of key volatile and refractory elements with different depletion factors (e.g., Zn, O, P, S, Si, Mg, Mn, Cr, and Fe) can be measured to derive the total column density of each element, and then compared with the hydrogen column density measured from the strength of the damping wings of Ly$\alpha$ or the H~I 21-cm emission line along the same line of sight (see Sect.~\ref{sec:depletion} for possible caveats). Dust inferred using this method has been reported in the Milky Way Fermi Bubbles \citep{SAVAGE2017}, the galactic fountain of NGC~891 \citep{BREGMAN2013,QU2019}, 
some DLAs at $z \lesssim 1$ \citep{PEROUX2006,RAHMANI2016},  and a few $z \lesssim 1.5$ LLS with $N_{\rm H} \simeq 10^{19} - 10^{20}$ cm$^{-2}$ \citep[``sub-DLAs'';][]{MEIRING2009b,MEIRING2009a}, although not all of them \citep[e.g.,][]{QUIRET2016}. On the other hand, dust depletion in $z \lesssim 1$ LLS with $N_{\rm H} < 10^{19}$ cm$^{-2}$ has been found to be modest, largely consistent with dust-free gas \citep[e.g.,][]{LEHNER2013,LEHNER2019}.  Most of these LLS are metal-poor, mostly ionized, and not believed to be directly associated with cool winds.

The results presented so far have largely focused on star-forming systems. However, it is clear that dust is also associated with AGN-driven outflows on all scales. We have already discussed the case for dust shielding to explain the AGN-driven Na~I outflows detected on small scales in FeLoBALs (e.g., Mrk~231; Sec. \ref{sec:Mrk231}) and on kpc scales in several AGN-dominated ULIRGs (sec. \ref{sec:lowz_neutral_atomic}). The location of the dust in LoBALs in general is uncertain, which makes the location and energetics of the outflows equally uncertain \citep[Eq.~(\ref{eq:U_H});][]{DUNN2010}.  In a few low-z cases, the same dust that is responsible for the anomalously red continuum colors of LoBALs and FeLoBALS seems to affect the kpc-scale narrow-line emission in these objects so it must be located outside of the outflow region \citep{DUNN2015}. This is not to say that none of the dust takes part in the FeLoBAL outflows, but this dust may have only subtle effects on the overall spectrum of these objects \citep[e.g., UV extinction without significant reddening;][]{VEILLEUX2016,HAMANN2017}. 

For completeness, we should mention that evidence for dust in the \emph{ionized} winds of AGN outside of the dust sublimation radius is also quickly growing. Mid-infrared interferometry of nearby Seyfert galaxies with VLT/MIDI has revealed that most of the 12 $\mu$m emission on scale of tens to hundreds of parsecs originates from optically thin dust in the polar region of the inner torus, likely entrained in a ionized wind \citep{HONIG2012,HONIG2013,BURTSCHER2013,TRISTRAM2014,LOPEZ-GONZAGA2014,LOPEZ-GONZAGA2016,ASMUS2011,ASMUS2016,HONIG2017,STALEVSKI2019,LEFTLEY2019}. 
Recently, \citet{MEHDIPOUR2018} reported the presence of dust in the X-ray wind of the nearby Seyfert galaxy IC~4329A, based on reddening measurements and the strengths of X-ray edge features from O, Si, and Fe that indicate dust depletion.  \citet{BARON2019b,BARON2019a} have also recently discovered a direct connection between ionized [O~III] outflows and the presence of dust emission at 10--30 $\mu$m in the SED of SDSS AGN. Using Eq.~(\ref{eq:r_subl}), the measured dust temperature $T_{\rm dust} = 100 - 200$ K is used to derive the location of this dust, $r \sim$ 500 pc, hence the size and energetics of the [O~III] outflows in AGN. Some of these dusty AGN-driven outflows may carry dust to CGM scales and help explain the large ($r \gtrsim 5-10$ kpc) scattering cones detected at $\sim$3000 \AA\ and in polarized light around obscured quasars \citep{ZAKAMSKA2005,OBIED2016}. These cones have opening angles 
and directions that match the blueshifted side of the [O~III] outflows. The scattering medium is inferred to be dusty and photoionized by the quasar.

\subsection{Driving Mechanisms in AGN-dominated Systems} \label{sec:lowz_driving_mechanisms}

The most powerful, most massive, and fastest cool outflows are associated with quasars. As discussed in Section \ref{sec:thermal_energy}, the most effective driving mechanism is the so-called ``blast-wave'' energy-driven mode. In this scenario the very fast ($v \sim$ 0.01-0.1~c) and energetic nuclear wind generated by the AGN radiation pressure produces a blast wave that expands with little radiative losses, i.e. energy-conserving \citep[e.g.][]{KING2010,ZUBOVAS2012,KING2015,FAUCHER-GIGUERE2012a}. In this model the cold molecular/atomic phase is accelerated either as a result of the galaxy ISM being entrained/shocked in/by the hot, fast outflow, or as direct cooling and fragmentation of the hot phase (see Sec. \ref{sec:origin}). In this scenario the kinetic power of the outflowing gas is expected to be of the order of a few percent of the AGN radiative luminosity (typically $\rm \dot{E}_{out}\approx 0.05~L_{AGN}$). Moreover, as mentioned in Section \ref{sec:thermal_energy}, the energy conserving nature of the outflow implies that the outflow undergoes a momentum boost, with respect to the momentum of the driving radiation radiation pressure, by a factor of $f v_{\rm wind}/v_{\rm out}$ from eq.\ (\ref{eq:pdot_outflow}), i.e. typically a few tens unless the efficiency $f$ to transform the kinetic energy of the nuclear wind into bulk motion of the large-scale outflow is $<< 1$ (most popular models expect $\rm \dot{p}_{out}\approx 10$-$20~L_{AGN}/c$).

Various studies have attempted to test the blast-wave scenario by constraining the kinetic power and momentum rate in AGN-driven outflows \citep[][]{CICONE2014,TOMBESI2015,NARDINI2015, FERUGLIO2015,FERUGLIO2017,VEILLEUX2017,RUPKE2017,FLUETSCH2019,BISCHETTI2019,LUTZ2019}. Large uncertainties in both quantities make a clear, solid assessment difficult to achieve. 
Moreover, in several galaxies the outflow is also partly, or mostly, driven by star formation, which complicates the interpretation. Despite all these caveats, the emerging picture is that AGN-driven outflows span a broad range in terms of $\rm \dot{E}_{out}/L_{AGN}$ and in terms $\rm \dot{p}_{out}/(L_{AGN}/c)$. However, most of the objects seem to lie below the prediction expected by the blast-wave, energy-conserving scenario. Figures \ref{fig:oh} and \ref{fig:co} show recent compilations of momentum and energy rates in galaxies in which the molecular outflow is mostly/partly AGN-driven. While a few galaxies are consistent with the expectation of the blast-wave scenario, the majority of them are below the blast-wave scenario expectations, some by orders of magnitude, which is difficult to reconcile in terms of erroneous assumption on the outflow geometry or even by assuming that a significant fraction of the outflow is in the ionized phase (for a significant fraction of these systems however the ionized phase is measured and found not to contribute significantly or, anyhow, not by the large factor needed to reconcile observations with the blast-wave scenario).

\begin{figure}[htb]
\begin{center}
\includegraphics[width=0.95\textwidth]{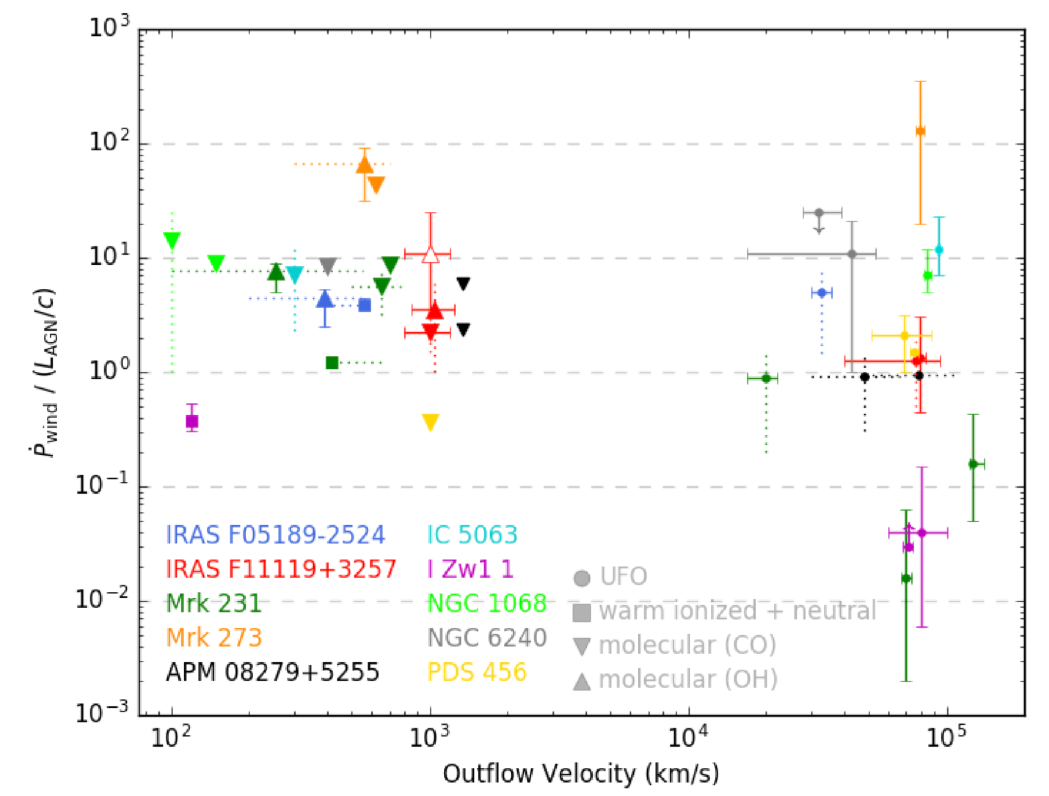}
\end{center}
\caption{Driving mechanisms of cool outflows in nearby AGN-dominated systems. Compilation of the momentum outflow rates normalized by the AGN radiation pressure for objects with both large-scale cool outflows (left) and nuclear X-ray winds (``UFO = ultra-fast outflow''; right). The color of each symbol identifies the object and the shape of the symbol indicates the method used to derive the plotted quantities. The open triangle of F11119+3257 is the instantaneous momentum rate from \citet{TOMBESI2015}; all other data points are time-averaged momentum rates. Images reproduced with permission from \citet{SMITH2019}, copyright by AAS.}
\label{fig:lowz_agn_driving_mechanisms}
\end{figure}

For a limited number of galaxies with measured molecular or atomic cold outflows, the momentum rate of the nuclear driving wind is also known through X-ray observations (Fig.~\ref{fig:lowz_agn_driving_mechanisms}). In these few cases, the momentum boost can be obtained through a direct comparison between the  momentum rate of the large-scale outflow and the nuclear wind. The first and still the best case of a fast quasar accretion-disk wind in a system with a massive molecular outflow was reported in the ULIRG F11119$+$3257 by \cite{TOMBESI2015} and confirmed by \cite{TOMBESI2017}.  As mentioned in Section \ref{sec:Mrk231}, in Mrk~231 the momentum rate of the large-scale cold outflow appears consistent with the nuclear wind boosted by $v_{\rm wind}/v_{\rm out}$ \citep{FERUGLIO2015}, i.e. with the (blast-wave) energy conserving scenario. A few additional sources seem consistent with this scenario \citep{NARDINI2015}. However in other cases the momentum rate of the large scale outflow is significantly lower than the expectations from the energy conserving scenario and nearly consistent with the momentum-conserving scenario \citep[Fig.~\ref{fig:lowz_agn_driving_mechanisms};][]{VEILLEUX2017,FERUGLIO2017,BISCHETTI2019,SMITH2019}.

One likely explanation for these results is that, although energy-driven, the nuclear AGN-powered wind does not couple efficiently with the galaxy ISM (i.e.\ small values of $f$ in eq.\ (\ref{eq:pdot_outflow})) and that most of the energy escapes along paths of least resistance (e.g. perpendicular to the galaxy disk in the case of regular late-type galaxies) without affecting the bulk of the ISM in the galaxy (especially the dense, star-forming component of the molecular gas). This (not really unexpected) result is obtained by most of the detailed numerical hydrodynamical simulations \citep{COSTA2014,GABOR2014,ROOS2015,COSTA2015,COSTA2018a,HARTWIG2018,HENDEN2018,KOUDMANI2019,NELSON2019}. Note that these scenarios do not imply that the AGN-driven wind does not have any major effect on the star formation in the host galaxy. Indeed, the energy injected into the galaxy halo results into heating of the CGM, which reduces or can even suppress cold accretion onto the galaxy. In the latter case star formation in the galaxy would quench as a consequence of `starvation', i.e. lack of fresh gas delivered to the galaxy to fuel further star formation. This is essentially the so-called `delayed' or `preventive' feedback mode, in which the energy injected into the halo does not happen through radio-jets but through wide-scale outflows. Indirect observational evidence of this mode being responsible for the quenching of star formation in galaxies has been suggested by various studies \citep[e.g.][]{WOO2017,PENG2015,TRUSSLER2018}.

An alternative scenario to explain the observed trends is that these AGN-driven outflows are primarily caused by direct radiation pressure onto the dusty clouds of the galaxy \citep{FABIAN2012,THOMPSON2015,ISHIBASHI2017a,ISHIBASHI2017b,COSTA2018b,ISHIBASHI2018a,ISHIBASHI2018b}. This scenario implies lower outflow momentum rates and lower energetics, and can potentially account for the scatter and lower values of  these quantities with respect to the energy-driven predictions. As discussed in Section \ref{sec:radiation}, this scenario was initially deemed implausible because of the Kelvin-Helmholtz and Rayleigh-Taylor instabilities that are expected to develop in this scenario and which would disrupt the outflowing clouds \citep{SCANNAPIECO2015,BRUGGEN2016,FERRARA2016b}. However, more recent works have shown that if the clouds are dense enough (as observed in many outflows now) their core can survive \citep{DECATALDO2019}  and additional cold gas can also  potentially cool in the tail of the outflowing clouds \citep[Fig.~\ref{fig:sims_clouds}d;][]{GRONKE2018b,GRONKE2019}. However, even if the gas clouds survive radiation (or ram) pressure acceleration, 
they have difficulties accounting for outflows that have large momentum rates and high kinetic powers. Multiple scattering of IR photons in a very dense, optically thick medium can potentially boost the momentum rate up to $\rm 2-5~L_{AGN}/c$ \citep{THOMPSON2015,ISHIBASHI2015,ISHIBASHI2018a,COSTA2018b}. Recent versions of these models predict a super-linear relation between outflow rate and $\rm L_{AGN}$, which seems supported by recent observations \citep{FLUETSCH2019}. However, these models still struggle to produce the large kinetic powers observed in some of the most powerful quasar-driven outflows (in these models typically $\rm \dot{E}_{out} < 0.01~L_{AGN}$).

As discussed in Sect.~\ref{sec:jetted_agn} and \ref{sec:lowz_molecular} (see also Sect.~\ref{sec:CONs} below), ram pressure from radio jets also contributes to driving cool outflows in some objects. However, the lack of a correlation between outflow rate and radio loudness in AGN with known molecular outflows suggest that radio jets are not a primary mechanism for most of the molecular outflows \citep{FLUETSCH2019}. This does not mean that when a radio jet is present in a galaxy it is not the primary driver of an outflow. The few detailed studies carried out so far indicate that, when a radio jet is present, there are no clear galaxy or AGN properties defining whether the molecular outflow is primarily driven by thermal pressure, or radiation pressure, or the radio jet (or even by SNe associated with star formation); each individual case has its own specific peculiarities.

\subsection{Compact Obscured Nuclei (CONs): A Special Evolutionary Phase?}\label{sec:CONs}

\begin{figure}[htbp]
\begin{center}
\includegraphics[width=1.1\textwidth]{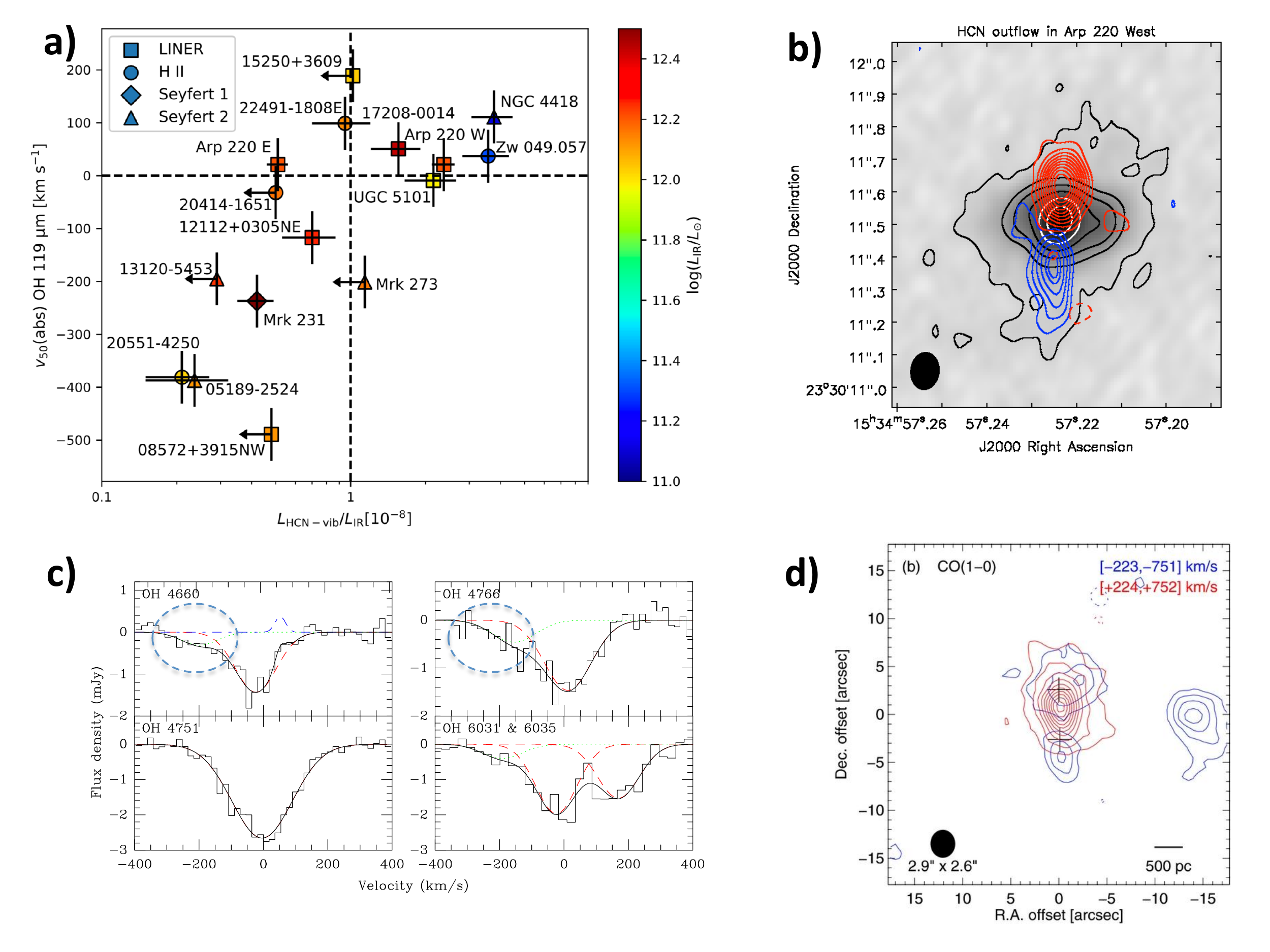}
\end{center}
\caption{Compact Obscured Nuclei (CONs). (a) Median far-IR OH absorption velocity as a function of the mm-wavelength HCN-vib luminosity relative to the total infrared luminosity. Squares, circles, diamonds, and triangles represent LINER, HII, Seyfert 1, and Seyfert 2 optical spectral types, respectively. Colors indicate the total infrared luminosity of each system. (b) Greyscale (+black and white contours) 92 GHz continuum image of the western nucleus of Arp~220. The red and blue contours show the red- and blueshifted HCN1-0 high-velocity emission. (c) VLA OH absorption lines at cm wavelengths revealing a fast outflow in the LIRG Zw049.057. (d) Collimated molecular outflow emerging from the southern nucleus of the LIRG merger NGC~3256. This nucleus is likely not a CON, but a dormant AGN where the outflow serves as a relic indicator of its past activity. Images reproduced with permission from (a) \citet{FALSTAD2019} and (c) \citet{FALSTAD2017}, copyright by ESO; and (b) \citet{BARCOS-MUNOZ2018} and (d) \citet{SAKAMOTO2014}, copyright by AAS.}
\label{fig:obscured_flow}
\end{figure}

{\em Herschel}-based studies of local molecular outflows \citep[Sect.\ \ref{sec:lowz_molecular};][]{VEILLEUX2013a} have shown that OH 119 $\mu$m outflows are ubiquitous among dusty U/LIRGs with deep OH and 9.7 $\mu$m silicate-dust absorption features (Sect.\ \ref{sec:ext_red}). This is consistent with a scenario where molecular outflows with an average opening angle of $\sim$145$^{\circ}$ are present in essentially all obscured systems,
but subside once the AGN clears a path through the obscuring material. \citet{GONZALEZ-ALFONSO2017b} 
have found that the most deeply buried sources (based on the depth of OH 65 $\mu$m) split into two subsets (groups I and III in their Fig. 4), where group III galaxies do not show fast OH 119 outflows, despite having the same $W_{\rm eq}$(OH 65) as the galaxies in group I.
\citet{FALSTAD2019} have recently shown that the most obscured galaxy nuclei, the CONs (Sect.\ \ref{sec:obscured}), in fact constitute the group with high $W_{\rm eq}$(OH 65), but without OH 119 outflows (Fig.~\ref{fig:obscured_flow}a).  The CONs are more obscured than the galaxies with fast OH 119 outflows, showing that $W_{\rm eq}$(OH 65) alone cannot identify the most deeply obscured sources. Instead, their identification requires using highly excited lines at longer, mm wavelengths (such as vibrationally excited HCN; Sect.~\ref{sec:obscured}).  There is evidence for inflows of molecular and atomic gas into the central regions of some CONs \citep[e.g.,][]{GONZALEZ-ALFONSO2012,COSTAGLIOLA2013,SAKAMOTO2013,AALTO2015a,FALSTAD2015,AALTO2019}.  It has therefore been suggested that CONs represent a brief evolutionary phase before mechanical feedback from the obscured starburst or AGN becomes important and creates a detectable outflow. Another possible explanation of these results is that the morphology (collimation), size, and orientation of the molecular outflows in CONs are different from those in galaxies of less extreme obscuration \citep{FALSTAD2019}. Since the CONs are extremely obscured, it is also possible that their outflows are obscured even in the far-infrared. To understand the difference in outflow structure with varying degree of obscuration (and nature and strength of the buried activity) requires multi-wavelength observations at high spatial resolution. 

Recent mm-wave studies of the CON ULIRG Arp~220 have revealed the presence of a fast (840 km~s$^{-1}$) collimated molecular and dusty outflow in its western nucleus \citep[Fig.~\ref{fig:obscured_flow}b;][]{SAKAMOTO2017,BARCOS-MUNOZ2018}. Interestingly, the collimated molecular outflow was found in HCN, not (as usual) in CO. The outflow is 120 pc long with a dynamical time scale of 10 Myr and a mass outflow rate $>$70 \msun\ yr$^{-1}$. It is oriented almost perpendicular to the line-of-sight and also shows opaque dust emission at its base. It is not clear if the driving source is primarily a compact starburst or a dusty AGN.  A slower outflow has been detected before in absorption \citep[e.g. in HCO$^{+}$ and SiO;][]{SAKAMOTO2008, TUNNARD2015}, and a faint, fast ($-$ 700 km s$^{-1}$) 119 $\mu$m OH absorption wing may also be present in this object \citep{VEILLEUX2013a}, but the connection between the collimated outflow and these various blueshifted absorption structures will require further investigation. Another intriguing case is the CON ULIRG IRAS~17208$-$0014, where the OH 119 $\mu$m absorption profile does not show any extended blue wing, but a fast (750 km s$^{-1}$) CO outflow is detected \citep{GARCIA-BURILLO2015}.  It appears collimated but its morphology is not yet fully mapped and hence the origin of the discrepancy between the OH and CO results is not clear.  The mass, energy, and momentum budgets of the molecular outflow in IRAS~17208$-$0014 seem to require the existence of a hidden AGN.  Other very obscured galaxies and CONs with outflows revealed at longer wavelengths include Zw049.057, ESO 320-G030, IC~860 and NGC~4418 \citep{BAAN1989,PEREIRA-SANTAELLA2016,FALSTAD2017,FLUETSCH2019,AALTO2019}. NGC~4418 and IC~860 have wide-angle, large-scale v-shaped dust lanes along the minor axis, and NGC~4418 shows a slow hot minor-axis flow similar to the galactic wind in M82 \citep{OHYAMA2019}. The link, if any, between the molecular outflows and these optical features is not yet understood. In the case of NGC~4418, the galactic wind may be powered by an extended 10 Myr old starburst.

In summary, several of the CONs show fast molecular outflows revealed by high-resolution mm-wave observations of CO and/or HCN. In some cases, their morphologies indicate highly collimated outflows, without obvious radio counterparts. 
The CON outflows may also be physically smaller than those seen in less obscured dusty galaxies, consistent with an evolutionary sequence. 
Note, however, that collimated, apparently radio-quiet, outflows are not restricted to CONs; they are also found in objects of less extreme obscuration. One example is the collimated molecular outflow associated with the southern nucleus of the merger NGC~3256 \citep[see Fig.~\ref{fig:obscured_flow}d;][]{SAKAMOTO2006,SAKAMOTO2014}. It is a fast ($\sim$2000 km~s$^{-1}$) outflow likely powered by a (now dormant) AGN.

\subsection{Fate of the Outflowing Material} \label{sec:lowz_fate}

Three questions arise when discussing the fate of the cool outflowing material: (1) does it escape the host galaxy altogether? (2) Does it experience a phase transition between cold molecular, cool neutral-atomic, and warm/hot ionized?  (3) Does it collapse to form stars? The data at $z \lesssim 1$ provide the best constraints on these three important questions.

The first of these questions has been addressed in several studies by comparing the measured radial velocity of the cool outflowing gas with the minimum requirement that it should exceed the escape velocity, often assuming $v_{\rm esc} \simeq (2-3)~v_{\rm circ}$, where the lower multiplicative factor is for CGM-scale gas while the higher multiplicative factor is for kpc-scale material (see Sec. \ref{sec:escape}). The fraction of neutral-atomic and molecular outflowing material that meets this simple requirement generally is less than 20\%, often consistent with zero \citep[e.g.,][]{HECKMAN2000,RUPKE2005c,MARTIN2005,RUBIN2014,SCHROETTER2015,SCHROETTER2016,SCHROETTER2019,FLUETSCH2019,BOUCHE2012}. The few cases where the cool material has velocities that exceed the escape velocity often are those where the velocities are measured down-the-barrel and on small scales \citep[$r \lesssim 1$ kpc; e.g.,][]{TREMONTI2007,WEINER2009,VEILLEUX2013a,CHISHOLM2017,GONZALEZ-ALFONSO2017b}. This perhaps indicates that drag forces are subsequently important in slowing down the material as it travels to the halo. However, one should also keep in mind the strong observational bias against detecting escaping outflows on CGM scales since (1) the volume and column densities, covering factor, and surface brightness of the outflowing clouds likely decrease with increasing radius \citep[e.g.,][]{KRUMHOLZ2017a}, and (2) the escaping clouds at large radii will have only modest velocities, of order $\sqrt{2}~v_{\rm circ}$, for ballistic motions. If instead the material accelerates for some time as it moves away from the center of the galaxy, as in the cases of thermally and radiatively-driven winds (Sect.~\ref{sec:thermal_energy} and \ref{sec:radiation}), then the material at small distances from the galaxy with velocities below escape speed may nonetheless be able to escape because it is still being accelerated. On CGM scales, this accelerated gas may be faster than the ballistically launched material but will remain hard to detect for the reasons listed earlier under (1).

While the current kinematic data, taken at face value, suggest that the cool material generally does not escape the galaxy, this does not necessarily mean that none of the outflowing material makes it out of the galaxy and into the IGM. Indeed, one very real possibility is that the cool outflowing material becomes ionized as it travels to the halo, so it is also important to look at the kinematics of the other warmer gas phases in the outflows. The high wind velocity implied by the measured X-ray temperature of the entrained gas in galactic winds ($v_{\rm X} \simeq 500 - 900$ km s$^{-1}$) has long been used as an argument that a significant fraction of the hot ionized gas may be escaping from dwarf galaxies \citep{MARTIN1999}, including perhaps M~82 \citep{LEHNERT1999}.  The measured velocities of the warm ionized phase (traced by emission and absorption lines of e.g. H$\alpha$, [O~III] 5007, Si~IV, C~IV, and O~VI) are generally smaller than $v_{\rm X}$, but they often -- though not always -- exceed the velocities of the cooler material \citep{VEILLEUX2013a,RUPKE2017,RODRIGUEZDELPINO2019}. The fraction of ionized gas that escapes the host galaxy is therefore higher than that of the cool gas, although it is still modest. Moreover, one should take into account that semi-ballistic motions are not the only way for the gas to escape the halo. Indeed, recent models and simulations have shown that under certain conditions the main effect of the AGN- and SN-driven outflows is to create hot bubbles that become buoyant within the galaxy halo, hence the gas ejected by the galaxy is transported to the outer halo primarily as a consequence of buoyancy rather than because of its velocity \citep{BOWER2017}.

The expectation that the outflowing molecular clouds may dissolve into the warm/hot ionized phase has been inferred from various studies \citep[e.g.][]{SCANNAPIECO2015,FERRARA2016b} in which the outflowing molecular clouds, subject to ram or radiation pressure, should undergo Kelvin-Helmholtz or Rayleigh-Taylor instabilities, and dissolve on timescales shorter than the outflow dynamical time (see Sect.~\ref{sec:entrainment} for more details). Moreover, photoevaporation has been identified as an additional process that may erode outflowing molecular clouds to the point that they can completely dissolve \citep[e.g.][]{DECATALDO2017}.

Direct observational evidence for gas phase transition from molecular to neutral-atomic or from neutral-atomic to warm/hot ionized is rare. Arguably the best case for molecular to neutral-atomic transition is that of the increasing neutral-to-molecular gas fraction with increasing heights above the disk in the nearby system M~82, where the linear resolution is only $\sim$30 pc arcsec$^{-1}$ \citep[Sect.~\ref{sec:M82};][]{LEROY2015b}.  In contrast, NGC~253, an object at basically the same distance as M~82 and with data of similar quality, does not show any obvious gas phase transitions in the outflowing gas \citep[Sect.~\ref{sec:N253};][]{BOLATTO2013b,WALTER2017}. The similarity of the velocity profiles of Si~II, Si~III, Si~IV, and O~VI in nearby starbursts also does not support phase transition within the inner $\sim$ kpc of these outflows. Detailed multi-phase comparisons have been conducted in a few, more distant infrared-bright galaxies using 3D data on the outflowing molecular, neutral-atomic, and ionized gas phases with matched sub-arcsecond resolutions, but $\sim$10$\times$ poorer linear resolution than for M~82 and NGC~253 \citep[e.g.,][]{RUPKE2013b,RUPKE2017}.  The conclusions of these comparisons have been mixed. In some cases, the gas phases overlap spatially and kinematically with each other, suggesting that they move together as a single entity without mixing or phase transition. In other cases, the cold molecular gas is slower and more concentrated in the nucleus than the neutral-atomic component, and the ionized gas is faster on average than the neutral-atomic gas and also more diffuse and extended \citep{RUPKE2017}.  The observational evidence for phase transitions on CGM scales is even weaker since most of the information on these scales comes from transverse spectroscopic studies of background quasars that do not spatially sample the absorbers. The use of galaxies as background sources is slowly changing this and promises to map cloud structures of different phases \citep{RUBIN2018b,RUBIN2018a}.

While some of the models mentioned above predict the molecular clouds to dissolve in the harsh environment of outflows, other models predict  the molecular clouds to not only survive but to actually cool even further, evolving in even denser molecular phases \citep[Sect.\ \ref{sec:in-situ}, e.g.][]{RICHINGS2018b,ZUBOVAS2014,DECATALDO2019,GRONKE2019}. In some of these models, the expectation is that the molecular clouds should become gravitationally unstable, fragment, and form stars inside the outflow \citep{ZUBOVAS2013,ZUBOVAS2014,ISHIBASHI2012,ISHIBASHI2013,ISHIBASHI2017b, EL-BADRY2016,WANG2018,DECATALDO2019}. In these models, the expected star formation inside the outflow could potentially reach hundreds of solar masses per year. This new mode of star formation would have remarkable consequences. Indeed, once formed in the outflow, the stars would have the kinematic imprint of the outflowing molecular gas out of which they have formed, and would start moving ballistically on quasi-radial orbits. If their velocity does not exceed the escape velocity, then they would start oscillating around the galactic center, hence contributing to the spheroidal component of galaxies (bulge, halo, or possibly even contributing to the  formation of elliptical galaxies). High velocity stars, stellar shells, and intracluster supernovae are some of the other properties that could be explained by this scenario, according to some authors. Finding observational evidence for star formation in galactic outflows is not simple as the signature of star formation can be easily overwhelmed by the presence of shocks and, even more likely, by the AGN photoionization \citep{GALLAGHER2019,HINKLE2019}. Moreover, young stars formed in the outflow may be difficult to disentangle from star formation in the host galaxy; indeed, as soon as stars form in the galactic clouds they respond only to gravity and are therefore expected to be rapidly decelerated to velocities that are hardly distinguishable from the stars in the host galaxy \citep{MAIOLINO2017}. 

Despite these issues, recent studies have presented observational indication of star formation in some galactic outflows. \cite{MAIOLINO2017} have investigated in detail the outflow in the ULIRG F23128-5919 and found that the both optical and near-IR nebular diagnostics of the outflowing gas are consistent with star formation. A similar analysis was extended by \citet{GALLAGHER2019} and by \citet{RODRIGUEZDELPINO2019} to samples of tens of local galaxies by exploiting the MaNGA integral field spectroscopic survey. About one third of the outflows have nebular line diagnostics consistent with excitation by star formation. The question is whether the excitation is due to star formation inside the outflow or externally from star formation in the disc of the host galaxy. The ionization parameter ($U_H \propto r^{-2} n_H^{-1}$; Eq.\ \ref{eq:U_H}) should be
different in the two cases. Indeed, in the case of external photoionization (e.g., young stars in the host galaxy), the ionization parameter should be lower in the outflowing gas, primarily because the gas is located at larger distances from the ionizing source, which is not compensated by a decrease in the gas density (as discussed in Sect.~\ref{sec:lowz_molecular}, the density of the gas in the outflows is actually systematically higher than in the disk). The finding that in those outflows with SF-like diagnostics the ionization parameter is similar, or even higher, than in their host galaxies favour the scenario in which star formation is occurring \textit{in situ}, i.e. inside the outflow, therefore bringing support to the various theoretical models discussed above.

Assessing the occurrence of this phenomenon is still difficult and sometimes controversial. Based on scaling relations, \citet{GALLAGHER2019} suggest that the phenomenon of star formation inside outflows may be even more prominent in more powerful and more massive outflows typically occurring in the more active phases at high redshift. However, the most powerful winds are driven by quasars, which can easily overwhelm any star-forming diagnostic in the outflow, even if the SFR in the outflow is relatively high \citep[see more detailed, quantitative discussion in][]{GALLAGHER2019}. \cite{HINKLE2019} investigate the ionization properties of four PG quasars and find that they are indeed dominated by AGN-like or shock-like excitation. They do find a region of one of the outflows dominated by SF-like excitation, but ascribe it to external photoionization based on the proximity, in projection, with star-forming regions in the host galaxy. This result underlines the need to accurately model the 3D position of the outflowing gas with respect to all of the star-forming regions in this object and all other galaxies with suspected \textit{in situ} star formation.

If star formation inside outflows is a channel contributing to the formation of the spheroidal component of galaxies, one may wonder whether the Milky Way's halo should show the signatures of such stars formed in past outflows. Within this context it is interesting that, by exploiting the latest GAIA data release, \cite{BELOKUROV2019} have recently discovered a significant population of metal-rich halo stars with purely radial orbits, i.e. with the same properties expected by models of stars formed in galactic outflows \citep[see e.g. sect. 4.1 of ][]{ZUBOVAS2013}. Indeed \cite{BELOKUROV2019} suggest that this might be one of the possible scenarios for the formation of such stars. It is also interesting that, according to their age, all these metal-rich MW halo stars with radial orbits were formed between z$\sim$1 and z$\sim$4, which is when the higher central star formation and AGN activity is expected to have generated the strongest outflows.

It should be noted that there are several other instances of ``positive feedback'' in galaxies in which
outflows or jets boost star formation in galaxies by compressing their ISM or CGM. Beautiful, clear examples of this phenomenon are seen in Cen A \citep[Sec.\  \ref{sec:CenA};][]{CROCKETT2012,SANTORO2016} and, thanks to new observing techniques (especially thanks to the new generation of wide-field IFS) more examples of outflow/jet-induced star formation occurring in other nearby galaxies are being found \citep[e.g.][]{CROFT2006,ELBAZ2009,CRESCI2015,SHIN2019}. However, these are cases in which the star formation is triggered in the host galaxy by the compression of the outflow/jet and the newly formed stars tend to preserve the kinematic properties of their host galaxy, hence a process fundamentally different from the star formation \emph{inside} outflows discussed above. While these findings support several recent models of jet/outflow-induced star formation \citep{MUKHERJEE2018a,GAIBLER2012,SILK2013,BIERI2015,BIERI2016,DUGAN2017,NAYAKSHIN2012,SILK2017}, as well as older models predating all of these observations \citep[e.g.,][]{DEYOUNG1981}, the inferred star formation rates associated with this mode are typically low (generally less than $0.1$--$1\,M_{\odot}$ yr$^{-1}$), hence they probably do not have a major impact on the overall evolution of galaxies. However, recent evidence for positive feedback associated with the outflow of a powerful type 2 quasar at $z\sim1.5$ has been detected, and in this case the inferred induced star formation rate is about $200\,M_{\odot}$ yr$^{-1}$. Moreover, star formation induced by the radio jets of high-z radio galaxies has been known for a long time \citep{MCCARTHY1993,CIMATTI1998,DISEREGOALIGHIERI1997,FEAIN2007}. Therefore, the effect of  star formation induced by outflows and jets may be important in the early Universe although its relevance relative to other mechanisms of galaxy formation has yet to be assessed properly.

\section{Cool Outflows in the Distant Universe} \label{sec:highz}

\subsection{Molecular Gas Component} \label{sec:highz_molecular}

While the detection of ionized outflows at high redshift has been obtained for large samples of star-forming galaxies and AGNs \citep[e.g.][]{STEIDEL2010,GENZEL2014,CANO-DIAZ2012, CARNIANI2015,CARNIANI2017,HARRISON2017,FORSTERSCHREIBER2019}, the detection of cold molecular outflows in high-redshift galaxies is still limited to a handful of objects, due to the observational difficulty in detecting the weak signatures of outflows in these distant systems and also to disentangle them from the dynamically quiescent ISM in the host galaxies. Moreover, even when detected, the limited projected spatial resolution often prevents a proper assessment of the extension and geometry of the outflow.

\begin{figure}[htbp]
\begin{center}
\includegraphics[width=0.95\textwidth]{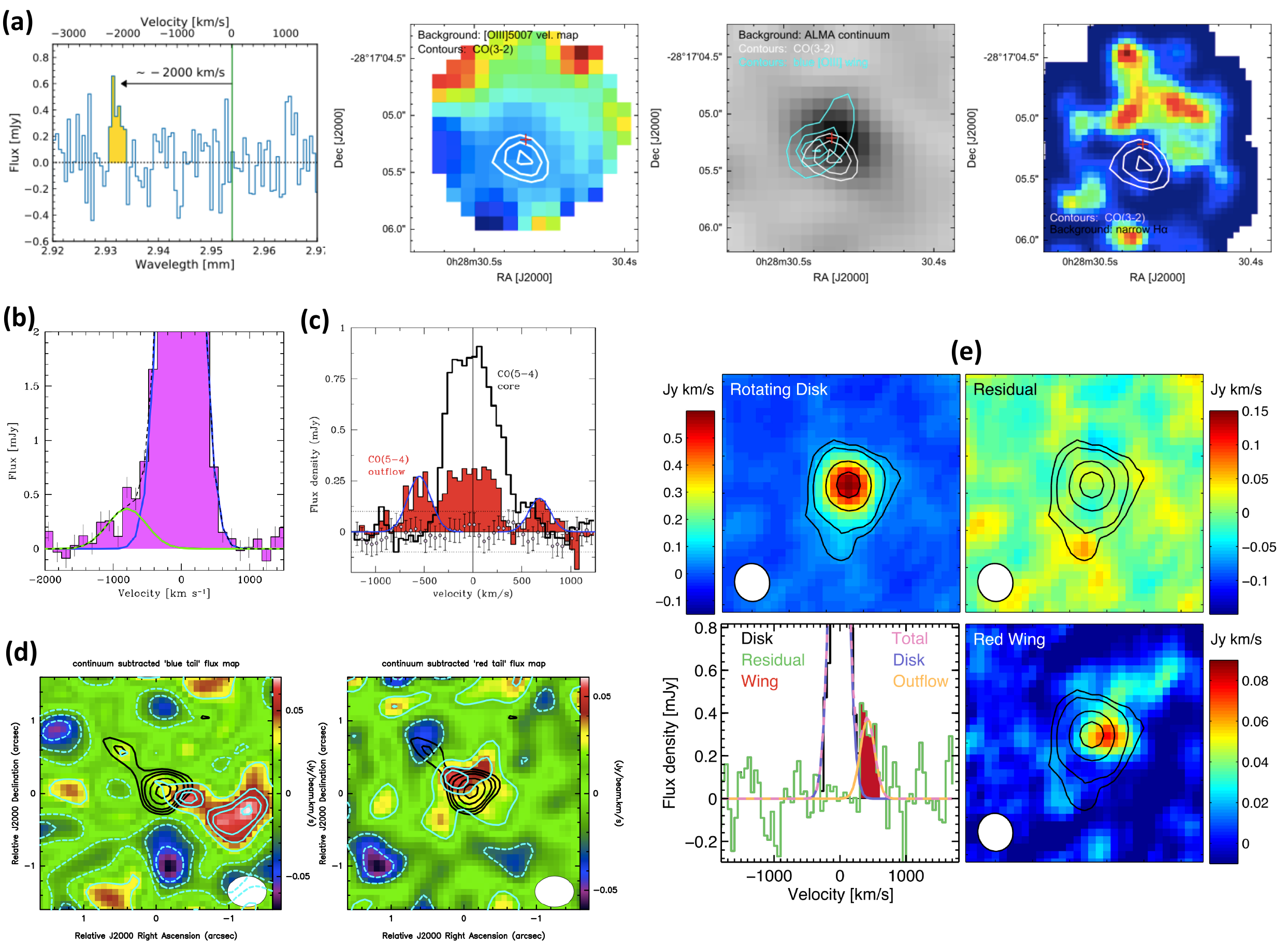}
\end{center}
\caption{Examples of molecular outflows in distant galaxies detected via emission lines. (a) From left to right: blueshifted CO (3-2) line emission from $z = 2.4$ quasar 2QZJ0028, flux map of the blueshifted CO (3-2) line emission (white contours) overlaid on the [O~III] velocity map, [O~III] blue wing (blue contours), ALMA 3-mm continuum map (greyscale), and narrow H$\alpha$ emission (color scale) tracing star formation in the host galaxy. (b) Asymmetric CO (4-3) line emission from $z = 3.9$ quasar APM 08279+5255. (c) CO (5-4) outflow (red histogram) and core (black histogram) in quasar XID2028. (d) Maps of the blueshifted ($v < -$350 km s$^{-1}$; left panel) and redshifted ($v > 350$ km s$^{-1}$; right panel) CO (5-4) line emission shown in panel (c), color-coded according to the flux. (e)  AGN-driven CO (3-2) outflow in the massive main-sequence galaxy zC400528 at $z$ = 2.3. The emission maps of the three components identified in the integrated spectrum shown in the lower left panel are presented in the three other panels. Images reproduced with permission from (a) \citet{CARNIANI2017}, (b) \citet{FERUGLIO2017}, (c, d) \citet{BRUSA2018}, copyright by ESO; and (e) \citet{HERRERA-CAMUS2019a}, copyright by AAS.}
\label{fig:highz_molecular_emission}
\end{figure}

High-sensitivity observations of CO transitions have revealed high-velocity molecular gas ascribed to outflows in about half a dozen high-redshift galaxies \citep[Fig.~\ref{fig:highz_molecular_emission};][]{WEISS2012,FERUGLIO2017,CARNIANI2017,VAYNER2017,FAN2018,BRUSA2018,HERRERA-CAMUS2019a}, although in some of these it is not yet totally clear whether the high-velocity molecular gas is truly associated with outflow or actually result from merging or irregular dynamics. In all of these cases the outflow is driven by an AGN, typically a powerful quasar, but also sometimes a lower luminosity AGN residing in a normal main-sequence galaxy \cite[Fig \ref{fig:highz_molecular_emission}e,][]{HERRERA-CAMUS2019a}, and also molecular outflows driven by radio-jets \citep{VAYNER2017}. These molecular outflows in distant galaxies have velocities up to 1000 km~s$^{-1}$ and extend from a few kpc to several kpc.

\begin{figure}[htbp]
\begin{center}
\includegraphics[width=0.9\textwidth]{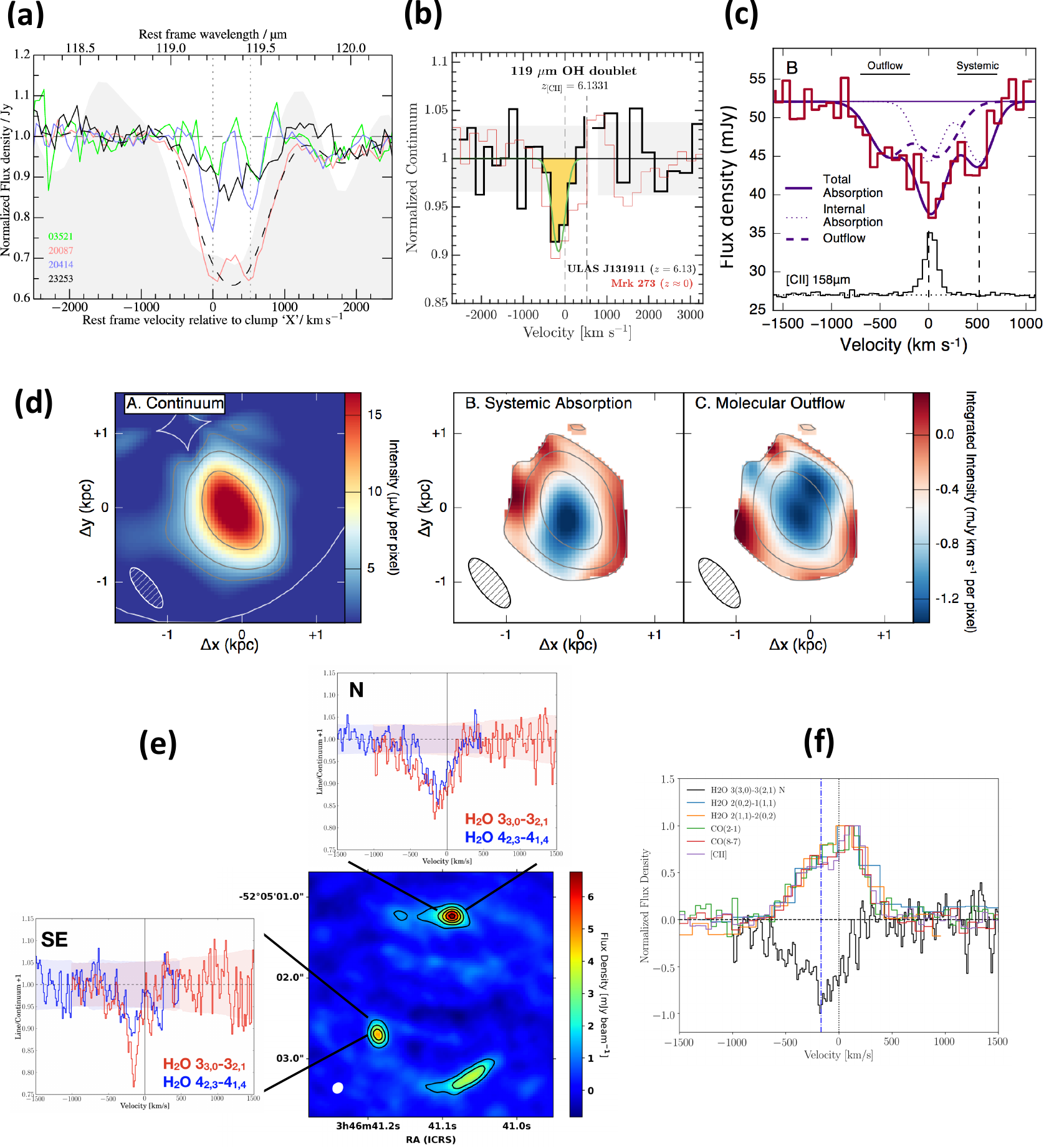}
\end{center}
\caption{Molecular outflows in distant galaxies detected via absorption lines. (a) Normalized OH 119 absorption profile of $z$ = 2.3 ULIRG  SMM J2135$-$0102 (in grey) compared to those of nearby starburst-dominated ULIRGs. (b) Normalized OH 119 absorption profile of $z$ = 6.13 quasar ULAS J131911+095051 compared with that of local ULIRG Mrk~273. (c) OH 119 absorption profile of $z$ = 5.3 lensed ULIRG SPT J231921$-$5557.9 compared with the profile of the [C~II] 158 $\mu$m emission line. (d) Lensing reconstruction of the continuum emission (left), systemic OH absorption (middle), and OH outflow (right) in the object shown in panel (b). (e) H$_2$O absorption spectra extracted at two locations in the continuum map of the lensed starburst ULIRG SPT-S J034640$-$5204.9. (f) Water absorption profile for the northern lensed image in panel (d) compared with the profiles of the [C~II] 158 $\mu$m, CO (2-1), and two different water emission lines. Also shown is the two-Gaussian fit to the H$_2$O absorption spectrum. Images reproduced with permission from (a)  \citet{GEORGE2014}, copyright by the authors; (b) Herrera-Camus et al. (2019c), copyright by ESO; (d) \citet{SPILKER2018}, copyright by the authors; (e, f) \citet{JONES2019b}, copyright by ESO.}
\label{fig:highz_molecular_absorption}
\end{figure}

The detection of molecular outflows through blueshifted far-IR molecular  transitions in absorption (redshifted to the millimeter and submillimeter bands, observable from the ground) has been limited largely to extremely powerful  starburst galaxies, which are also strongly lensed, as the detection of these features in absorption requires a strong continuum (typically in excess of 100 mJy). The only exception is quasar J131911+095051 \citep[][]{HERRERA-CAMUS2019c}. There are currently only four outflow detections, from redshift 2.3 to 6.1 (Fig.~\ref{fig:highz_molecular_absorption}), three of these obtained through the detection of OH transitions \citep[][]{GEORGE2014,SPILKER2018,HERRERA-CAMUS2019c} and one of them through the detection of blueshifted high-level transitions of water \citep[][]{JONES2019b}. The latter is a particularly interesting result, as it implies very dense ($\sim 10^5-10^6~{\rm cm}^{-3}$) and warm (300-500~K) molecular outflowing gas, with large column densities ($\sim 10^{24}~{\rm cm}^{-2}$ or even higher).

The fact that the high-z molecular outflows are detected only in AGN/quasar host galaxies when searching for CO high-velocity emission, while outflows are detected almost exclusively in starburst galaxies when searching for blueshifted far-IR OH or H$_2$O absorption, does not really have any underlying, intrinsic physical reason. The reason for this dichotomy is most likely a consequence of observational selection effects. Indeed, on the one hand AGN-driven molecular outflows tend to be more vigorous (in terms of mass, outflow rate and kinetic power) and to have higher velocities with respect to starburst-driven outflows, hence easier to detect in the sensitivity-limited CO searches of outflows at high-z. On the other hand, the detection of blueshifted OH and H$_2$O absorption at high-z requires strongly lensed galaxies with strong far-IR continuum, and there are statistically many more lensed starburst galaxies with these characteristics (typically detected by the SPT and {\em Herschel} surveys) than lensed quasars.

\begin{figure}[htb]
\begin{center}
\includegraphics[width=0.7\textwidth]{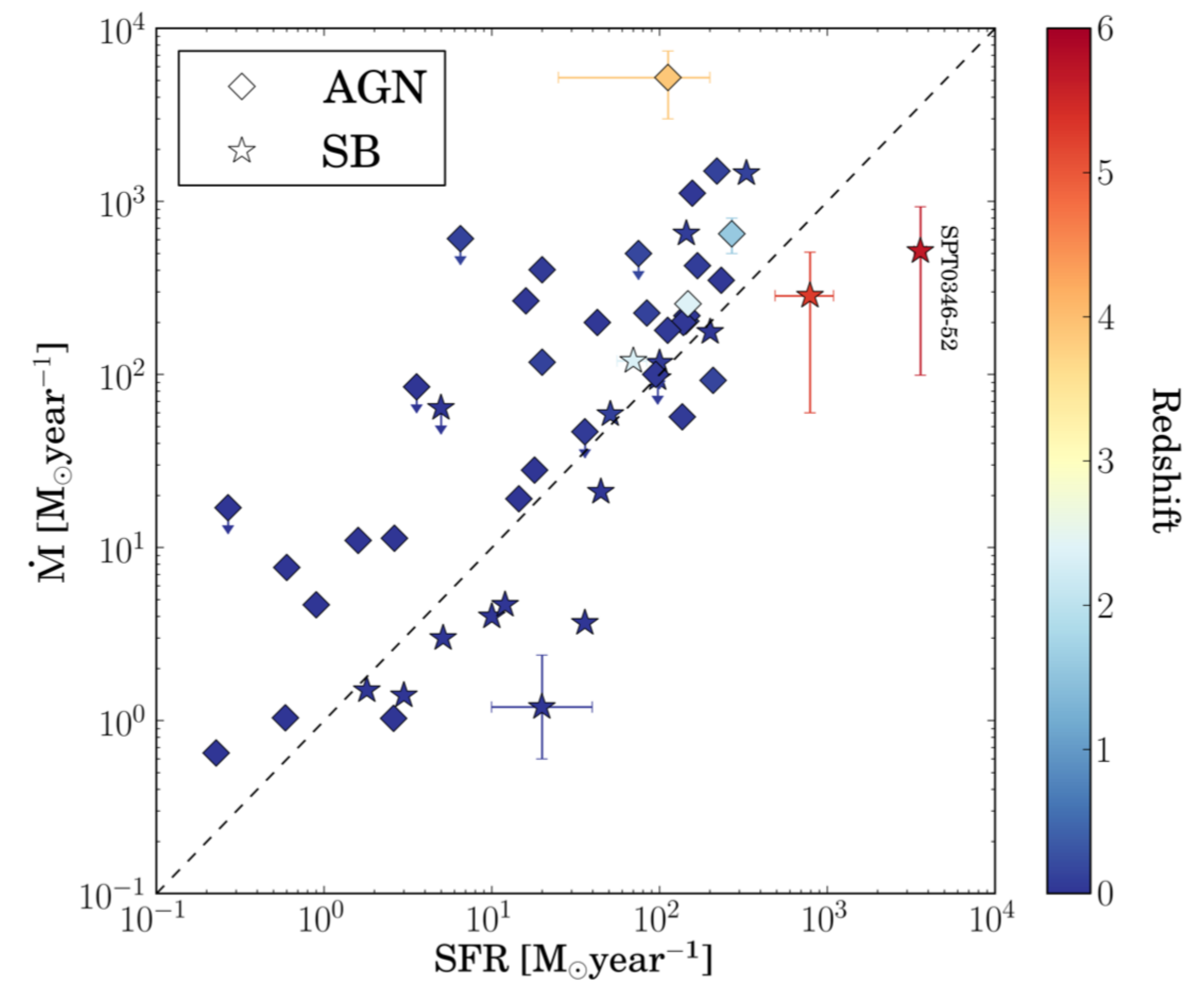}
\end{center}
\caption{Molecular mass outflow rates in distant galaxies plotted as function of star formation rates. The dashed line shows the 1:1 equality line. The symbols differentiate between AGN and starbursts and are color-coded by redshifts \citep{JONES2019b}, copyright by ESO.}
\label{fig:jones2019_summary}
\end{figure}

Overall, the quantitative findings of these high-redshift studies illustrate that the properties of the molecular outflows in distant galaxies follow similar trends as local galaxies, although the outflow properties seem to saturate at the more extreme and powerful tail (Fig.~\ref{fig:jones2019_summary}). More specifically, high-redshift star forming galaxies are characterized by a mass loading factor of about unity, as found for their local counterparts. However, for the most extreme starbursts found at high redshift, this implies outflow rates even exceeding 1000~$\rm M_{\odot}~yr^{-1}$. For instance, the most extreme  galaxy for which the molecular outflow has been detected (the lensed galaxy SPT-S J034640$-$5204.9 at z=5.6; Fig.~\ref{fig:highz_molecular_absorption}d-e) is characterized by an extremely powerful and extremely compact starburst, resulting in an impressive surface density of star formation rate of 10,000~$\rm M_{\odot}~yr^{-1}~kpc^{-2}$, an order of magnitude larger than the so-called ``maximal starburst'', i.e. the maximum Eddington-limited star formation surface density beyond which radiation pressure is expected to rapidly blow away the ISM \citep{THOMPSON2005,CROCKER2018}. This galaxy is clearly in a highly transient phase. The outflow rate inferred from the water P-Cygni profile is only a few hundred $\rm M_{\odot}~yr^{-1}$, much lower than the star formation rate in the same galaxy, 3000~$\rm M_{\odot}~yr^{-1}$, implying an outflow mass loading factor much less than unity, i.e. much lower than observed for low-z galaxies (it is the rightmost data point in Fig.~\ref{fig:jones2019_summary}). Apparently, being so highly super-Eddington, the outflow capacity has saturated; in this case the outflow cannot regulate the star formation rate (as in less extreme, low-z galaxies) and result into a runaway process in which star formation deplete gas in a few Myr, likely resulting into a compact massive quiescent galaxy, unless replenished with fresh gas from the IGM \citep[][]{JONES2019b}. The second most extreme galaxy among those with blueshifted molecular absorption detection (another SPT lensed galaxy at $z$ = 5.3; Fig.~\ref{fig:highz_molecular_absorption}b-c) displays similar properties, i.e. a very compact starburst (although not as extreme as SPT-S J034640$-$5204.9) and a low loading factor, although the latter is still consistent with unity within the uncertainties (it is the second data point from the right in Fig.~\ref{fig:jones2019_summary}).

Even in these high-z systems, the presence of a powerful AGN does boost the outflow rate by a significant factor, as in local systems, therefore confirming the important role of AGNs, and especially powerful quasars, in regulating or even quenching star formation in galaxies (Fig.~\ref{fig:jones2019_summary}). The energetics of these quasar-driven outflows at high-z are typically well below the $\sim$5\% of L$_{AGN}$ expected by the blast-wave, energy-driven scenario \citep[e.g.,][]{BISCHETTI2018}. The momentum boost of these outflows also seems generally well below the factor expected by the energy-driven scenario \citep[i.e. typically of the order of L$_{\rm AGN}$/c or a few times L$_{\rm AGN}$/c, in contrast to the expectation of $>$10~L$_{AGN}$/c expected in the energy-driven scenario;][]{FERUGLIO2017,BISCHETTI2018}. These findings suggest that, as expected by numerical simulations, in these early systems the bulk of the hot nuclear wind does not couple efficiently with the galaxy ISM and mostly escapes into the IGM \citep{COSTA2015}. Alternatively, a significant fraction of the outflow may be accelerated by radiation pressure, which results in a much lower momentum boost \citep[e.g.][]{COSTA2018b,ISHIBASHI2018c}.

\subsection{Neutral Atomic Gas Component} \label{sec:highz_neutral_atomic}

\begin{figure}[htbp]
\begin{center}
\includegraphics[width=0.95\textwidth]{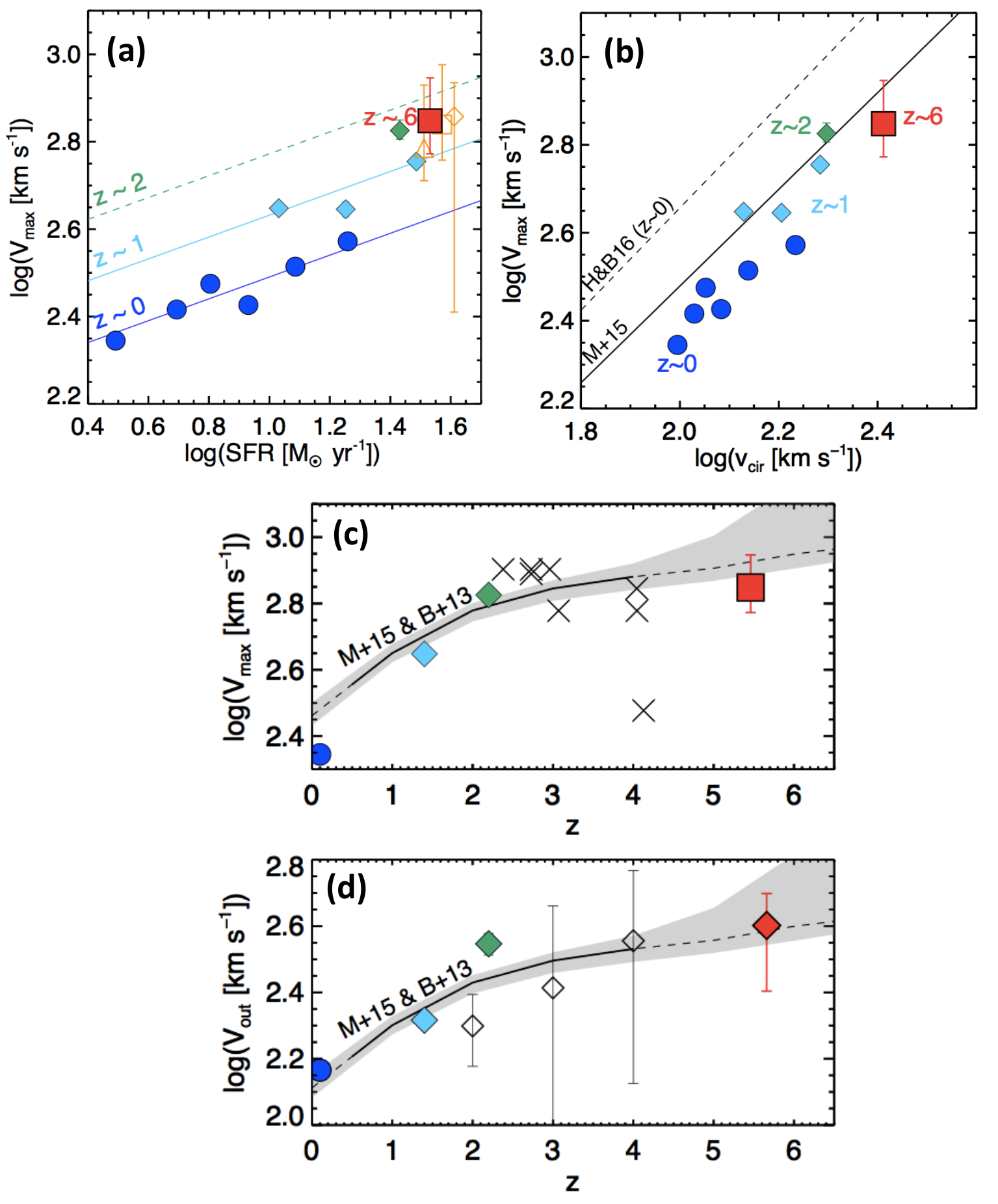}
\end{center}
\caption{Neutral-atomic outflows in distant star-forming galaxies. Trends of increasing maximum outflow velocities with (a) SFR, (b) circular velocity and (a)-(d) redshift are present in the data. Images reproduced with permission from \citet{SUGAHARA2019}, copyright by AAS.}
\label{fig:highz_sf_neutral_atomic}
\end{figure}

\begin{figure}[htbp]
\begin{center}
\includegraphics[width=0.8\textwidth]{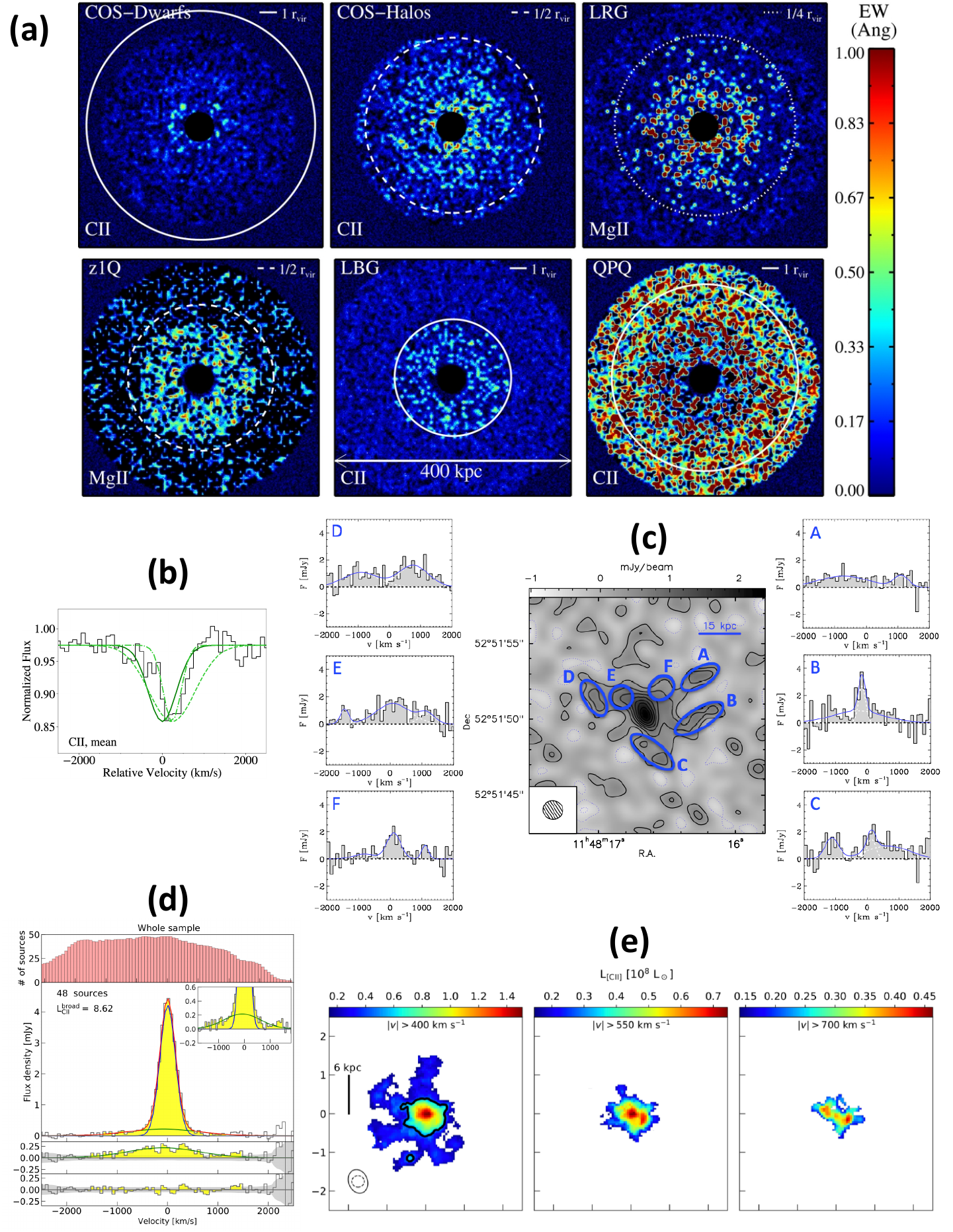}
\end{center}
\caption{Neutral-atomic outflows in distant quasars. (a) UV C~II 1334 and Mg~II 2796 absorbers around $z \simeq 2$ massive galaxies hosting quasars (lower-right panel) compared with those of $z \simeq 0$ dwarf and luminous galaxies (COS-Dwarfs and COS-Halos, respectively), $z \simeq 1$ Luminous Red Galaxies (LRG), $z \simeq 1$ quasars (z1Q), and $z \simeq 2$ Lyman-Break Galaxies (LBG). (b) Velocity shift between the mean stack of C~II absorbers around $z \simeq 2$ quasars (black histogram) and the expected kinematics purely from clustering (solid green line). The dashed and dashed-dotted lines are absorption profiles centered on the data but with $\sigma_v$ = 554 and 214 km s$^{-1}$, respectively. (c) Spatially resolved [C~II] 158 $\mu$m outflow around the quasar SDSS J1148+5251. (d) Stacked integrated [C~II] 158 $\mu$m spectrum of 48 quasars at 4.5 $< z <$ 7.1. (e) Luminosity maps of the high-velocity [C~II] emission derived from the spectrum in panel (d). Images reproduced with permission from (a) \citet{PROCHASKA2014}, (b) \citet{LAU2018}, copyright by AAS; and (c) \citet{CICONE2015}, (d, e) \citet{BISCHETTI2018}, copyright by ESO.}
\label{fig:highz_quasar_neutral_atomic}
\end{figure}

There are many more reported cases of neutral-atomic outflows at high redshifts than molecular outflows. Down-the-barrel and transverse spectroscopic surveys of neutral/low-ionization UV absorption lines (C~II, O~I, Al~II, Si~II, and Fe~II in absorption) and emission lines (Si~II$^*$, especially common at $z \sim 4$) from $z \gtrsim 2$ main-sequence star-forming galaxies have revealed ubiquitous, fast ($v_{50} \simeq 100 - 300$ km s$^{-1}$, $v_{98}$ up to $\sim$1000 km s$^{-1}$),
neutral-atomic outflows extending to at least $\sim$ 50-100 kpc \citep[$\sim 0.5 - 1~r_{\rm vir}$; e.g.,][]{SHAPLEY2003,ADELBERGER2005,ERB2012,FOX2008,STEIDEL2010,RUDIE2012,RUDIE2019,SUGAHARA2017,SUGAHARA2019,DU2018,GATKINE2019}. 
The outflow velocities tend to increase with redshifts up to $z \sim 2$, above which they seem to level off \citep[Fig.~\ref{fig:highz_sf_neutral_atomic},][]{SUGAHARA2017,SUGAHARA2019,DU2018}, as predicted by some models \citep[e.g.,][]{MURATOV2015}. This turning point at $z \sim 2$ may also be due to a systematically higher SFR surface density in galaxies at $z \gtrsim 2$ (due to selection biases) or a change in outflow geometry (a more spherical outflow geometry at high $z$ will boost the detection fraction of outflows and increase the average/median outflow velocities in the stacked galaxy spectra). 

In most cases, these results rely on the measurements of the strongest interstellar lines, which are typically saturated and therefore only trace the covering fraction and kinematics of the outflowing gas but not the column density (mass). Over the years, down-the-barrel studies of a total of $\sim$ 20 gravitationally lensed systems at $z \simeq 2-4$ have addressed this weakness by measuring the fainter unsaturated lines of several elements (from H to Zn) in a variety of ionization stages \citep{PETTINI2002,QUIDER2009,DESSAUGES-ZAVADSKY2010,STARK2013,RIGBY2018a,RIGBY2018b,JONES2018}. These studies have confirmed the ubiquity of fast low-ionization outflows in high-$z$ main-sequence star-forming galaxies, and reported mass outflow rates of the low-ionization gas alone that typically match or exceed the star formation rates.

The presence of powerful AGN at the center of star-forming galaxies likely enhances the velocity of cool outflows on small scales \citep{TALIA2017}, although in some cases the AGN photoionize the cool gas that is located too close from the center \citep{PERROTTA2016}. In the most extreme cases of $z \simeq 2$ LoBALs and FeLoBALs, the outflowing low-ionization material reaches velocities of several $\times$ 1000 km s$^{-1}$ on scales of 1-3 kpc \citep[based on Eq.~(\ref{eq:U_H}) and densities derived from Fe II$^*$/Fe II and Si II$^*$/Si II ratios;][]{MOE2009,DUNN2010}. The implied outflow mass rates and kinetic luminosities are a few $\times$ 100 M$_\odot$ yr$^{-1}$ and 0.1--1\% $L_{\rm BOL}$, respectively.
On CGM scales, transverse studies of C~II and Mg~II absorbers around powerful $z \simeq 2$ quasars \citep[Fig.~\ref{fig:highz_quasar_neutral_atomic}a;][]{PROCHASKA2014,LAU2016,LAU2018}  have revealed large reservoirs ($\sim 2 \times 10^{11} M_\odot$) of cool gas with $\sim$0.7 covering fraction, $\sim$1/4 solar metallicity, and super-solar O/Fe ratios, indicative of enrichment primarily by core-collapse SNe. To avoid Fe dilution by SNe Ia, the metals must be injected into the CGM of these quasars in less than $\sim$1 Gyr through a burst of star formation of $\gtrsim$50 M$_\odot$ yr$^{-1}$. \citet{PROCHASKA2014} have argued that the large amount of metals in the CGM of these quasars may simply reflect their large halo masses ($\gtrsim 10^{12.5}$ M$_\odot$) rather than the influence of the quasars. Indeed, the broad velocity distribution of the C~II absorbers 
can largely be explained by virial motion within the galactic halos. However, the observed velocity offset of $+$200 km s$^{-1}$ of the cool absorbing system with respect to systemic velocity (Fig.~\ref{fig:highz_quasar_neutral_atomic}b) seems to require a net outflow of $\sim$ 400-450 km s$^{-1}$, combined with anisotropic and/or intermittent ionizing radiation from the quasars \citep{LAU2018}.

Photoionization by internal or external energy sources is also an important factor when studying the cool CGM in line emission. For instance, CGM-scale Ly$\alpha$ nebulae have been detected around several high-$z$ star-forming and active galaxies through stacking analyses \citep[e.g.,][]{STEIDEL2011,MOMOSE2016}   and deep integral-field spectroscopy \citep[e.g.,][]{BORISOVA2016,WISOTZKI2016,WISOTZKI2018,ERB2018,ARRIGONIBATTAIA2019,CAI2019,O'SULLIVAN2019},  but in most cases, these nebulae are photoionized by the central quasars or starbursts \citep[e.g.,][]{LEIBLER2018} or the cosmic UV background \citep[e.g.,][]{WISOTZKI2018}, and thus they trace largely the warm-ionized component of the CGM rather the cool neutral-atomic gas phase. Moreover, the severe radiative transfer effects of this strong resonant line complicate the interpretation of the broad Ly$\alpha$ line profiles, making this line an ambiguous kinematic tracer of outflows in general (Sec. \ref{sec:lya}).  Much of the recent effort has instead been directed at mapping the high-$z$ CGM using cool-gas tracers that are not affected by these effects.

Being one of the brightest lines in the spectrum of galaxies, the [C~II] 158 $\mu$m transition has been extensively used to search for outflows in galaxies at $z >$ 4, i.e. at redshifts high enough that this transition is shifted to the submm/mm-wave bands with good atmospheric transmission. As discussed in Section \ref{sec:m_and_mdot}, this transition is expected to trace predominantly the atomic neutral component of the ISM and outflows, although contribution from the ionized phase may be significant. [C~II] is currently detected in a few hundred galaxies at high redshift, however, as for CO, the detection of the outflow component has been much more difficult, as it implies detecting broad wings, typically much fainter than the core of the line, and often distributed in a more diffuse component. The most prominent detection of cold atomic outflow at high-z has been obtained for the famous quasar J1148+5251 at $z$ = 6.4 \citep[Fig.~\ref{fig:highz_quasar_neutral_atomic}c;][]{MAIOLINO2012,CICONE2015}, which was observed for over 30 hours with NOEMA, and where [C~II] is detected out to velocities in excess of 1000 km~s$^{-1}$, extending on scales of 30 kpc, matching expectations from zoom-in cosmological simulations and theoretical models specifically aimed at intepreting feedback in this quasar \citep{VALIANTE2012,COSTA2015}. In this case  the outflow has been mapped in such detail that it was possible  to infer the geometry and temporal evolution of the outflow. More specifically, the outflow appears to have a biconical structure along our line of sight; indeed detailed modelling \citep{COSTA2015} shows that this is the only geometry that can match the nearly symmetric, double-peaked distribution of the [C~II] high-velocity component on large spatial scales. Moreover, the distribution of the dynamical timescale of the various outflowing clumps indicates that the outflow has been ongoing for at least $\rm 10^8$~yr and in a bursty mode.
 
However, J1148+5251 seems to be an extreme case. Detections of atomic cold outflows through [C~II] wings has so far been achieved only for a few additional individual quasars \citep[][Carniani et al. in prep.]{DIAZ-SANTOS2016}. Yet, the combination of ALMA [C~II] data of nearly 50 quasars at z$\sim$5-7 (corresponding a total of nearly 40 hours of integration) has revealed clear broad wings in the stacked cube \citep[Fig. \ref{fig:highz_quasar_neutral_atomic}d-e;][]{BISCHETTI2018}, indicating that these primeval quasars on average drive high-velocity cold outflows as expected by models and cosmological simulations \citep[e.g.][]{DIMATTEO2005,VALIANTE2012,COSTA2015,COSTA2018a}. The stacked cube also suggests that these quasar-driven outflows have extensions of several kpc. Stacking analyses no doubt miss a significant fraction of faint outflows due to orientation effects  \citep{STANLEY2019}, so these results should be considered conservative lower limits. 

Early attempts to detect cool winds in distant ``normal'' star-forming galaxies through the [C~II] line and through stacking of 10-20 galaxies have resulted in tentative detections of broad wings \citep{GALLERANI2018,FUJIMOTO2019}. A clearer and more definitive result has been obtained through the ALPINE ALMA large program, which has enabled the stacking of nearly 100 galaxies \citep[][]{GINOLFI2019}. A clear detection of broad wings, extending to velocities of several 100~km~s$^{-1}$ and on kpc scales, is however obtained only when the stacking is restricted to the most star forming galaxies (SFR $>$ 100  $M_{\odot}~{\rm yr}^{-1}$).

\subsection{Dust Component} \label{sec:highz_dust}

Direct evidence for dust in cool outflows based on the detection of spatially extended dust emission coincident with extended broad CO or [C~II] line emission is rare \citep[possibly seen only in SDSS J1148+5251;][]{CICONE2015}. The same is true for dust depletion in individual low-ionization outflows. In the few cases of  $z \simeq 2-3$ gravitationally-lensed star-forming systems where this type of detailed down-the-barrel analysis is possible, the results are ambiguous. The outflowing gas often shows super-solar alpha-to-iron ratios consistent with dust depletion, but these anomalous ratios may also be explained if the bulk of the stars in these galaxies formed recently \citep[$\lesssim$ 0.3 Gyr;][]{PETTINI2002,DESSAUGES-ZAVADSKY2010,JONES2018}.

Arguably the most convincing argument for dust in cool outflows at $z \gtrsim 1$ comes from reddening measurements of quasars with intervening strong Mg~II absorbers. \citet{MENARD2012}  have deduced that strong Mg~II absorbers at $z = 0.4 - 2.0$ carry about half of the total amount of dust outside of galaxies (see Sec. \ref{sec:lowz_dust}), or nearly all of the circumgalactic dust within the virial radii of galaxies. These absorbers have a constant MW-like dust-to-gas ratio that is consistent with an outflow origin, where dust is neither added through stellar processes nor destroyed by sputtering over the inferred long (several Gyr) effective lifetime of these clouds.

Dust is also present in most $z \gtrsim 1.5$ DLAs with $N_{\rm H} > 10^{20.3}$ cm$^{-2}$ and sub-DLAs with $N_{\rm H} = 10^{19 - 20.3}$ cm$^{-2}$, based on background quasar reddening \citep{FUKUGITA2015} and absorption-line measurements of differential elemental depletion
\citep{WOLFE2005,DESSAUGES-ZAVADSKY2006,JENKINS2009,KULKARNI2015,QUIRET2016,DECIA2016,DECIA2018}. However, these systems generally have $3-10\times$ lower metallicity than that of Mg II absorbers and their dust-to-gas ratio, metallicity, and line width all decrease with increasing $N_H$, contrary to that of Mg~II absorbers. The metals in DLAs may have been ejected from galaxies in cool winds as with Mg II clouds, but this gas must have subsequently mixed with primordial gas on intergalactic scales to produce the observed metallicity and dust-to-gas ratio, and their dependence on $N_{\rm H}$ \citep{FUKUGITA2015}. Note also that the MW-like 2175 \AA\ graphite feature is detected in some of these systems, but not in Mg~II absorbers \citep{MA2017}.

Most high-redshift AGN, particularly those with red or IR-bright SEDs, contain large quantities of dust, but a firm association between dust and outflows in these systems is still lacking. The case for entrained dust in LoBAL and FeLoBAL quasars has already been discussed in Section \ref{sec:lowz_dust}, when discussing low-redshift systems, so it is not repeated here. Many more quasars with very red optical-IR colors, remarkably blue or flat colors in the UV, and fast powerful warm-ionized winds seen in both line emission and absorption have been discovered by cross-correlating the sources in the WISE mid-infrared catalog of galaxies with those from other large ground-based extragalactic surveys. These so-called ``extremely red quasars'' or ERQs represent a promising new class of objects to search for dust in outflows \citep{BANERJI2015,ASSEF2015,HAMANN2017,PERROTTA2019,COATMAN2019,TEMPLE2019}.  A recent spectropolarimetric follow-up study of a small subset of these objects by \citet{ALEXANDROFF2018} has revealed high polarization ($\sim$ 15\%) in the UV continuum, indicating that most of the continuum emission at these wavelengths is due to anisotropic scattered light, most likely off of a dusty medium in proximity to the AGN. \citet{ALEXANDROFF2018} have also argued that the strength and swing in position angle of the polarization across the profiles of the emission lines can most simply be explained with a geometrically thick equatorial dusty scattering outflows of several thousand km s$^{-1}$ that extends on scales of tens of parsecs. On the other hand, \citet{TEMPLE2019}  recently concluded that most of the dust in extremely red quasars lies on galactic scales \emph{outside} of the ionized [O~III] outflows. Clearly, the jury is still out on this issue, and the possibility that dust may be present on both nuclear and galactic scales, as in FeLoBALs, is not excluded.

\subsection{Fate of the Outflowing Material} \label{sec:highz_fate}

It is difficult to estimate the escape velocity in high-redshift systems. However, most of the molecular outflows detected at these high redshifts seem hosted in very massive galaxies, characterised by high SFR and high AGN power (this is simply a consequence of the current selection  biases), so even velocities as high as 1,000 km~s$^{-1}$ would not enable the cold outflowing gas to escape the halo of these galaxies. Hence the gas is likely to rain back onto the host galaxy unless heated and prevented to cool.

Nevertheless, it is interesting to note that quasar-driven outflows seem to have an immediate quenching effect on those regions of the host galaxy hit by the outflow: indeed in these regions star formation appears to be suppressed relative to other regions of the host galaxy \citep{CANO-DIAZ2012,CRESCI2015,CARNIANI2016}. However, overall the elevated star formation rate across the rest of the quasar galaxy seems unaffected, which has raised questions on the effectiveness of quasar-driven outflows in quenching star formation at high redshift \citep[e.g.][]{BALMAVERDE2016,SCHULZE2019}. In fact, there are indications that, in some of these high-z galaxies characterized by quasar-driven outflows, star formation can actually be boosted as a consequence of gas compression in  regions surrounding the fast outflow \citep{CRESCI2015,CARNIANI2016}. In these cases, the net feedback effect seems to be more positive than negative.

The story for the outflowing neutral-atomic material is seemingly quite different. As mentioned in Sec. \ref{sec:highz_neutral_atomic} and shown in Fig\ \ref{fig:highz_sf_neutral_atomic}, the outflow velocities of the neutral-atomic gas at $z \gtrsim 2$ appear to be larger on average than those measured at low redshifts. More importantly, the stellar masses of the high-$z$ host galaxies are typically 1$-$2 dex smaller than those at low redshifts \citep[e.g.,][]{BEHROOZI2013}, so their escape velocities are correspondingly 0.3-0.6 dex smaller (Sec. \ref{sec:escape}). As a result, the bulk of the cool outflowing material in these high-$z$ systems often has a large enough velocity to escape from the host galaxies altogether, 
contrary to the cases at low redshifts (Sec. \ref{sec:lowz_neutral_atomic}, \ref{sec:lowz_fate}). If unimpeded by drag forces, this outflowing neutral-atomic material will contribute to the enrichment of the CGM and IGM \citep[e.g.,][]{MURATOV2015,MURATOV2017}, and may re-accrete onto neigboring galaxies \citep{ANGLES-ALCAZAR2017a}.

The difference between the molecular and atomic gas phases in high-$z$ outflows no doubt is due to selection biases rather than a physically meaningful difference between the phases. High-$z$ cold molecular and [C~II]-based outflows have so far been detected only in massive galaxies, for which the escape velocity is very high, while the neutral outflowing phase observed through absorption line studies is mostly detected in lower mass main-sequence star-forming galaxies which have smaller escape velocities. So the current results reflect the different populations of galaxies being probed by different outflow tracers rather than differences in outflow phases.

\section{Open Issues and Future Directions} \label{sec:open}

\subsection{Theoretical Challenges} \label{sec:open_theoretical}

\subsubsection{Wind--ISM Interaction} \label{sec:open_wind-ism}

Fundamentally, the issue here is whether cool ISM clouds entrained in a hot wind can reproduce the observed characteristics of cool outflows. The first basic question is whether ram-pressure forces exerted by a fast low-density  wind on a dense ($\sim$10$^4$ cm$^{-3}$), cold ($\sim$10-100 K) cloud initially at rest can accelerate this cloud to $\sim$ 200 (1000) km s$^{-1}$ over a scale of $\sim$ 200 (1000) pc, without shredding it into pieces by erosion processes. These densities, velocities, and distance scales are representative of the outflowing molecular gas in NGC~253 and Mrk~231, respectively (Sect.~\ref{sec:N253} and \ref{sec:Mrk231}). The fast ($\sim$ 450 km s$^{-1}$) and very dense ($\sim$ 10$^7$--10$^5$ cm $^{-3}$) clouds seen outflowing on parsec scale in NGC~1068 (Sect.~\ref{sec:N1068}) are equally if not more challenging to explain in this entrainment scenario. Recent high-resolution 2D and 3D simulations \citep[][respectively]{MCCOURT2018,SPARRE2019} of radiatively cooling clouds interacting with a supersonic, hot wind indicate that shattering / fragmentation of clouds larger than the cooling length is unavoidable. However, the quantitative differences between the results from these simulations\footnote{ As pointed out by \citet{SPARRE2019}, instabilities have a smaller effect in 3D than in 2D because a 3D flow has the freedom to use the $z$-direction to avoid disturbing dense clouds. As a result, the level of fragmentation is lower in 3D than in 2D, and the increase in covering fraction for large clouds in 3D is less than that seen in 2D.} underscore the need to strive for the most realistic state-of-the-art simulations and address the issue of numerical convergence to make substantial progress answering even this simple question. The importance of extending both the parameter space (e.g. $t_{\rm cool, mix}/t_{\rm cc}$) and the simulation domain to follow up the evolution of the stripped material mixing with the hot fluid has also recently been emphasized by \citet[][see Fig.~\ref{fig:sims_clouds}]{GRONKE2018a,GRONKE2019}. Other forces (e.g., radiation, cosmic rays, magnetic fields) may be added later to try to reproduce the substantial momentum boost ($\dot{p} > L/c$) observed in energy-driven cool outflows, but the treatment of these other forces will require additional refinements to the simulations (see below) that are computationally costly and may compromise the accuracy of the overall results. Ultimately, the outcome of these simulations should be used to design observational diagnostics that distinguish between these various processes.

\subsubsection{Production and Survival of Dust and Molecules} \label{sec:open_dust}

As discussed in Sec. \ref{sec:thermal_energy} and \ref{sec:in-situ}, the molecular material in cool outflows may be produced \textit{in situ} from the shocked ISM itself as long as the molecular gas manages to (re-)form within $\lesssim 10^6$ yrs.  This short timescale favors H$_2$ formation catalyzed by dust grains but it is at present unclear to what extent dust is able to survive the reverse and forward shocks produced by the violent interaction of the fast inner wind with the ISM (Sec. \ref{sec:dust_cycle}). In recent years, simulations and models of reprocessing of dust grains through shocks in SN remnant environments have shown that large ($a \gtrsim 0.1 \mu$m) grains may survive or ``escape'' the shocks and decouple from the gas 
\citep{SLAVIN2004,SILVIA2010,BISCARO2016},  but the results of these models do not necessarily apply to the more extreme conditions of cool-gas outflows (characterized by higher shock velocities and stronger UV and X-ray radiation fields). Similarly, the failure of diffuse ejecta to reform the dust once the dust is destroyed by shocks \citep{BISCARO2014}  may not be relevant to the denser conditions of cool outflows. Moreover, dust growth in the more diffuse cold neutral medium may also be significant \citep{DRAINE2009a,FERRARA2016b,ZHUKOVSKA2016}, but requires tracking the evolution of nano-grains in this lower density environment (since $t_{\rm acc} \propto a/n$; Eq.~(\ref{eq:t_acc})). Modeling of the reprocessing (destruction, formation, and evolution) of both the dust and molecular gas needs to be done self-consistently assuming the special time-dependent conditions of cool outflows to track the non-equilibrium molecular chemistry, and fully assess the validity of the scenario where the molecules form \textit{in situ} within the wind.

\subsubsection{Coupling the Radiation to the Cool Gas} \label{sec:open_radiation}

Coupling of the radiation to the cool gas is inherently a difficult numerical problem since it requires in principle to resolve scales as small as the photon mean free path or dust destruction radius to capture all of the relevant physics, even in the simpler single-scattering optical thin limit \citep[although promising workarounds have been developed in recent years;][]{KRUMHOLZ2018,HOPKINS2019}. In the optically thick limit, the numerical treatment of radiation Rayleigh-Taylor instabilities and frequency diffusion (due to $\kappa_{\rm eff, IR} \propto \nu^2$ for IR radiation) is critically important and as a consequence considerable debate still remains regarding the efficacity of radiation to drive cool outflows (Sec. \ref{sec:radiation}). Future numerical simulations that utilize the latest methods to solve the frequency-dependent radiative transfer equation and interpolate the solution back onto fluid elements will need to be tailor-made to reproduce the conditions of local U/LIRGs, where the most extreme momentum boosts $\dot{p}/(L/c)$ are observed, and should emphasize the distinguishing physical features of outflows accelerated by radiation pressure over those driven by thermal processes. Given the (tentative) evidence for elevated dust-to-gas ratios in the extraplanar material of some galaxies (Sec. \ref{sec:lowz_dust}) and decoupled dust/gas velocities in the ionized wind of M82 \citep[][]{YOSHIDA2011,YOSHIDA2019},
the possibility of the dust grains decoupling dynamically from the gas due to the stronger radiation forces acting on them should also be investigated, keeping tract of the sizes and charges of dust grains and the collisions, Coulomb interactions, and Lorentz forces with the molecular, neutral-atomic, and ionized gas making up the outflow \citep{DRAINE1979a,DRAINE1979b,KRUMHOLZ2013,THOMPSON2015}.

\subsubsection{Importance of Cosmic Rays} \label{sec:open_crays}

Tremendous progress have been made in recent years in modeling the effects of CRs as an additional source of pressure to drive warm and cool outflows (Sec. \ref{sec:cosmic_rays}). The results are sensitive to the treatment of the microphysical processes associated with CR transport, e.g. relative importance of advection, streaming, and anisotropic diffusion of CRs along the magnetic fields.  Future theoretical efforts should be directed at making \emph{quantitative} morphological and kinematic comparisons between cosmic-ray driven outflows and those that do not include the effects of cosmic rays over a broad range of realistic initial ISM conditions (multi-phase density distribution, magnetic field strength and orientation), energy source distribution (SNe in a starburst, AGN), and detailed physics of the CR-plasma interaction. The importance of cosmic rays as a source of ionization of the cold molecular gas should also be investigated using the physical conditions appropriate to cold-outflow hosts (e.g., gas-rich U/LIRGs) to allow direct comparisons with the few existing observational constraints on partially ionized molecular outflows (e.g., Mrk~231; Sec. \ref{sec:Mrk231}). 

\subsection{Observational Challenges} \label{sec:open_observational}

\subsubsection{Unbiased Census of Local Molecular Outflows} \label{sec:open_census}

Current statistics on local molecular outflows are not representative, since they are highly biased in favor of special, highly active or gas-rich galaxies, such as ULIRGs, quasars, AGN, and starbursts where outflows may be brighter and easier to observe. The census of neutral atomic outflows also suffers from biases, but the situation is much better due to SDSS and large long-term investments of telescope time in optical and near-infrared spectroscopic surveys (Sect.\ \ref{sec:lowz_neutral_atomic}). An unbiased census of local molecular outflows can in principle be carried out with current facilities (NOEMA, ALMA), given enough observing time. Future infrared and radio facilities ({\em SPICA}, the {\em Origins Space Telescope}, Square Kilometer Array, Next Generation VLA), however, have the sensitivity and throughput to significantly improve the picture. The {\em Origins Space Telescope}, for example, will be an efficient survey machine for low and high redshift \citep{BONATO2019}, capable of fast, very sensitive mid and far-infrared spectroscopy. As an interferometer, the Next Generation VLA (ngVLA) will provide high-resolution spectral imaging with very high sensitivity for the ground transitions of CO, HCN, and HCO$^+$ at $\lambda\sim3$~mm, and several other powerful diagnostics of cold molecular outflows \citep{BOLATTO2018}. The phase 1 mid-frequency version of the Square Kilometer Array (SKA), which should start construction soon, is optimizing for surveying at $\lambda\sim 21$~cm, and will be an extremely capable instrument to image outflows in neutral hydrogen emission and absorption \citep[e.g.,][]{MORGANTI2012}.

\subsubsection{Outflow Physical Properties} \label{sec:open_energetics}

The measurements of the outflow physical properties,  specifically mass loss rate, kinetic power, and momentum rate, are uncertain. The main issues are the geometry, velocity projection effect, and the determination of the mass in the outflow. There is a need for higher angular resolution and sensitivity (especially to diffuse low-surface brightness emission) to be able to apply several independent diagnostic tools on the same targets to reduce systematic errors and constrain the key quantities needed to calculate the energetics. Use of the fainter, higher level mm-wave CO, [C~I], HCN, and HCO$^+$ transitions on a large sample of objects promises to provide a more complete picture of molecular outflows (especially a more accurate assessment of the gas mass in the outflow), while combined down-the-barrel and transverse multi-line rest-frame UV and optical absorption-line studies and deep 3D emission-line mapping of statistically significant samples of galaxies should help answer questions about in-flight gas phase transition and gaseous extent of the neutral-atomic outflows. Soon, {\em JWST} will help probe the warm H$_2$, predicted to be the dominant molecular component according to some quasar feedback models \citep{RICHINGS2018b,RICHINGS2018a}.  Next-generation UV and X-ray missions will provide valuable constraints on the kinematics, ionization state, chemical composition, and dust content of the energetically important warm-hot and hot wind fluids, the ``piston'' in thermally driven cool outflows. The importance of CRs in driving these outflows will also be assessed quantitatively with LOFAR \citep[e.g.,][]{HEESEN2018,HEESEN2019,MULCAHY2018} and ngVLA.

\subsubsection{Negative Feedback: Zone of Influence and Escape Fraction} \label{sec:open_negative}

The true impact of cool outflows can only be fully assessed by mapping the entire zone of influence of these outflows since negative feedback limited to small scales would not prevent the cool CGM from accreting back onto the galaxy host and forming new stars. Current absorption- and emission-line measurements are still severely limited by the sensitivity of current observations, particularly the surface brightness limits of emission-line mapping. On the short term, this can be improved by increasing the exposure times on a few key targets using large-format 3D spectroscopic instruments such as MUSE on the VLT and KCWI on Keck to provide full three-dimensional coverage of the outflows on CGM scales and allow reliable decomposition of the outflowing material from the gravitationally bound material \citep[e.g.,][]{RUPKE2019}. However, robust conclusions based on a statistically significant sample of galaxies will have to wait for the next generation of ground-based facilities including IFS on ELTs and ngVLA, and space-borne facilities such as {\em SPICA}, the {\em Origins Space Telescope}, {\em LUVOIR}, the {\em Advanced Telescope for High-ENergy Astrophysics} ({\em Athena}), and the {\em Lynx} X-ray telescope.

\subsubsection{Positive Feedback: Induced and In-Situ Star Formation} \label{sec:open_positive}

The evidence for positive feedback is still sparse, even in the local universe. There is a need to carry out a systematic search for positive feedback in a statistically significant sample of objects. Actively star-forming hosts may need to be excluded from this sample to avoid any confusion with the source of ionization (internal vs external to the outflow), unless the geometry of the outflowing gas with respect to the host material is unambiguous. The technical challenge of disentangling the different sources of excitation in the gas (AGN, star formation, shocks) will still remain, but great strides have been made in recent years to resolve this problem by using the full suite of sophisticated photoionization and shock models in combination with kinematic predictions \citep{DAVIES2014,DAVIES2016,KEWLEY2019}. Yet, the fact that the presence of an AGN easily overwhelms the signature of star formation remains an issue difficult to overcome. Near-IR diagnostic nebular lines are promising alternatives to the classical line diagnostic diagrams in the optical \citep[e.g.][]{MAIOLINO2017} and are less affected by the AGN dominance; currently these near-IR diagnostics are  observable with good S/N only for a small number of outflows, but the sensitivity and resolution of {\em JWST} will enable to trace these diagnostics in essentially all local outflows and also
at intermediate redshifts. The ultimate test will be directly detecting the signature of young stars and their kinematics; within this context deep observations with {\em HST}-COS or even with future facilities such as {\em LUVOIR} will help to unambiguously disentangle positive feedback phenomena.

\subsubsection{Evolution with Look-Back Time} \label{sec:open_evolution}

Current observations indicate that warm-ionized and cool-atomic galactic winds are common at high redshifts, where main-sequence galaxies are very active and experiencing rapid growth.
However, very little is known about their cold components, except in a handful of spectacular examples \citep[e.g.,][]{HERRERA-CAMUS2019a}.
Because of the biases and selection effects introduced by using different diagnostics to study the evolution of cool outflows with look-back time, it is preferable to use the same wind diagnostics at all redshifts. A fair comparison also requires matching linear resolution and flux limits of the observations. Ground-based 8-10 meter class OIR telescopes, ALMA, and VLA have already proven very useful to search for the rest-frame UV/optical, FIR, and mm-wave signatures of cool winds at $z$ $\simeq$ 2-6. {\em JWST} should soon extend our grasp to higher redshifts using optical/UV diagnostics without compromising on sensitivity and linear resolution. But large, sensitive, systematic surveys of molecular outflows reaching out to epochs where most galaxies are rapidly evolving are necessary to establish the importance of outflows in the cosmological context. Future facilities such as the ELTs, the ngVLA, {\em SPICA}, the {\em Origins Space Telescope}, and {\em LUVOIR} should be able to detect and characterize cool winds during the epoch of galaxy formation and fast growth of black hole seeds 
\citep[e.g.,][]{GONZALEZ-ALFONSO2017a}. Combining this together with studies of the growth of black holes at these redshifts, enabled by next-generation X-ray facilities such as {\em Athena} and {\em Lynx}, 
will shed new light on some of the key processes that shape galaxies.

\begin{acknowledgement}
The authors thank the two referees of this review, particularly Mark Krumholz who provided a thorough report and made several constructive suggestions which improved the review. S.V. thanks the Editor, Joel Bregman, for the invitation to write this review. He thanks NASA for partial support of this research through NASA grant ADAP NNX16AF24G. S.V. also acknowledges support from a Raymond and Beverley Sackler Distinguished Visitor Fellowship and thanks the host institute, the Institute of Astronomy, where most of this review was written. He is also grateful for support by the Science and Technology Facilities Council (STFC) and by the Kavli Institute for Cosmology, Cambridge. S.V. also appreciates the hospitality of the Space Telescope Science Institute, where this review was completed. R.M. acknowledges ERC Advanced Grant 695671 `QUENCH' and support by the Science and Technology Facilities Council (STFC). A.B. acknowledges partial support from NSF-AST1615960. S. A. acknowledges funding from the European Research Council (ERC) under the European Union's Horizon 2020 research and innovation programme, grant agreement No ERC-2017-ADG-789410. S.A. also acknowledges the Swedish  Research Council grant 621-2011-4143.
\end{acknowledgement}

\bibliographystyle{spbasic}
\bibliography{full-library-v5.bib,added-library-v2.bib,added-library2.bib,added-library3.bib}

\end{document}